# Diffusion of the Adoption of Online Retailing in Saudi Arabia

**Rayed Abdullah AlGhamdi**

*Bachelor of Computing, Master of ICT*

School of Information & Communication Technology

Science, Environment, Engineering and Technology Group

Griffith University

Submitted in fulfillment of the requirements of the degree of Doctor of Philosophy

July 2012

# STATEMENT OF ORIGINALITY

This work has not been previously submitted for a degree or diploma in any university. To the best of my knowledge and belief, the thesis contains no material previously published or written by another person except where due reference is made in the thesis itself.

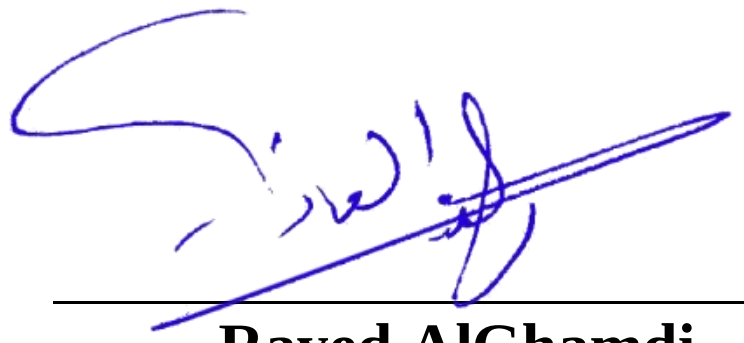

**Rayed AlGhamdi**

# Dedication

***To***

My beloved parents for their endless love, support and care since I was born and whose encouragement guided me throughout my study.

My beloved wife for her patience and support

My lovely three kids

My brothers and sisters

&

My country, The Kingdom of Saudi Arabia

# Acknowledgment

*Thanksgiving to the Almighty Allah for his guidance, grace, mercy and blessing that have empowered me throughout my life.*

**I highly appreciate the enormous support by Associate Prof. Marilyn Ford guiding me in this critical stage of my PhD journey to extensively revise this thesis, *very thanks Marilyn!***

Special thanks to **King Abdulaziz University** for giving me this opportunity and providing me with sponsorship including full financial support to my family and I to complete my postgraduate studies. In addition, thanks to the Saudi Arabian Cultural Mission in Canberra for their services and support.

Very thanks to **Dr. Steve Drew**; the supervisor, friend and brother, the one have accompanied me in this journey from the beginning to the end.

I also would like to take this opportunity to express my appreciation to all of those provided support and assistance, and those I have known and worked with; including: Dr. Ann Nguyen, Dr. Sue Nielsen, Mrs. Mary Pavlaki, Dr. Osama Alfarraj, Dr. Mohammed Alshehri, Prof. Tom Nguyen, Dr. Jeremy Nguyen, Dr. Thamer Alhussain, Dr. Salem Alkhalaf , Mr. Adel Bahadad, Dr. Luke Houghton, Dr. Peter Grimbeek, Dr. Bill Metacalf, Ms Heidi Perrett and Mr. Attiat Allah AlBishri.

Last but not least, thanks to Griffith University for providing excellent learning environment and assist the students to achieve their goals.

# LIST OF PUBLICATIONS

## Journal Publications

## International Conference (Refereed)

4. **AlGhamdi, R**, Nguyen, A, Nguyen, J & Drew, S 2011, 'Factors Influencing Saudi Customers' Decisions to Purchase from Online Retailers in Saudi Arabia: A Quantitative Analysis', in P Kommers & P Isaías (eds), *IADIS International Conference e-Commerce 2011*, Roma, Italy, pp. 153-161

5. **AlGhamdi**, **R**, Nguyen, A & Jones, V 2012, 'Wheel of Online Retailing Development in Saudi Arabia', *The First International Conference on Robot Intelligence Technology and Applications (RiTA 2012)*, Gwangju, South Korea, pp. 110-121.

## Articles

These articles were published in Arabic in the main Saudi newspaper 'Alriyadh'. The purpose of publishing these articles was to create an awareness of the study. The articles summarize the main findings and highlight the priority needed.

1. **AlGhamdi, R**. (2010, 23 Dec). Between the Ministry of Commerce and Ministry of IT: E-commerce in the Kingdom and the missing hope! *Alriyadh newspaper*. <http://www.alriyadh.com/2010/12/23/article588081.html> مابين وزارة التجارة ووزارة الاتصالات: التجارة الإلكترونية في المملكة والأمل المفقود!

2. **AlGhamdi, R**. (2011, 12 Jun). E-commerce in Saudi Arabia: the absence of plans for the future! *Alriyadh newspaper*. <http://www.alriyadh.com/2010/12/23/article588081.html> التجارة الإليكترونية في السعودية: غياب الخُطط المستقبلية!

3. **AlGhamdi, R**. (2011, 12 July). The five wings model to enhance confidence in online shopping environment in the Kingdom, *Alriyadh newspaper*. <http://www.alriyadh.com/2011/07/12/article649928.html> نموذج الأجنحة الخمسة لتعزيز الثقة في بيئة التسوق الإليكتروني في المملكة

## Co-authored papers

1. Alfarraj, O, Drew, S & **AlGhamdi, R** (2011) 'EGovernment Stage Model: Evaluating the Rate of Web Development Progress of Government Websites in Saudi Arabia', *International Journal of Advanced Computer Science and Applications (IJACSA)* 2: 9, pp. 82-90 Sep

2. Alshehri, M., Drew, S., Alhussain, T., & **AlGhamdi, R**. 2012. The Impact of Trust on E-Government Services Acceptance: A Study of Users' Perceptions by Applying UTAUT Model. *International Journal of Technology Diffusion (IJTD),* vol. 3, no. 2, pp. 50-61.

## Submitted Papers

## Papers to be Submitted

### Papers in progress

# ABSTRACT

Despite the fact that the Kingdom of Saudi Arabia (KSA) is a leading producer of oil and natural gas (CIA 2012), a member of G-20, has the largest and fastest growth of ICT marketplaces in the Arab region (U.S. Commercial Services 2008, Alfuraih 2008), including computer hardware and software, and is very wealthy, online retailing activities are not progressing at the same speed of its growing ICT marketplace (CITC 2011). For this reason, this thesis explores the factors influencing the slow progress of online retailing in Saudi Arabia. While several studies exist investigating factors influencing e-commerce adoption by organizations, this thesis is different in that it suggests that the stage of e-commerce maturity of companies is important when considering why companies do or do not adopt e-commerce and how companies can be encouraged to move to a higher stage of maturity. Thus, the present thesis considered the factors influencing e-commerce adoption by retailers in Saudi Arabia based on stages of e-commerce maturity of each organization. To evaluate the stage of e-commerce maturity among retailers in Saudi Arabia, the Stage Of Growth E-business (SOG-e) model of Mckay, Marshall and Pranato 2000 was adopted.

The research started with the identification of the research problem and questions, and a review of the literature. The data collection includes two stages. The first data collection stage consisted of interviews with 16 retailers covering the four stages of e-commerce maturity. The second stage of data collection was designed based on the results of the interviews to further test and explore the qualitative findings with a wider sample of retailers; 153 retailers, covering different sizes, ages, business categories, and stages of e-commerce maturity. The outcomes showed striking differences between retailers as a function of the stage of growth of their companies in regard to factors relating to their perceptions of consumers and to organization factors, while at the same time showing agreement between retailers in companies at different stages in relation to environmental factors. The environment was recognized as the same by respondents in all four stages. This investigation helps to identify the incentives and barriers for each stage of maturity. These findings show that it is important to consider stage of e-commerce growth when considering barriers/incentives for organizations to adopt e-commerce. Based on the findings, recommendations are provided in terms of stages of e-commerce maturity in order to contribute to e-commerce development in Saudi Arabia. These recommendations include e-commerce educational and awareness programs, with each stage targeted with specific programs, assistance in getting the lower stage companies online, introduction of more secure online payment systems such as SADAD, enhancing the legislation system for e-commerce, and more effort to building trust with the Saudi mailing and addressing system, and enhance the uses of home mailboxes. The thesis shows that it is important for researchers to consider the stage of e-commerce maturity of companies when considering incentives and barriers to e-commerce development. What one group sees as an incentive, another group may see as a barrier, or as neutral.

# Table of Contents

# List of Tables

# List of Figures

# List of Abbreviations

| | |
|---|---|
| B2B | Business to Business |
| B2C | Business to Customers |
| B2E | Business to Employee |
| B2G | Business to Government |
| C2C | Customer to Customer |
| C2G | Consumer to Government |
| CDSI | Central Department Of Statistics & Information |
| CEO | Chief Executive Officer |
| CITC | Communication and Information Technology Commission |
| CSC | Council of Saudi Chambers |
| DOI | Diffusion of Innovation |
| E-commerce | Electronic commerce |
| E-government | Electronic government |
| EIU | Economist Intelligence Unit |
| E-readiness | Electronic Readiness |
| GCC | Gulf Cooperation Council |
| ICT | Information and Communication Technology |
| IS | Information System |
| IT | Information Technology |
| KACST | King Abdulaziz City for Science and Technology |
| KSA | Kingdom of Saudi Arabia |
| MCIT | Ministry of Communication and Information Technology |
| M-commerce | Mobile Commerce |
| MENA | Middle East and North Africa |
| MENA | Middle East and North Africa |
| MFA | Ministry of Foreign Affairs |
| MHE | Ministry of Higher Education |
| MoC | Ministry of Commerce |
| MoL | Ministry of Labour |
| P2P | Peer to Peer |
| PKI | Public Key Infrastructure |
| SADAD | The Saudi national electronic payment system |
| SOG-e | Stage of Growth e-business |
| UAE | United Arab Emirates |
| UN | United Nation |
| WiMAX | Worldwide Interoperability for Microwave Access |

# CHAPTER 1: INTRODUCTION

This thesis is concerned with the diffusion of the adoption of online retailing in Saudi Arabia. The initial chapter presents the research problem, questions, objectives, and the significance and contribution of the research. A background to e-commerce is also presented along with a justification of why this research is conducted in Saudi Arabia. An overview of the e-commerce scenario in Saudi Arabia and the research methodology are presented. Finally, the chapter ends with an overview of how this thesis is structured.

## 1.1. Research problem

The uptake of e-commerce by retailers in Saudi Arabia has been sluggish (Alfuraih 2008; CITC 2011). Despite the fact that the Kingdom of Saudi Arabia (KSA) is a leading producer of oil and natural gas in the world (CIA 2012), a member of G-20, has the largest and fastest growth of ICT marketplaces in the Arab region (U.S. Commercial Services 2008, Alfuraih 2008), including computer hardware and software, and is very wealthy, e-commerce activities are not progressing at the same speed of its growing ICT marketplace (CITC 2011). The latest report issued by the Saudi Arabia's Communication and Information Technology Committee (CITC) demonstrated that only 6% of residents in Saudi Arabia have purchased online, only 16% of the internet users are planning to buy online, and only 15% have plans to sell online (CITC 2011). Further, only 8% of businesses in Saudi Arabia sell their products/services online, B2B was estimated at 54%, B2C at 28%, and B2G at 18% (CITC 2011). The case is similar with the other Gulf Cooperation Council (GCC) countries. The countries of this regional interest group include Saudi Arabia, the United Arab Emirates, Oman, Qatar, Kuwait, and Bahrain. This region is classified as one of the richest regions in the world as it produce almost a half of the world's supply of oil. However, online shopping in this cooperative group of countries is still in its early stages (IMRG 2011).

A large body of literature already exists identifying factors that influence businesses to adopt and use e-commerce, and consumers to purchase online. A thorough review of the literature that has investigated e-commerce in Saudi Arabia, other Gulf countries, and various developing countries, shows that most studies have concentrated on environmental factors, such as ICT infrastructure, online payment mechanisms, legislative and regulatory framework, logistics infrastructure, and education and awareness (Al-Rawi, Sabry and Al-Nakeeb 2008; Alrawi and Sabry 2009; Albadr 2003; Sait, Al-Tawil & Hussain 2004; Al-Solbi and Mayhew 2005; Alfuraih 2008; Alwahaishi, Nehari-Talet and Snasel 2009; Aleid, Rogerson and Fairweather 2010a; Aleid, Rogerson and Fairweather 2010b; Al-Hudhaif & Alkubeyyer 2011; Eid 2011). While certain inhibitors and facilitators are recognized, there has been no attempt to give a broader framework that shows the relationship between facilitators and inhibitors and the level of e-commerce maturity that a company has reached, nor has there been a breakdown in terms factors other than environmental factors, many of which could be very important. The puzzle of why Saudi Arabia lags behind other wealthy countries in e-commerce together with recommendations for how the situation can be changed so that more companies can move to higher maturity levels can hopefully be addressed more satisfactorily by using a broader framework than that used in the past. Such an approach, which would yield more than a simple list of inhibitors and facilitators, could also be used, in the future, by other researchers to study similar situations in other countries.

## 1.2. Research questions and objectives

The following research questions were articulated:

- What are the differences between retailers at different stages of e-commerce maturity in term of factors that influence their adoption or non-adoption of e-commerce activities?
- Apart from environmental factors, what else influences retailers' decision in Saudi Arabia on whether or not to adopt e-commerce activities?

- Given the findings, what recommendations can be made to help companies move to a higher stages of e-commerce maturity?

In order to answer these questions, the following objectives were formulated. Critically review the literature in the area of e-commerce in general and e-commerce and online retailing in Saudi Arabia in more detail. Second, adopt a maturity model, in particular the Stage of Growth (SOG-e) maturity model of (Mckay, Marshall and Pranato 2000), and using four levels of e-commerce maturity develop criteria to distinguish different levels of e-commerce maturity. Third, by interviewing retailers from companies at different levels of maturity, investigate the factors that influence Saudi retailers' decision on whether to adopt or not to adopt e-commerce and in particular allow the interviews to be broad enough to allow factors of all kinds, not just environmental, to surface. Fourth, use the factors identified in the interview stage of the research to develop a survey to investigate the factors further with a wider sample of companies at the four levels of e-commerce maturity. Finally, to differentiate between the different levels of e-commerce maturity in terms of the rating of the factors in order to provide suitable recommendations for companies to advance to higher maturity levels.

## 1.3. Research contribution

A great deal of literature has identified factors that influence businesses on whether to adopt or not to adopt e-commerce (Kendall et al. 2001; Tigre 2003; Drew 2003; Andersen, Björn & Dedrick 2003; Grewal, Iyer & Levy 2004; Wymer and Regan 2005; Gibbs, et al 2006; Hafez 2006; Sleem 2006; Shalhoub & AlQasimi 2006; To & Ngai 2006; Kraemer, Dedrick, and Melville 2006; Ho, Kauffman & Liang 2007; El Said and Galal-Edeen 2009; Scupola 2009; Colton, Roth & Bearden 2010; Nair 2010). These studies dealt with various factors that might influence businesses to adopt and use e-commerce concentrating mainly on organizational and business factors, and, to a lesser extent, customer practices and environmental factors. The concentration on the organizational factors possible relates to the fact that the e-commerce infrastructure environment is already well established in the countries studied. In studies held within Saudi Arabia there is greater emphasis on environmental factors; such as, ICT infrastructure, logistics, and online payment mechanisms (Aladwani 2003; Albadr 2003; Sait, Al-Tawil & Hussain 2004; Al-Solbi & Mayhew 2005; Alfuraih 2008; Alwahaishi, Nehari-Talet & Snasel 2009; Alrawi & Sabry 2009; Aleid, Rogerson & Fairweather 2010). Thus, for example, logistic infrastructure issues, such as the lack of mailing addresses and the cost of delivery in Saudi Arabia, have been highlighted as inhibiting factors for businesses to sell online. Little attention has been given to non environmental factors. Moreover, no studies have investigated how the Saudi businesses that already sell online and have reached a high level of e-commerce maturity overcame these issues. A study of companies at different levels of e-maturity could be invaluable in suggesting what could be done to increase the likelihood of other company's increasing their maturity stage. The businesses that have reached a high stage of e-commerce maturity should provide practical evidence of positive and optimistic attitudes and practices regarding use of e-commerce, whereas the businesses that have not reached higher levels of maturity should provide practical evidence of more negative and pessimistic attitudes and practices. The study, therefore, should contribute to efforts leading to greater e-commerce development.

Before moving on to the actual research conducted for the present thesis, it is useful to first briefly consider e-commerce in general and Saudi Arabia in particular. This short consideration will give the reader greater background on the research topic to begin with, though more detailed discussions will be given later.

## 1.4. E-commerce and Saudi Arabia

At the most basic level, commerce is the trade of goods for money, and electronic commerce (e-commerce) is commerce enabled by the Internet (Whiteley 2000). Comprehensively, e-commerce includes pre-sale and post-sale activities across the supply chain over the Internet (Chaffey 2004). It can be defined as "all electronically mediated information exchanges between an organization and its external stakeholders" (Chaffey 2004, p. 9). In some definitions, e-commerce is made equivalent to electronic business (e-business). However, in this study, e-commerce is considered as a subset of e-business in line with Davis & Benamati (2003). E-business can be defined as business processes using “all electronically mediated information exchanges, both within an organization and with external stakeholders” (Chaffey 2004, p. 10). As the present study investigates what would encourage online retailing in KSA, the definition that defined e-commerce as commerce enabled by Internet including pre-sale and post-sale activities was adopted in this research. This definition facilitates the explanation of this process.

E-commerce encompasses a broad area and is classified into different subtypes or “models”. Figure 1.1 presents a classification encompassing Business to Customer (B2C), Business to Business (B2B), Customer to Customer (C2C), Peer to Peer (P2P), and Mobile Commerce (M-commerce). Davis & Benamati (2003) categorize e-commerce into B2C, B2B, C2C, and Business to Employee (B2E). There are also Business to Government (B2G) and Consumer to Government (C2G) models to consider. The focus of this study falls under B2C.

B2C is categorized into seven models (Laudon & Traver, 2010); portal, online retailer, content provider, transaction broker, market creator, service provider, and community provider. Table 1.1 demonstrates these classifications in further detail. To narrow the research topic further down, this research is involved with online retailing. Online retailing can be defined as an “online version of traditional retail; [which] includes virtual merchant (online retail store only), Bricks-and-Clicks e-retailers (online distribution channel for a company that also has physical store), Catalog Merchants (online version of direct mail catalog), Manufacturers selling directly over the web” (Laudon & Traver 2010). Figure 1.1 demonstrates how the research topic is narrowed down from the general topic, e-commerce, to specific area, i.e. online retailing; they are highlighted in grey coloring.

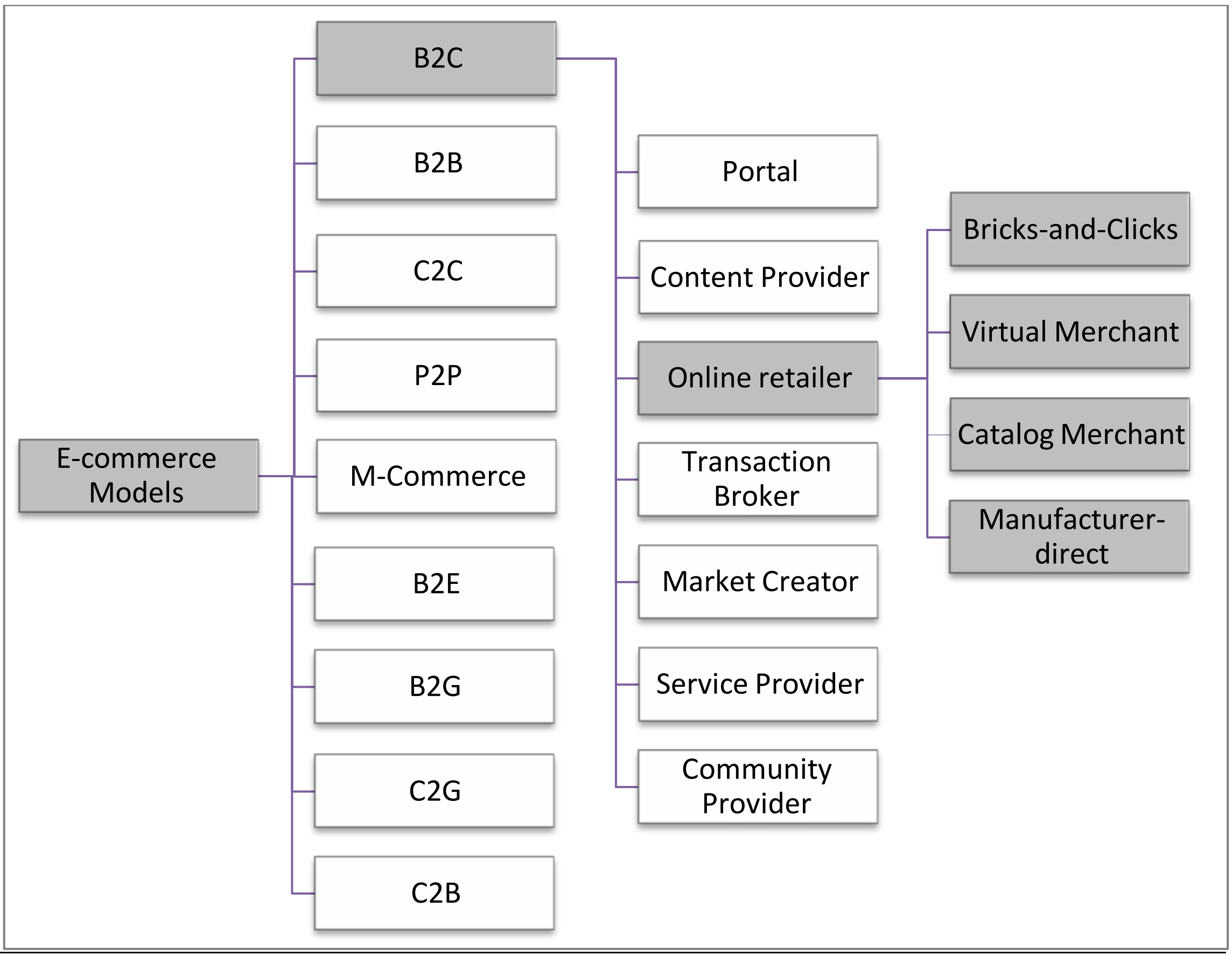

Figure 1.1: E-Commerce Models and B2C Sub-models

Table 1.1: Business to Customers (B2C) E-commerce Models

| Business Model | Variations | Examples | Description | Revenue Model |
|---|---|---|---|---|
| Portal | Horizontal/General | Yahoo.com AOL.com MSN.com | Offers an integrated package of content and content search, services, news, e-mail, chat, music downloads, video streaming, calendars, etc. Seeks to be a user's home base. | Advertising, subscription fees, transaction fees |
| | Vertical/ Specialized (Vortal) | Sailnet.com | Offers services and products to specialized marketplace | Same |
| | Search | Google.com AskJeeves.com | Focused primarily on offering search services | Advertising, affiliate referral |
| Online Retailer | Virtual Merchant | Amazon.com | Online version of retail store where customers can shop at any hour of the day or night without leaving their home or office | Sale of goods |
| | Bricks-and-Clicks | Walmart.com Sears.com | Online distribution channel for a company that also has physical stores | Same |
| | Catalogue Merchant | LLBean.com Lillian.com | Online version of direct mail catalogue | Same |
| | Manufacturer-direct | Dell.com Mattel.com Sony.com | Manufacturer uses online channel to sell direct to customers | Same |
| Content Provider | | WSJ.com Sportlines.com CNN.com ESPN.com RealRhapsody | Information and entertainment providers such as newspapers, sports sites, and other online sources that offer customers up-to-date news and special-interest, how-to guidance and tips, and or information sales | Advertising, subscription fees, affiliate referral fees |
| Transaction Broker | | Etrade.com Expedia.com Monster.com Travelocity.com Hotels.com Orbitz.com | Processors of online transactions, such as stakeholders and travel agents, that increase customers' productivity by helping them get things done faster and more cheaply | Transaction fees |
| Market Creator | | eBay.com Priceline.com | Web-based businesses that use Internet technology to create markets that bring buyers and sellers together. | Transaction fees |
| Service Provider | | VisaNow.com xDrive.com myCFO.com BlueFlag.com | Companies that make money by selling users services, rather than products. | Sales of services |
| Community Provider | | IVillage.com Friendster.com About.com | Sites where individuals with particular interests, hobbies, common experiences, or social networks, can work together and meet online. | Advertising, subscription, affiliate referral fees |

Adapted from (Laudon & Traver, 2010, p. 69)

Many businesses around the world have introduced e-commerce tools into their businesses to gain a competitive advantage. The adoption of e-commerce systems, especially in the developed continues, is growing quickly (Kamaruzaman, Handrich & Sullivan 2010). The period 1995-2000 saw a notable proliferation of e-commerce start-ups and online retailing systems in the USA (Dedrick et al. 2006; Dinlersoz & Hernández-Murillo 2005). Since 2000, the rapid growth of e-commerce activities has been obvious in the developed world. Global e-commerce spending is worth about US$10 trillion at present, compared to US$0.27 trillion in 2000. The USA accounts for the largest share (about 79%) of the current total, followed by Europe (Kamaruzaman, Handrich & Sullivan 2010). By comparison, the Middle East and African region has a very small share (around 3%) (Kamaruzaman, Handrich & Sullivan 2010).

The USA followed by the UK accounts for the world largest market for online retailing. Online retail in USA accounted for 3.6% ($142 billion) of total retail sales in 2008 (U.S. Census Bureau 2010) and in UK accounted for 10.7% (almost $74 billion/52 billion EUR) of their retail trade in 2010 (Centre for Retail Research 2010). According to the Nielson (2010) report, the top 10 products/services globally sold online are books, clothing/accessories/shoes, airline ticket/reservations, electronic equipment, tours/hotel reservations, cosmetics/nutrition supplies, event tickets, computer hardware, videos/DVDs/games, and groceries.

It would seem that Saudi Arabia, as a leading world oil producer, would take advantage of e-commerce tools and applications. However, despite the fact that Saudi Arabia has the largest growing market for ICT products in the Arab region (Saudi Ministry of Commerce 2001, Alotaibi & Alzahrani 2003, U.S. Commercial Services 2008, Alfuraih 2008), e-commerce activities are not showing notable growth (Al-Otaibi & Al-Zahrani 2003, Albadr 2003, Aladwani 2003, CITC, 2007, Agamdi 2008). The Saudi Government's decision to introduce e-commerce in Saudi Arabia started in 2001. In response to the fast development of e-commerce around the world, the Saudi Ministry of Commerce established a permanent technical committee for e-commerce. However, this Committee has ceased to exist, and that the roles of e-commerce supervision and development were transferred to the Ministry of Communications and Information

Technology (MCIT) in 2006. Since 2006, the efforts by MCIT to support e-commerce have not yielded results. When confronted with the lack of success thus far, MCIT simply claimed that they are still in the early stages of studying e-commerce (Ghawanny 2011).

Unlike western nations, Saudi Arabian electronic systems have been concentrated on Saudi e-government development with e-commerce taking second place. In contrast, the development of e-commerce and indeed the Internet in Western countries has been driven by organizational and financial competition for market share – with the focus being on independent businesses. Typically, KSA and most Gulf states (Oman, UAE, Qatar, Kuwait and Bahrain) are late adopters of Internet technology, and while there is a rapid technological diffusion of internet technology there is a lag in the development of e-commerce adoption based on low level of public knowledge (and trust) about electronic models of business.

## 1.5. Research method overview

The research starts with the identification of the research problem and questions, and a review of the literature. The data collection includes two stages: interviews to obtain qualitative data and surveys to obtain quantitative from a larger number of participants. The first data collection stage consisted of interviews with 16 retailers. Then an analysis of the qualitative data from the interviews led to a list of factors influencing retailers on whether or not to adopt e-commerce and a classification of these factors into groupings. The second stage of data collection was designed to further tests and explores the qualitative findings with a wider sample of retailers; 153 retailers, covering different sizes, ages, business categories, and e-commerce maturity stages. To measure e-commerce maturity stages, the Stage of Growth e-business (SOG-e) model of (Mckay, Marshall and Pranato 2000) was adopted to evaluate the e-commerce maturity among retailers in Saudi Arabia. The sample was categorized into four groups based on their e-commerce maturity stages. The quantitative analysis identifies the factors significantly influencing retailers on whether or not to adopt e-commerce, with differences in the influence of factors being shown between the four groups at different stages of e-commerce maturity.

## 1.6. Ethical considerations

The application for ethical clearance for this project has been approved by Griffith University Human Research Ethics, and the ethical certificate was issued in September 2009 with protocol number ICT/06/09/HREC (appendix A, page 183). It ran from 20 Sep 2009 to 03 Nov 2012. An extension was granted on 16 August 2013 (appendix E, page 194). Accordingly, all ethics guidelines were ensured and followed.

## 1.7. The structure of the thesis

In the following chapter, Chapter 2, the nature of ICT infrastructure in Saudi Arabia, the Saudi retail business, the current state of the nation's online retail sector development, and online retailing in other countries are reviewed. First, an overview about Saudi Arabia, its location, population, economy and social system are all discussed. Second, all available data about Information and Communication Technology (ICT) in Saudi Arabia is reviewed, including the Internet, its history, and current statistical data. Third, a background of the retail business in Saudi Arabia including its nature and volume in the economy are presented. Fourth, a discussion about e-commerce in Saudi Arabia and e-commerce statistics are provided. The last section in this chapter discusses the global online retailing history, volume and trend.

Chapter 3 reviews the literature that defines Business to Customers (B2C) electronic commerce drivers and challenges. This section covers the range of issues that might influence this type of business. A number of studies that have been made regarding e-commerce development in Saudi Arabia are also reviewed. This review highlights the challenges that electronic commerce encounters in Saudi Arabia. In addition, Chapter 3 reviews e-commerce maturity models, with an emphasis on the use of the SOG-e model of (Mckay, Marshall and Pranato 2000). Criteria for classifying businesses into four levels of e-commerce maturity are developed.

Chapter 4 provides a comprehensive description of the methodology used for the interviews. The chapter presents detailed descriptions of the processes used for qualitative data collection and analysis. Chapter 5 presents the qualitative data analysis of the interviews of 16 retailers in Saudi Arabia. The main outcome of this chapter is an identification of factors influencing retailers' decisions on whether or not to adopt and use e-commerce. This outcome is the basis for the survey used to obtain quantitative data from 153 retailers from companies identified as belonging to one of four levels of e-commerce maturity.

Chapter 6 provides a comprehensive description of the quantitative study methodology used. The chapter presents detailed descriptions of the quantitative data collection and analysis processes. Chapter 7 presents the analysis of the survey data. Analysis of variance, pairwise t-tests with Bonferroni corrections, and multiple linear regression were performed to identify significant factors influencing e-commerce adoption or lack of adoption, differences between companies at different maturity levels, and interactions between factors and the four levels of maturity.

In Chapter 8, the results are discussed and recommendations to develop e-commerce for retailers in Saudi Arabia and help retailers to move to higher stages of e-commerce maturity are suggested.

Finally, Chapter 9 gives a summary to what have been done in this thesis and summarizes the answers to the research questions. Following this, research contributions are presented. Finally, this study's limitations and directions for future research are highlighted.

# CHAPTER 2: CONTEXTUAL BACKGROUND

Having introduced the research problem and topic, it is good here to give the reader detailed background of the research context. In this chapter, the background of Saudi Arabia, including the nature of its ICT infrastructure, the Saudi retail business sector, the current state of the nation's online retail sector development are presented. First, an overview about Saudi Arabia, its location, population, economy and social system are all discussed. Second, all available data about Information and Communication Technology (ICT) in Saudi Arabia, including Internet, its history, and current statistical data are reviewed. Third, background of the e-government project in Saudi Arabia is provided as well as background of the retail business in Saudi Arabia, including its nature and volume in the economy are presented. Finally, official discussions regarding e-commerce in Saudi Arabia and e-commerce statistics are provided.

## 2.1. The Kingdom of Saudi Arabia (KSA)

This research involves online retailing development in Saudi Arabia. Saudi Arabia is the centre of the Islamic world hosting the two Islamic holy cities, Makkah and Maddenah. Islam is the basic system for the country, which plays a major role in Saudi peoples' daily lives (Saudi Ministry of Foreign Affairs 2009). Saudi Arabia covers an area of approximately 2,240.000 square kilometers (Commercial Office 1996). Its population reached 27,136,977 with an annual growth rate of 3.2% between 2004 and 2010. Saudi citizens represent 68.93% (18,707,576) of the whole population and the rest are foreigners (CDSI 2010).

Saudi Arabia is the leading largest oil producer and exporter in the world and is one of the globe's 20 largest economies (CIA 2012). Today, Saudi Arabia "represents a unique country and convergent blend of social conservatism and technological prowess, a tremendous transformation from an isolated, desert land that it was over 50 years ago" (Sait, Altawil, & Hussain 2004).

Believing in the importance of follow-up with global developments, scientific and technological progress, Saudi Arabia has sponsored tens of thousands of Saudi students, over the last six years, to study in the developed countries, qualifying them to play their roles in the development of various fields (Saudi Ministry of Higher Education 2009a). The interdisciplinary scholarships include study in medical sciences, engineering, information technology, natural sciences (Biology, chemistry, physics and mathematics), law, and business fields. From the business fields, e-commerce is considered one of the most important specializations needing development in Saudi Arabia (Saudi Ministry of Higher Education 2009b).

Table 2.1: Summary of relevant information about Saudi Arabia

| | Number | | | Percentage |
|---|---|---|---|---|
| Population[1] | 27,136,977 | Men | | |
| | | Women | | |
| | | Saudis | 18,707,576 | 68.93% |
| | | Foreigners | 8,429,401 | 31.07% |
| | | | | |
| Area[2] | 2,240,000 sq km | | | |
| Retail Sales[3] | SAR160 billion | | | |
| Telephones[4] | 4.49 million | | | 16% |
| Mobile[4] | 54.8 million | | | 195 % |
| Internet Users[4] | 12.5 million | | | 44% |
| ICT Development[5] | # 52 (out of 159) | | | |
| Logistics Index[6] | # 40 (out of 155) | | | |
| Ease of Business[7] | # 12 (out of 183) | | | |
| e-Readiness[8] | # 52 (out of 70) | | | |
| e-Government[9] | # 41 (out of 193) | | | |

Sources: [1] (CDSI, 2010), [2] (Commercial Office 1996), [3] (Habtoor 2011a), [4] (MCIT 2011b), [5] (IMRG 2011), [6] (TLF 2010), [7] (IFC 2012), [8] (EIU 2010), [9] (UN 2012)

The Saudi Ministry of Labor has started a Saudisation project (replacing foreign labour with Saudi nationals, especially in the private sector) recently. Given that Saudis generally prefer to work in the public sector due to the feeling that job security is better there, the private sector is occupied with 85-90% of foreign workers. The unemployment rate among Saudis reached 12.10% in 2012 and among females is about 28% (CDSI 2012a). There are also changes underway in the recruitment of more women (MOL 2012). Another reason for the private sectors employing more foreign workers is because of their salaries is relatively cheaper compared to Saudi workers. The Saudi Ministry of Labor is working on making the cost of having foreign worker close or higher than having Saudi workers to encourage companies to employee more Saudis with an acceptable minimum level of salary (MOL 2012b). All these circumstances work together putting pressure on companies to find solutions to offer jobs to Saudis and minimize the number of their workers. E-commerce seems a solution here.

## 2.2. Information and Communication Technology (ICT) in KSA

Information and Communication Technology (ICT) plays a significant role in the countries' economy. Attempts to build a strong ICT infrastructure in KSA took place in the early years of the 21st century (AL-Shehry 2008). Over the last decade, the Saudi Government has paid high consideration to this field and has become the largest and fastest growing ICT marketplace in the Arab region (Saudi Ministry of Commerce 2001, Alotaibi & Alzahrani 2003, U.S. Commercial Services 2008, Alfuraih 2008, CITC 2010). ICT spending reached SR 83 billion in 2011 (CITC 2012). Strong growth rates are to be maintained over the next few years, with IT spending set to expand at a compound annual growth rate of 11.4% through 2015, and this rapid growth is fuelled mostly by increased spending on hardware and IT services (CITC 2012). Recently, the Saudi Government has concentrated on improving the Information Technology (IT) infrastructure by opening the telecommunication sectors to privatization (AL-Shehry 2008).

In 2001, the Saudi Government created the Communications and Information Technology Commission (CITC) to take over the main role for ICT regulation in the country. CITC provides the basis and regulatory framework of the ICT sector in Saudi Arabia. "The Act includes a number of objectives, including: provision of advanced, sufficient and affordable communications services; creating the proper climate to encourage fair competition; utilizing frequencies efficiently, transferring telecommunications technology and keeping breast with its developments, and realizing clarity and transparency in processes procedures, in addition to achieving the principles of equality and non-discrimination and protecting the public interest as well as the interests of users and investors" (CITC 2011a).

In 2003, the Saudi Government decided to create an independent ministry called the Ministry of Communications and Information Technology (MCIT). The role of MCIT is, "the supervision of the activities of communications and information technology sector in the Kingdom of Saudi Arabia and ensuring the comprehensiveness of the sector and its spread as vital services that contribute in the socio-economic development and consequently in the development and prosperity of the homeland and citizens" (MCIT 2011a).

Saudi Arabia is globally classified as one of the largest mobile telecommunications markets. By the end of 2012, the total number of mobile subscriptions grew to around 52 million, with penetration standing at 176.9 % of population (CITC 2013) which is higher than the world average of 67%, the developing countries average of 57% and the developed countries average of 114% (Alhussain et. al 2013). The majority of these mobile subscriptions are prepaid which represent 87% in line with the trend in other similar markets around the world (MCIT 2011b). The increased adoption of mobile services in Saudi Arabia caused a slight decline in residential fixed telephone lines. The number of fixed telephony lines stand at around 16% of the population (4.49 million) which is "slightly lower than the world average of 17.8%, but is higher than the Arab World average of 10.5% and the developing countries average of 13.5%" (MCIT 2011b). The Internet services market has undergone notable development. The number of Internet users reached 16.2 million (55%) by the end of the first quarter of 2013 (MCIT 2013). As the extent of Internet access is a key indicator for e-commerce readiness (Sait, Altawil & Hussain 2004), further details about the Internet and its usage in Saudi Arabia are provided in the following paragraph.

All the website requests made by local Internet users pass through a content filtering system, which is a practice put into place when the Internet was introduced into Saudi Arabia in 1997. (Algedhi 2002, Saudi Internet 2007b). Connecting to the Internet in Saudi Arabia has to go through King Abdulaziz City for Science and Technology (KACST)1; see Figure 2.1.

[1] "King Abdulaziz City for Science and Technology (KACST) is an independent scientific organization administratively reporting to the Prime Minister. KACST is both the Saudi Arabian national science

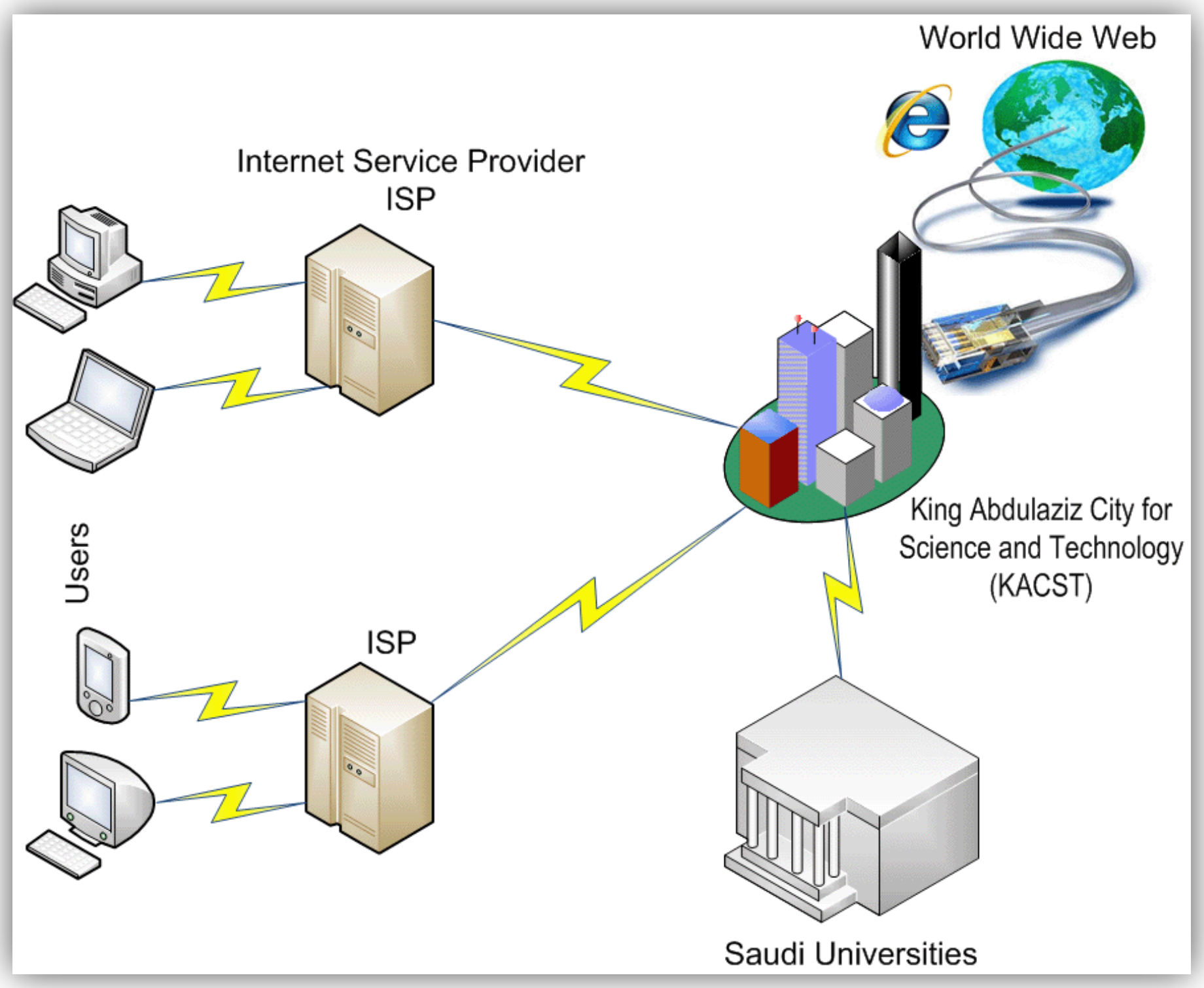

Figure 2.1: Graphical design shows how Saudi users connect to the Internet in KSA

"The vast majority of the web pages that are blocked by the filtering system (95 %) are pornographic" (Saudi Internet 2007b). In addition, websites that directly conflict with, abuse Islamic religion or Saudi laws and regulations are blocked (e.g. pages that involve drugs, gambling, alcohol, terrorism, account theft, and copyright infringement, etc.) (Algedhi 2002, Saudi Internet 2007b, CITC 2010). When a user tries to access a website that is blocked, he/she will then be redirected to a page that informs him/her that the requested website/webpage is not available; see Figure 2.2 below (Saudi Internet 2007b, CITC 2010).

agency and its national laboratories. The science agency function involves science and technology policy making, data collection, funding of external research, and services such as the patent office." (KACST, 2009)

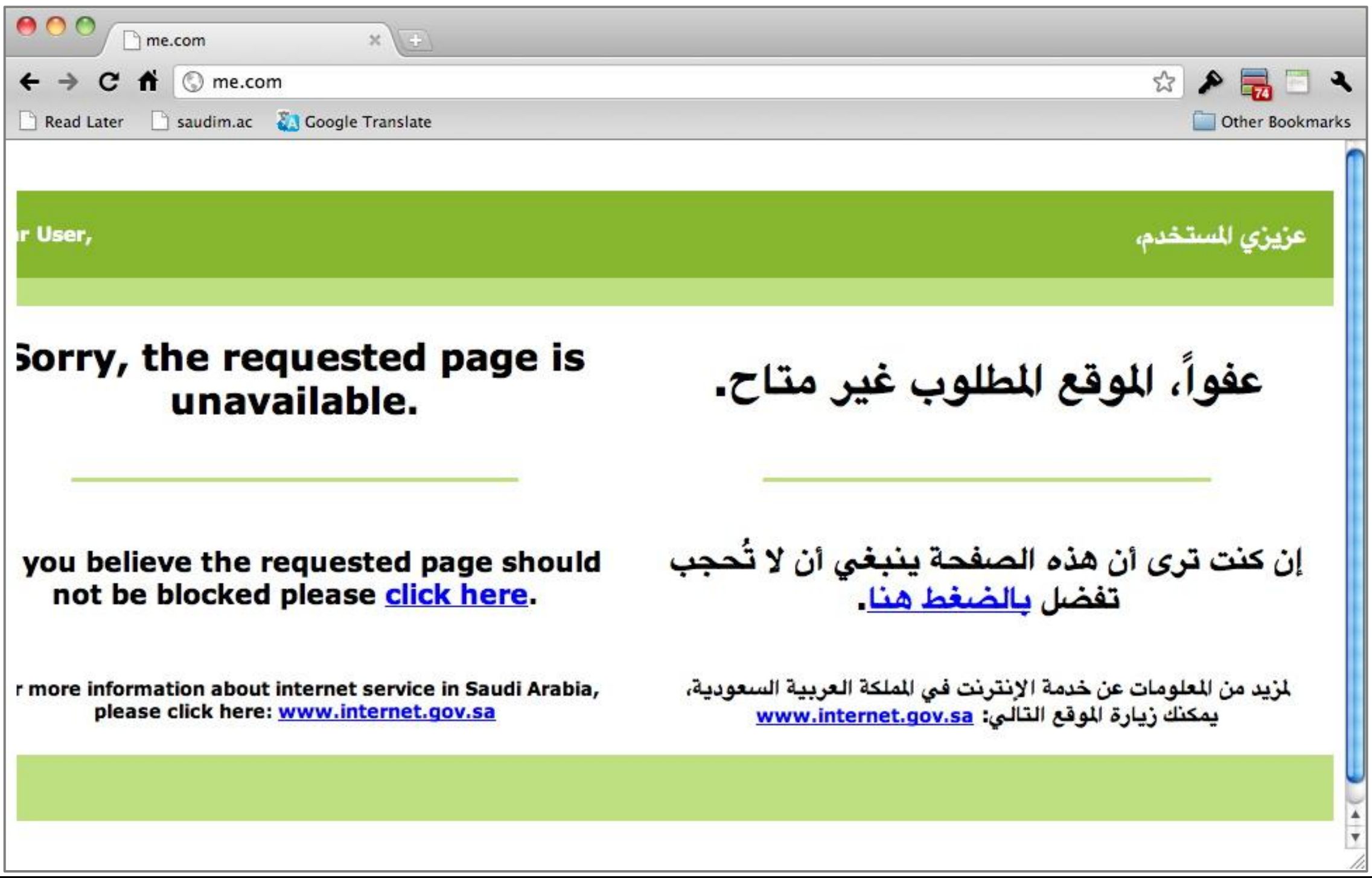

Figure 2.2: Blocked Page (Access not allowed, filtered by CITC)

Any Internet user can communicate with the Saudi Communications and Information Technology Commission (CITC) to contribute by making a request to block or unblock a website or page. Special forms need to be filled out and then a team of full-time employees studies these requests with approval only given if justified. (Saudi Internet 2007b, CITC 2010)

The number of Internet users in Saudi Arabia grew from around 1 million to 16.2 million users with a population penetration of 55% in the period between 2001 and the end of first quarter 2013 (MCIT 2013); see Figure 2.3. The growth in broadband availability, decreased costs of personal computers and laptops, increased public awareness and IT literacy, availability of local content on the Internet, increase of e-services such as online banking and e-government applications are all notable contributors to the growth of Internet users in Saudi Arabia (MCIT 2011b).

Broadband subscriptions in Saudi Arabia have grown to around 2.69 million for fixed broadband lines with penetration rate stood at around 42.9% of households and reached 14.59 million for mobile broadband representing a penetration percentage of 49.6% of the population at the end of first quarter 2013 (MCIT 2013). Fixed broadband services include DSL, fixed Wireless (Wimax) and other fixed lines. The penetration rate of fixed broadband stood at around 33% of households and mobile broadband represents a penetration of 39% of the population. The mobile broadband services include "subscriptions to voice SIMs with use of data communications at broadband speeds, as well as subscriptions to dedicated data SIMs" (MCIT 2011b).

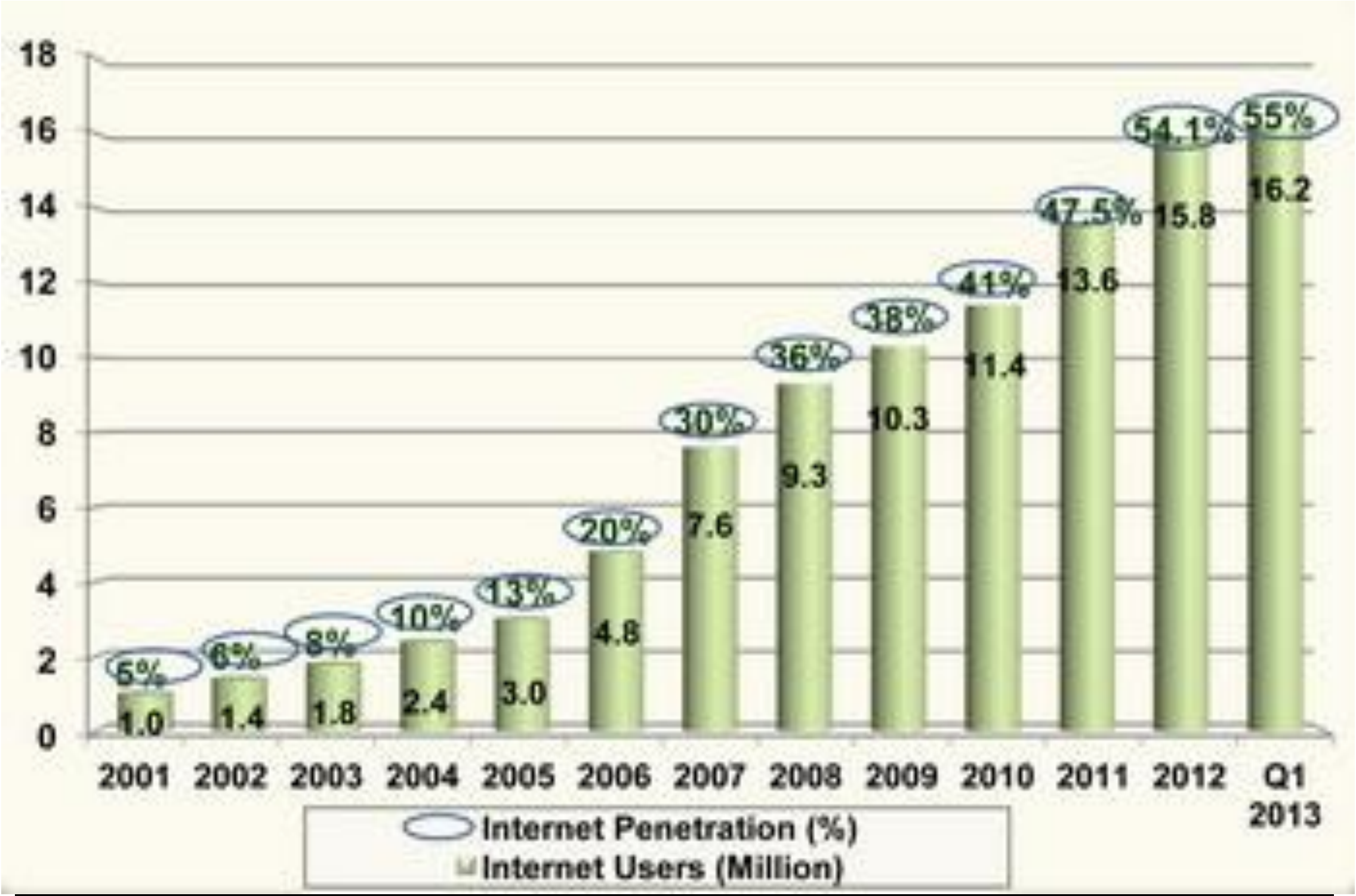

Figure 2.3: The growth of the Internet usage in KSA
Source: (MCIT 2011)

The Saudi Communication and Information Technology Commission (CITC) confirm that the Internet has become an integral part of Saudi society. Around 96% of residential Internet users believe that the Internet is an important source of information and entertainment. Young people who are keen on using IT technologies are a major factor in the growth of the Internet in Saudi Arabia. In particular, "the highest frequency of Internet use can be seen among students, who are encouraged to use the Internet at their schools" (CITC 2010, p. 18). Figure 2.4 below demonstrates the Saudis' usages of the Internet.

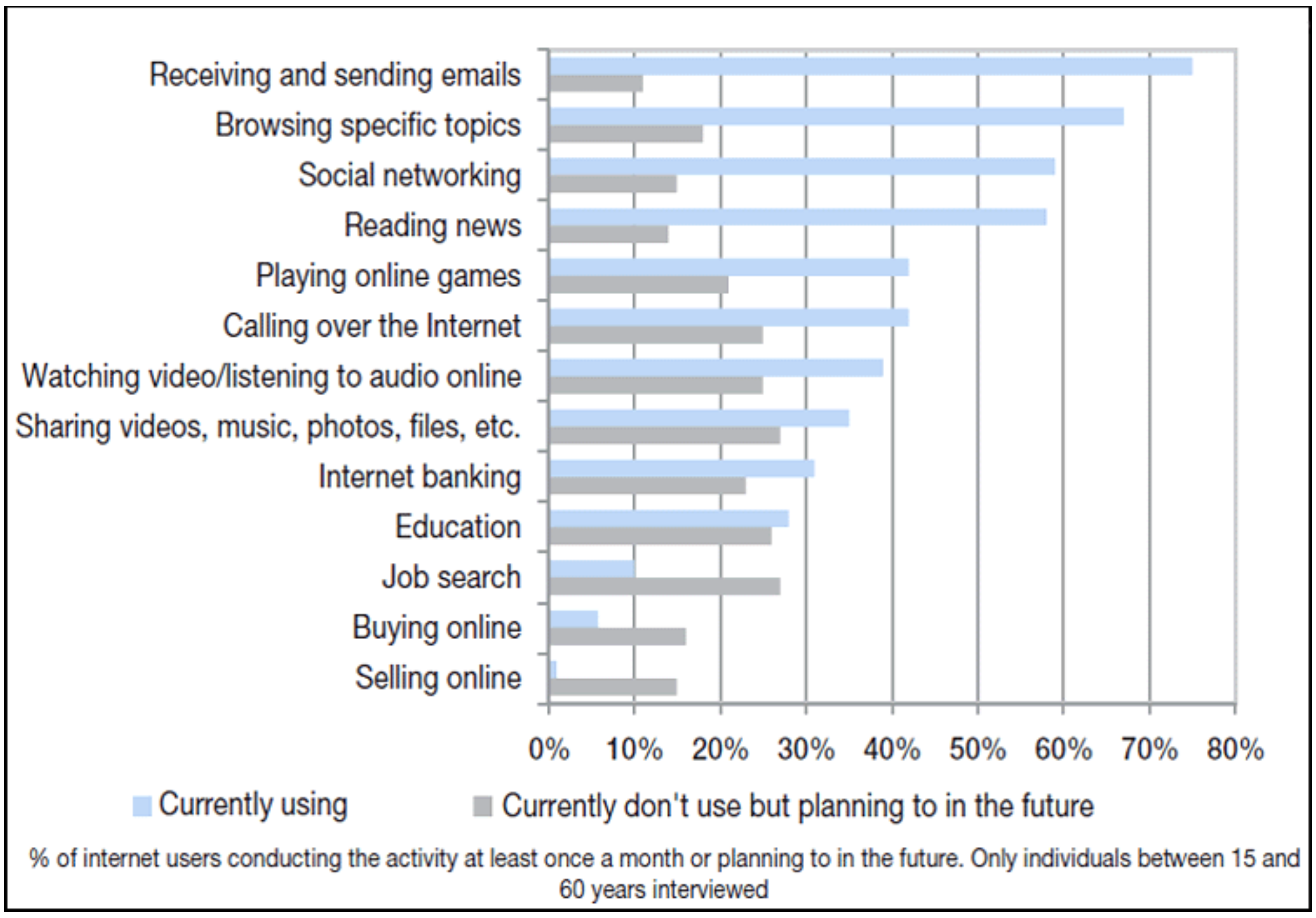

Figure 2.4: Saudi Usage of Internet Content
Source: (CITC 2010, p. 18)

On the other hand, businesses in Saudi Arabia most often use the Internet to communicate with government authorities and access banking and financial services. “As partner ecosystems mature and the availability of online content increases, activities such as buying online from suppliers, HR functions, training and education, and market research can expect to see significant increases in usage. These increases in usage are in line with the potential business benefits the internet ecosystem can provide to companies, especially in terms of time and cost savings” (CITC 2010, p. 19).

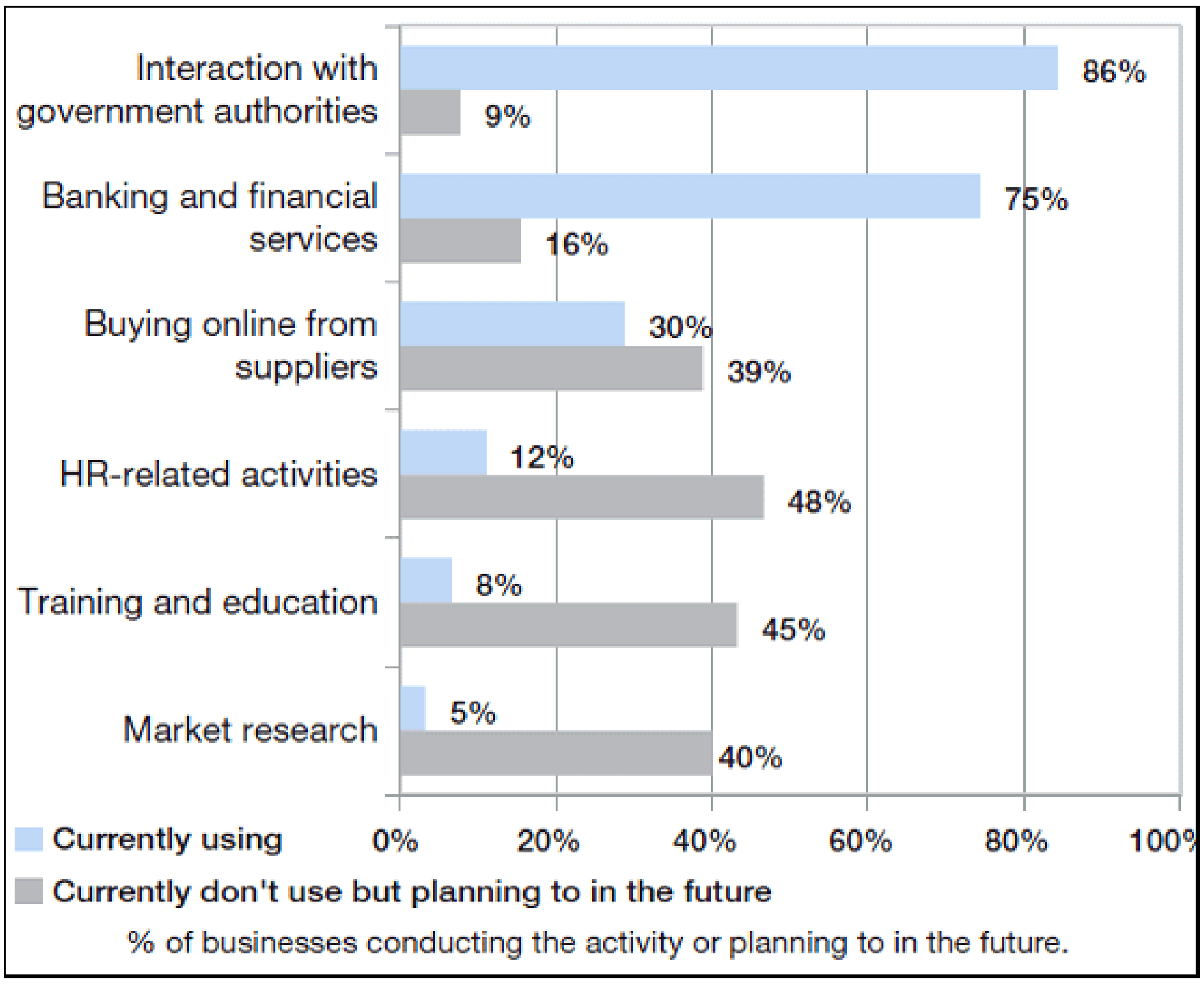

Figure 2.5: Business Usage of the Internet
Source: (CITC 2010, p. 19)

## 2.3. E-government in Saudi Arabia

In 2003, a decision was taken by the Saudi Government to start work on e-government; however, a committee for e-government was established and the actual work started in 2005 (Alfarraj, Nielsen & Vlacic 2010). With cooperation of three government entities (The Ministry of Communication and Information Technology (MCIT), the Ministry of Finance and the Communication and IT Committee (CITC)), an e-government program called 'Yasser' was launched in 2005 (Al-Shehry et al. 2006). This program acts as an umbrella for all e-government activities, procedures, legislations and all related issues (Alfarraj, Nielsen & Vlacic 2010). An e-government plan was set up with the following vision "By the end of 2010, everyone in the Kingdom will be able to enjoy from anywhere and at any time – world class government services offered in a seamless user friendly and secure way by utilizing a variety of electronic means" (Al-Shehry et al. 2006). However, this vision has not been achieved as set up in a timely manner, which means the plan was not realistic (Alfarraj, Nielsen & Vlacic 2010). The main problem, which was not taken into consideration, was the ICT infrastructure and assessing the e-readiness of the different government departments (Alfarraj, Drew & AlGhamdi 2011,

Alshehri, Drew & Alfarraj 2012). As a result, an e-government second action plan with the vision: “Enable use of efficient, integrated customer friendly and secure multiple e-Government services” (covering the period 2012-2016) has been launched; considering human resource training and development, promote cooperation and innovation culture, and maximizing efficiency of e-services provided by government agencies, (Yasser eGov Program 2012).

E-government and e-commerce share some similarity in terms of transaction requirements and development in e-government can serve as an engine to power e-commerce development as well (Blakeley & Matsuura 2001). The similarity between e-government and e-commerce is that both of them depend on ICT infrastructure, online payment systems, and mailing/post systems to reach their users/customers and deliver their services/products (Al-Shehry et al. 2006). They are also similar in that both required a legal system to regulate the online business environment (Blakeley and Matsuura 2001 and Barzilai-Nahon & Scholl 2007). In addition, both e-commerce and e-government will benefit from online business educational programs and building the electronic services awareness in a country (Blakeley & Matsuura 2001, Barzilai-Nahon & Scholl 2007, Alwahaishi, Nehari-Talet & Snasel 2009). In contrast, the commercial driver for e-commerce, which is different from e-government, needs to be taking into consideration.

## 2.4. Retail commerce in Saudi Arabia

The average annual population growth rate in the Kingdom of Saudi Arabia has reached 3% over the past 10 years. This seems to be playing a role in making Saudi Arabia a dynamic retail market and most notably in the Middle East (AMEinfo 2008). Saudi Arabia is the fastest and largest growing retail market in Arab region (ACG 2009, Ventures Middle East LLC 2010). According to the latest report issued by Saudi Alhokair group, the wholesale and retail trade grew at a compound annual growth rate of 5.8% in the past 10 years. In 2010 the retail trade volume exceeded SAR90 billion (US$1= SAR3.75), although it was only expected to reach up to 70 billion. In 2011, the size of the retail market in Saudi Arabia estimated greater than SAR160 billion

dominated by small and medium size companies accounting for more than 85% of market share (Habtoor 2011a). By 2015, retail sales in Saudi Arabia will grow to almost SAR375 billion (IMRG 2001).

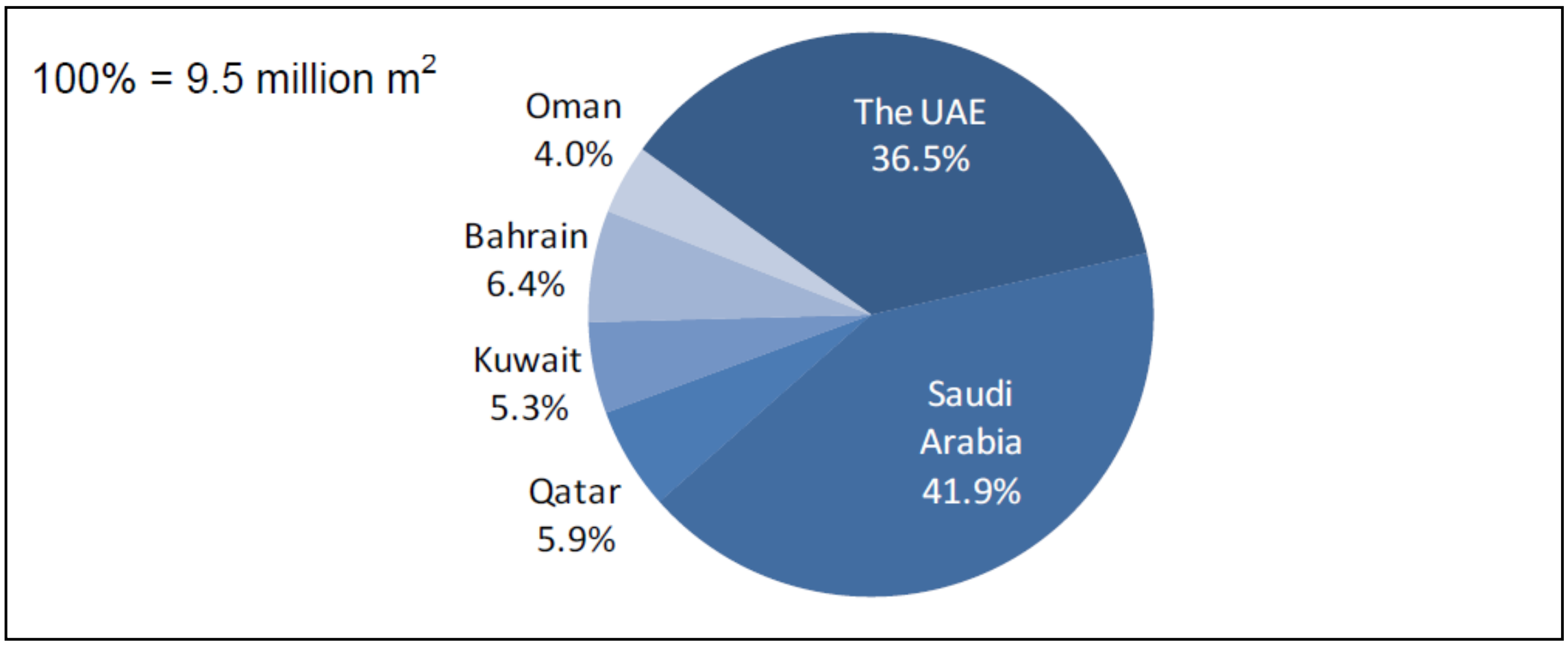

Figure 2.6: Completed Gross Leasable Area in the Gulf Countries, 2008 (%)
Adapted from (ACG 2009, p 22)

"While the global economic slowdown in 2009 had mixed effects on the countries across the Middle East, the Kingdom of Saudi Arabia (KSA), despite the contraction in its GDP, emerged as a buoyant economy in terms of retail growth and attractiveness for global investors. Adding to these inherent macroeconomic drivers to the growth of the industry, the country is also the traditional hub of religious tourism, housing two of the world's largest shrines, Mecca and Medina, which further add fillip to the growth of retail spend per capita in the region. Retail expenditure per capita in the Kingdom grew from US$ 1,825 in 2005 to US$ 2,439 in 2009. Food and grocery, Apparel, Automotive and Furniture have been the major recipients of this increased spending" (Ventures Middle East LLC 2010).

The retail sector in the Saudi market is fragmented and dominated by sales in individual stores, despite the emergence of a number of large retail chains in recent years (AMEinfo 2008). Unorganized retailers like baqalas (neighborhood stores) dominate most of the retail industry in Saudi Arabia with the top five retailers accounting for only 13.9% of the total market. However, the opportunity for the organized retailers to increase their market share is therefore vast (ACG 2009); See Table 2.2.

Table 2.2: Trade structure in Saudi Arabia and United Arab Emirates (UAE)

| No of Outlets (% industry turnover) | Saudi Arabia | The UAE |
|---|---|---|
| Hyper/supermarket | 430 (39%) | 280 (85%) |
| Other self-service | 3490 (20%) | 260 (3%) |
| Large grocery | (10860) 22% | 1550 (6%) |
| Baqalas | 18150 (18%) | 5040 (7%) |

Adapted from (ACG 2009, p. 23)

Baqalas (neighborhood stores) are in almost every main street, generally located within residential/business buildings. A typical retail street scene is depicted in Figure 2.7.

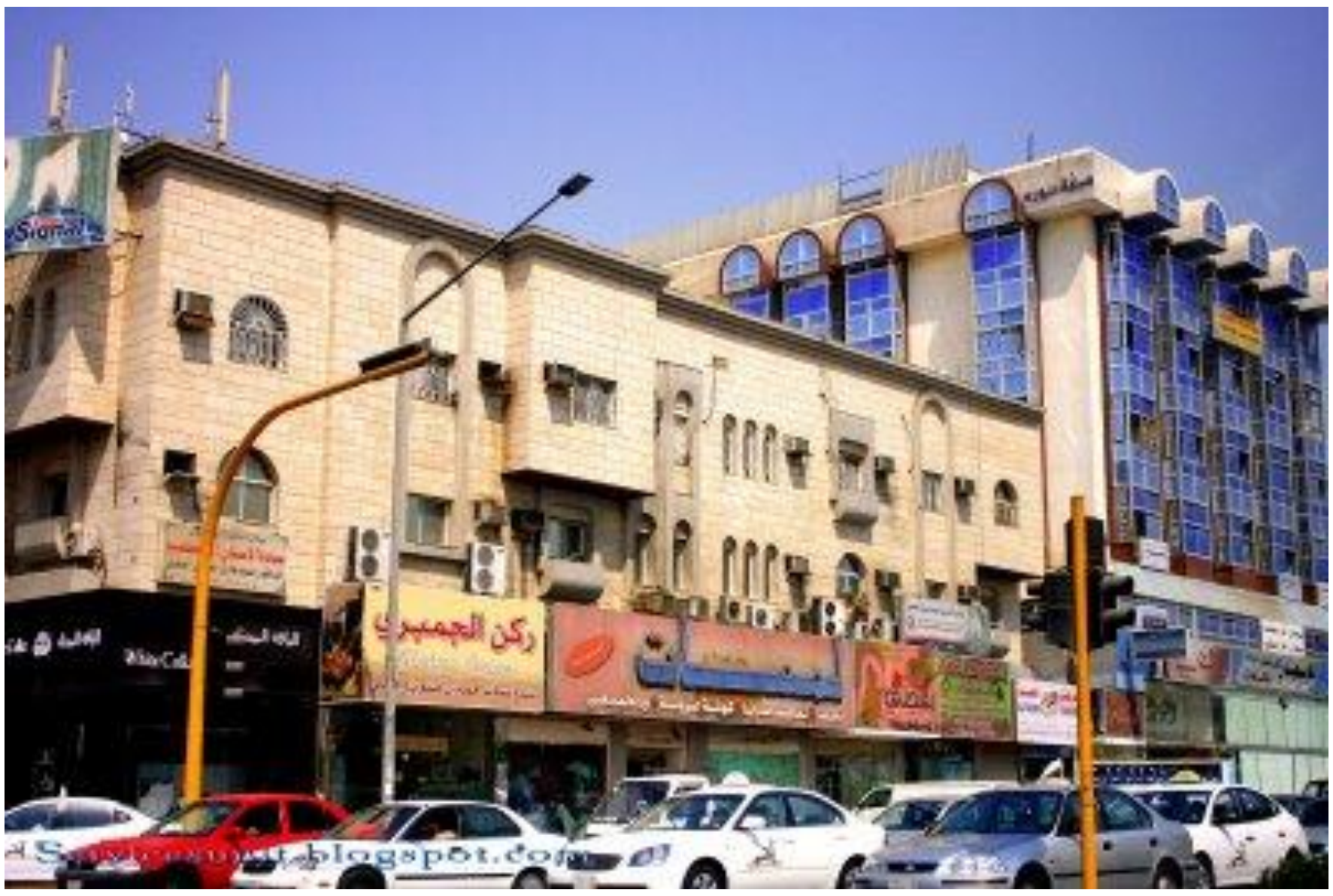

Figure 2.7: Baqalas (neighborhood stores) along Jeddah Road
Source: (BWN 2008)

According to the 2010 global retail development index, Saudi Arabia ranked number four. Saudi Arabia has a booming retail market with a stable economy. The hypermarket and supermarket sectors have changed significantly over the past six years. This change is due to the entrance of foreign players and the expansion of local players. International apparel retailers have entered Saudi Arabia through local partnerships over the past decade (A.T. Kearney 2010). However, "Government regulations still constrain international retailers. Saudi Arabia's foreign investment rules require a minimum of 25 percent local capital. The government also regulates hours of operations during religious periods, often requiring stores to close or dim lights during prayer times. Nevertheless, international retailers will be pivotal toward shaping this vibrant market." (A.T. Kearney 2010)

Business hours vary from one company to another and from one area to another; however, mostly working hours are divided in two periods: 8 am – 2:30 pm and 5 pm – 9 pm. During Ramadan (the fasting month) these periods changed to 9 am – 3 pm and 8 pm – 1 am. The weekdays are Saturday to Wednesday and the weekend Thursday and Friday. (CSC 2009)

## 2.5. E-commerce in Saudi Arabia

The results of e-commerce development have not yet reached the originally stated aspirations, falling behind the world's expectations of a country like Saudi Arabia with its level of importance and weight in the global economy. While the Saudi Government has played a major role in promoting the rapid growth of ICT in general, it appears to have placed rather less emphasis on e-commerce than on e-learning and e-government.

The official decision to introduce e-commerce in Saudi Arabia started in 2001. In that year, Saudi Ministry of Commerce established a permanent technical committee for e-commerce including members from the Ministries of Commerce, Communication and Information Technology and Finance. The committee also includes members from the Saudi Arabian Monetary Authority (SAMA) and King Abdulaziz City for Science and Technology (KACST) (Saudi Ministry of Commerce 2001). The Committee prepared a general framework and plans for improving factors that influence e-commerce development – such as IT infrastructure, legislation and regulation, payment systems, security needs, delivery systems, education and training. Apart from a booklet (Saudi Ministry of Commerce 2001) issued by the Ministry of Commerce in 2001, little information is publicly available regarding the Committee's further work or impact. Indeed, the Committee has ceased to exist, and that the roles of e-commerce supervision and development have been transferred to the Ministry of Communications and Information Technology (MCIT) in 2006.

Since 2006, the responsibility of e-commerce has transferred to the MCIT. This information was gained from an exploratory phone call made in December 2010 to the Saudi Ministry of Commerce. So far, the efforts of e-commerce support by MCIT are hapless. Direct contact with the MCIT in April 2011, regarding information about e-commerce support and development, revealed that the MCIT is still in its early stages of studying e-commerce.

In 2007, the Saudi Communication and Information Technology Commission (CITC) carried out extensive studies of various aspects of Internet usage in Saudi Arabia, one of which was e-commerce awareness and activity (CITC 2007). It reported that only 9% of Saudi commercial organizations, mostly medium and large companies from the manufacturing sector were involved in e-commerce and only four out of ten private companies had their own websites. As for the customers, while 43% were aware of e-commerce, only 6% had ever bought or sold products online, mainly airline tickets and hotel bookings (CITC 2007). CITC’s IT Report 2010 reiterated that e-commerce in Saudi Arabia is still in its early stages. In particular, most Saudi retail chains have yet to establish an online channel, and only 8% of Saudi businesses sell online (CITC 2011).

Based on the CITC’s 2010 survey, only 6% of residents in Saudi Arabia have purchased online. This figure is low compared to the 13% in the Czech Republic and 58% in the USA. However, 16% of the Internet users in Saudi Arabia are planning to buy online and 15% have plans to sell online. “Increased plans to buy online create opportunities for Saudi online sellers. They can emulate the success of existing regional and domestic companies, such as saudi.souq.com or Saudi Airlines, which are popular among the internet [sic] users surveyed” (CITC 2011). Similarly with businesses in Saudi Arabia, they are increasing their online purchases. In contrast, online/e-commerce sales by companies and business organizations are still low with 8% of the businesses currently selling products/services online.

Firms in Saudi Arabia seem not to be following the developed countries' rapid progress towards global e-commerce. Despite the fact that consumers in Saudi Arabia are not showing strong movement towards purchasing online, the young generation are spending more on ICT, are using the Internet, and using Smartphones (MENA), (E-commerce Facts 2012; Qasem, G 2012). According to the 2012 Google report, Saudi Arabia is one of the five countries which has more than 50% of their population on Smartphones, UAE 62%, Saudi Arabia 60%, Norway 54%, Australia 52%, UK 51%, and Sweden 51% (Google report 2012). Online shoppers in Saudi Arabia are increasing as access to technology and communications infrastructure increases. The Arab Advisory Group carried out an extensive survey in mid-2006, targeting Internet users in four Arab countries (Saudi Arabia, UAE, Kuwait and Lebanon). The survey covered Internet usage and, e-commerce activities in these countries. While UAE ranked first in the rate of annual spending on e-commerce per capita, Saudi Arabia ranked first in the overall money spent on e-commerce activities. As for the prevalence of e-commerce activities among the population, UAE ranked first at 25.1%, Saudi Arabia second at 14.3%, Kuwait third at 10.7% and Lebanon last at 1.6% (AAG 2008). A recent survey of Saudi Arabia's 11.4 million Internet users (representing 41% of population) found that around 3.1 million Saudis have purchased online. Airline tickets and hotels bookings take the largest percentage of these purchases (ACG 2009, AAG 2011).

In addition to e-commerce scenario in Saudi Arabia, since 2005 the Saudi Post carried out a notable development. Postal deliveries to homes and buildings have been made since 2005 (Saudi Post 2008; Alfuraih 2008). However before 2005, individuals had no uniquely identifying addresses and mail was not delivered to homes and offices but, instead, was collected from postal offices (Saudi Post 2008). Nevertheless, a significant proportion of the public has remained without home addresses, largely as matter of old habit and personal choice. Moreover, the mail services provided by the Saudi Post do not cover all cities of Saudi Arabia. Up until now, mail services have only been rolled out in the main cities. Home postal addresses in towns, villages, and some parts of some cities are still absent (Aleid, Rogerson & Fairweather 2010a). The number of subscribers of the home mail delivery service by the Saudi Post has reached approximately half a million so far (Alriyadh 2010) which means that just under 2% of the population owns an individual house mailbox.

Another issue related to e-commerce is that there is a shari'ah block on the notion of interest in KSA. Further, in Saudi Arabia many consumers are reluctant to use credit cards, both because of a lack of trust and because some consumers are culturally averse to carrying out transactions linked with conventional interest rates (Shalhoub & AlQasimi 2006, Alwahaishi, Nehari-Talet & Snasel 2009). From a cultural perspective, Saudis are still an overtly cash-oriented society (Toland 2006, Aleid, Rogerson & Fairweather 2010a, IMRG 2011). A proper idea which may contribute overcoming the online payment issues is the growing popularity of the national payment system, in Arabic called SADAD, among citizens. This is interesting and can contribute to e-commerce development since it is trusted and widely used for e-government services. SADAD is the government payment gateway but it was only scaled to a low number of large business connections. To address this limitation, SADAD have worked on Biller Base Expansion, which will increase its biller options from around 100 to 20,000 (CITC 2010). With growing citizens' satisfactions and use of SADAD, it becomes the second largest electronic payments system in the world serving around 29 million citizens (Saudi Gazette 2013).

## 2.6. Online retailing in Saudi Arabia

At the end of 2010, Saudi Post launched an electronic mall, the first online marketplace in Arabic and English (E-mall 2010), giving Saudi retailers the chance to sell their products online and benefit from cheap delivery fees. In March 2011, e-mall administrators revealed that to that date, there were 50 sellers and 50,000 buyers, which is a 10% increase in buyers and sellers since 2010. 2,000 deals have taken place, totaling 2 million KSR, with the preferred payment method being SADAD, which is an online payment system similar to Bill PAY in Australia (Al-Mohamed 2011). It seems

Saudi Post adopting an online mall is to encourage more citizens to subscribe to their services, including having a home mailbox.

The spending on online retailing in Saudi Arabia is growing. The online retail sector is estimated to be around SAR 3 billion (US$800 million, US$1= SAR 3.75). This figure represents 20% of the total electronic trading in Saudi Arabia. The average online purchase is around SAR400/US$106 (Hamid 2011). This growth in spending of the community of online customers combined with the slowness of retailers in introducing an online sales channel in Saudi Arabia is an indicator that retailers in Saudi Arabia are not realizing the importance of online retail yet.

Encouragingly, however, “roughly half of businesses in Saudi Arabia have plans to adopt e-commerce in the near future. In terms of their target customers’ revenue share, B2B was estimated at 54%, B2C at 28%, and B2G at 18%”. (CITC 2010)

## 2.7. Conclusion

The Saudi Government has begun to realize the importance of building a strong ICT infrastructure and as a result have increased their effort to expand the technology within the country. These efforts mean that Saudi Arabia has the largest and fastest growth of ICT marketplaces in the Arab region. The penetration of mobile phones adoption has reached 195%. Similarly with the growing ease of Internet access, there has been a notable increase in the number of Internet users. Internet users reached 44% in 2011 where it was only 5% in 2001. Despite the substantial investments made to date in ICT infrastructure and training, Saudi Arabia ranked 52nd out of the 70 countries listed in a 2010 e-readiness report, which assessed the quality of each country’s ICT infrastructure as well as the ability of its government, businesses, and people to use ICT.

The ICT market in Saudi Arabia not only has the largest and fastest growth in the Arab region; the retail market has the same tendency and holds a very good position globally. The Saudi market ranked fourth in the 2010 global retail market. Saudi Arabia has a booming retail market with a stable economy. In 2010 the retail trade volume exceeded

SAR90 billion (US$1= KSR3.75). In 2011, the size of the retail market in Saudi Arabia estimated greater than KSR160 billion dominated by small and medium size companies accounting for more than 85% of market share. Retail expenditure per capita in the Kingdom grew from US$ 1,825 in 2005 to US$ 2,439 in 2009. Food and grocery, apparel, automotive and furniture have been the major recipients of this increased spending.

While the Saudi Government has played a major role in promoting the rapid growth of ICT in general, it appears to have placed rather less emphasis on e-commerce than on e-learning and e-government. Since the responsibility of e-commerce has transferred to the Saudi ministry of Communication and Information Technology (MCIT) in 2006, the efforts of e-commerce support by MCIT have been hapless. Data was collected in 2011, to establish the extent of e-commerce support and development done by MCIT. When confronted with their inadequate efforts thus far, MCIT explained that they are still in the early stages of studying e-commerce. Only a very small number of Saudi businesses sells online. In particular, most Saudi retail chains have yet to establish an online channel, and only 8% of Saudi businesses sell online. Firms in Saudi Arabia seem not to be following the developed countries' rapid progress towards global e-commerce. In contrast, online shoppers in Saudi Arabia are increasing with access to technology and communications infrastructure. A recent survey of Saudi Arabia's 11.4 million Internet users found that around 3.1 million Saudis have purchased online. The largest percentage of these purchases involves Airline tickets and hotels bookings (AAG 2011).

The following chapter reviews the extant literature that highlights the challenges and drivers of Business to Customers (B2C) e-commerce globally and locally in Saudi Arabia. The chapter also reviews the current role of the Saudi Government in terms of supporting electronic government and electronic commerce. The types of roles that governments around the world have enacted in order to promote online commerce are also reviewed.

# CHAPTER 3: LITERATURE REVIEW

Although the present approach takes a broader approach to e-commerce adoption than simply determining a list of possible inhibitors and facilitators, it is still important to consider the factors found in the literature that influence businesses on whether or not to adopt e-commerce. While this research focuses on the retailers' side, it is still important to review the literature regarding consumers' factors because business involves two parties; sellers and buyers. In addition, two theoretical models of technology adoption at organizational level, Diffusion of Innovation (DOI) and Technology-Organization-Environment (TOE) framework, will be reviewed together with studies that have used them to provide good understanding about the research phenomena. These two models give a good understanding of looking to different aspects (e.g. organization, external environment, and technology) of an organization that might be influencing whether to adopt or not to adopt. Finally, the chapter provides a background about the e-commerce maturity models and detailed reviews the Stage of Growth e-business (SOG-e) model (Mckay, Marshall and Pranato 2000). The criteria to evaluate businesses' maturity of e-commerce were developed based on the SOG-e with some adaption to fit the research purpose.

## 3.1. E-commerce Maturity

The previous sections reviewed the factors influencing on whether to adopt or not do adopt e-commerce. These factors were in general for businesses and consumers. However organizations differ in terms of rating these factors based on their level of e-commerce maturity. For example, an organization which has website and interact with its consumers online may significantly differ in rating the lack of knowledge to use e-commerce in contrast to an organization that has no online presence at all. To find these differences, organizations are needed to be evaluated and classified into different levels based on their e-commerce maturity levels. Mature e-commerce is the state of having full development. Several models exist to help assessing the level maturity in organizations. Maturity models include stages from an initial state to maturity to help organizations assess as-is situations, to guide improvement initiatives, and to control progress (Röglinger, Pöppelbuß & Becker 2012). Morais, Gonçalves & Pires (2007) reviewed the principal models of e-business maturity and present the most cited models. Table 3.1 below demonstrates a comparison of the most cited maturity models.

Table 3.1: Comparison of the maturity models

| Model | Perspective | Development | Emphasis | Verification | Focus | Source | Stages |
|---|---|---|---|---|---|---|---|
| KPMG | Business | Linear | Non-specific | No | E-commerce | Private Sector | 3 |
| Model of Grant | Business | Linear | SME | Yes | E-business | Academia | 5 |
| Model of McKay | Technology | Linear | Non-Specific | No | E-business | Academia | 6 |
| Model of Earl | Business | Linear | Non-Specific | No | E-business | Academia | 6 |
| SOG-e | Technology | Linear | Non-Specific | Yes | E-business | Academia | 6 |
| Model of Rayport & Jaworski | Technology | Linear | Non-Specific | No | E-business | Academia | 4 |
| Model of Rao | Technology | Linear | Non-Specific | No | E-business | Academia | 4 |
| Model of Chan & Swatman | Business | Linear | Non-Specific | Yes | E-business | Academia | 4 |

Reproduced form (Morais, Gonçalves & Pires 2007)

Among these models, Stage of Growth e-business (SOG-e) model is the more validated (Morais, Gonçalves & Pires 2007) and accounts for both the Internet-based IT activity in organizations alongside with traditional information technologies and systems (Mckay, Marshall and Pranato 2000). It was developed in 2000 by Mckay, Marshall and Pranato (2000) and verified in using empirical research assessing the progression of e-business maturity in Australian Small and Medum Enterprises (SMEs), (Prananto, McKay & Marshall 2003).

The SOG-e consists of 6 stages (Mckay, Marshall and Pranato 2000):

- Stage 1: no online presence
- Stage 2: static online presence
- Stage 3: interactive online presence
- Stage 4: Internet commerce
- Stage 5: integrated organization
- Stage 6: extended enterprise

In stage 1, an organization has no clear direction for e-business. In stage 2, an organization is considering the importance of e-business; however, definite plan for moving forward does not exist. In stage 3, an organization is considering the significance of e-business and has definite plan for moving forward; however, the focus is on technology-centric perspective and not influenced by needs of business. In stage 4, an organization is making e-business adoption and development more business-focused, moving towards integrating and coordinating e-business components and the business processes. In stage 5, an organization Integrates between processes and activities of normal business and e-business. E-business initiatives, in this stage, provide strategic benefits through the building of strategic systems. In stage 6, an organization strongly integrates the e-business components and business processes within organization and with business partners making e-business involved with every aspect of the organization. E-business initiatives, in this stage, establish and hold strategic advantage. (Prananto, McKay & Marshall 2003)

Each of SOG-e stages is assessed based on four layers: e-business strategy, e-business system, staff arrangement, and impact on business processes (Prananto, McKay & Marshall 2003). The SOG-e is built on Galliers and Sutherland (G&S) model (1994) which has six stages to assess the maturity of traditional IS/IT within an organization (Mckay, Marshall and Pranato 2000). In addition to the assessment of the maturity of traditional IS/IT within an organization, six stages are added/cooperated to assess the maturity of e-commerce and all together assess the maturity of e-business within an organization (Mckay, Marshall and Pranato 2000). The SOG-e has, therefore, four stages to assess e-commerce maturity and four stages to assess traditional IS/IT maturity and the meeting point for these two stages groups is in stages five and six which serve to assess e-business maturity, see Figure 3.1 (Mckay, Marshall and Pranato 2000).

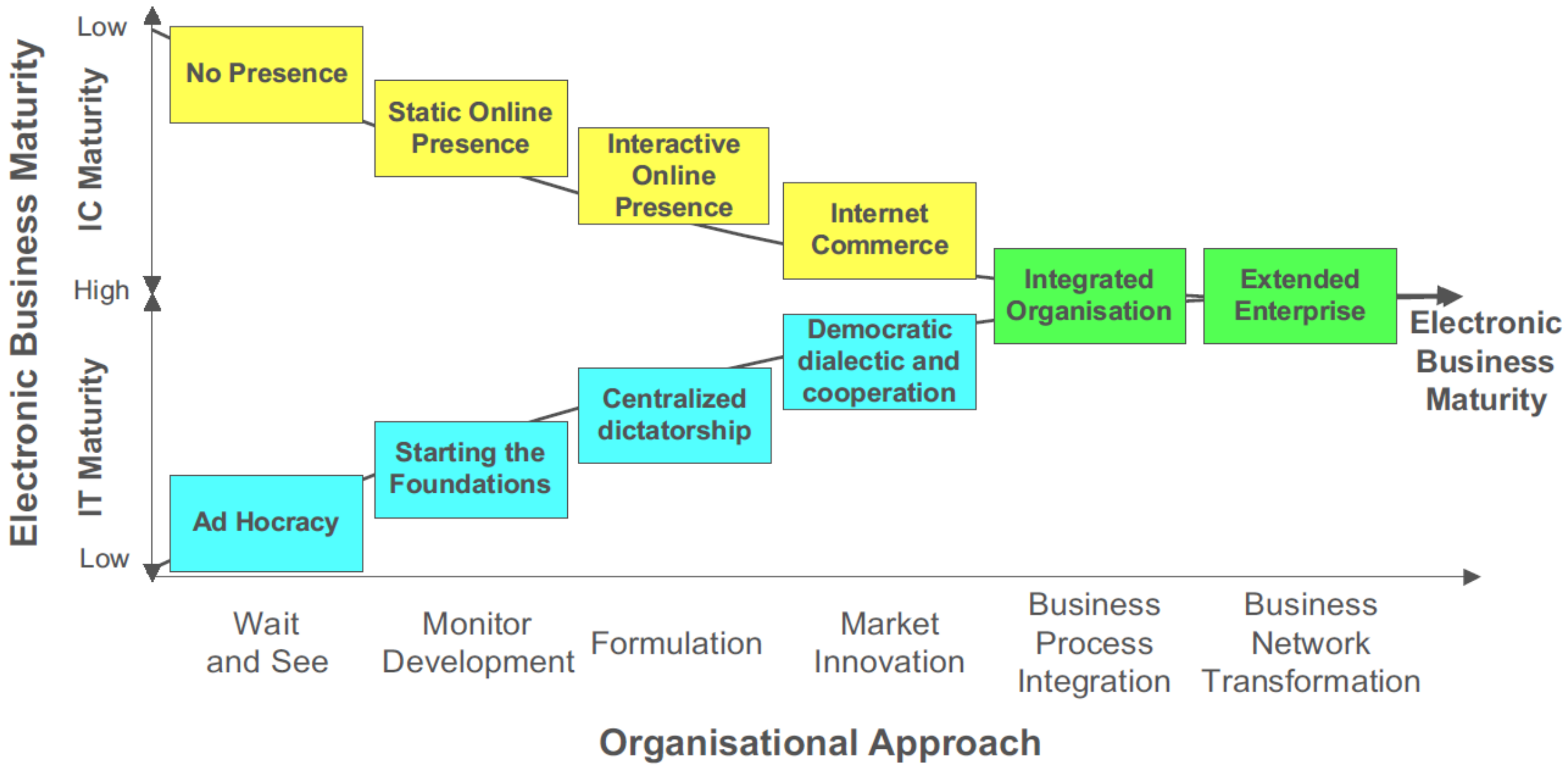

Figure 3.1: The SOG-e model of e-business maturity (Mckay, Marshall and Pranato 2000)

In this case, e-commerce is regarded as subset of e-business (Davis & Benamati 2003) and this is the same definition which is the current study adopt. The current study adopt the definition of e-commerce that says e-commerce is commerce enabled by Internet including pre-sale and post-sale activities whereas e-business as supporting the range of business processes using all electronically mediated information exchanges, both inside and outside an organization (Chaffey 2004). As the focus of this study is on e-commerce, therefore, the stages of SOG-e to assess e-commerce maturity will be adopted, see Figure 3.2. Furthermore, due to the KSA is still in its early stage of e-commerce, only the first four stages of e-commerce maturity will be used.

The criteria for each stage of SOG-e are more descriptive which give choice to the participant to choose from descriptive options. The participants may be not aware of their selections and assessing themselves. For this reason, the current study used SOG-e and the literature to develop specific criteria for each stage.

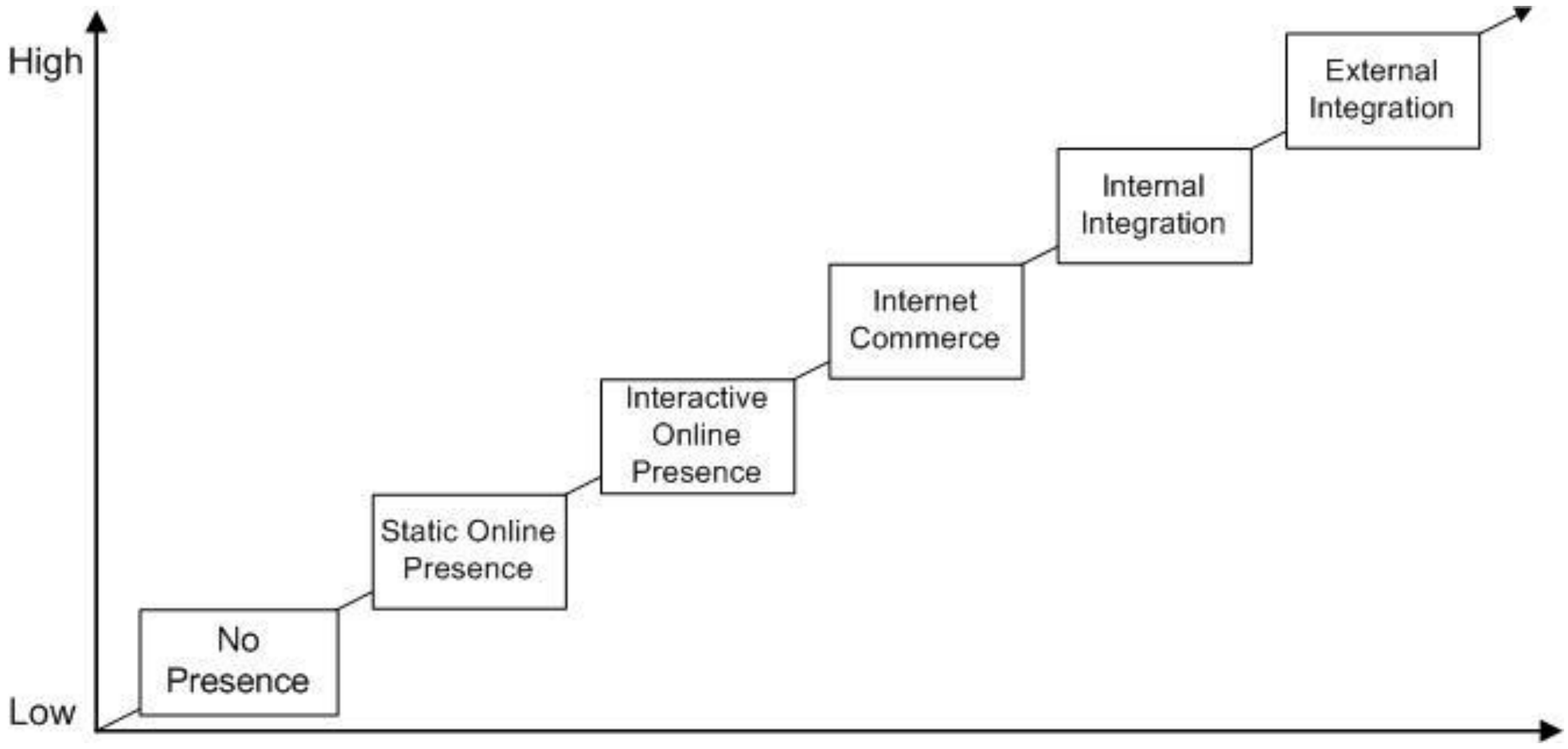

Figure 3.2: SOG-e Stages of e-commerce maturity (Prananto 2004)

Molla & Licker (2004, 2005a, 2005b) developed a framework based on SOG-e and literature to assess the actual and planned implementation of e-commerce within organization. They placed companies into six stages of e-commerce based on the following criteria as illustrated in Table 3.2. Attention is drawn here that Molla & Licker (2004, 2005a, 2005b) identified organizations that sell online in Stage five whereas in SOG-e companies which sell online are classified in stage four. These criteria of the six stages of e-commerce maturity (Molla & Licker 2004, 2005a, 2005b) are close to the ones established for the stages of e-government model which developed by United Nation to assess the maturity of e-government (UN 2000) and this model can be used to assess e-commerce maturity as well (Ghasemzadeh & Ghasemzadeh, F 2003).

Table 3.2: Criteria of the six stages of e-commerce maturity

| Stage | Criteria |
|---|---|
| 1 | Not connected to the Internet and no e-mail |
| 2 | Connected to the Internet with e-mail but no Web site |
| 3 | Static Web, that is publishing basic company information on the Web |
| 4 | Interactive Web presence, that is, accepting queries, e-mail, and form entry from users |
| 5 | Transactive Web, that is, online selling and purchasing of products and services such as customer service |
| 6 | Integrated Web, that is, a Web site integrated with suppliers, customers, and other back-office systems allowing most business transactions to be conducted electronically |

Source: Molla & Licker 2004, 2005a

The e-government stage model has, in addition to no presence stage, five stages: emerging presence, enhanced presence, interactive presence, transactional presence, and seamless or connected (United Nations 2008), Table 3.3.

Table 3.3: Criteria of the e-government stage model

| Stage | Criteria |
|---|---|
| No presence | No website |
| Emerging presence | Official website for the country containing information about it and at there must be links for the country's ministries and departments, such as health, education and so on. |
| Enhanced presence | The government provides on its website more information to citizens on public policy and the government as well as other information such as reports and regulations, all of which is easily and continuously accessible through archives. |
| Interactive presence | The government provides on its website downloadable forms for other services in order to enhance the ease and convenience of the service requester. Simple two-way communication with the ability to post comments online is also offered. |
| Transactional presence | The government has started to provide online services and allows citizens to access these services 24/7 in order to represent Government to Citizens (G2C) interactions. Examples of these services are applications for ID cards and online license renewals. |
| Seamless/connected | The governments activate back offices; that is, they have transformed themselves into an online entity that meets their citizens' needs and can respond to their citizens in easy and modern ways. Thus, it represents the most developed level of online government initiatives and has the following characteristics: horizontal connections (among government agencies), vertical connections (central and local government agencies), infrastructure connections (interoperability issues), connections between governments and citizens, and connections among stakeholders (government, private sector, academic institutions, Non-Government Organizations NGOs and civil society). |

Source: United Nations 2008; Alfarraj, Drew & AlGhamdi 2012

### 3.1.1. Criteria for e-commerce maturity stages

Based on the discussion above, the criteria presented in Table 3.4 were developed to determine organizations' level of e-commerce maturity.

Table 3.4: Adopted criteria determine organizations' level of maturity

| Stage | | Criteria |
|---|---|---|
| 1 | No presence | • No website<br>• No defined plan for e-commerce<br>• E-commerce is seen as having no impact on their business |
| 2 | Static Online Presence | • Static Web, that is publishing basic company information on the Web<br>• There is some sense of direction for e-commerce but there is no proper planning<br>• E-commerce is seen as having little impact on their business |
| 3 | Interactive Online Presence | • Interactive Web presence, that is, accepting queries, e-mail, and form entry from users<br>• The website is used for an organization's advertisements<br>• Online catalogue is used<br>• There is a clear direction for e-commerce development, but it is still focused much on technology-centric perspective and not influenced by business needs.<br>• E-commerce is seen as having considerable impact on their business |
| 4 | Internet Commerce | • Transactive Web, that is, online selling of products and services<br>• Consumers are able to place their orders and pay online using shopping cart<br>• Clear planning and there are attempts to integrate and coordinate e-commerce initiatives with business strategy<br>• Dedicated staff with technical expertise with the help of, or together with, business-oriented staff<br>• E-commerce is seen as a driver of business process reengineering |

## 3.2. Theoretical models of technology adoption and use

A number of theories have been used by Information System (IS) researches to describe the adoption and use of technology. At the organizational level, two popular theories have been used widely either alone or with other theories to describe the adoption and use of technologies (including e-commerce). These two theories are Diffusion of Innovation (DOI) theory (Rogers 2003) and technology-organization-environment (TOE) framework (Tornatzky and Fleischer 1990).

DOI theory was developed by Rogers in 1962. It provides an inclusive view of the process of innovation decision-making (Rogers 2003), and undertakes the explanation

how an idea, practice, or object that is perceived as new by an individual, or another unit of adoption, is spread (Eastin 2002). Rogers (2003, p. 5) defines diffusion as "the process during which an innovation is communicated through certain channels over time among members of a social system". Rogers sets out four main elements for the diffusion of innovation. These elements are (1) the innovation, (2) communication channels, (3) time, and (4) social systems (Rogers 2003), see Figure 3.3. An innovation is "an idea, practice, or object that is perceived as new by an individual or other unit of adoption" (Rogers 2003, p. 12). Communication is "the process by which participants create and share information with one another in order to reach a mutual understanding" (Rogers 2003, p. 18). "The inclusion of time as variable in diffusion research is one of its strengths" (Rogers 2003, p. 20). The time dimension is "involved in diffusion in (1) the innovation-decision process by which an individual passes from first knowledge of an innovation through its adoption or rejection, (2) the innovativeness of an individual or other unit of adoption (that is, the relative earliness/lateness with which an innovation is adopted) compared with other members of a system, and (3) an innovation's rate of adoption in a system, usually measured as the number of members of the system who adopt the innovation in a given time period" (Rogers 2003, p. 20). A social system is defined "as a set of interrelated units that are engaged in joint problem solving to accomplish a common goal. The members or unit of a social system may be individual, informal groups, organizations, and/or subsystems." (Rogers 2003, p. 23)

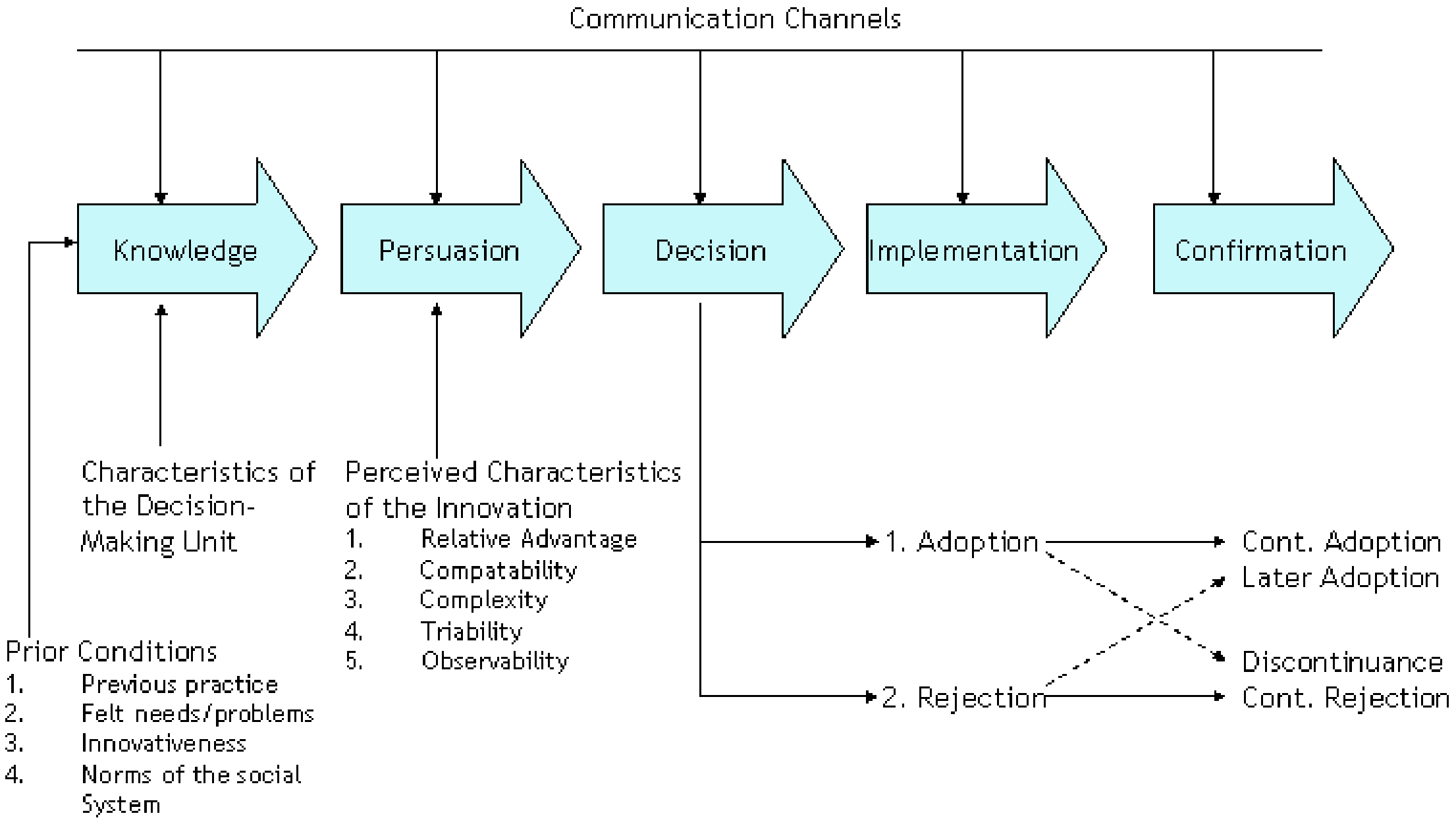

Figure 3.3: Model of the five stages in the innovation-decision process (Rogers 2003, p. 170)

Rogers (2003) discussed the DOI at two levels: individuals and organizations. At organizational level, innovativeness of organization is related to three independent variables: management characteristics and attitude toward change, internal characteristics of organizational structure and external characteristics of the organization, Figure 3.4. It includes six aspects: centralization, complexity, formalization, interconnectedness, organizational slack and size.

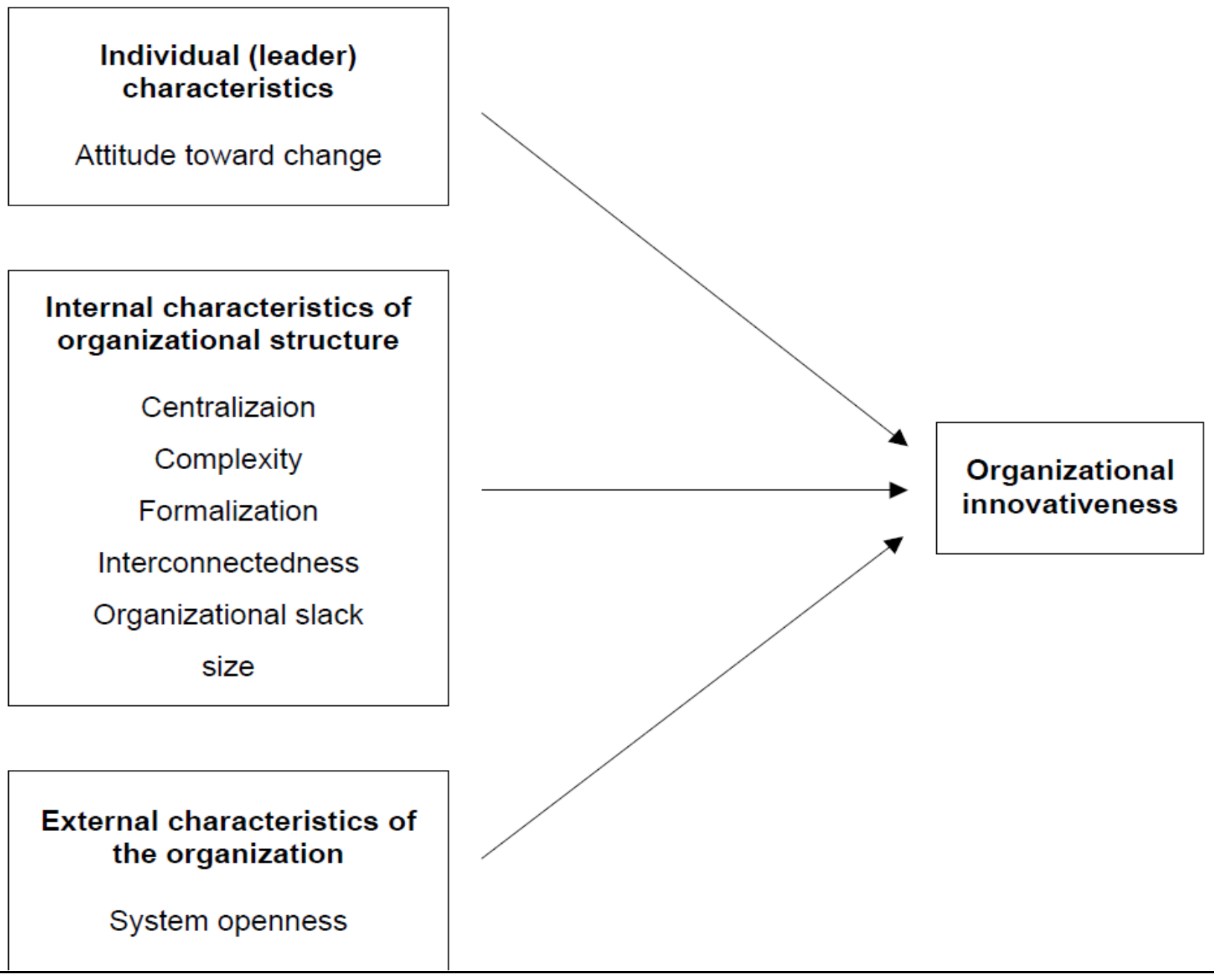

Figure 3.4: Independent variables related to organizational innovativeness (Rogers 2003)

Rogers (2003) categorizes adopters of an innovation to five types including innovators, early adopters, early majority, late majority, and laggards. Figure 3.5 illustrates the theoretical percentages of each adaptor classification as Rogers (2003) defines. Reaching the diffusion stage is when the early majority complete the adoption. These categorizations will guide the current study in categorizing the research sample, as it will be discussed in details in the methodology chapter.

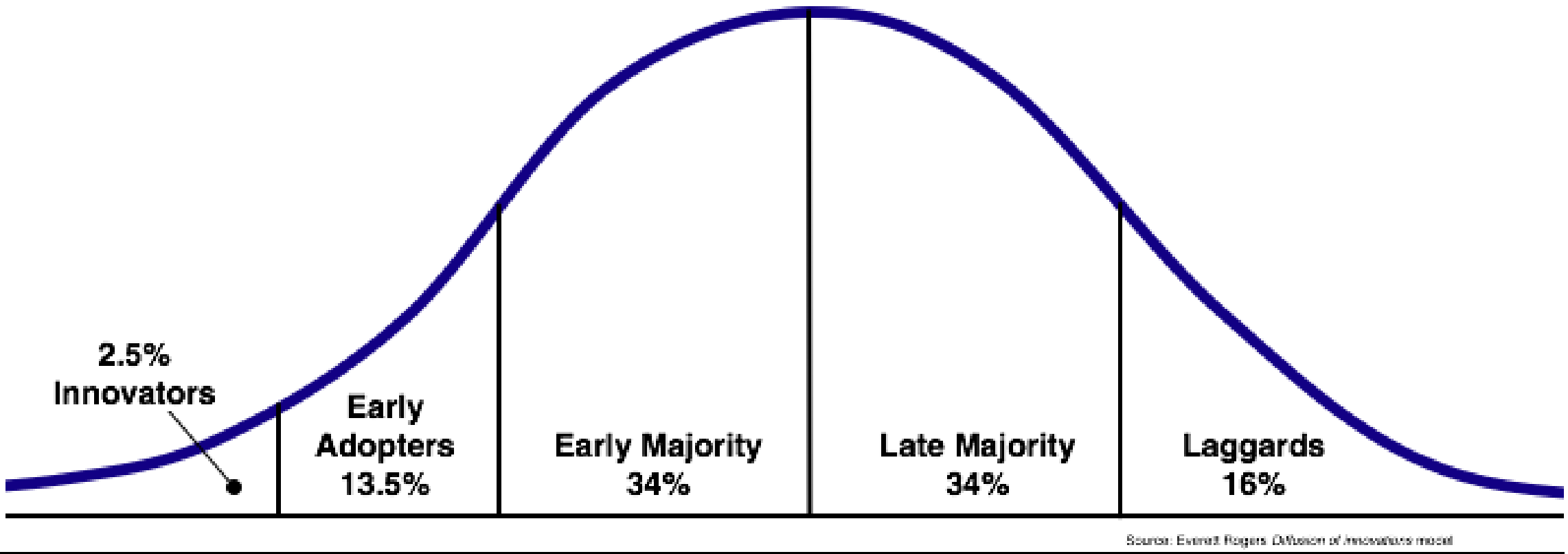

Figure 3.5: Adaptor categorization on the basis of innovativeness (Rogers 2003, p. 281)

TOE framework was developed by Tornatzky and Fleischer (1990) to describe technology adoptions by organizations. It identifies three principle contexts affect organization adoption and use of new technology. These areas include the technological context, the organizational context, and the environmental context (Tornatzky and Fleischer 1990), Figure 3.6. The technological context includes the all the technologies both internal and external that are having influence to the organization’s processes. The organizational context includes “the organizations size, the centralization, formalization, and complexity of its managerial structure; the quality of its human resources; and the amount of slack resources available internally” (Tornatzky and Fleischer 1990, p. 153). The environmental context refers to the field in which an organization conducts its business. This field includes industry, competitors, access to resources supplied by others and interactions with the government (Tornatzky and Fleischer 1990).

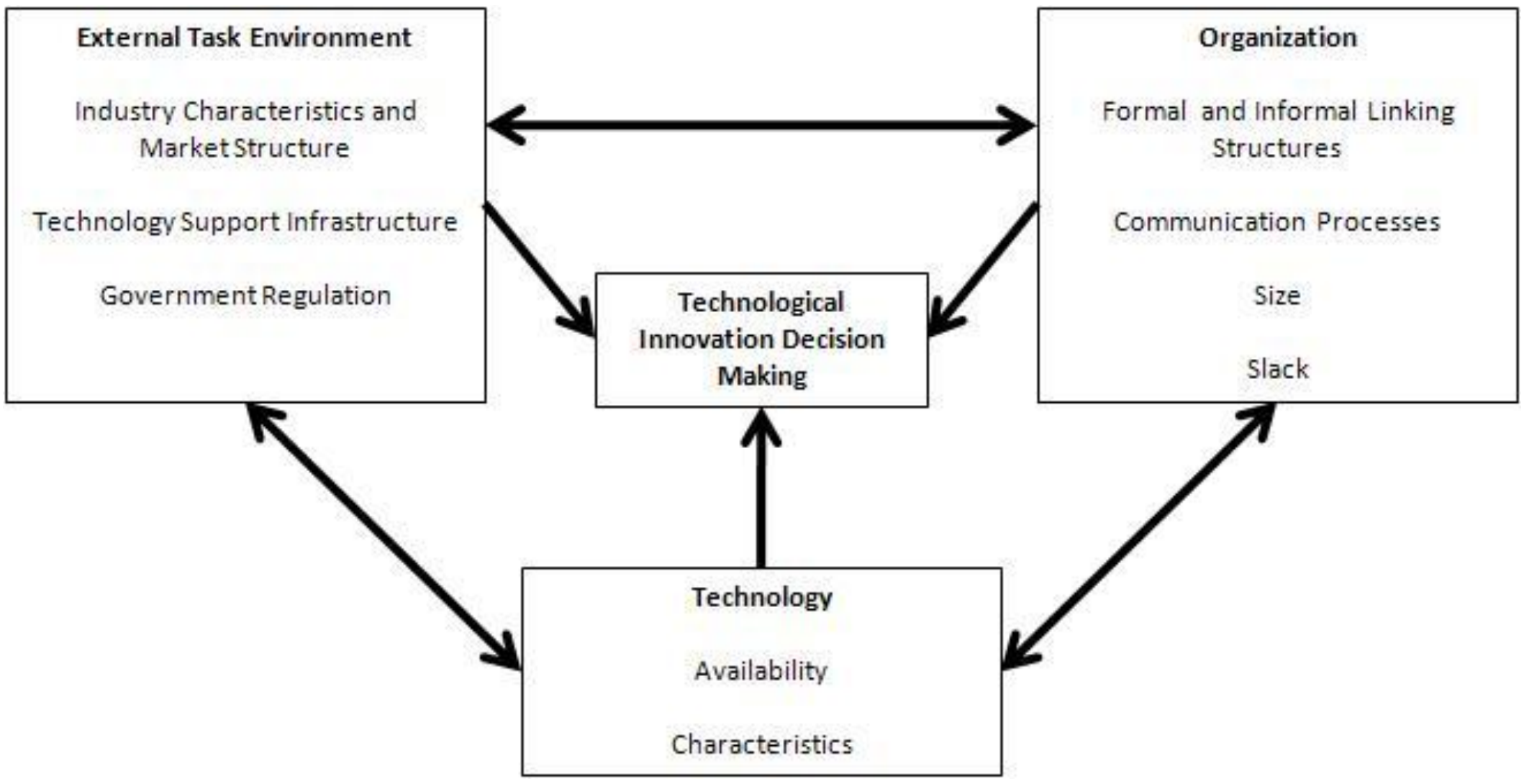

Figure3.6: The context of Technological Innovation (Tornatzky and Fleischer 1990)

A significant number of empirical studies have used DOI (e.g. Ling 2001; Al-Gahtani 2003; Pease & Rows 2005; Hsu, Kraemer & Dunkle 2006; Sparling, Toleman & Cater-Steel 2007; Chong & Pervan 2007; Li 2008); TOE (e.g. Kuan & Chau 2001; Hong & Zhu 2006; Zhu & Kraemer 2005; Lippert & Govindarajulu 2006; Lin & Lin 2008; Oliveria & Matins 2011; Awa, Ukoha & Emecheta 2012), and both DOI and TOE (e.g. Yu-hui 2008; Chong et al. 2009; Zhu, Dong, Xu & Kraemer 2006; Wang, Wang & Yang 2010) as a theoretical basis for studying contextual factors that influence the process of technologies (including e-commerce) adoptions and uses by firms.

Oliveria & Matins (2011) review the literature of information technology adoption models at firm level. They found that most empirical studies were deduced from the TOE and DOI. Using both because the DOI quite applicable to an investigation of innovation adoption and the TOE includes the environment context (which is not included in DOI) and both describe the organizational and technological contexts for better explanation of intra-firm innovation adoption (Hsu Kraemer & Dunkle 2006; Oliveria & Matins 2011). Table 3.5 summarises the reviewed studies that used DOI and TOE to studying contextual factors that influence the technology adoption by businesses.

Table 3.5: Literature used DOI and TOE to studying contextual factors that influence the technology adoption by businesses

| Author | Topic | Methodology | Variables under consideration |
|---|---|---|---|
| Yu-hui 2008 | E-procurement Adoption in Chinese Manufacturing Enterprises | Survey questionnaire was developed based on literature review and refined based on the responses of the respondents of the pilot test Data were gathered through a telephone survey of 120 respondents who agreed to participate.<br>Logistic regression is used to examine the relationships between the technological, organizational and environmental characteristics and the enterprises' adoption of e-procurement. | **Technological context**<br>• Relative advantage<br>• Complexity<br>• Compatibility<br>**Organizational context**<br>• Financial slacks<br>• Top management support<br>• Technological readiness<br>**Environmental context**<br>• External pressure<br>• External support |
| Chong et al. 2009 | adoption level of c-commerce | A survey instrument was developed for the testing of the hypothesis developed using data gathered from a review of current literature combined with in-depth case studies with two major E&E companies in Malaysia together with ten of their suppliers. A self-administered questionnaire was distributed to 400 Malaysian organizations and 109 usable questionnaires were returned. | **Innovation Attributes**<br>• Relative advantage<br>• Compatibility<br>• Complexity<br>**Environmental**<br>• Expectations of market trends<br>• Competitive pressure<br>Information Sharing Culture<br>• Trust<br>• Information Distribution<br>• Information Interpretation<br>**Organization readiness**<br>• Top management support<br>• Feasibility<br>• Project champion characteristics |
| Zhu et al. 2006a | post-adoption digital transformation | Tests the proposed model using a dataset of 1415 companies from six European countries. The measurement items were developed based on a comprehensive review of the literature as well as expert opinion. | **innovation characteristics**<br>• relative advantage<br>• compatibility<br>• costs<br>• security concern<br>**contextual factors**<br>• technology competence<br>• organization size<br>• competitive pressure<br>• partner readiness |
| Hsu Kraemer & Dunkle 2006 | E-Business Use in U.S. Firms | The research model was tested with a questionnaire designed to collect data on the variables from 294 firms. The initial questionnaires were pilot tested on 20 firms randomly selected from the sample frame, and some items were revised for clarity.<br>Confirmatory factor analysis (CFA) using AMOS 4.0 structural equation modeling software was conducted to check the reliability and validity of the measurement model. | Perceived Benefits<br>Organizational readiness<br>• Firm size<br>• Technology resources<br>• Globlization level<br>**External pressure**<br>• Trading partner pressure<br>• Government pressure<br>**Environment**<br>• Regulatory concerns<br>• Completion intensity |
| Wang, Wang & Yang 2010 | RFID adoption in the manufacturing industry | Data collected from 133 manufacturers in Taiwan is tested against the proposed research model using logistic regression.<br>The principal construct measures were based on existing instruments in the literature. | **Technology**<br>• relative advantage<br>• compatibility<br>• complexity<br>**Organization**<br>• top management support,<br>• firm size<br>• technology competence,<br>**Environment**<br>• information intensity<br>• competitive pressure trading partner pressure |
| Yuan & Chau 2001 | EDI adoption in SMEs | Case-based and survey-based approaches were used to test a model against data collected from 575 small firms in Hong Kong. Six factors are tested using logistic regression. | **Technology**<br>• Perceived Direct Benefits<br>• Perceived indirect Benefits<br>**Organization**<br>• Perceived financial Cost<br>• Perceived Technical Competence<br>**Environment**<br>• Perceived Industry Pressure<br>• Perceived Government pressure |
| Lin & Lin 2008 | e-business diffusion (internal & external org) | Data gathered from 163 IS executives in large Taiwanese firms were employed using survey to test the relationships between the research model constructs using a structural equation modeling (SEM) approach. Data analysis utilized a two-stage approach to assess the reliability and validity of the measures before their use in the research model. Confirmatory factor analysis (CFA) was applied to assess the reliability and validity of the proposed constructs. | **technological context**<br>• IS infrastructure<br>• IS expertise<br>**organizational context**<br>• organizational compatibility<br>• expected benefits of e-business<br>**environmental context**<br>• competitive pressure<br>• trading partner readiness |
| Liu 2008 | e-commerce development in firms | 156 cross-sectional samples in 16 industries, covering 11 cities and regions in Shaanxi in China, were collected by email survey, online survey and telephone interviews. Analysis - By using statistical package (SPSS11.5), reliability analysis and factor analysis were employed to identify underlying factors and dimensions of the model; multiple regressions were exploited to test the hypotheses. | **Technological:**<br>• Support from technology<br>• Human<br>• capital<br>• Potential support from technology<br>**Organizational:**<br>• Management level for Informatization<br>• Firm size<br>• Controls : Firm property<br>**Environmental**:<br>• User satisfaction<br>• EC security |
| Hong & Zhu 2006 | e-commerce adoption and migration | Survey data of 1036 firms in a broad range of industries were collected and used to test model. Analysis based on multi-nominal logistic regression<br>A combination of exploratory and confirmatory factor analysis methods was used in validating the measurement of the multi-item constructs. | • Technology Integration<br>• Web Spending<br>• Web foundation<br>• EDI Use<br>• Partner Usage<br>• Perceived Obstacles<br>• Controls: Firm size & Industry type |
| Awa, Ukoha & Emecheta 2012 | E-Commerce Adoption by SMEs | Literature review | **Technology:**<br>• Perceived Usefulness<br>• Perceived Ease of Use<br>• Perceived Behavioural Control<br>• Perceived Service Quality<br>**Organization:**<br>• Scope of Business Operations<br>• Firm's Size<br>• Organization Mission<br>• Facilitating Conditions<br>• Individual Difference Factors<br>• Social Influence or Subjective Norms<br>**Environment:**<br>• Consumer Readiness<br>• Competitive Pressure<br>• Trading Partners'Readiness<br>• Perceived Trust |
| Aguila-Obra & Padilla- | Internet technology adoption | Used the TOE for the grouping (contexts) and identified factors for each context from t the literature. | **Technological factors**<br>External communication (e-mail);<br>Obtaining information from suppliers; |
| Mele´ndez 2006 | based on the different phases | Questionnaire- based data collected from 280 companies which using Internet technology, applying factor and clustering analysis. | offering information to consumers; contact with governmental agencies; internal communication; sending purchase orders to suppliers; product and market research; receiving orders from customers; ability to reach out to international markets form and extend business; networks; operational efficiency; management effectiveness; competitive advantage; improve organization image; new business opportunities<br>**Organizational Factors**<br>Internal technical support; top management support; IT experience; IT in use; IT knowledge by top; management; IT expertise among employees; IT expertise among supervisors; IT training; Positive attitude to IT use; organizational structure<br>**External Factors**<br>Outside consultants; use of IT by trading partners; organization's image; internet image |
| Zhu & Kraemer 2005 | Post-Adoption Variations in Usage and Value of E-Business | The theoretical model tested by using structural equation modeling (SEM) on a dataset of 624 firms across 10 countries in the retail industry.<br>To empirically assess the constructs of the model, CFA was conducted using structural equation modeling as implemented in partial least squares (PLS). | **Technology context**<br>• Technology competence<br>**Organization context**<br>• Size<br>• International scope<br>• Financial commitment<br>**Environ. Context**<br>• Competitive pressure<br>• Regulatory support |
| Lippert & Govindaraj ulu 2006 | Web Services Adoption | Literature review<br>A model developed based on TOE and has not been tested. The stdu suggest that, designs to test the model may include the use of survey techniques and analytical protocols such as structural equation modeling (SEM). | **Technological factors**<br>• security concerns<br>• reliability<br>• deployability<br>**organizational factors**<br>• firm size<br>• firm scope<br>• technological knowledge<br>• perceived benefits<br>**environmental factors**<br>• competitive pressure<br>• regulatory influence<br>• dependent partner readiness<br>• trust in the web service provider |
| Zhu, Kraemer & Xu 2006 | The Process of Innovation Assimilation: Technology Diffusion Perspective on E-Business | Developing an integrative model to examine three assimilation stages: initiation→ adoption→ routinization.<br>To test the conceptual model and the associated hypotheses, a large-scale survey of firms in three major industries (manufacturing, retail/wholesale, financial services) was conducted across 10 countries. A unique data set of 1,857 firms from 10 countries is used to test the conceptual model and hypotheses. | **Technological context**<br>• Technology Readiness<br>• Technology integration<br>**Organizational context**<br>• Firm size<br>• Global scope<br>• Managerial obstacles<br>**Environmental context**<br>• Competition intensity<br>• Regulatory environment |

## 3.3. Factors influencing businesses on whether to adopt or not to adopt e-commerce

The literature investigated various issues that influence businesses on whether to adopt or not to adopt e-commerce. The issues highlighted in this review include organizational electronic readiness (e-readiness), competitive pressure, set-up and maintenance cost, product brand strength, relative advantages of using e-commerce, consumer purchasing power, privacy and security, and type of products. National environmental factors also play significant role in the adoption decision and are reviewed.

*Organizational e-readiness* - An organization's electronic readiness refers to the ability of the ICT infrastructure and relevant information systems to meet the business' electronic needs and abilities (EIU 2010). This includes the availability of IT staff or consultants with web-skills (Wymer & Regan 2005). Empirical results in a study done by To & Ngai (2006) shows that having a higher caliber of IT resources will positively affect a retail company's likelihood to adopt e-commerce. Wymer & Regan (2005) finds that a major reason for Small and Medium Enterprises (SMEs) in the USA delaying the adoption of e-commerce is a lack of resources to initiate and support E-business and E-commerce Information Technology (EEIT). The lack of having sufficient support for employees and skilled IT professionals in a company is a barrier to e-commerce adoption (Andersen, Björn-Andersen & Dedrick 2003, Wymer & Regan 2005, To & Ngai 2006). Finding and maintaining skilled staff is a greater challenge especially for larger businesses and multichannel retailers (Forrester Research Inc 2010). The lack of experience in online marketing and selling leads to a lack of success in the Internet business environment (Grewal, Lyer & Levy 2004).

*Competitive pressure* - Competitive pressure refers to the situations of marketplace where there are firms running similar businesses in competition with each other. In relation to e-commerce, this refers to the situation where selling online successfully and profitably can affect market share in the traditional marketplace (To & Ngai 2006). To & Ngai (2006) argue that for online retailing that competitive pressure is influential on the likelihood of retailers to adopt e-commerce. "A competitive environment makes a company respond more quickly" (To & Ngai 2006, p. 1136). Wymer & Regan (2005) found the competitive pressure stood out strongly as an influencing factor when looking at businesses that are already adopting e-commerce. They called them 'adopters'. The pressure of competitive marketplace is the most important factor for adopter's companies (Wymer & Regan 2005). In contrast, the reviewed literature for developing countries and Saudi Arabia do not mention competition as an influencing factor in the adoption of e-commerce. Because of the lack of maturity of online business environments in developing countries, and the convenience of traditional shops nearby, in some countries, there is no competitive pressure on the Internet. (Shalhoub & AlQasimi 2006).

*Setting up and maintaining costs* - The cost of setting up and maintaining electronic business, including the website and need of skilled IT staff, may represent a substantial obstacle for businesses wishing to adopt e-commerce. In all reviewed literature, only Wymer & Regan (2005) mentions this factor as a barrier to the adoption of e-commerce by SMEs in USA. Wymer & Regan (2005) included in their study's a sample mixture of businesses already adopting e-commerce (46%) and non-adopters (54%). They studied the factors that influence the e-commerce adoption by SMEs. They found that the only consistent factor that came out significant for both two groups (adopters and non-adopters) is the cost of setting-up and maintaining an e-commerce website, skilled IT employees, online marketing and so on. Both groups considered this factor as a barrier to e-commerce adoption and use (Wymer & Regan 2005).

*Relative advantage* - Commonly, relative advantage is linked to money, saving/gaining more money; however, it also refers to expediency, effort and time, and reduced inconvenience in using or adopting an innovation (Al-Ghaith, Sanzogni & Sandhu 2010). Relative advantage is defined as "the degree to which an innovation is perceived as being better than the idea it supersedes" (Rogers 2003, p. 229). So, to what extent is a company advantaged through the adoption of e-commerce? This is a major influencing factor for many companies in deciding whether to accept or reject e-commerce adoption (Kendall et al. 2001, Gibbs et al. 2003, Wymer & Regan 2005, To & Ngai 2006, Kraemer, Dedrick, & Melville 2006, Aleid, Rogerson & Fairweather 2010b)

"Firms view e-commerce as an additional channel for doing business, as a means of reducing cost, as a vehicle for improving operational performance, and/or as a whole new platform for doing business with great prospects for achieving these gains" (Gibbs et al. 2003, p. 10). When a company perceives that online retailing would benefit their business, the adoption is more likely to be accepted (Wymer & Regan 2005). "The greater the number of these perceived advantages, the higher the likelihood of adopting the innovation will be" (To & Ngai 2006).

Kendall et al. (2001) explored factors that influenced Singaporean small and medium enterprises (SMEs) on whether to adopt or not to adopt e-commerce. They used Diffusion of Innovation (DOI) theory as a framework; however, their usage of this theory was incomplete because they focused only on technology attributes and ignored the other contributing factors that affect the diffusion of a new technology. In any case, the relative advantage of using e-commerce in Singaporean SMEs emerged as the most significant factor influencing their willingness to adopt such technology. Kraemer, Dedrick & Melville (2006) analyzed the e-commerce in eight countries: Brazil, China, France, Germany, Japan, Mexico, Taiwan and the USA. They found that the most significant driver for e-commerce adoption and use is the advantages that it would bring including expanding markets, improving coordination with consumers and suppliers, and entering new markets (Kraemer, Dedrick & Melville 2006).

*Consumer purchasing power* - The increase in demand from consumers purchasing online is a key motivator for businesses to sell online (Gibbs et al. 2003). Kraemer, Dedrick & Melville (2006, p. 40) state that, "firms are more likely to respond to customers pressure than to supplier pressure, suggesting that e-commerce adoption is pulled rather than pushed through the value chain". This factor is the main influencer for businesses in Saudi Arabia on whether to adopt or not to adopt e-commerce. Businesses in Saudi Arabia consider consumer purchasing power in Saudi Arabia as weak and not supporting them to run their businesses online. Therefore, they are waiting to see how the online marketplace develops (Aleid, Rogerson & Fairweather 2010b).

*Privacy and security issues* - “The control over one’s personal data and the attempted access to data by unauthorized others” are two critical problems for both businesses and consumers (Ackerman & Davis 2003). These concerns become greater in an ineffective legal and regulatory environment. This statement is confirmed by Kraemer, Dedrick, & Melville (2006) that privacy and security concerns are rated the number one inhibitor for businesses to adopt e-commerce in all eight countries in their study except with three countries “where strong privacy protections have been enacted”. Wymer & Regan (2005) find that security was a significant barrier, especially with the businesses that decided not to adopt e-commerce solutions. Concern for security is the most significant barrier for non-adopters (Wymer & Regan 2005).

*Type of product* - Type of product plays a role for some companies to take the decision on whether to adopt or not to adopt e-commerce. For example, companies that sell food or fragile products are concerned about selling online, as the goods may not be received in the condition in which they were sent. Some other types or categories of products (e.g. computer products, books, music CDs and videos) have strong consumer acceptance for online purchase, which has a positive influence on retailers to adopt e-commerce solutions (Grewal, Lyer & Levy 2004). Products with brand name are also another inflecting positive factor. A brand is “a name, term, design, symbol, or any other feature that identifies one seller's good or service as distinct from those of other sellers” (American Marketing Association Dictionary). Strong brand is a drive to online retailing competitive advantage. “Strong brands pull online shoppers to retail Web sites and help generate customer satisfaction and e-tail [online retail] loyalty” (Colton, Roth & Bearden 2010). Companies selling brand name products may not encounter difficulty in gaining the trust of customers. Their reputations help them to build online businesses more quickly and easily than new or unwell-known retailers (Schneider 2002).

*ICT infrastructure* - Information and Communication Technologies (ICT) infrastructure is a pillar for the Internet economy (Sleem 2006, Alfuraih 2008). The availability and well-developed ICT infrastructure within a country should be associated with faster e-commerce diffusion (Gibbs et al. 2003, Ho, Kauffman & Liang 2007). In fact, “ICT plays a major role in all aspects of human life; in politics, economic, social and cultural development" (Jimoh & Salawu 2009, p. 150). ICT refers to all technology forms that are used for information transition, storage, creation, share or exchange (Jimoh & Salawu 2009). The e-commerce literature places high emphasis on the wide availability of broadband Internet as the backbone of ICT infrastructure strength (Andersen, Björn-Andersen & Dedrick 2003).

With regards to the availability of the technological prerequisites for e-commerce, many developing countries sit behind the developed countries. The gap in telephone and computer availability (the two main requirements for Internet) highlights the difference (Sleem 2006). For instance, Nair (2010) refers to Malaysia as having a lag in e-commerce development because the ICT infrastructure is less developed than in the more developed countries; slow deployment of broadband; "diffusion of computers, Internet, ICT services and online facilities in Malaysia are relatively lower than that in more developed countries". Furthermore, "the cost of the Internet and other ICT services are relatively higher in Malaysia than in developed countries" (Nair 2010, p. 200). The low speed and high cost of Internet access is a critical factor influencing national e-commerce development (Ho, Kauffman & Liang 2007). To contribute to solving this problem, Hafez (2006, p. 39) suggests for developing countries "(1) increases in investments in infrastructure which lead to increased networking capacity, thus improving geographical coverage; and (2) increases in competition amongst commercial Internet service providers (ISPs) thus serving to decrease access and service costs for users, which in turn further increases user access and business volume".

A number of studies have been made regarding e-commerce development in Saudi Arabia. For example, Albadr (2003), Aladwani (2003) and Sait, Al-Tawil & Hussain (2004) highlighted the availability, speed, and cost of Internet connection as potentially critical determinants of the participation by individuals and organizations in online activities, as of the early 2000s. While these studies are a little dated, the issues they examined remain of some relevance. Despite substantial investments made to date in ICT infrastructure and training, Saudi Arabia ranked 52nd out of the 70 countries listed in a 2010 e-readiness report, which assessed the quality of each country's ICT infrastructure as well as the ability of its government, businesses, and people to use ICT (EIU 2010). Use of the Internet has been available through domestic servers in Saudi Arabia since 1998 (Alfuraih 2008). In 2001, the Saudi Government liberalized the telecommunication sector opening the market to foreign investment and therefore competition (Alfuraih 2008). At the present in Saudi Arabia, individuals, private and public organizations (except universities), must subscribe to Internet services through one of 45 currently licensed Internet Service Providers (ISPs), (CITC 2011). The Internet users in Saudi Arabia increased from 5% of the population in 2001 to an estimated 46% at the end of 2011 quarter three (MCIT 2011). Mobile broadband subscriptions reached 11.5 million at the end of Q3 2011, representing a penetration of 40.5% of the population and the fixed broadband penetration rate stood at 30.6% of households at the same period (MCIT 2011). However, most of the services provided by these companies are mainly from the main cities. Most small towns and villages are still not well served by Internet connections or have no Internet connections at all (Aleid, Rogerson & Fairweather 2010a). "The main communications companies in Saudi Arabia agreed that there is a lack of infrastructure in rural regions. However, they justified this by the lack of profitability in these areas. They reported that they paid huge license fees to enter the Saudi market and their main aim is to seek profit. This has led to a shortage of communication investment and, therefore, a lack of Internet and communication services in rural areas" (Aleid, Rogerson & Fairweather 2010b, p. 381). Internet connection fees in Saudi Arabia are considered high compared to leading developing countries and developed countries (Alfuraih 2008). Prices need to be reviewed to make it attractive for most of the households to connect to the Internet (Aleid, Rogerson & Fairweather 2010a).

*Online payment infrastructure* - Providing e-commerce services with the necessary trusted and secure financial transactions requires trustworthy online payment mechanisms (Gibbs et al. 2003, Sleem 2006). "Among the variety of electronic payment tools, credit cards play a special role in online transactions" (Ho, Kauffman & Liang 2007, p. 241). In the West, credit card is the most popular method to conduct online purchases (Nielsen 2008). By contrast, consumers in the developing countries are reluctant to use credit cards due to privacy and security concerns (Shalhoub & AlQasimi 2006). Security and privacy concerns regarding using credit cards online are greater in countries where the consumer protection laws are less established (Gibbs et al. 2003, Kraemer, Dedrick &Melville 2006). This suggests that to ensure greater confidence in online business environment, development for online payment mechanisms is needed (Gibbs et al. 2003).

The situation in Saudi Arabia is not different to those in the developing countries; credit cards are not widely used as a payment method (Alfuraih 2008). The credit card penetration rate in Saudi Arabia is about 15% comparing to 140% and 273% in Korea and Singapore for example (Alwahaishi, Nehari-Talet & Snasel 2009). Low credit card usage contributes to the low diffusion rate of B2C e-commerce in the developing countries including Saudi Arabia (Alwahaishi, Nehari-Talet & Snasel 2009, Shalhoub & AlQasimi 2006, Ho, Kauffman & Liang 2007). As such, providing alternative, trustworthy and easy-to-use payment options is a critical need for the industry (Alwahaishi, Nehari-Talet & Snasel 2009). “While various solutions to the problem of payments for e-commerce transactions are being developed, it is essential that banks in developing countries must participate to make these attributes of payment systems as a criterion of helping the firms to get access Saudi market” (Alwahaishi, Nehari-Talet & Snasel 2009, p. 229).

*Legislative and regulatory framework* - The set of protocols, laws and regulations that affects the conduct of online business as well as the relationships between government, businesses, and customers are critical in terms of e-commerce promotion (Sleem 2006, Alrawi & Sabry 2009). The lack of legal framework for e-commerce in a country leads consumers to have significant reservations about purchasing online (Gibbs et al. 2003; Andersen, Björn-Andersen & Dedrick 2003). The consumers’ concerns regarding trust in business practices, privacy/security, “resistance to using credit cards, and preferences for in-store shopping and inspection of products” become greater in an environment where the legal framework is less established (Gibbs et al. 2003, p. 11).

“In most of the developing countries, the existence national laws do not contemplate the use of modern means of communication” (Sleem 2006, p. 355). A lack of clear legislation, regulation, rules and procedures, in KSA, to protect the rights of all parties involved in e-commerce transactions has been cited as a major issue by commercial organizations (Al-Solbi & Mayhew 2005, Aleid, Rogerson & Fairweather 2010b) and also consumers (Aleid, Rogerson & Fairweather 2010a). Although Saudi Arabia contributes to the efforts of UNCITRAL (United Nations Commission on International Trade Laws), (Saudi Ministry of Commerce 2001), many stakeholders regard the lack of national legislation and regulation in this area as a key obstacle to further and more rapid development (Albadr 2003, Al-Solbi & Mayhew 2005). Consumers in Saudi Arabia raised the following issues, in terms of e-commerce legal system. Statutory rights, customers' rights protection act, return and refund policy, and a known government body for legal issues does not currently exist in Saudi Arabia (Aleid, Rogerson & Fairweather 2010a).

*Logistics infrastructure* - Logistics is "the process of planning, implementing, and controlling the efficient, effective flow and storage of goods, services, and related information from point of origin to point of consumption for the purpose of conforming to customer requirements" (Finch 2007). Effective logistics infrastructure is a key for e-commerce success (Grewal, Lyer & Levy 2004, Sait, Al-Tawil & Hussain 2004). In Saudi Arabia, a significant obstacle, as highlighted by Al-Solbi & Mayhew (2005), has been the lack of individual home addresses. Before 2005, individuals had no uniquely identifying addresses, and mail was not delivered to homes and offices but, instead, was collected from postal offices (Saudi Post 2008). Postal deliveries to homes and buildings have been made since 2005 (Saudi Post 2008), (Alfuraih 2008). Nevertheless, a significant proportion of the public has remained without home addresses, largely as matter of old habit and personal choice. The mail services provided by the Saudi Post do not cover all cities of Saudi Arabia. Up until now, it has only been rolled out in the main cities. Homes postal addresses in towns, villages and some parts of cities are still absent (Aleid, Rogerson & Fairweather 2010a). The number of subscribers of the home mail delivery service by the Saudi Post has reached more than half a million so far (Alriyadh 2010). This number shows that almost 2% of the population own an individual house mailbox. Reasons for the low use of the service might include the lack of awareness of the growing importance of having a mailbox. They might also include the citizens' distrust in the security of receiving their mail through this new service or its providers (Aleid, Rogerson & Fairweather 2010a).

*Government intervention* - Government intervention for e-commerce adoption and diffusion takes various forms from country to country; however, government regulation can be critical to supporting e-commerce (Kraemer et al. 2006, Wymer & Regan 2005). Gibbs et al (2003) note that government promotion commonly comes in the form of technical support, training, and funding provided by the government to businesses, particularly Small and Medium Enterprises (SMEs). Further, they argue that "government regulation, such as privacy protection, can be critical to supporting e-commerce" (Gibbs et al 2006). In the USA, the government's role in this area was mainly to develop an effective infrastructure (Dedrick, et al. 2006). Andersen, Björn-Andersen & Dedrick (2003) identify four governance initiative sets that promote e-commerce diffusion. These sets are knowledge diffusion, economic incentives, regulation and legislation, and electronic government. With regards to electronic government, Tiger (2003) states that the Brazilian government's use of Internet providing e-services and e-procurement has stimulated e-commerce activities in the private sector. Scupola (2009) finds that the government could mostly influence through knowledge deployment including training programs, subsidies, and educational and informational campaigns to build the awareness in a country. Government can influence most common areas such as "promotion of IT and e-commerce in businesses, especially SMEs, by providing them with technical support, training, and funding for IT use" and "building consumer acceptance of IT and promoting IT literacy, [which] most notably in Singapore, and more recently in Brazil, Denmark, France, and Mexico" (Gibbs et al. 2006).

Comparing developing countries to developed countries, it is apparent that government incentive is most important in the developing world (Kraemer et al. 2006). The role of government in developing and industrializing economies (such as Taiwan) was often far more prominent and extensive (Kraemer, Dedrick & Melville 2006). Al-Rawi, Sabry & Al-Nakeeb (2008) reviewed and explored e-commerce growth in the Gulf region (Saudi

Arabia, Oman, United Arab Emirates, Qatar, Kuwait and Bahrain). Based on their review, a number of factors that may contribute to e-commerce growth in the Gulf region have been drawn up. These factors include all those mentioned above. The thing that makes this study unique is mentioning, in particular, the governments' roles for the Gulf region. They found that the governments' roles in promoting e-commerce is essential in terms of ensuring coherent policy for IT and e-commerce for consumer protection; secure, transparent, predictable, and enabled environment; support, coordination, collaboration, and cooperation (Al-Rawi, Sabry & Al-Nakeeb 2008). By contrast, in Saudi Arabia, there is an absence of government responsibility for e-commerce (Aleid, Rogerson & Fairweather 2010b). The Saudi Ministry of Commerce reported that they are not responsible for e-commerce because this is the role of the Ministry of Communication and Information Technology (MCIT). On the other hand, MCIT stated that they are only responsible for the ICT infrastructure. Therefore, it can be said that there are no long and short-term strategic plans for the e-commerce environment development as there is no government agency specifically allocated to this (Aleid, Rogerson & Fairweather 2010b).

To sum up this section, Table 3.6 lists the factors believed to be the most important in the decision making on whether or not to adopt e-commerce.

Table3.6: The identified factors influencing businesses on whether or not to adopt e-commerce in the literature

| **Factors** | **Supporting references** |
|---|---|
| E-Readiness (technologies, skilled staff etc) | (Stockdale & standing 2006 ; Drew 2003; Simpson & Docherty 2004; Grewal, Iyer & Levy 2004; To & Ngai 2006; Andersen, Björn & Dedrick 2003; Wymer and Regan 2005; Tan & Teo 1998; Aladwani 2003 |
| Setup and maintenance costs | (Kaynak, Tatoglu & Kula 2005; Wymer and Regan 2005) |
| Security issues | (Kaynak, Tatoglu & Kula 2005; Simpson & Docherty 2004; Wymer and Regan 2005; Kraemer, Dedrick, and Melville 2006) |
| Type of products | (Simpson & Docherty 2004; Grewal, Iyer & Levy 2004; Colton, Roth & Bearden 2010; Schneider 2002) |
| Relative advantage | (Rogers 2003; Kaynak, Tatoglu & Kula 2005; Drew 2003; Simpson & Docherty 2004; Poon & Swatman 1999; To & Ngai 2006; Kendall et al. 2001; Gibbs, et al 2006; Wymer and Regan 2005; Kraemer, Dedrick, and Melville 2006; Tan & Teo 1998; Aleid, Rogerson and Fairweather 2010) |
| Consumer purchasing power | (Kaynak, Tatoglu & Kula 2005; Drew 2003; Simpson & Docherty 2004; Gibbs, et al 2006; Kraemer, |
| | Dedrick, and Melville 2006; Aladwani 2003 Aleid, Rogerson and Fairweather 2010) |
| Competitive pressure | (Drew 2003; Simpson & Docherty 2004; To & Ngai 2006; Wymer and Regan 2005; Tan & Teo 1998) |
| E-commerce law / legislative infrastructure | (Kaynak, Tatoglu & Kula 2005; Simpson & Docherty 2004; Grewal, Iyer & Levy 2004; Kraemer, Dedrick, and Melville 2006; Andersen, Björn & Dedrick 2003; Sleem 2006; Alrawi and Sabry 2009; Albadr 2003; Al-Solbi and Mayhew 2005; Aleid, Rogerson and Fairweather 2010; Al-Hudhaif & Alkubeyyer 2011) |
| ICT infrastructure | (Stockdale & standing 2006; Simpson & Docherty 2004; Gibbs, et al 2006; Tan & Teo 1998; Aladwani 2003; Sleem 2006; Hafez 2006; Nair 2010; Alrawi and Sabry 2009; Albadr 2003; Sait, Al-Tawil & Hussain 2004; Alfuraih 2008; Aleid, Rogerson and Fairweather 2010; Al-Hudhaif & Alkubeyyer 2011) |
| Logistic infrastructure | (Sait, Al-Tawil & Hussain 2004; Al-Solbi and Mayhew 2005; Alfuraih 2008; Alwahaishi, Nehari-Talet and Snasel 2009; Aleid, Rogerson and Fairweather 2010; Al-Hudhaif & Alkubeyyer 2011) |
| Online payment infrastructure | (Gibbs, et al 2006; Sleem 2006; Shalhoub & AlQasimi 2006; Alrawi and Sabry 2009; Alfuraih 2008; Al-Hudhaif & Alkubeyyer 2011; Alwahaishi, Nehari-Talet and Snasel 2009) |
| Government intervention | (Tigre 2003; Gibbs, et al 2006; Andersen, Björn & Dedrick 2003; Wymer and Regan 2005; Kraemer, Dedrick, and Melville 2006; Scupola 2009; Al-Rawi, Sabry and Al-Nakeeb 2008) |

## 3.4. Factors influencing consumers to purchase from businesses online

Although this thesis focus on retailers' side, it is important to look at consumers' side as well because business has two parties; sellers and buyers. Therefore, this section reviews the literature which discusses various issues that influence consumers to purchase online. The highlighted issues in this review include lack of trust due to security and privacy issues, reluctance to use credit cards online, language barriers, preferences for in store shopping, user interface quality of e-commerce websites, lack of product trail, relative advantage, familiarity of products, and seller's good reputation.

*Lack of trust due to security/privacy concerns* - Trust is central to "interpersonal and commercial relationships because it is crucial wherever risk, uncertainty, or interdependence exist" (McKnight & Chervany 2001, p. 42). Sellers mostly refuse using a particular system or technology when the requirements of keeping their information private and transactions secure are not met (Ackerman & Davis 2003). Similarly with consumers, concerns for privacy and security regarding credit card and other personal information cause significant reservations about purchasing online (Gibbs et al. 2003, Ackerman & Davis 2003, Aladwani 2003, Grewal, Lyer & Levy 2004). Lack of trust due to security and privacy concerns becomes a strong barrier in an ineffective legal and regulatory environment (Gibbs et al. 2003).

Fraud, concerns with security risks and illegal access to personal and financial information, are major concerns for consumers wishing to purchase online in the developing countries (Alrawi & Sabry 2009, Nair 2010, Aleid, Rogerson & Fairweather 2010a). In Saudi Arabia, these concerns are among the first rated barriers (Aleid, Rogerson & Fairweather 2010a, Eid 2011). These concerns become greater in Saudi Arabia because a law to protect privacy does not exist (Aleid, Rogerson & Fairweather 2010a). This confirms the stance the previous literature takes in regard to ineffective legal and regulatory environment, which creates distrust due to security and privacy concerns and becomes a strong barrier (Gibbs et al. 2003).

*Reluctance to use credit cards* - In the West, using credit cards to pay is the most popular method to conduct online purchases (Nielsen 2008). However, there is a reluctance to use credit cards for purchasing online in the developing countries (Shalhoub & AlQasimi 2006, Sleem 2006). This reluctance, along with other reasons makes the prevalence of using credit cards for online purchases in the developing countries, including Saudi Arabia, low (Shalhoub & AlQasimi 2006). This suggests a need to provide e-commerce services with the necessary trusted and secure financial transactions, which requires a strong financial infrastructure and more options for online payment (Sleem 2006).

The reluctance to using credit cards for purchasing online seems to be more likely linked with cultural issues than security issues. In Saudi Arabia, many consumers are reluctant to use credit cards, both because of a lack of trust and because some consumers are culturally averse to carrying out transactions linked with conventional interest rates (Shalhoub & AlQasimi 2006, Alwahaishi, Nehari-Talet & Snasel 2009). From a cultural perspective, Saudis are still an overtly cash-oriented society (Toland 2006, Aleid, Rogerson & Fairweather 2010a, IMRG 2011).

*Language barriers* - Literature identifies language as an influencing factor for purchasing online (Gibbs et al. 2003). It is mentioned as an inhibitor amongst non-English speaking consumers. This is because of the fact that large amount of the Web content is in English (Gibbs et al. 2003). This suggests that having an e-commerce website that uses the local language (maybe as well as in English) may have a higher chance of attracting customers than one that uses English only. This plays a significant role, especially in Asian countries (e.g. China, Japan, and Taiwan), where language differences and difficulties with English inhibits B2C e-commerce (Gibbs et al. 2003). The reviewed literature regarding Saudi Arabia does not support the difference in the language as a barrier. This may partly be due to the widely adopted teaching of English as a second language in that country.

*Preferences for in-store shopping/ convenience of shops nearby* - For several reasons people prefer the traditional shopping in stores such as the ability to test products by hand, convenience of shops nearby, social outing, shopping habit and so on (Andersen, Björn-Andersen & Dedrick 2003, IMRG 2011). Some people enjoy visiting malls and stores for shopping. They like to interact with others, mingling with crowds, the sights, sounds and smells. To them this type of shopping is part a social outing. Although online retailers try to create an enjoyable shopping environment, “it is still perceived by many to be rather sterile in comparison to a traditional store” (Grewal, Lyer & Levy 2004, p. 710). The convenience of shops nearby and the ingrained buying habits may lead consumers to prefer normal shopping, which causes a slower adoption of e-commerce by businesses (Andersen, Björn-Andersen & Dedrick 2003).

*Good quality of e-commerce websites* - Determining the quality of an e-commerce website refers to the system quality, information quality, and service quality (Lin 2007). Gibbs et al. (2003) found that the lack of valuable and useful content on B2C e-commerce websites is considered a major inhibitor for consumers to conduct online purchases, however, the usability and interactivity of B2C websites contributes to e-commerce success in the Arab world (Alrawi & Sabry 2009). Eid (2011) found that providing greater level of integrity, reliability, and/or credibility in B2C websites has a positive influence on Saudi online consumers; they perceive those e-commerce websites as trustworthy.

*Lack of product trial / inspection by hand* - Online retailing, despite allowing for some multimedia presentations (e.g. photos and videos), cannot provide physical inspection to products. For example, perfumes need to be smelled, and clothes, shoes or toys need to be tried on or examined, and so on. By contrast, “the products most frequently purchased through online retailing remain those with standardized formats, such as books and music. Consumers often require high sensory evaluation and/or trial for products such as clothing, but these cannot be adequately represented digitally”. (Grewal, Lyer & Levy 2004, p. 711)

*Relative advantage (prices, convenience etc)* - Relative advantage is not only involved with money but also involved with saving effort and time and reducing inconvenience and anything else that brings a benefit (Al-Ghaith, Sanzogni & Sandhu 2010). As stated previously, Rogers (2003) defines relative advantage as “the degree to which an innovation is perceived as being better than the idea it supersedes”. When consumers perceive online shopping will be advantageous to them, they will then be more likely to purchase online. The reviewed literature only mentioned the cost in this regard. Grewal, Lyer & Levy (2004) mention that online retailers should offer competitive prices in order to draw consumers from traditional shops.

In many cases the convenience is about having the product delivered to a customer’s home, at no extra cost, so that she/he does not have to transport it and install it her/himself. It is about not having to find a car park. It is about not having to have a car. It is about placing an order if the goods are not in stock so that they are delivered as soon as they are available. It is about being able to compare the same goods (or comparable goods) from a number of online suppliers whilst on the computer rather than walking from shop to shop. By offering comparably lower prices, consumers would have the choice of either spending less money, or purchasing a higher quality item online, for a price comparable to a lesser item in a bricks and mortar establishment.

*Familiarity of products / seller's good reputation* - Product familiarity refers to how a particular product is familiar to a consumer (Josiassen, Lukas & Whitwell 2008). A seller's reputation refers to the opinions and evaluations of consumers about products, services, customer support, after sales services, and so on (Melnik & Alm 2005). El Said & Galal-Edeen (2009) found that the perceived familiarity and reputation of online stores plays a significant role in consumers' trust and, as result, on their decision to purchase online. The relationship between trust and these two factors (familiarity and reputation) are found to be culturally sensitive (El Said & Galal-Edeen 2009). In high uncertainty-avoidance cultures, such as the Arab countries, familiarity of products and reputation of sellers needs to have relatively more attention paid to it as they play highly significant roles on the decision of purchasing online (El Said & Galal-Edeen 2009, Aladwani 2003).

To sum up this section, Table 3.7 lists the factors believed to be the most important in the decision making by consumers on whether or not to purchase online.

Table 3.7: The identified factors influencing consumers on whether or not to purchase online in the literature

| **Factors** | **Supporting references** |
|---|---|
| Lack of trust due to security/privacy concerns | Ackerman & Davis 2003; Gibbs et al. 2003, Ackerman & Davis 2003; Aladwani 2003, Grewal, Lyer & Levy 2004; Alrawi & Sabry 2009; Nair 2010; Aleid, Rogerson & Fairweather 2010a; Eid 2011 |
| Reluctance to use credit cards | Shalhoub & AlQasimi 2006; Sleem 2006; Alwahaishi, Nehari-Talet & Snasel 2009; Toland 2006; Aleid, Rogerson & Fairweather 2010a; IMRG 2011 |
| Lack of product trial / inspection by hand | Grewal, Lyer & Levy 2004; Aleid, Rogerson & Fairweather 2010a |
| Preferences for in-store shopping/ convenience of shops nearby | Andersen, Björn-Andersen & Dedrick 2003; IMRG 2011; Grewal, Lyer & Levy 2004 |
| quality of e-commerce websites | Lin 2007; Gibbs et al. 2003; Alrawi & Sabry 2009; Eid 2011 |
| Relative advantage (prices, convenience etc) | Al-Ghaith, Sanzogni & Sandhu 2010; Rogers 2003; Grewal, Lyer & Levy 2004; |
| Familiarity of products / seller's good reputation | Josiassen, Lukas & Whitwell 2008; Melnik & Alm 2005; El Said & Galal-Edeen 2009 |
| Language barriers | Gibbs et al. 2003 |

## 3.5. Limitations in the existing literature

A great deal of literature has identified factors that influence businesses on whether to adopt or not to adopt e-commerce. However, little attention has been paid to classifying these factors into groups and evaluating them based on e-commerce maturity stage for organizations. To my best knowledge, no studies have been found to evaluate the factors influencing retailers on whether to adopt or not to adopt e-commerce from the perspectives of each on e-commerce maturity stages; form no use of e-commerce to the higher uses. Although there are slow progress of e-commerce adoption in Saudi Arabia, there are Saudi retailers who have managed to create successful online businesses, so the question is, how do they do it, what can be learnt from these companies, how could that information be used to help other companies and countries with similar context etc.

Most of the existing studies investigating e-commerce in Saudi Arabia pay high emphasis on consumers and environmental factors. For example, Al-Solbi and Mayhew (2005) highlighted the absence of clear e-commerce regulations, legislation, and rules; Alfuraih (2008) highlighted weaknesses of the e-commerce infrastructure including ICT, mail system, and payment system; Alwahaishi, Nehari-Talet & Snasel (2009) highlighted the technical and cultural issues; Alrawi & Sabry (2009) highlighted the government support and ICT infrastructure; and Aleid, Rogerson & Fairweather (2010) emphasis on cultural, technical, delivery, legal and government issues. Plenty of studies that investigate e-commerce in Saudi Arabia from consumers' perspectives exist. For example, Al-maghrabi & Dennis (2009) and ALMowalad & Putit (2013) investigated the female consumers' behaviours on online shopping; Algaith (2011) and Alsuwat (2013) also focused on consumers' behaviours and their impacts on online shopping in Saudi Arabia; Eid (2011) investigated the determinants of e-commerce customer satisfaction, trust, and loyalty.

Therefore, the current research contributes to address these limitations by doing the following. Practically, the study contributes to the knowledge by evaluating the factors that influence businesses' decision on whether to adopt or not to adopt e-commerce based on four e-commerce maturity stages. These four levels of e-commerce maturity are:

- Stage 1: businesses with no presence online/ do not have website
- Stage 2: businesses with static website
- Stage 3: businesses with interactive website
- Stage 4: businesses with e-commerce websites

This classification of businesses is to distinguish between the factors that influence each stage. This distinction is expected to help find out which level would benefit more from e-commerce adoption initiatives provided by government and/or industry. The puzzle of why Saudi Arabia lags behind other wealthy countries in e-commerce together with recommendations for how the situation can be changed so that more companies can move to higher maturity stages can hopefully be addressed more satisfactorily by using a broader framework than that used in the past. Such an approach, which would yield more than a simple list of inhibitors and facilitators, could also be used, in the future, by other researchers to study similar situations in other countries.

The following Chapter discusses in details how this research will be conducted. Chapter 4 details the qualitative study methodology, followed by Chapter 5, the interviews analysis. Then the quantitative data methodology is presented in Chapter 6 and the questionnaire survey analysis is presented in Chapter 7.

# CHAPTER 4: QUALITATIVE STUDY METHODOLOGY

This Chapter starts with describing the research design. The research design followed the exploratory mixed method. In the exploratory mixed method design, a qualitative study is conducted first for the purpose of exploring the phenomena and then a quantitative study is conducted based on the findings of the qualitative study for the purpose of testing in a wider sample. In this Chapter, the qualitative study methodology is detailed.

## 4.1. Research design

Morse (2003) and Creswell (2007, 2008, and 2011) classified mixed methods design into four types (see Figure 4.1):

1. **Triangulation** – mixed methods design is used to "simultaneously collect both quantitative and qualitative data, merge the data, and use the results to understand a research problem". Both quantitative and qualitative data collected separately and at the same time (Creswell 2008, p. 557).
2. **Embedded** – mixed methods design is used to "collect quantitative and qualitative data simultaneously but to have one form of data play a supportive role to the other form of data" (Creswell 2008, p. 558).
3. **Explanatory** – mixed methods design "consists of first collecting quantitative data and then collecting qualitative data to help explain or elaborate on the quantitative results" (Creswell 2008, p. 560).
4. **Exploratory** – mixed methods design is "the procedure of first gathering qualitative data to explore a phenomenon, and then collecting quantitative data to explain relationships found in the qualitative results" (Creswell 2008, p 561). This design is used to explore a phenomenon, develop a model, "identify themes, design an instrument", and then test it (Morse 2003, Creswell 2008).

The exploratory mixed method design was adopted.

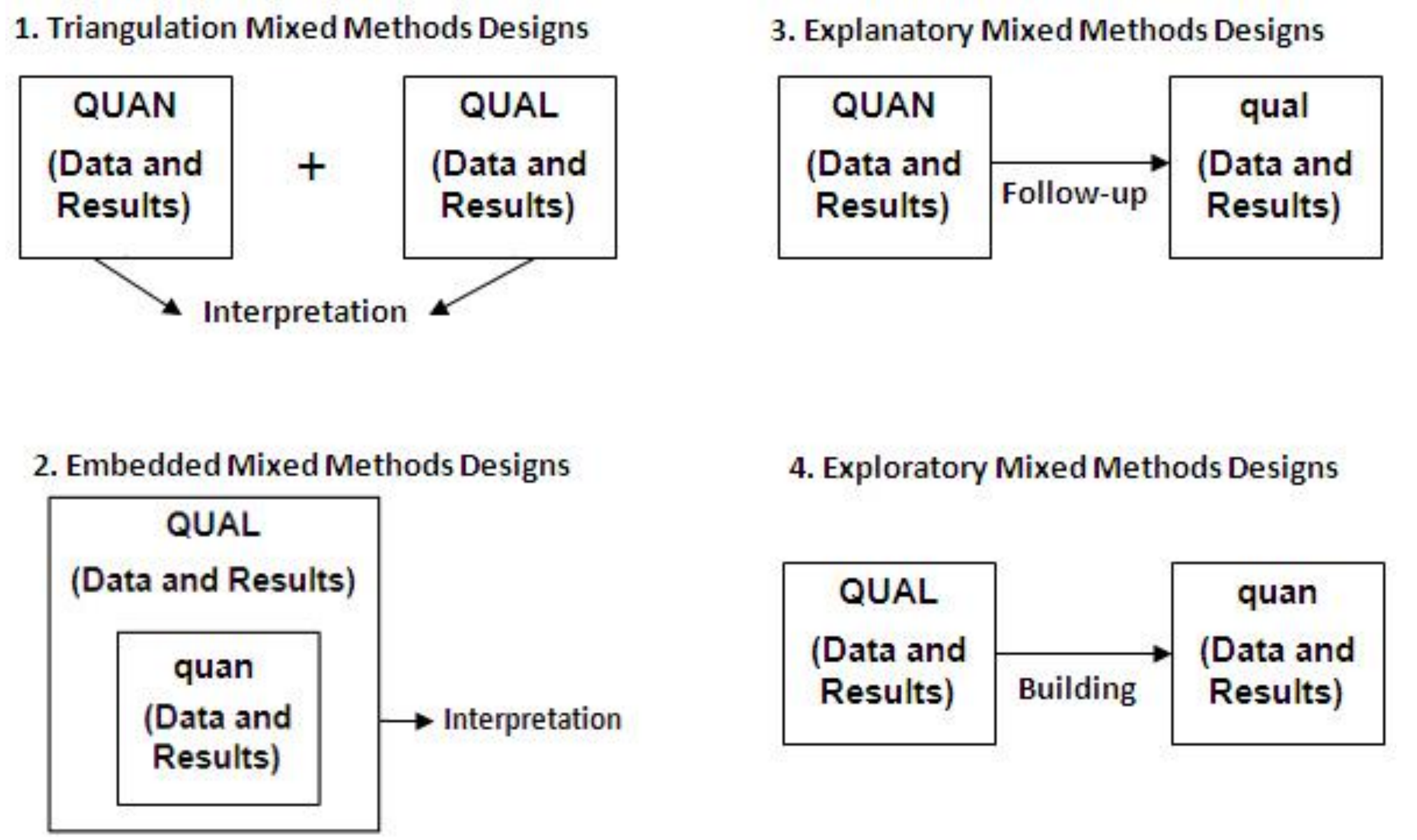

Figure 4.1: Types of Mixed Methods Designs (Morse 2003; Creswell 2007)

## 4.2. Process of the qualitative data collection

Rogers (2003) identified five attributes determining an innovation's rate of adoption. He highly recommended that each diffusion research should adopt and develop the measures of the five perceived attributes. These five variables are

(1) perceived attributes of innovations (relative advantage, compatibility, complexity, trialability, and observability),
(2) type of innovation-decision (optional, collective, authority),
(3) communication channel diffusing the innovation at various states in the innovation-decision process (mass media, interpersonal),
(4) nature of the social system (norms, degree of network interconnectedness, etc), and
(5) extent of change agent's promotion efforts (Rogers 2003, p. 222).

Based on these five attributes of Rogers (2003), interview questions were developed. The researcher formulated the questions and discussed them with an academic colleague. The questions then were formulated in English and translated into Arabic. In addition, to ensure the accuracy, relevance, and clarity of the interview questions, pilot interviews were conducted with two PhD students who were familiar with the research topic and who were Arabic-English speaking. They both were interviewed face to face and they were asked to act as real interviewees and raise comments or feedback for the question clarity, relevance and translation. Then, the researcher and the two PhD students worked together to reformulate questions that would be used to guide the real interviews. Table 4.1 presents the final interview questions which were developed to guide the interviews.

Table 4.1: The interview questions to guide the interviews

| | | |
|---|---|---|
| General Questions | | What are your targeted customers/ marketplace?<br>What do you market/what are your products?<br>What are the most popular goods that you sell? |
| | | Does your organization have access to the Internet?<br>Does your company have a website?<br>If so, what kind of information does your organization's website provide? |
| Perceived Attributes of Innovations | Relative advantage | Will doing business over the Internet lower your business cost?<br>Is setting up online channel and ongoing maintenance cost cause concern for your organization?<br>Will online channel put your company in better competition position?<br>Do you think the online channel will support traditional retailing?<br>What is your perception about the importance of online channel for your business in the future? |
| | Compatibility | What changes are needed to have an e-commerce system?<br>Do you think multichannel retailing will be compatible in your company?<br>Has any of your suppliers sell online? If not, would you consider using online channel to sell to your customers cause headache to your company's relationship with the suppliers?<br>Is changing of company's policy and organisational structure are required in order to do business on the Internet? |
| | Complexity | Do you find it difficult/confronting to use online system?<br>What makes it difficult advertising your product online?<br>What are the most difficult tasks for your organization selling products online?<br>Does your company have IT department/professionals?<br>Do you have concern about the security of payment over the Internet?<br>Do you have concern that information involved in a transaction over the Internet is not private? |
| | Trialability | Have you bought online?<br>If a sample of a system was available to try, would you use it?<br>Do you want to try a sample before applying e-commerce?<br>Do you see that it is important for Chamber of Commerce/ Ministry of Commerce to/e-commerce solutions companies offer free/cheap deals for trail purposes? |
| | Observability | Have you observed online sales activities?<br>When you see those who use the Internet to do business, will this help you to take to decide if your company should go into it as well?<br>How would you see doing business over the Internet generating the desired returns in terms of profit? |
| Communication Channels | | What channels do you use to market your products?<br>What are your favourite channels for marketing products?<br>How do you most often communicate personally (telephone, in person, e-mail, etc)? |

| | |
|---|---|
| Nature of the Social System | Do you have any friends in business that do regular online purchasing and selling of goods?<br>What does your social/professional group think about the risks and benefits of e-commerce?<br>If you/when you offer online channel sales, how do you evaluate the reaction of your customers?<br>How would you describe the effects of social system in KSA on online retailing? |
| Extent of change Agents' promotion efforts | Are you part of commercial network?<br>Do you receive any business future information/guidelines from Jeddah Chamber of Commerce for example?<br>How do you describe the efforts of chamber of Commerce/ Ministry of Commerce in terms of promoting e-commerce activities in the county? |

The researcher delivered, in person or by mail, official letters to more than 40 retailers in Jeddah, Saudi Arabia. However, no responses were received. The researcher then contacted the relevant departments in government and business to ask for help in order to contact retailers. These departments were not very helpful, simply referring to their websites as a means of finding retailers' contact details. For this reason, the researcher engaged personal contacts to coordinate the selection of participants for interviews. The selection took into consideration covering different types and sizes of retailing businesses. Sixteen interviews were carried out, which is an appropriate number for a qualitative study according to Creswell's (2013) guidelines.

The study's sample covered various categories of the retail market. Interviews were conducted with 16 retail decision-makers: 5 owners, 2 Chief Executive Officer (CEO), 2 regional directors, 5 directors, 1 marketing manager, and 1 IT manager. All were in Jeddah, the second largest and main economic city of Saudi Arabia. The sample covered 6 small, 5 medium, and 5 large organizations. The size of the businesses in Saudi Arabia are identified according to the number of employees; a micro size company has 1-5 employees, a small size company has 6-50 employees, a medium size 51-200 employees, and a large size more than 200 employees (Ministry of Commerce 2009). The sample in the present study also covered different types of retail businesses: telecommunications, computers, sports, supermarkets, restaurants, printing services, Internet services, electrical and electronic products, beauty and body cares, books, watches and clocks, and chocolate and biscuit manufacturing. Nine out of sixteen of the participating businesses have a company website and only three of them sell online. Each organization's e-commerce maturity stage was determined based on whether no website existed (stage 1), a static website existed (stage 2), the existence of an interactive website (stage 3), or the existence of an e-commerce website which allows consumers to place their orders and pay online (stage 4). In terms of e-commerce maturity, three retailers were classified in stage 4, two retailers in stage 3, five retailers in stage 2, and 6 retailers in stage 1. The classification criteria for e-commerce maturity model were based on whether they have or do not have existence of static, interactive or e-commerce website, see Table 3.4 in section 3.1.1 in the previous chapter.

Each meeting was held in the participant's organization, in their office. It started with giving the participant the interview information sheet about the study and the interview consent form to sign. See the Appendix C page 188 for the English version of the interview information sheet and Appendix D page 191 for the English version of the interview information sheet, and Appendix E page 194 for the English version of the interview consent form and Appendix F page 196 for the Arabic version of the interview consent form. All the interviews began with the general questions given in Table 5.1 followed by the other questions, though not necessarily in the given order, with the order varying depending on the interviewee's answers. Also, sometimes extra questions emerged as an interviewee responded. Furthermore, free time was given at the end of each interview session to allow the interviewees – if they wished – to add further information. . Each interview session took approximately 30 minutes. Questions were asked verbally and responded to verbally. Fourteen of the interviews were in Arabic and two were in English because the two interviewees did not speak Arabic. All the interview sessions were recorded using a recorder application on a Smartphone. The interviews were transcribed and then translated into English by the researcher with the help of translation experts. The 16 interviews can be found in Appendix G, page 198, with each having a pseudonym not a real name. Thus, for example, the notation *(Salem, medium, 4)* means the participant pseudonym is Salem, his organization size is medium, and the level of e-commerce maturity of his organization is 4.

Table 4.2 demonstrates the interviewees' positions and their organization profile including business category, company size, and e-commerce maturity stage.

Table 4.2: The interviewees' organizations profiles

| Pseudonym | Business Category | Participant's Position | Company Size | E-commerce Maturity Stage |
|---|---|---|---|---|
| Mohammed | Appliances and home improvement | CEO | Large | 3 |
| Ahmed | Beauty products | Marketing Manager | Large | 2 |
| Moneer | Telecommunications | Director | Large | 4 |
| Thamer | Sporting Goods | Regional Director | Large | 4 |
| Osam | Telecommunications | Regional Director | Large | 3 |
| Nasir | Books | CEO | Medium | 2 |
| Salem | Chocolates | IT Manger | Medium | 4 |
| Fadi | Groceries | Director | Medium | 2 |
| Tameem | Printing services | Director | Medium | 1 |
| Naif | Watches | Owner | Medium | 2 |
| Ali | Appliances and home improvement | Owner | Small | 1 |
| Waleed | Audio/Video | Owner | Small | 1 |
| Saleh | Computer related | Director | Small | 1 |
| Saeed | Groceries | Owner | Small | 1 |
| Hassan | Kitchen utensils | Director | Small | 1 |
| Talal | Mobile phones | Owner | Small | 2 |

## 4.3. Process of the qualitative data analysis

The qualitative data obtained from the interviews was analysed using content analysis. Content analysis is a research technique for studying texts, breaking it into small component units and finding relationships between these units to give a meaning (Denscombe 2010). In inductive content analysis (Elo and Kyngäs 2008) a researcher reads the content and defines the codes as they emerge during data analysis; that is, the codes are derived from the data (Hsieh & Shannon 2005). The current study used inductive content analysis because the study did not aim to test pre-identified concepts but rather used a more exploratory approach to investigate issues influencing retailers on whether to adopt or not adopt e-commerce.

All 16 interview transcripts were loaded into nVivo software version 8, which is a tool that assists in undertaking an analysis of qualitative data (Bazeley 2010).

After preparing data, the coding process started. The process of coding involved combining the text into small categories of information, and then assigning a label to the code. The coding process can be defined as “aggregating the text or visual data into small categories of information, seeking evidence for the code from different databases being used in a study, and then assigning a label to the code” (Creswell 2013, p. 184). In the coding process, transcribed interviews were read line by line looking for segments that can give meaning and these were labeled with codes. The outcome of this process provides a long list of codes, in the present case 82 codes. Table 4.3 presents the produced codes.

Table 4.3: list of the codes produced form the first coding process

| | | |
|---|---|---|
| Aware of offering competitive advantage | lack of e-commerce knowledge in our company | Issues related to products types |
| Consumers distrust Arabian online businesses | Level of demand in buying online | Personal feeling or experience affect negatively |
| Consumers Familiarity with e-commerce | level of offering competitive advantage | Personal feeling or experience affect positively |
| Consumers knowledge to buy online | local bank has to do something | Phone or e-mail order |
| Consumers level of demand of buying online | Main products | Phone order and delivery |
| Consumers lived or studied in developed countries | Management Attitude toward e-commerce | Planning to use e-commerce |
| Consumers preferences | Marketing channels | Products availability in local market |
| Consumers reluctance to use credit cards | most popular products | Protection system |
| Consumers trust | National environment | Purpose of the website |
| Consumers understanding of e-commerce benefits | nature of business | Readiness of the consumers |
| Culture or habit of online shopping | Need for Educational programs | Required action by government |
| Delivery issues | Not considering Internet for Ads | Required changes in our organization to use e-commerce |
| Developing Trust with consumers | Not profitable | Required changes to sell online |
| Dominance of traditional thinking of doing business | Not useful for Small retailers | Research is required |
| E-commerce difficulty | online payment system | SADAD_ national online payment system |
| E-marketing required hard efforts | Our business familiarity or preference | Secure network |
| experience buying online | Our business priority | security concerns |
| Familiarity of Saudis with e-commerce | Our marketing Strategy | Sellers distrust online payment |
| Favourite marketing channels | Own delivery system | Setting up cost |
| Free trails for e-commerce software | Own delivery system | Size of online or e-commerce population |
| future of E-commerce in KSA | PayPal | Skilled Staff |
| General Negative Perception | Perception about e-commerce | Targeted consumers |
| Idea of online catalogue | Perception about e-commerce profits | technology available |
| infrastructure to support e-commerce | Perception about e-commerce users | Trust and security |
| Issues relate to Saudi mailing and addressing system | Perception about Internet users in KSA | Using Internet increase seller popularity |
| Issues related to nature of the business | Perception about Saudi mailing system | Wi-fi & broadband services availability |
| Issues related to offering low prices | Personal feeling or experience | Willingness to pay for delivery services |

Another step was required to reduce the number of these codes by combining similar codes together. This step is referred to as 'focused coding'. The purpose of focused coding was not only to reduce the large number of codes by combining the similar codes, but also to produce a structured list of codes. Re-reading of the texts and consideration of all 82 codes in Table 4.3 led to six categories encompassing 55 codes: consumer issues, organizational issues, environmental issues, market issues, retailer perceptions, and general issues. Table 4.4 presents the 55 codes organized into these six categories. The fully coded transcripts can be found in Appendix H, page 225.

Table 4.4: The codes organized into six categories after the focused coding process

| **Consumers issues** | **Organizational issues** |
|---|---|
| Consumers distrust Arabian online businesses | E-commerce difficulty |
| Consumers Familiarity with e-commerce | experience buying online |
| Consumers knowledge to buy online | Issues related to nature of the business |
| Consumers level of demand of buying online | Issues related to offering low prices |
| Consumers lived or studied in developed countries | Issues related to products types |
| Consumers preferences | lack of e-commerce knowledge in our company |
| Consumers reluctance to use credit cards | Management Attitude toward e-commerce |
| Consumers trust | Our business familiarity or preference |
| Consumers understanding of e-commerce benefits | Our business priority |
| Culture or habit of online shopping | Our marketing Strategy |
| **Environmental issues** | Own delivery system |
| Delivery issues | Personal feeling or experience |
| Free trails for e-commerce software | Phone or e-mail order |
| Issues relate to Saudi mailing and addressing system | Planning to use e-commerce |
| Need for Educational programs | Research is required |
| Need for researches | security concerns |
| online payment system | Sellers distrust online payment |
| PayPal | Setup cost concern |
| Protection system | **General** |
| Required action by government | Favourite marketing channels |
| SADAD_ national online payment system | Main products |
| Wi-fi & broadband services availability | Required changes in our organization to use e-commerce |
| **Retailer perceptions** | most popular products |
| future of E-commerce in KSA | Purpose of the website |
| Perception about e-commerce | Marketing channels |
| Perception about e-commerce profits | Targeted consumers |
| Perception about e-commerce users | **Market issues** |
| Perception about Internet users in KSA | E-marketing required hard efforts |
| Using Internet increase seller popularity | level of offering competitive advantage |
| | Availability products in local market |

Elo & Kyngäs (2008) suggest using a panel of experts to support the content analysis outcomes of interviews. Byrne (2001) also suggests that using people other than the primary researcher to assess the findings is a good strategy for conducting credible qualitative research. Thus, an independent academic colleague went through all the interview transcripts and assessed the coding. Given the feedback, to organize the codes under three main categories: consumers related factors, environment related factors, and organization related factors; all the codes were then organized under these three categories. For further qualitative analysis quality assurance, all the code statements were organized under the three main categories together with examples of statements by the participants that the researcher believed exemplified the code statement and were shown to another academic. This led to the final code statements as presented in Table 4.5 together with examples of statements considered as relevant to that code. It can be seen that there are 22 code statements. These code statements can be considered suggested factors that influence retailers' decision on whether to adopt or not to adopt e-commerce. Notice that these factors result from interviews where the guiding questions were developed based on the five attributes influencing the rate of a new technology of Rogers' theory (Rogers 2003).

Table 4.5: Evidence from the participants' words/phrases to support the codes statements

| **Examples of participant words/phrases** | **Code Statement** |
|---|---|
| **Consumers related factors** | |
| Culture of the people to accept e-shopping is not encouraging us to open the electronic market. | Cultural attitudes or habits regarding online shopping |
| Unfortunately the culture of our customers is not encouraging us to go ahead with this idea. | |
| there is no difficulty [to sell using online system] but this issue refers to the culture of people to accept and use. | |
| Culture of people plays a role [in discouraging online shopping]. | |
| the culture of people is not encouraging, so it is not useful for us. | |
| People culture to buy online discourages us to think of providing option to buy online | |
| we already know that this region doesn't give you more sales on the Internet because that habit of the people here in this region used to go outside and buy | |
| We have not reached the stage to sell and buy online yet | |
| the selling with customers online is not useful for us due to the habit of people. When the culture of people changed and accept to buy online we will apply e-commerce system. | |
| This is because the nature of people in this country especially businessmen. We have wholesales too but I don't think businessmen, in this country, search for products on the Internet. | |
| We make %10 off on online sales however the problem with people culture to accept this new way of buying. | |
| our culture buying and selling online is very weak. Whoever sells or buys should be professional. | |
| The absence of the habit to buy and sell online makes it very difficult to progress in this field. | |
| people here understand the benefit of e-commerce 100% | Consumers' understanding of e-commerce benefits |
| I seen people there the indigenes original Saudi people superbly well educated and they know exactly the benefit of e-commerce and everything | |
| If there is people understand the benefit of e-commerce I agree with you it help to gain more profits | |
| Customers do not trust to buy online and they only accept with strike conditions, because it requires credit cards payment | Consumers' trust |
| I advice the main supporter in this field is to find out a way to build the | |
| honesty/trust between customers and sellers | |
| we think if there is a similar system [i.e. PayPal] available in Saudi Arabia and in Arabic language will encourage to build trust which is very important factor to buy online | |
| There are people when they pay online remain in doubt and not sure of receiving their purchased orders | |
| There is fear, because ignorance breeds fear. There is a doubt to deal by money on the Internet | |
| our customers don't trust [online payment] too | |
| We are working now to have certificates from trusted organizations to build the trust with our customers | |
| Naturally the trust needs to be built. No one will trust you with the first impression | |
| Trust is required. The most important is trust, if there is no trust you cannot sell and buy | |
| If people know that the one who run an e-commerce website is Arabian or the company owned by Arabian, they will be unmotivated to buy | |
| people in general tend not to use credit cards to buy online except those who used to do | Consumers' reluctance to use credit cards |
| Also there is another problem that people are afraid to buy online using credit cards | |
| People are reluctant to use credit Cards | |
| we notice that people feel difficult to put their credit card information | |
| They [consumers] are afraid to put their credit cards number on the internet | |
| We have tried to market on the Internet But there is no demand | Consumers knowledge to buy online |
| we have not seen strong foothold of e-commerce in this part of the world | |
| Because currently there is no rush to e-shopping in Saudi Arabia | |
| there is no good demand to buy online. For example, in western region of Saudi Arabia we receive 3-5 orders buying online every week and these are nothing compared to normal way of business | |
| those who sell or buy is very small | |
| The majority of people in our country do not know how to buy and sell on the Internet | Current level of consumer knowledge of buying online |
| Also there is Ignorance in the community. Most of them don't know the meaning of e-commerce as result they will not venture | |
| After that, when people become well educated, we can benefit from applying such idea like this | |
| The perceptions among people are that buying through the Internet cause to steal your money! ... The rumor here spreads quickly | |
| This subject requires education to make it easy for people understand how it works and what its benefits are | |
| . It depends on your learning and understanding. There is no education, no understanding, and therefore the user feels that there is a difficulty. You must educate people first, how to buy / use / get used to purchase online | |
| Illiterate today is not the one who cannot read and write but the one who don't know how to deal with technologies | |
| You need to educate people first and spread good understanding for e-commerce | |
| People here are not familiar with e-commerce and this is not something that you can apply it in a day | Consumers' familiarity with e-commerce |
| customers are not familiar with this idea | |
| The familiarity, Sellers and customers are not familiar with online sales and purchases | |
| Because the idea of [e-shopping] popularity is not coming from the customers, once the businessmen or the ones who run the business in KSA get the confident that the people are more keen to go online and visit the places they will be more than happy to jump to the bandwagon and put their business in e-commerce channel | |
| In general people have to become familiar with this system [before we | |
| offer online sale system] | |
| We [Saudi community] are still fairly new in the use of electronic commerce | |
| When people accept then we will do | |
| We are growing community, and the idea of e-commerce is new to us, and we need time to understand these things | |
| I can see also one of the inhibitors is delivery fees or insurance | Consumers willingness to pay for delivery fees |
| we have concern regarding delivering the purchased products and its fees | |
| the customer maybe will not accept to pay delivering fees | |
| With delivery there is a problem if there is delivery fees, customer maybe not paying these fees except those from specific class or busy | |
| There might be extra fees to buy online [for delivery] | |
| If you request from the customer to pay extra 10 SR for such a fees like this and if the price in total comes more than the price in the local shops, he/she may buy it from them without buy it from you online | |
| **Environment related factors** | |
| But we need to have the infrastructure right. In order to have the infrastructure right we must make sure that the Wi-Fi is available in better cases and more places. We must ensure that broadband is available for other people in remote places | Wi-fi & broadband services availability |
| But I think someone needs to kick the whole things of having the access of Wi-Fi, of having the access of broadband, having the access in every house hold. Once this done, obviously you will see the results of that very soon | |
| The percentage of users who browse the Internet is high | Internet users in Saudi Arabia |
| Today almost every home has access to the Internet, many people browse it and of course this brings customers to your shop | |
| Nowadays people aware and as I said to you almost %90 of people use the Internet and people become well educated in this matter | |
| Today almost every house has Internet connection and I believe Internet usages will increase in the future | |
| The system that protect customer right when he/she pays, he/she will receive the purchased products/services | Protection system |
| Yes sure there is no clear system to protect sellers and customers rights | |
| It has to be done from early stage, customers have to be ensured there are rules and legislations protect their right | |
| With no clear e-commerce law and legislations applying e-commerce systems are not encouraging | |
| It has to have a clear [protection] system, i.e. from government, Imposed on all involved parties | |
| Citizens will have more trust if this subject sponsored by the government because we, in Saudi Arabia, have great confidence in anything that comes through the government. | |
| Like this it should be there is a certification body from the government itself to say that this company is a certified company by local government and you can buy from them. This is good to build the customers trust with the certified companies as the government trust them | |
| it should be there is a certification body from the government itself to say that this company is a certified company by local government and you can buy from them. This is good to build the customers trust with the certified companies as the government trust them | Required action by government |
| Citizens will have more trust if this subject sponsored by the government because we, in Saudi Arabia, have great confidence in anything that comes through the government | |
| there has to be a mechanism whereby the customers feel confident and will be a %100 trusting to put their credit cards details on the Internet | Online payment systems |
| There are secure payment systems; you just need to incorporate that into your system | |
| We provide two options [for online payment], credit cards and SADA system | |
| not affording options [e.g. PayPal]for online payment [cause a difficulty using online system] | |
| Having more options for online payment other than credit cards only is an Encouraging factor to buy online | |
| Banks should provide easy options to have two credit cards, one with large amount of money and another one with small amount to be used in online payment | |
| This idea [debit cards] makes people get used to and this will remove their fear | |
| We provide two options for online payment: credit cards and direct debit. However the second option is less accepted | |
| providing more online payments because not everybody has credit card, It should be there is another ways of online payments | |
| I suggest for local banks to offer other easy options for online payment to encourage people to purchase online | |
| SADAD system is more secure system and great | SADAD, national online payment system |
| SADAD is great idea and more secure than credit cards and encourage people to buy online | |
| SADAD before was very expensive solution, it is good for large companies but it's not for middle size company like us | |
| The problem is mail address; still some people do not have mail address | Issues relate to the Saudi mailing and addressing system |
| We depend on the mobile phone number; if the mobile phone number is not correct then it's difficult to deliver the product. This is not only for us but also for other companies and home delivery people they are struggling with these things. But there are certain locations (e.g. companies buildings) you can know the address clearly, you know which street and block, building, floor, flat etc. But with community houses is not very clear. | |
| **Organization related factors** | |
| Yes [it is difficult], you have to design a commercial website, train your employees, deliver goods | E-commerce difficulty |
| The website needs cost and even we go ahead to build an e-commerce website, it is not guaranteed to be successful. Also there are other issues like online payment [difficulty], goods delivery [difficulty] | |
| Selling online is difficult to show all the details of the products unless you have customers understand the details of the products because they have known this product before with normal buying and dealt with you normally | |
| I feel it is difficult and case a headache. The normal way is better and easy | |
| E-marketing is very important and requires a lot of efforts to keep customers buying from you. | |
| The majority of similar companies, %99, do not have e-commerce websites and shops are widespread where we cannot have competitive advantage to sell online | Issues related to the nature of products |
| Same products [electrical and electronics] are widespread in the market | |
| our products related to smell, shape, and the customer wants to see and touch physical product | |
| Some customers are interested to buy from you but when they see Fragile products they will go away, it is better for them to buy from local market instead of buying online where they cannot ensured the product will shipped safely | |
| this [delivery in satisfying condition] is an inhibitor for us because some of our products are made from glass and need very high care in shipment | |
| our stuff don't know about marketing and selling on the Internet where they required training programs | Business familiarity and knowledge with e-commerce |
| Honestly there is Ignorance inside our company in terms of e-commerce, and that is why our website only display products and not receive orders | |
| I don't know the procedures to sell online | |
| We are familiar with the way of normal selling where a customer has to come to our shop and see the products | |
| We prefer personal marketing it is much better for us. When you explain to a customer face to face is totally different from leaving him alone to read the features of a product on the Internet | |
| Our employees are familiar with direct sales | |
| The normal way of selling is the best for us and we do not want to sell online at all. | |
| This [selling online] is not similar to normal way of selling where a customer comes to the shop, see a product, and make sure he/she selects the right product | |
| So the normal way of business is the best for us | |
| this [e-commerce] does not work with our customers. We, currently, don't depend on new customers but loyal customers | |
| this way is cannot be compared with the normal way of selling and buying which requires physical meeting (seller, customer, product) | |
| When a customer come to buy everything is finished with payment and giving him/her a product but on the Internet you have to take care of many things | |
| I personally like to do move toward e-commerce system. This will actually help your customers to browse, search, check the prices for competitor, and check the quality of products in terms of features, advantages and benefits and then make the right choice. It makes comfort and less time to move around | Management Attitude toward e-commerce |
| We believe that most of the company's services should be available on its website to save a customer's time and effort visiting the shop. Visiting shops are required for our customers when there are physical products need to be purchased. Even physical products can be sold online but it requires us more careful thinking | |
| However we will continue providing this option (selling online) to encourage people to use it | |
| using the Internet to order our products is not our priority at the moment | Business priority |
| Our e-commerce website is not targeting consumers in Saudi Arabia and we don't have a priority to do so | |
| sometimes I receive orders by e-mail but I don't consider them as serious orders. I give a priority to those who come to the shop to buy | |
| we have a different marketing strategy which is having branches/shops in the main malls to be directly close to customers | |
| these issues [security and trust] cause concern to us | Security and trust concerns |
| this [security of online transactions] cause a difficulty | |
| We do not trust online payment, the money must paid cash | |
| Online payment maybe represents an inhibitor | |
| we don't trust online payment | |
| The website needs cost and even we go ahead to build an e-commerce website, it is not guaranteed to be successful | Setup and maintenance cost concern |
| Building the website is very easy, You can buy cheap stuff/software from the Internet, there are e-commerce packages available cost around $100 to buy a package and publish your e-commerce website. Before it was very expensive, but know things become cheaper | |
| It is the same thing paid advertisements instead of paying money and efforts for e-commerce website and you don't know there is people look at it or not | |
| Financial costs involved in something that does not bring profit | |

Having explained the qualitative study methodology, a detailed discussion of the qualitative data can be given. Thus, in the next chapter, an analysis and discussion of the interview data is presented. It is based on the main three categories explained above; consumers, organization, and environment related factors, issues or concerns.

# CHAPTER 5: ANALYSIS AND DISCUSSION OF THE INTERVIEW DATA

In the previous Chapter, the process of the interview data analysis was detailed. The process started with open coding, resulting 82 codes, followed by focused coding, resulting 55 codes, and ending up with 22 code statements. These 22 codes statements were used in this chapter to formulate the discussion around them under three themes: consumers related factors, environment related factors, and organization related factors. Table 5.1 presents the codes statements resulting from the interviews data analysis which can be called factors influencing retailers' decisions to adopt e-commerce in Saudi Arabia.

Table 5.1: factors influencing retailers' decisions to adopt e-commerce in Saudi Arabia

| Consumers related factors |
|---|
| • Cultural attitudes or habits regarding online shopping |
| • Consumers' familiarity with e-commerce |
| • Consumers' understanding of e-commerce benefits |
| • Consumers' trust |
| • Consumers' reluctance to use credit cards |
| • Consumers' level of demand of buying online |
| • Consumers' knowledge to buy online |
| • Consumers' willingness to pay for delivery fees |
| Environment related factors |
| • Internet users in Saudi Arabia |
| • Wi-fi & broadband services availability |
| • Protection system |
| • Required action by government |
| • Online payment systems |
| • SADAD, national online payment system |
| • Issues relate to the Saudi mailing and addressing system |
| Organization related factors |
| • E-commerce difficulty |
| • Issues related to the nature of products |
| • Business familiarity and knowledge with e-commerce |
| • Management Attitude toward e-commerce |
| • Business priority |
| • Security and trust concerns |
| • Setup and maintenance cost concern |

The discussion will be grouped around the 22 factors developed in chapter 4 and will also highlight comments of retailers in companies at different stages of e-commerce maturity. It is interesting to see how retailers in companies of different stages of e-commerce maturity might differ in their attitudes regarding the 22 factors. Where comments of interviewees are given, the pseudonym is given, together with a number in brackets which indicates the e-commerce maturity stage of the participant's company. As discussed earlier in the earlier methodology chapter, the meaning of these numbers are as follow: stage 1 indicates that a company has no online presence/ no website; stage 2 indicates that a company has a static website that gives some information about a company; stage 3 indicates that a company has an interactive website that accepts feedback, comments, and communication from consumers; stage 4 indicates that a company has an e-commerce website that allows consumers to place their orders and that accepts their payment online.

## 5.1. Consumers' factors/issues/concerns

Interestingly, the retailers' interviews raised several issues relating to consumers that could discourage them from adopting e-commerce in Saudi Arabia. These issues include the retailers' perceptions of: cultural attitudes or habits regarding online shopping; consumers' familiarity with e-commerce; consumers' understanding of e-commerce benefits; consumers' trust in online shopping; consumers' reluctance to use credit cards; consumers' level of demand of buying online; consumers' knowledge of how to buy online; and consumers' willingness to pay for delivery fees.

### 5.1.1. Cultural attitudes or habits regarding online shopping

From the statements made by interviewees it appears that some retailers believe that the culture of the people in Saudi Arabia to buy or sell online is such that it discourages retailer to adopt and online retailing system. The retailers believe that people in the KSA prefer traditional shopping, visiting the shops, and inspecting the quality and uses of the products before making a decision to buy.

Saeed (1), Saleh (1) and Waleed (1) all stated that online shopping has not been a habit for Saudis and that they would not benefit from such selling online. Their comments are all close to what Saeed (1) says: "*the culture of people is not encouraging, so it is not useful for us*".

Participants whose companies only have a static website did not seem different from participants with companies at stage 1. They raised the same issues that because consumers in Saudi Arabia do not have a habit of online shopping they do not sell online. Ahmed (2) and Nasir (2) made statements close to what Talal (2) said: "*the selling with customers online is not useful for us due to the habit of people. When the culture of people changed and accept to buy online we will apply e-commerce system*".

Their thinking is all about whether they will gain more profit from selling online. From their perspective they think that e-commerce is not useful for them unless consumers accept it and it becomes a habit for them.

What about companies that do sell online in Saudi Arabia; do they think about consumers in KSA and their shopping habits? Two participants, Salem (4) and Thamer (4), agreed that online sales cannot compare to normal sales in KSA. They stated that they know that selling online in Saudi Arabia does not give more profit. However, their companies want to be active and take a positive role or to experiment in the new market. The main reason motivating them to sell online is to open up a new marketing channel and establish a good reputation for their companies' e-commerce in the region. Some believe that e-commerce in Saudi Arabia has a bright future. Thamer (4) stated that "*I believe that the Saudi market is bullish and a fertile ground for investments, but –in my opinion- even if we start thinking about e-commerce we need time, not less than 10 years, to reach maturity in this field*".

### 5.1.2. Consumers' familiarity with e-commerce

Interestingly, retailers who raised the issue that consumers are not familiar with e-commerce have no experience selling online. Ali (1) stated that "*the familiarity, sellers and customers are not familiar with online sales and purchases*". This retailer has no experience buying online, does not know friends who have experience buying online, and his business is not involved with e-commerce at all. So, Ali's thoughts about consumers are based on his surrounding environment, and are not based on practice.

On the other hand, , Muhammed (3) seems more optimistic, saying "*once the businessmen or the ones who run the business in KSA get the confident that the people are more keen to go online and visit the places they will be more than happy to jump to the bandwagon and put their business in e-commerce channel*". Muhammed's company, which sells electronics and home appliances, is planning to sell online in the near future. They have almost everything ready to go ahead. So it seems that his impression about his consumers' familiarity with e-commerce is positive.

The situation here is like a chicken and egg dilemma; which should start first? Should companies start selling online and hope the consumers will follow or should companies wait until consumers show stronger movement towards online shopping. Osam (3) is of the opinion that it is best to wait until people become familiar with online shopping, stating that "*in general people have to become familiar with this system before we sell online"*. Similarly with Talal (2), "*when people accept to buy online we will apply e-commerce system*". By contrast, Thamer (4) and Salem (4) did not agree to wait for consumers to be mature in purchasing online. They knew that they will not gain more profits or many online consumers similar to online retailers in western countries; "*we already know that this region doesn't give you more sales on the Internet because that habit of the people here in this region used to go outside and buy*", Salem (4) stated. However, they "*continue providing this option* [selling online] *to encourage people to use it*", Thamer (4) said.

### 5.1.3. Consumers' understanding of e-commerce benefits

Two participants raised the issue of consumers' understanding of e-commerce benefits and they are totally different in their opinions. Talal (2) saw it negatively, that consumers in Saudi Arabia do not understand the benefit of e-commerce. "*If there are people who understand the benefit of e-commerce I agree with you it help to gain more profits*". . For this reason, it seems Talal believes that there is no point in using e-commerce becaue people do not understand its benefits and would not buy from business online. Mohammed (3) totally disagrees with that judgment about consumers. He said: "*People here understand the benefit of e-commerce 100%... I've seen people there are the indigenous original Saudi people superbly well educated and they know exactly the benefit of e-commerce and everything*". .

Again, personal feeling or experience seems to play a role here in these two different opinions. The first participant, Talal (2), has no experience purchasing online and his company has no engagement in e-commerce activities. Therefore, his judgment is not based on practice. By contrast, Mohammed (3) has lived in western countries and has engaged in buying online many times. In addition, Mohammed's company has an interactive website gaining consumers' feedback and comments regarding their products. His company has done almost everything to move forward for the next step and to sell online.

### 5.1.4. Consumers' trust

The issue of consumers' trust has received high attention from the retailers who were interviewed. Interviewees often raised the issue that they believe that consumers are sensitive and uncomfortable about online shopping because it involves paying online and there is an absence of inspecting the product in one's hands. While participants whose companies are classified in level 1 and 2 in e-commerce maturity raised their concerns that there is no trust in online shopping; participants whose companies' e-commerce maturity levels are 3 and 4 seem to have moved beyond raising the concerns and instead suggest solutions to overcome this issue and build trust with online shoppers as it is explained in the following passages.

Tameem (1) clearly stated that he does not trust buying online from Arabian businesses. Later when he was asked why his business was not taking steps to sell online, he replied: "*if people knew that the one who run an e-commerce website is Arabian or the company owned by Arabian, they will be unmotivated to buy*"! Similarly in Nasir's (2) case, he has negative feelings regarding dealing with money over the Internet, even with using Internet banking he is sensitive. The interesting thing here is to know that Nasir's company accepts orders using phone calls or Fax and receives payment from consumers using bank transfer, and ships ordered products with well known international shipment companies such as DHL and FedEx. Nasir (2) commented regarding selling online that "*there are people when they pay online remain in doubt and not sure of receiving their purchased orders*". Interestingly, he thinks his consumers remain in doubt when they purchase online and but that do not have doubts when they order using the phone or when they transfer money using a bank account.

Moving beyond raising the concerns of consumers trust, other participating retailers suggested solutions to overcome this issue and build trust with online shoppers. Most of the participating retailers who suggested these solutions are involved in e-commerce activities.

Most of the suggestions to overcome consumers trust issues were concerned with providing alternative ways of paying online to help customers to choose the one they feel safe with. It seems that some retailers believe there consumers in Saudi Arabia have an exaggerated of using their credit cards online as Osam (3) commented, "*people are reluctant to use credit cards, I think that this fear is exaggerated*". The national payment system SADAD was mentioned as a good solution here to gain consumers' trust. It is used for online purchases in Moneer's (4) company and he commented that "*SADAD is great idea and more secure than credit cards and encourage people to buy online*". The PayPal system for paying online was also mentioned as a method that helps gain trust with online consumers. Naif (2) commented regarding the process of PayPal saying: "*with this procedure customers feel happy to deal with this intermediary e-payment option which protects their rights and also build the trust with companies that deal with PayPal*". The online payment systems are further discussed in the environment related factors section.

Other solutions that may work in Saudi Arabia to build trust with online consumers were the involvement of the government and existence of e-commerce protection system. Thamer (4) suggested that "*citizens will have more trust if this subject sponsored by the government because we, in Saudi Arabia, have great confidence in anything that comes through the government*". Salem's (4) company is "*working to have certificates from* *trusted organizations to build the trust with our customers*". Salem (4) suggested that "*it should be there is a certification body from the government... this is good to build the customers trust with the certified companies as the government trust them*". Ahmed (2) also suggested that the government build clear legislation system for e-commerce. Ahmed (2) stated that "*customers have to be ensured there are rules and legislations protect their right. This is very important to build the trust with customers... I advice the main supporter in this field is to find out a way to build the honesty/trust between customers and sellers*". These two issues of government involvement and are further discussed in the environment related factors section.

### 5.1.5. Consumers' reluctance to use credit cards

Interestingly, some participants who raised the issue of their belief that consumers are reluctant to use credit cards for online payments do sell online. Their statements were based on experience. For example for Salem's (4) company they notice that their consumers feel difficult to put their credit cards information on their e-commerce website, so they may contact them for arranging other options for payment as he stated "*we notice that people feel difficult to put their credit card information... we have another way here especially for local customers, we have call centre and the customer can call us (free calls) and make the order by phone and go to pick it up from the branch*". Salem's company is working to provide more options to pay for online purchased ordered to overcome the issue of consumers' reluctance of using credit cards as he considered the importance of "*providing more online payments because not everybody has credit card, it should be there is another ways of online payments*".

Similarly with Moneer's (4) and Thamer's (4), they partly refer the low acceptance of online shopping among Saudis to the consumers' reluctance to use credit cards. Thamer (4) stated that consumers "*are afraid to put their credit cards number on the internet*". And he suggested "*for local banks to offer other easy options for online payment to encourage people to purchase online*". Providing other options for consumers to pay online seems very important for these retailers. For example, Thamer's company provide the option of Cash on delivery. Cash on delivery seems acceptable for Thamer's company because the company has its own delivery system and purchased products delivered by the company's staff using its own trucks. Other option for online payment that has become popular and trusted for Saudis to pay for e-government services is the national payment system SADAD. Recently, SADAD has become available for commercial businesses and private companies to use allowing their consumers to pay online. Moneer's (4) company provide SADAD for their consumers to pay for their online purchases, "*we provide two options, credit cards and SADAD system*", Monner siad. Moneer (4) commented on using SADAD in his company that "*SADAD system is more secure system and great. With this system there is no need to entre personal payment details on our website*". These two empirical examples suggest that for companies in Saudi Arabia, using more options for online payment is critical to overcome the issue of consumers' reluctance using credit cards.

### 5.1.6. Consumers' level of demand of buying online

Participating businesses considered consumers purchasing power in Saudi Arabia as weak and not supporting them to run their businesses online. The issue of low level of demand raised by participants in all different four groups of e-commerce maturity levels. Saleh (1) commented that "*the current situation here is not encouraging us to go ahead with this idea*" and Ahmed (2) stated that they "*have tried to market on the Internet but there is no demand*". While it is not based on practice to hear from businesses which are not involved in e-commerce activities, i.e. Saeed (1), Saleh (1) and Ahmed (2), that the level of online purchase demand is low; this was confirmed by businesses which are involved in e-commerce activities. Thamer (4) and Salem (4) confirmed that the online sales cannot be compared to the normal way of selling. Very few consumers choose to buy from their businesses online as they stated. Thamer (4) commented "*there is no good demand to buy online. For example, in western region of Saudi Arabia we receive 3-5 orders buying online every week and these are nothing compared to normal way of business*". Similarly with Salem (4) as he stated that "*we are not in Europe or America where people easily buy from the Internet, culturally we are different than them that's why e-commerce is not getting much more businesses*".

It is again about the chicken and egg dilemma, which should start first companies sell online or consumers buy online. While many businesses seem they adopt wait to see strategy, there are other businesses despite they mentioned the issue of low level of online purchases in Saudi Arabia, they choose to be among those explorers, motivators and developers of e-commerce in the region. For example, while Thamer (4) and Salem (4) acknowledge the low level of consumers online purchases demand as Thamer (4) stated "*there is no good demand to buy online*", they choose to go ahead and "*continue selling online to encourage people to buy online*", Thamer (4) said. It is not only that the main reason motivating them to sell online but rather other factors contributed to this. For example, these companies have well established e-readiness and sell online was considered to them as additional marketing channel. Furthermore, they want to have strong foothold in e-commerce in Saudi Arabia. They see that e-commerce has promising future in the region, "*I believe that Saudi market is bullish and a fertile ground for investments, but –in my opinion- even we start thinking about e-commerce we need time not less than 10 years to reach the maturity in this field*", Thamer (4) said. It seems correct that the level of online purchase demand in Saudi Arabia is low compared to western countries; however, Saudi Arabia has huge potential for e-commerce market **(**Gabr 2013). This issue is further discussed in the discussion chapter.

### 5.1.7. Consumers' knowledge to buy online

Interestingly, participants who have perceptions or raised the issue of consumers' knowledge to buy online are the ones whose companies do not sell online, and interact online with consumers. Their given statements are like prejudgments that consumers do not have the knowledge to buy online! Similar to what have been said earlier in consumers' familiarity and benefits understanding of e-commerce sections, again personal feeling or experience plays a significant role here in these perceptions. It is not logical to judge the consumers' knowledge to buy online while you have not interacted with them online! Taking Ali's (1) statement as an example, "*the majority of people in our country do not know how to buy and sell on the Internet*". Including the word 'sell' in his statement is a key to what he meant. He included consumers being not knowledgeable to buy online as same as he/his company was not knowledgeable to sell online.

Similarly with Nasir (2) as he commented that "*there is ignorance in the community. Most of them don't know the meaning of e-commerce*". It seems and can be understood from these generalised statements they based on their own/ or their company knowledge and experience. For example, Nasir (2) stated that "*honestly there is ignorance inside our company in terms of e-commerce*". A question may be asked here, how for a person judge others knowledge about something he/she is not knowledgeable about it? In addition, none of these participants, (i.e. Ali (1), Saleh (1), Nasir (2), and Talal (2)), has experience buying online, they never purchased online. They also answered that they have no friends or relatives who do have experience purchasing online.

While it is not logical to accept this judgments, it still worthy to consider this perception that they have about consumers as significant factor influencing their decision to use e-commerce. Education and building awareness programs may be useful in this regard. Sellers may need to be aware of consumers behaviours buying online based on research reports and actual financial spending. This may add value changing this negative perception about consumers and as result helps to positively influence their decision toward using e-commerce.

### 5.1.8. Consumers' willingness to pay for delivery fees

The same as the previous issue, participants who have perceptions or raised the issue of consumers' willingness to pay for delivery fees are the ones whose company do not sell online, and interact online with consumers. Their given statements in this regards are based on as they perceived not on empirical evidence. Again, it seems they have this perception based on their own experiences. None of these participants, Saleh (1) and Hassan (1) except Ahmed (2), has experience buying online. Ahmed (2) only has bought one time an airline ticket.

> *I can see also one of the inhibitors is delivery fees*, Ahmed (2)

*We have concern regarding delivering the purchased products and its fees,* Hassan (1)

*If you request from the customer to pay extra 10 SAR for such a fees like this and if the price in total comes more than the price in the local shops, he/she may buy it from them without buy it from you online,* Saleh (1)

By contrast, businesses that involved in selling online had no concern regarding the delivery fees. This is because they have known and gained experiences on how to deal with delivery companies and reduce the cost of shipments. For example, Salem's (4) company has had contracts with FedEx well known and trusted international shipment company, "*we have done is having very strong contract with express and trusted shipment company, FedEx*", Salem (4) said. They ship for them for very competitive prices. Salem also mentioned that if a consumer or his employee has a contract with any shipment company, a consumer is able to provide them with his/her shipment company reference number when completing the order online, so the order is delivered with no cost. Salem (4) commented "w*e have also another shipment type, the customers who have accounts with the shipment companies they can provide us their account numbers with anyone of these companies and we will arrange with these companies to come and pick up your products and they will charge you as the same agreement you have with them*". These are examples demonstrate how these companies came up with solutions to reduce the cost.

However, reducing the cost and looking for competitive prices are not always what consumers looking for. This is what retailers in the lower stages of e-commerce maturity not aware of. In many cases the convenience is about having the product delivered to a customer's home, at no extra cost, so that she/he does not have to transport it and install it her/himself. It is about not having to find a car park. It is about not having to have a car. It is about placing an order if the goods are not in stock so that they are delivered as soon as they are available. It is about being able to compare the same goods (or comparable goods) from a number of online suppliers whilst on the computer rather than walking from shop to shop. By offering comparably lower prices, consumers would have the choice of either spending less money, or purchasing a higher quality item online, for a price comparable to a lesser item in a bricks and mortar establishment.

## 5.2. Environment factors/issues/concerns

Environmental factors refer to the environment of technological readiness and other involved factors to serve the running of e-commerce. The literature discusses various environmental issues that influence rate of e-commerce growth in a particular environment. The raised factors, issues or concerns by the participants regarding the e-commerce environment in Saudi Arabia include Internet services and users, e-commerce protection system, required action by government, online payment system, national payment system, and the Saudi mailing and addressing systems. Here is the discussion of all these issues with sentences being quoted from the retailers' comments to support the discussion.

### 5.2.1. Internet users in Saudi Arabia

There is no doubt that the low percentage of internet users in an environment is an inhibitor to the growth of e-commerce. On the contrary, high percentage of internet users plays significant role in the development of online services and attracts more businesses to be online. The internet users in Saudi Arabia are seen high by some participating businesses and may play a motivational role for e-commerce success. "*Today almost every home has access to the Internet, many people browse it and of course this brings customers to your shop*", Hassan (1) said.

As discussed earlier in the background chapter, the internet users in Saudi Arabia is increasing rapidly. The Saudi market has becomes attractive for more online retailers in the Arab world **(**Gabr 2013). To what extent this would play a motivational role for Saudi businesses to sell online? Despite the high numbers of Internet users in Saudi Arabia, some participants recognised the activities of e-commerce are very low but acknowledge that will be changed in the near future and e-commerce in Saudi Arabia has promising future. Thamer (4) believed that "*the Saudi market is bullish and a fertile* *ground for investments, but –in my opinion- even if we start thinking about e-commerce we need time, not less than 10 years, to reach maturity in this field*".

### 5.2.2. Wi-fi & broadband services availability

The e-commerce literature places great emphasis on the wide availability of broadband Internet as the backbone of ICT infrastructure strength. Despite the recent development in broadband services compared to previous years in Saudi Arabia, still more efforts are needed as suggested by some retailers. The CEO of a large-size company that runs a business selling complete home and electronic solutions, Mohammed (3), see that the current situation of Internet services and access in Saudi Arabia needs to be improved in order to create more attractive environment for e-commerce. Mohammed (3) commented "*I think someone needs to kick the whole things of having the access of Wi-Fi, of having the access of broadband, having the access in every house hold. Once this done, obviously you will see the results of that very soon*".

### 5.2.3. Protection system

Regulations and rules, in commercial transactions, are very important in terms of systemizing the work, and protecting the rights for all involved parties. Some retailers, in the current study, urge for e-commerce consumers protection system. For example, Ahmed (2) saw that "*there is no clear system to protect sellers and customers rights... the system that protects customer right when he/she pays, he/she will receive the purchased products/services must be ensured*". Interesting is to hear that businesses urge for consumers protection system. It is understandable why these retailers urge for having legislation system for e-commerce. With the lack of legislative system for e-commerce, trust is difficult to be built with consumers. Businesses which raised this issue they stated that setting up legislative system is required action by government. As discussed in earlier section, consumers trust this type of procedures of having protection system and enhanced by the government helps to build trust with consumers. This is confirmed by Thamer (4) and Salem (4) whose companies already involved in selling online. They have not clearly indicated that a clear e-commerce law should be exist but they mention the involvement of the government as a key to develop trust with consumers. Their views were discussed in the next section.

### 5.2.4. Required action by government

Government support for e-commerce growth takes various forms from country to country such as technical support, training, funding provided ensuring coherent policy for IT and e-commerce for consumer protection; secure, transparent, predictable, and enabled environment; support, coordination, collaboration, and cooperation.

As discussed in the earlier section, businesses needs government to be involved in e-commerce as a third party which helps businesses to build trust with consumers. The businesses which urge for government involvement were the ones which they are sell online. Salem (4) and Thamer (4) saw that government involvement helps them to build trust with their consumers. Salem (4) suggested that "*it should be there is a certification body from the government itself to say that this company is a certified company by local government and you can buy from them. This is good to build the customers trust with the certified companies as the government trust them*". Similarly with Thamer (4), it is all about contributing to building trust with consumers, "*citizens will have more trust if this subject sponsored by the government because we, in Saudi Arabia, have great confidence in anything that comes through the government*".

### 5.2.5. Online payment systems

Providing e-commerce services with the necessary financially secure transactions requires trustworthy online payment mechanisms. Participating businesses in the current study which raised the issue of online payment systems were all agreed that providing more options making them available for consumers to choose from the option that they feel safe with is significant incentive for consumers. On the businesses side, the wide available online payment options for businesses to implement for their e-commerce systems, the higher they accept to sell online. It is again the matter of providing consumers with means that contributes to build trust with businesses. For that reason, Thamer (4) emphasis on providing local payment systems that helps consumers to get used to use. Credit cards are fine and widely use world-wide, however in Saudi Arabia credit cards seem yet not popular. Thamer (4) suggested that "*for local banks to offer other easy options for online payment to encourage people to purchase online*", similarly with Salem (4) and Osam (3). Osam (3) suggested that "*banks should provide easy options to have two credit cards, one with large amount of money and another one with small amount, called debit cards, to be used in online payment*". This suggestion is good; however, local banks in Saudi Arabia charges fees for having debit cards or intrant purchases cards and that is why there is a need for improving this situation.

While some businesses urge for having local solutions for online payments which helps to build trust with consumers and encourage them to buy online, there are some other businesses that use international payment system such as PayPal. However, PayPal still not popular payment method to be used in Saudi Arabia. One popular method that is widely used in Saudi Arabia for paying bills online is SADAD. This system seems to be very good solution for providing trustworthy online payment method that is accepted by consumers. SADAD is discussed in the next section.

### 5.2.6. SADAD, national online payment system

Following from the previous point that businesses are looking for online payment systems that help to build trust with their online consumers. The national payment system SADAD seems great for providing mean to pay online and build trust with online consumers. SADAD is a national electronic bill presentment and payment service provider for the Kingdom of Saudi Arabia, and was launched on October 3rd, 2004. The core mandate for SADAD is to facilitate and streamline bill payment transactions for consumers through all channels of the Kingdom's Banks. It relies on existing banking channels (such as Internet banking, telephone banking, ATM transactions and even counter transactions) to allow bill payers to view and pay their bills via their banks. It has been found that consumers are comfortable and tend to have more trust with using SADAD in e-government services.

SADAD may represent good solution to overcome the issue of consumers trust paying online. In the current study, a personal experience of one participant purchasing online using SADAD has leaded him to influence his company providing SADAD for their consumers as an option for online payment. Moneer's (4) company provide two options for its consumers to pay online: credit cards and SADAD system. Moneer (4) commented on using SADAD in his organization to complete online purchase orders saying "*SADAD is great idea and more secure than credit cards and encourage people to buy online*". Moneer (4)

At the time of conducting this study, SADAD was limited for use to 100 billers, mostly large and government organizations. For that reason, SADAD was not considered as online payment method for e-commerce. Salem (4) commented "*SADAD before was very expensive solution, it is good for large companies but it's not for middle size company like us*". However, recently this limitation have been addressed and the government working on billers base expansion, which will increase its biller options from 100 to 20,000 (CITC 2011).

### 5.2.7. Issues relate to the Saudi mailing and addressing system

There are some retailers in Saudi Arabia that have their own delivery system. There are some others who have no problem with the delivery of phone orders. They organize with local shipment companies to deliver the goods, however, this is not considered as a professional delivery service because these normal delivery companies do not get a clear home address, so they deliver to their offices only and then contact the customer to come pick up his/her order. When a customer wants the products to be delivered to their home, they have to arrange this with the delivery company and pay an extra fee. They have to explain to the driver where their home is located.

With the recent development in the mailing and addressing systems in Saudi Arabia that each building has address and mailbox, still some companies work using the old system relying on phone numbers. For example, Salem's (4) company "*depend on the mobile phone number; if the mobile phone number is not correct then it's difficult to deliver the product*". It seems part of the problem with the international delivery companies that they do not follow the update and new addressing system in Saudi Arabia which carried by the Saudi post. The Saudi post delivers to homes. In addition, Some businesses have lack of knowledge of the efforts made by the Saudi Post since 2005. For example, Salem

(4) thought that "*there are certain locations (e.g. companies' buildings) you can know the address clearly, you know which street and block, building, floor, flat etc; but with community houses this is not very clear*". Similarly with Osam (3) where he commented on this issue saying "*the problem is mail address; still some people do not have mail address*".

There have been recent developments in the mailing and addressing systems carried out by the Saudi Post. Huge efforts have been done naming streets and numbering all residential subdivision inside cities in Saudi Arabia. One possible factor, pertaining to the low use of building addresses, might be the lack of awareness either by retailers or consumers of these recent changes in the Saudi mailing and addressing systems.

## 5.3. Organization factors/issues/concerns

The participating businesses has raised several issues relate to their organizations that affect their decision to adopt e-commerce. These issues include e-commerce level of difficulty; issues related to the nature of products; management attitude toward e-commerce; e-commerce familiarity and knowledge; business priority; security and trust concerns; setup and maintenance costs. Below is the discussion of all these issues with sentences being quoted from the retailers' comments to support the discussion.

### 5.3.1. E-commerce difficulty

The difficult tasks of e-commerce are different based on each company e-commerce maturity level. For example, participants whose companies in level 1 and 2 of e-commerce maturity model raised issues regarding creating e-commerce website, train employees, online payment and products delivery. For example, Saeed (1) commented that "*e-commerce is difficult; you have to design a commercial website, train your employees, deliver goods, etc*". The raised concerns regarding creating e-commerce website, online payment and products delivery are not seen difficult by the participants whose companies already sell online. Of course this is because the strong e-readiness of these companies. However, part of their success is their knowledge and experience to find solutions. For example, creating e-commerce website is not seen an issue by Salem (4). He commented that "*building the website is very easy. You can buy cheap stuff/software from the Internet, there are e-commerce packages available cost around $100 to buy a package and publish your e-commerce website*". Similarly with Thamer (4), "*there are complete e-commerce solutions that you can easily use*". It is the same thing that was discussed in paying delivery fees issue that somebody may have negative perception bout something and consider it difficult, expensive or not useful while that in reality is not right. The issue here is about knowledge, lack of knowledge cause a person to pre-judge.

Other issues regarding the difficulty of e-commerce relate to the market when sell online. These issues are all made around the difficulty to compete online while products are available in the local market. Mostly, businesses in level 1 and 2 raised the difficulty to compete online while products are available in the local market. They consider e-commerce is not useful for products that can be found in the local market, e-commerce is useful for products that is unique and does not exist in the local market. Here are examples of what these businesses thought.

> *I agree to go ahead with this idea if there is products are not available in the local market where we can have competitive advantage.* Ali (1)
> *Computer shops are widespread in Jeddah, so to go ahead with this idea is not benefiting us,* Saleh (1)
> *Having no similar products in the local market help to achieve high profits, but if the products are available in the market, I believe that profits will be very weak.* Waleed (1)

The raised issue regarding the difficulty to compete online is an interesting and important point for discussion. The usual way around this is to think of different online business models, with efficiencies to reduce the cost of doing business. It is much like KSA moving from petrol to an information economy as retailers must reduce the reliance on transport (cost of petrol and time) and storage of stock (rental). For example, Amazon can sell books cheaper than many local bookstores because it does not have to rent shop space or run a delivery service itself. Order details are sent directly to publishing houses for direct delivery by a specialist courier or postal service. There are also models that provide certain items only by Internet sales so customers have to use that channel if they want that product. Similarly, access to specialist items direct from manufacturers or wholesalers is a price reduction model. Part of the problem may be that Saudi retailers are not aware of the different online business models, and that they can take advantage of different forms of disintermediation – removing service costs from the retail price. This lack of education is part of the problem surrounding e-commerce – not just for retailers, but for the customers as well. Saudis are not adept at creating a competitive advantage online yet. It is a new way of thinking about supply chains.

### 5.3.2. Issues related to the nature of products

According to the statements made by interviewees, it appears that the decision of whether or not to adopt an online retailing system depends on the type of business or product. For example, companies that sell food and fragile products are reluctant to sell online due to their concerns that the goods may not be received in the same condition in which they were shipped.

Some of these issues relate to the ability of businesses to deliver or know delivery company that deliver in satisfactory situation. For example, Saeed's (1) business involves with selling grocery. He commented on delivery saying that "*delivering goods need special care with some of our products because they require being stored in a specific temperature to delivered safely and healthy*". That is the case that Saeed's (1) company does not have the ability to have refrigerated trucks to deliver their product and do not know delivery company to do so. It is not like other products that do not need special care. However, Salem's (4) company sell chocolates and biscuits which need to be stored in cool area. Although their products fragile and need special cares, they manage well to deliver their products in satisfactory conditions. Despite Salem (4) raised concern because customers are afraid that the fragile product may not be delivered in one piece, his company is managing well with FedEx, delivery company, and have had no complains form their consumers. Salem (4) commented "*some customers are interested to buy from you but when they see fragile products they will go away, it is better for them to buy from local market instead of buying online where they cannot ensured the product will shipped safely. To reduce this thing what we have done is* *having very strong contract with express and trusted shipment company, FedEx*". Therefore, it is again about knowledge and gaining experience to know more options and alternative.

Furthermore, delivery of special products is not the only area of consumer concern as indicated by retailers; there are other issues involved with the type of product e.g. a beauty company selling: perfumes, makeup, cosmetics and body lotions, shampoos and skin care. Ahmed's (2) business involves selling beauty products. He commented that it is difficult for them to sell online, consumers will not accept it because "*our products related to smell, shape which need to be physically seen*". While it is agreed that some consumers need smell, touch and inspect products in their hands, many successful businesses exist in selling beauty products online, e.g. the Body Shop. It is about consumers get used to know products. The products that have high demand purchased from normal shop, they are the ones have more chance to be successful online. This is because consumers they already know that product, know smell, shape etc and they will not have any problem buying it online once the trust is established. Again trust is mentioned here. It is more likely to be trust issue not the type of products issue.

### 5.3.3. Management Attitude toward e-commerce

The management attitude towards the changes is a key for organizations to adopt e-commerce. Rogers (2003) identified the attitude of an organization's management towards change as critical factor in terms of an organizational innovativeness.

The current study highlights positive and negative attitudes of management toward e-commerce and how they influence the adoption decision. Mohammed (3) has positive attitude toward e-commerce and has well experience using e-commerce as buyer. He is "*personally like to do move toward e-commerce system where customers will be able to browse, search, check the prices for competitor, and check the quality of products in terms of features, advantages and benefits and then make the right choice. It makes comfort and less time to move around*". This positive attitude led to that Mohammed's company is almost ready to start selling online. Their website currently display products, features and prices, and gains feedback and interacts with consumers on their website.

Another positive attitude for another case contributes to the continuance of selling online despite the low level of online purchase demand is with Thamer's (4) case. Despite his evaluation that consumers purchases online is very low, his company continue offer online channel for its consumers because they that the Saudi e-commerce market is bullish and a fertile ground for investments. Thamer (4) commented "*I believe that Saudi market is bullish and a fertile ground for investments... we will continue providing this option* [selling online] *to encourage people to use it*".

In contrast, negative attitude towards e-commerce influence negatively to the adoption decision. Despite Nasir's (2) company is accept orders by phones, payments by bank transfer and deliver products with DHL and FedEx, they do not accept this process to be complete on their website due to this negative attitude. The management attitude in this company have negative attitude toward using Internet to sell and buy and pay online. They think that "*there are thefts of credit cards numbers, and there are hackers penetrate your privacy. So this is a problem for the money dealing on the Internet*", Nasir (2) said. In this case it is clear that the negative attitude makes no room for thinking differently or looks for alternatives. Since Nasir's company has their website demonstrating the products descriptions and prices, accept orders by phones, payments by bank transfer and deliver products with DHL and FedEx, it seems not an issue to accept orders on website and accept the payment as they wished, i.e. bank transfer. However, this negative attitude makes no room for thinking differently.

### 5.3.4. Business familiarity and knowledge with e-commerce

It's normal to find resistance to the acceptance of a new technology or idea, as people need time to become familiar with. Regers (2003) showed an adoption timeline for a new technology/idea where there were few adopters in the beginning, and with time the percentage of people adopting its use increased. Some participants in this study don't want to change because they find the normal way of selling is much better and more profitable for them and they are not familiar with e-commerce. For example, Ali (1) commented "*we are familiar with the way of normal selling where a customer has to come to our shop and see the products*". Similarly with Hassan (1) where he clearly indicated that "*e-commerce is not similar to normal way of selling where a customer comes to the shop, see a product, and make sure he/she selects the right product*".

All the cases mentioned that they are not familiar with e-commerce and have no sufficient knowledge to use e-commerce were in the e-commerce maturity levels 1 and 2. Within company, insufficient knowledge to deal with e-commerce was significant inhibitor for these companies to use e-commerce. Hassan (1) mentioned that "*our staffs don't know about marketing and selling on the Internet where they required training programs*". Similarly with Nasir (2) when he was asked what prevent his company not to use e-commerce, he replied "*honestly there is Ignorance inside our company in terms of e-commerce, and that is why our website only displays products and not receives orders*". To go ahead with e-commerce, they are required to train their employees and employ other skilled people. In some cases, this is seen costly, not useful and not fitted with their business priorities.

### 5.3.5. Business priority

Business priority can be defined as "a statement of the level or urgency of important business needs" (Suppiah, not dated). Participating businesses which mentioned the business priority, they stated that e-commerce is not one of their current business priorities so they do not consider its implementation at the top their businesses. For example, Fadi's (2) company already has its website that show their products and prices, has telephone orders and own delivery system. It seems almost everything is ready for them to do e-commerce, the only thing that they need to activate receiving orders on their website. However, e-commerce is not considered in their business priorities at the moment. "*Using the Internet to order our products is not our priority at the moment*", Fadi (2) said.

Another example affecting business priority is consumers' purchasing power. A manufacturing company producing Islamic watches and clocks has e-commerce website mainly targets international market and has no priority to sell locally. This is because "*the purchasing power comes from overseas and that's why this website is in English*", Naif (2) said. They do not have priority to sell online locally because they considered that their products are plentiful in the Saudi market and there is no point to sell online locally. It is interesting case for discussion! While this business already has a channel selling online to international market, what prevent them to target local market on the same channel that they already they have? It seems that there are other factors other than business priority and products being available in the local market. Unfortunately, the researcher missed to emphasize on this interesting point to figure out what is really behind that decision.

A third cause for business priority not to do e-commerce is because the perceived low income that can be gained from selling online. That is the nature of business; business men are looking to maximize their profits. The idea or strategy that is seen bring more profits has higher priority and the opposite has less priority. Ahmed (2) replied on a question why not using e-commerce saying "*we have a different marketing strategy which is having branches/shops in the main malls to be directly close to customers*". For Ahmed's (2) company e-commerce is seen not bring more profits whereas having physical shops and contacted with consumers in person do maximize their profits, it has higher priority.

### 5.3.6. Security and trust concerns

Concerns regarding security of using technologies have received considerable attention in the literature. These concerns become greater in an ineffective legal and regulatory environment (Dedrick, & Melville 2006) and are the most significant barrier for technology non-adopters (Wymer & Regan 2005). Similarly with this study, participating businesses which have concern regarding security and trust of e-commerce are non-adopters, levels 1 and 2 of e-commerce maturity model. From the statements made by interviewees, it appears that there is a lack of trust regarding online activities and that is linked to the lack of understanding within the business community. This lack of trust is primarily directed at online payment. For example, Ali (1) commented that "*we do not trust online payment; the money must paid cash*". Participants who raised this issue had no experience selling and purchasing online, thus their concern is perceived. Some of these concerns are exaggerated. For example, Nasir (2) clearly stated that "*we don't trust online payment and our customers don't trust it either, the* *ignorance of something breeds fear from it*". With the lack of knowledge the issues of trust and security increase. Therefore, Nasir's perception about selling and buying over the Internet was not good because "*there are thefts of Credit Cards numbers, and there are hackers who penetrate your privacy*". On the Internet there are secure and trust payment methods that enormous business around the world using them. Again, the issue here is about the lack of knowledge and experience.

### 5.3.7. Setup and maintenance cost concern

The cost of setting up and maintaining electronic business, including the website and IT skilled staff may make it difficult for businesses to adopt e-commerce. Analysis of the data from interviewees suggests that there is some perceived difficulty and cost associated with setting up an online retailing system.

For businesses in levels 1 and 2 it seems they have higher concern regarding the setup cost. For example, Saleh (1) stated that "*the website needs cost and even we go ahead to build an e-commerce website, it is not guaranteed to be successful*". That is right that they have major concern regarding the results of using e-commerce, because the cost involved is not only with regards to build a website. Using e-commerce for businesses in stage 1 and 2 of e-commerce maturity involves training programs, computer network, dedicated staff to take care of e-commerce activity, etc, whereas businesses in stages 3 and 4 of e-commerce maturity have no concerns because they well established in internal e-readiness.

Again, businesses here link the cost of spending and profits that will be return. For the businesses that have concerns regarding setup and maintenance they do not see that e-commerce will bring profits. For that reason they choose not to be involved financially "*in something that does not bring profit*", Waleed (1) said.

This is one of the aspects where taxation law is useful in many countries. If e-commerce is a priority, then that country could do two things to make adoption less of a burden: create tax concessions, cover the initial losses that may be incurred. It appears that perception of set-up costs may be based on ignorance. There are several online companies that aid in getting a business online by providing business templates that can be quickly configured to the business' needs, and get the business online. Setup costs are generally replaced with ongoing costs in this model by paying for an online business hosting service that incorporates all of the payment, advertising, etc. For somewhere like the KSA, where there is no taxation, concessions can still be made to offset business losses during adoption. This is where government support is vital to the promotion and adoption of this technology.

## 5.4. Summary and direction to the quantitative study

We have seen that the interview data has led to the development of 22 suggested factors influencing retailers in Saudi Arabia on whether or not to adopt e-commerce. Interestingly, the analysis came up with findings that have not been identified in the literature. It looks as though part of the problem comes from the retailers' perceptions (or in some cases, prejudgment) about consumers in Saudi Arabia. The more interesting thing is that there seem to be differences between retailers in companies in different maturity stages in terms of having different attitudes regarding the issues of using e-commerce. Having said that the interview data suggests that retailers in companies in different stages of e-commerce maturity have different attitudes, it is necessary to carry out a, quantitative investigation using many more retailers to examine whether there are significant differences between retailers in companies at different levels of maturity.

To sum up, this chapter presented an analysis and discussion of the interview data and ended up with a list of 22 factors influencing retailers' decisions on whether or not to adopt e-commerce. This list of the 22 factors will be used to develop the design of the quantitative study. The following chapter details the quantitative study methodology.

# CHAPTER 6: SURVEY DEVELOPMENT AND SAMPLING FOR THE QUANTITATIVE STUDY

In the previous Chapter, the interview data was discussed and resulted in the suggestion of 22 factors that may influence retailers' decisions on whether or not to adopt e-commerce. It was suggested that retailers in different stages of e-commerce maturity have different attitudes toward the factors that influence their decisions. However, the data used in the qualitative study is not enough to generalise the findings. More quantitative investigation catering for more retailing participants is needed to explore the differences. This Chapter describes the processes that were followed to collect and analyse the quantitative data. Collecting quantitative data involved the use of a survey. The survey design, validity, translation accuracy, population and sampling, and survey distribution processes are described in the present chapter.

## 6.1. Building the survey, and ensuring validity and translation accuracy

The survey was built on the qualitative study findings. The design of the survey processes went through several processes to ensure its validity. The survey was designed through five steps.

***Step one*** involved reviewing the 22 factors that emerged from the interview data using two independent people. The researcher arranged the 22 factors in a table, see Table 4.5 in Chapter 4, and against each factor placed all the participants' phrases that the researcher considered to be related to each factor. Then the table with the participant phrases was sent to two academic colleagues to provide comments based on the following:

- Do the statements the researcher uses for the factors adequately capture the sentences/phrases provided by the participants?
- Are the factors formulated in a neutral way, so as not to be either positive or negative?

- If you do not agree with any wording of a factor, please suggest better wording.
- If you see that some statements may be combined, deleted or added, please write your suggestions on the document.

The outcome of this process suggested some factors rewording and combining. In addition, two factors, “*protection system*” and “*required action by government*”, were combined to be one named “*current level of safeguards to protect consumers when online shopping*”. This step resulted in 21 factors, with revised wording as shown in Table 6.1 below.

Table 6.1: The 21 factors resulted from the first step of validation

| No. | Factor statement |
|---|---|
| 1 | People’s current attitudes and habits regarding shopping in person |
| 2 | Current level of consumers familiarity with e-commerce |
| 3 | Current level of consumers understanding of e-commerce benefits |
| 4 | Current level of consumer trust in online shopping |
| 5 | Current level of consumer credit card usage |
| 6 | Current level of consumers’ demand of buying online |
| 7 | Current level of consumer knowledge of buying online |
| 8 | Level of willingness of consumers to pay delivery fees |
| 9 | The number of people using the internet |
| 10 | Level of availability of Wi-fi and broadband services |
| 11 | Current level of safeguards to protect consumers when online shopping |
| 12 | Current availability of online payment options |
| 13 | The available of SADAD for online payment for internet shopping |
| 14 | The current mailing and addressing system in Saudi Arabia |
| 15 | Level of easiness of doing e-commerce |
| 16 | Nature of our products we sell |
| 17 | Level of e-commerce knowledge and familiarity in our company |
| 18 | Management attitude towards e-commerce |
| 19 | Our current business priority |
| 20 | Our company’s level of trust in online selling processes |
| 21 | Setup and maintenance cost |

***Step two*** involved checking the further checking on the clarity of the new naming of the 21 factors. Two Saudi assistant professors, who are knowledgeable about the research topic, were consulted to provide feedback in this step. A random number generator (random.org) was used to select 2 statements from the interview participants' phrases for each the 21 factors. An example is shown in Figure 6.1.

**1. People's current attitudes and habits regarding shopping in person**

- *"People culture to buy online discourages us to think of providing option to buy online"*
- *"Our culture buying and selling online is very weak. Whoever sells or buys should be professional"*

Your comment:

Figure 6.1: Example of a factor and two random statements from those considered by the researcher to be related to that statement

The two consultants were asked the following:

- Do the statements given by the researcher capture the phrases provided by the interview participants?
- Are the researcher statements formulated in neutral ways?
- If you do not like any statement, please suggest better wording to describe the interview participants' phrases.

This step resulted in the rewording of some factors and the combining of two factors. The two combined were "*current availability of online payment options*" and "*the available of SADAD for online payment for internet shopping*". They were combined to a new factor named "*current availability of online payment options*". Therefore, the outcome of this process was 20 factors, as shown in Table 6.2. The 20 factors were then arranged for the third step.

Table 6.2: The 20 factors resulted from the second step of validation

| No. | Factor statement |
|---|---|
| 1 | People's current attitudes and habits regarding shopping online |
| 2 | Current level of consumers familiarity with e-commerce |
| 3 | The current level of consumer recognition of e-commerce benefits |
| 4 | The current level of consumer trust in online shopping |
| 5 | The current level of consumer usage of credit cards for online shopping |
| 6 | The current level of consumer demand for buying online |
| 7 | The current level of consumer knowledge of buying online |
| 8 | Level of willingness of consumers to pay delivery fees |
| 9 | The percentage of people using the internet |
| 10 | Level of availability of wi-fi and broadband services |
| 11 | Current level of e-commerce legislation for online shopping |
| 12 | Current availability of online payment options |
| 13 | The current mailing and addressing system in Saudi Arabia |
| 14 | Level of ease for our company to do e-commerce |
| 15 | Nature of the products our company sells |
| 16 | Level of e-commerce knowledge and familiarity in our company |
| 17 | Management attitude towards e-commerce in our company |
| 18 | The current business priorities of our company |
| 19 | Our company's level of trust in online selling processes |
| 20 | Setup and maintenance costs of e-commerce for our company |

***Step three*** involved a sorting process performed to further ensure that the factors wordings made by the researcher are appropriately capturing the statements made by the participants. For each of the 20 factors, a random number generator (random.org) was used to select 2 statements from a list of statements that the author considered to be related to the factor. Each of the resulting 40 statements was printed on a small piece of paper and each of the 20 factors on a larger piece of paper. In a face to face meeting, two academic colleagues were asked to sort the 40 statements under the 20 factors. The instructions are given here.

- In front of you there are 20 large pieces of paper giving factors that may influence whether companies in the Kingdom of Saudi Arabia have e-commerce sites to sell products online.

- You will also see 40 smaller pieces of paper. These are 40 comments that Saudi retailers made about what influences them to either have or not have an e-commerce site to sell products online.
- Please read the factors and the statements carefully and for each of the 20 factors find 2 statements that you think best relate to that factor.

All the statements were categorized the same as the researcher categorisation except for two factors. Statements under these two factors were not categorised as the researcher expected them to be. It was apparent that there were mixed concepts between some factors and as the result the two factors were combined to be one. The factor named "*current level of consumer familiarity with e-commerce*" was combined with the factor named "*people's current attitudes and habits regarding online shopping*". Therefore, the final total number of the factors is 19, and these are shown in Table 6.3.

Table 6.3: The 19 factors resulted from the third step of validation

| No. | Factor statement |
|---|---|
| 1 | People's current attitudes and habits regarding shopping online |
| 2 | The current level of consumer recognition of e-commerce benefits |
| 3 | The current level of consumer trust in online shopping |
| 4 | The current level of consumer usage of credit cards for online shopping |
| 5 | The current level of consumer demand for buying online |
| 6 | The current level of consumer knowledge of buying online |
| 7 | Level of willingness of consumers to pay delivery fees |
| 8 | The percentage of people using the internet |
| 9 | Level of availability of wi-fi and broadband services |
| 10 | Current level of e-commerce legislation for online shopping |
| 11 | Current availability of online payment options |
| 12 | The current mailing and addressing system in Saudi Arabia |
| 13 | Level of ease for our company to do e-commerce |
| 14 | Nature of the products our company sells |
| 15 | Level of e-commerce knowledge and familiarity in our company |
| 16 | Management attitude towards e-commerce in our company |
| 17 | The current business priorities of our company |
| 18 | Our company's level of trust in online selling processes |
| 19 | Setup and maintenance costs of e-commerce for our company |

***Step four*** involved checking the full draft of the survey and its Arabic translation. To ensure the quality assurance of translating the survey's questions into Arabic, two Saudi assistance professors, who are knowledgeable about the research topic, were asked to review and ensure the quality of the questions' translations. Based on feedback and comments, the Arabic version was revised for clarity.

***Step five*** involved testing the survey with two representatives of the research sample. Two retailing managers were given the Arabic version of the survey and encouraged to answer all its questions. They also were asked to highlight and provide comments if there were any ambiguous questions, sentences, or phrases. Based on their comments and feedback, the Arabic version of the survey was ready for distribution.

## 6.2. The survey design

The main section of the final survey can be seen in Figure 6.2. It can be seen that the 19 factors are listed, together with a seven-point Likert scale ranging from -3 to +3: (-3) major barrier, (-2) barrier, (-1) minor barrier, (0) neutral, (+1) weak incentive, (+2) incentive, and (+3) strong incentive.

| Factor | Level of Impact<br>*Barrier* ------------------------------------- *Incentive*<br>-3 | -2 | -1 | 0 | +1 | +2 | +3 |
|---|---|---|---|---|---|---|---|
| People's current attitudes and habits regarding shopping online | ○ | ○ | ○ | ○ | ○ | ○ | ○ |
| The current level of consumer recognition of e-commerce benefits | ○ | ○ | ○ | ○ | ○ | ○ | ○ |
| The current level of consumer trust in online shopping | ○ | ○ | ○ | ○ | ○ | ○ | ○ |
| The current level of consumer usage of credit cards for online shopping | ○ | ○ | ○ | ○ | ○ | ○ | ○ |
| The current level of consumer demand for buying online | ○ | ○ | ○ | ○ | ○ | ○ | ○ |
| The current level of consumer knowledge of buying online | ○ | ○ | ○ | ○ | ○ | ○ | ○ |
| Level of willingness of consumers to pay delivery fees | ○ | ○ | ○ | ○ | ○ | ○ | ○ |
| The percentage of people using the internet | ○ | ○ | ○ | ○ | ○ | ○ | ○ |
| Level of availability of wi-fi and broadband services | ○ | ○ | ○ | ○ | ○ | ○ | ○ |
| Current level of e-commerce legislation for online shopping | ○ | ○ | ○ | ○ | ○ | ○ | ○ |
| Current availability of online payment options | ○ | ○ | ○ | ○ | ○ | ○ | ○ |
| The current mailing and addressing system in Saudi Arabia | ○ | ○ | ○ | ○ | ○ | ○ | ○ |
| Level of ease for our company to do e-commerce | ○ | ○ | ○ | ○ | ○ | ○ | ○ |
| Nature of the products our company sells | ○ | ○ | ○ | ○ | ○ | ○ | ○ |
| Level of e-commerce knowledge and familiarity in our company | ○ | ○ | ○ | ○ | ○ | ○ | ○ |
| Management attitude towards e-commerce in our company | ○ | ○ | ○ | ○ | ○ | ○ | ○ |
| The current business priorities of our company | ○ | ○ | ○ | ○ | ○ | ○ | ○ |
| Our company's level of trust in online selling processes | ○ | ○ | ○ | ○ | ○ | ○ | ○ |
| Setup and maintenance costs of e-commerce for our company | ○ | ○ | ○ | ○ | ○ | ○ | ○ |

Figure 6.2: The identified factors from the qualitative study to be tested quantitatively, as they appeared in the survey design

Of course, it is also necessary to collect background information on the participating businesses. In addition, to categorizing the participating businesses, questions to measure the e-commerce maturity in the participants' organizations was included. As discussed in the literature review chapter, specific criteria were developed based on the SOG-e model (Mckay, Marshall and Pranato 2000) and the literature to determine organizations' level of e-commerce maturity. This section of the survey consisted of 10 items, for each item participants were given four choices regarding their company having certain responsibilities; (have it it now, within 3 years, probably never, and no defined plan). Figure 6.3 shows the statements used to measure e-commerce maturity in the survey. One more measurement for e-commerce maturity involves the impact of e-commerce on an organization. A question was given to describe the impact of e-commerce on an organization with four choices: no impact, little impact, considerable impact, and critical.

| **E-commerce Capabilities** | **Have it now** | **Within 3 years** | **Probably never** | **No definite plan** |
|---|---|---|---|---|
| Publish information about our organization on our own website | | | | |
| Use of the internet for marketing and advertisements | | | | |
| Publish product descriptions including prices on our own website | | | | |
| Communicate with consumers through our own website | | | | |
| Communicate with consumers using social networks (e.g. twitter and facebook). | | | | |
| Get consumer feedback or ratings of products on our website | | | | |
| Allow consumers to place orders on our website using a shopping cart | | | | |
| Accept payment from consumers using online payment systems | | | | |
| Have dedicated IT staff and business-oriented staff to manage e-commerce activities | | | | |
| Place orders with our suppliers over the Internet | | | | |

Figure 6.3: The criteria to measure e-commerce maturity, as appear in the questionnaire survey design

The other questions in the survey collected information about the participant's position in their organization, the organization's age, its primary business category, the number of employees, and the annual revenue. A question was given to the retailers that sell online to identify the percentage of total sales coming from their websites. Other general questions involved gaining better understanding of the participants' perceptions about the size of the e-commerce market in KSA; recent developments in the Saudi mailing and addressing systems; the national online payment system SADAD; and the latest report released by the Saudi Communication and Information Technology Commission (CITC) regarding the Internet usages and e-commerce activities. Finally, participants of the research were offered a copy of the result of the research if they completed and returned the questionnaire. The survey cover page included all the necessary information: research purpose, expected benefit of the research, voluntary participation, ethical conduct, value to the participants, and contact details. The participants were told they could keep the cover page for their future references.

Since Arabic is the official communication language in Saudi Arabia and English is commonly used in businesses, the survey questions were designed in English with an Arabic translation version also available, so that the participant could opt for the language with which they were most familiar. The English version of the survey is in Appendix K, page 252 and the Arabic version is in Appendix L, page 258.

## 6.3. Research population and sampling

The target research population was retailers Saudi Arabia.  The sample was selected from the targeted identified research population. In quantitative research there are two main sampling designs: probability sampling and non probability sampling. Probability sampling includes simple random sampling, stratified random sampling, cluster sampling, and systematic sampling whereas non probability sampling includes convenience sampling, snowball sampling, quota sampling, consecutive sampling, and purposive sampling (Al-Amoor 2012).

This research adopted snowball sampling. Snowball sampling is where a participant is asked to suggest other participants that might be willing to participate in the research and so on with other participants. Therefore, the sample is growing like a rolling snowball (Heckathorn 1997). The adoption of the snowball sampling was for the following reasons. It is highly difficult and takes long time to reach decision makers in the Saudi organizations using the official communications. Based on the researcher's previous experience plan to officially communicate and wait for participants' responses failed. The researcher then contacted the relevant departments (in government and business sectors) to ask for help in order to contact retailers. They were not helpful in this regard and simply referred to their websites as a means to find retailers' contact details. For this reason, the researchers engaged close personal contacts to coordinate the selection of participants and then potential participants were selected, via the snowball approach, where some participants were initially approached, and they would then be asked to recommend others who might be willing to participate, and so on.

Judgment was also exercised to ensure that the questionnaire forms were delivered to a wide range of businesses in terms of their size, the type of products or services that they offered, and stage of e-commerce maturity. Regarding the size of the businesses in Saudi Arabia, they are identified according to number of employees and annual revenue. Micro size company is with 1-5 employees, small size company is with 6-50 employees and its capital up to SAR 20 million, medium size is with 51-200 employees and its capital SAR 20-50 million, and large size is with more than 200 employees and its capital more than SAR 50 million (Ministry of Commerce 2009).

## 6.4. Survey distribution and response rate

The survey was distributed in two forms: in person and online. 137 businesses were contacted in person by the researcher and his association in Jeddah (the main economic city in Saudi Arabia), and Riyadh (the capital city) during September and October of 2013. Most of the contacted businesses, 87, filled the survey on time whereas the rest either refused to participate or asked the researcher to leave the survey and said they would contact the researcher when it is filled. No phone calls were received. The researcher contacted the businesses again, and only 5 were filled and ready for pick up. A total of 92 completed forms were returned, giving a response rate of around 67.2%.

Electronic copies of the questionnaire forms were also kept on the website of the Griffith University research survey centre (English version / Arabic version). The researcher collected the email addresses of 416 retail companies that are members of the Jeddah and Riyadh Chambers of Commerce, although approximately 100 were returned because of addresses that were no longer invalid. Associates of the researcher provided an additional list of around fifty business email addresses with which they were familiar.

The researcher also contacted 60 online retailers that were registered in the Mothoq website. Mothoq is a volunteer website in Saudi Arabia that registers online businesses as trusted. Thus, in total there were 426 retailers who potentially could choose to participate in the online survey via the Griffith website. Sixty-one did so, providing a response rate of 14.3%.

Therefore, the total number of the complete surveys for both paper and online is 153.

## 6.5. Description of the obtained sample

The 153 surveys were prepared for analysis. First, each of the 153 surveys was assisted based on the answers on the criteria of the e-commerce maturity for the purpose to categorise it into one of the four e-commerce maturity stages. The responses to the criteria questions were based on 'yes' they have now, plan to have it within 3 years, probably never or no definite plan. Only the first choice was considered to be accounted criteria item. Therefore, the sample was categorized based on the following criteria.

<u>Stage 1: No website</u>

- They do not have website
- E-commerce has no impact on our business

<u>Stage 2: Static website</u>

- They have website that publish information about the company
- E-commerce has little impact on our business

<u>Stage 3: Interactive website</u>

- Communicate with consumers through website
- Products info & prices
- Consumer feedback or ratings of products on website
- Communicate with consumers using Twitter and Facebook
- E-commerce has considerable impact on our business

<u>Stage 4: E-commerce website</u>

- Consumers are able to place orders on website using shopping cart
- Consumers are able to pay using online payment systems

- Detected IT and business staff to manage e-commerce activities
- E-commerce is critical to our business

Therefore, the sample covered almost equal number for each e-commerce maturity stages; 24.2% for stage one, 26.8% for stage two, 22.9% for stage three, and 26.1% for stage four. In addition, it has good spread on the company sizes, old and annual revenue. Companies with 1-5 employees (micro-size) represent 13.7% of the sample, compared to 27.5% for small-size (6-50 employees), 34.6% for medium-size (51-200 employees), and 24.2% for large (200+ employees), companies. Nearly one-third of the companies (39.9%) are 10-20 years old, compared to 20.3% less than 10 years, 24.8% 21-30 years, and 14.4% more than 30 years, old. Regarding individuals filling out the survey on behalf of their companies, more than one-third (41.2%) has the top responsibility; managing directors, general manger, CEO or owner. This is followed by department mangers (28.1%) and one-quarter (24.8%) has the second responsibilities in their companies; deputy managing director or deputy general manager. The remaining 5.9% reported having other positions which has no responsibility to e-commerce decision.

Table 6.4 presents, for each category of participating businesses, the number of those businesses in each of e-commerce maturity stage. Rather surprisingly, a higher percentage of micro-size businesses (33.3%) followed by small businesses (28.6%) than medium-sized (20.8%) or large businesses (27.0%) have e-commerce websites (in stage 4 of e-commerce maturity). The situation is opposite for the e-commerce maturity stage 3; a higher percentage of large businesses (37.8%) than medium-sized (20.8%), small businesses (14.3%), or micro businesses (19.0%). Small-size businesses possessing static websites, stage 2, are the highest (31.0%) followed by large businesses (29.7%), medium-sized (26.4%), and micro businesses (14.3%). For businesses not have website, stage 1, micro and medium-sized businesses are higher (33.3% and 32.1%) than small businesses (26.2%) and large businesses (5.4%).

Table 6.4: Company attributes and rate of e-commerce maturity stages

| **Category of retailers** | **Responses** | **E-commerce Maturity stage** | | | |
|---|---|---|---|---|---|
| | | 1 | 2 | 3 | 4 |
| | No. | No. | No. | No. | No. |
| All participating retailers | 153 | 37 | 41 | 35 | 40 |
| ***Participant's job position*** | | | | | |
| Managing Director/General Manager/CEO/Owner | 63 | 17 | 15 | 13 | 18 |
| Deputy Managing Director /Deputy General manager | 38 | 8 | 13 | 5 | 12 |
| Department Manager | 43 | 8 | 11 | 17 | 7 |
| Other | 9 | 4 | 2 | 0 | 3 |
| ***Company size*** | | | | | |
| 1-5 employees (micro-size) | 21 | 7 | 3 | 4 | 7 |
| 6-50 employees (small-size) | 42 | 11 | 13 | 6 | 12 |
| 51-200 employees (medium-size) | 53 | 17 | 14 | 11 | 11 |
| 200+ employees (large-size) | 37 | 2 | 11 | 14 | 10 |
| ***Company old*** | | | | | |
| Less than 10 years | 31 | 5 | 7 | 5 | 14 |
| 10-20 years | 61 | 24 | 13 | 15 | 9 |
| 21-30 years | 38 | 6 | 15 | 6 | 11 |
| more than 30 years | 22 | 2 | 6 | 9 | 5 |
| No answer | 1 | - | - | - | - |
| ***Company annual revenue*** | | | | | |
| Up to SAR 20 million | 62 | 20 | 16 | 10 | 16 |
| SAR 20-50 million | 44 | 10 | 13 | 9 | 12 |
| More than SAR 50 million | 37 | 0 | 10 | 16 | 11 |
| No answer | 10 | - | - | - | - |
| ***Business category*** | | | | | |
| Apparel, accessories, and footwear | 15 | 3 | 3 | 3 | 6 |
| Books and school needs | 9 | 2 | 1 | 2 | 4 |
| Building materials | 5 | 2 | 3 | 0 | 0 |
| Cars, auto parts, and accessories | 7 | 0 | 3 | 4 | 0 |
| Computers and accessories | 9 | 4 | 1 | 3 | 1 |
| Electronics | 12 | 4 | 5 | 1 | 2 |
| Furniture | 10 | 3 | 3 | 1 | 3 |
| Groceries | 14 | 3 | 2 | 7 | 2 |
| Jewellery & watches | 7 | 2 | 4 | 0 | 1 |
| Home appliances | 7 | 3 | 3 | 1 | 0 |
| Medicine and medical equipment | 7 | 2 | 4 | 1 | 0 |
| Perfumes and beauty products | 14 | 1 | 3 | 3 | 7 |
| Printing equipment and/or services | 3 | 0 | 1 | 2 | 0 |
| Sporting goods | 5 | 3 | 0 | 0 | 2 |
| Telecommunications services | 6 | 0 | 0 | 3 | 3 |
| Toys and video games | 8 | 2 | 1 | 0 | 5 |
| Travel and tourism | 5 | 2 | 1 | 1 | 1 |
| others | 10 | 1 | 3 | 3 | 3 |

Now after explaining the quantitative study methodology and describing the collected sample, detailed discussion and analysis of the quantitative data can be given. In the next chapter, an analysis and discussion of the survey data is undertaken to identify the significant factors influencing retailers' decision on whether or not to adopt e-commerce, and to find out the differences between the four groups of e-commerce maturity stages in terms of rating these factors, and the relative ranking of these factors for each group. The "R-project" software for statistical computing and Statistical Package for the Social Sciences SPSS version 20 were used to analyse the quantitative data collected from the survey responses. Analysis of variance was performed to determine whether there is a significant interaction between maturity level and perceptions of the 19 factors identified in the qualitative study. Single sample t-tests were performed to identify the significant factors influencing e-commerce adoption for all the population and for each of the four groups. Furthermore, pairwise t-tests using the Bonferroni correction for multiple tests was performed to determine in what way groups differ for each of the significant factors. Finally, multiple linear regressions were performed to test whether the attributes of the organizations other than maturity level were significant.

# CHAPTER 7: ANALYSIS OF QUANTITATIVE DATA

The data from the survey discussed in Chapter 6 were first prepared for analysis by placing the data from each participant into a spreadsheet, with notations for stage of maturity of their company and their responses on the 7-point scale for each of the 19 factors. Although the 7-point scale in the survey went from -3 (barrier) to +3 (incentive), for ease of analysis the data were transformed to go from 1 (barrier) to 7(incentive). The first thing was to determine whether maturity level interacted with responses for the 19 factors. That is, are there different responses on the 7-point scale to the factors, depending on maturity stage? An analysis of variance gives an answer to this question. To further investigate where any differences in factor responses for different maturity stages exist, t-tests were performed.

## 7.1. Analysis of variance

Analysis of variance was performed to determine whether the responses to the factors significantly differ from each other; whether, overall, the responses from participants in companies at different stages of maturity differ from each other, and whether there is an interaction between factors and stage of maturity. Table 7.1 presents the statistical results of the analysis of variance which was performed.

Table 7.1: Analysis of variance of the factors and stage of maturity

| | Df | Sum | Mean | F value | *p* value |
|---|---|---|---|---|---|
| factor | 18 | 547 | 30.4 | 19.03 | .000*** |
| maturity | 1 | 1065 | 1065.5 | 667.11 | .000*** |
| factor: maturity | 18 | 551 | 30.6 | 19.17 | .000*** |
| Residuals | 2869 | 4582 | 1.6 | | |

Signif. codes: 0 ‘***’ 0.001 ‘**’ 0.01 ‘*’ 0.05 ‘.’ 0.1 ‘ ’ 1

The statistical results show that: there are significant differences between the responses to factors, between the respondents in companies at different stages of maturity, and more importantly, that there is a significant interaction between factors and stage of maturity. To find out where these differences are, further statistical tests were performed as indicated in the following sections.

## 7.2. Identifying the significant factors for each stage of maturity

First, the responses of the sample in each stage of e-commerce maturity for each of the factors are considered. For ease of expression, the 19 factors were given codes F1 to F19. Table 7.2 presents the names of the factors and their codes. Note here that the factors in Table 7.2 are divided into three categories/ themes: consumer related factors (F1-F7), environment related factors (F8-F12), and organization related factors (F13-F19). These categories were identified earlier in the qualitative analysis, Chapter 5.

Table 7.2: Names and codes of the factors grouped into three categories

| Factor | code |
|---|---|
| **consumer related** | |
| People's current attitudes and habits regarding shopping online | F1 |
| The current level of consumer recognition of e-commerce benefits | F2 |
| The current level of consumer trust in online shopping | F3 |
| The current level of consumer usage of credit cards for online shopping | F4 |
| The current level of consumer demand for buying online | F5 |
| The current level of consumer knowledge of buying online | F6 |
| Level of willingness of consumers to pay delivery fees | F7 |
| **environment related** | |
| The percentage of people using the internet | F8 |
| Level of availability of wi-fi and broadband services | F9 |
| Current level of e-commerce legislation for online shopping | F10 |
| Current availability of online payment options | F11 |
| The current mailing and addressing system in Saudi Arabia | F12 |
| **organization related** | |
| Level of ease for our company to do e-commerce | F13 |
| Nature of the products our company sells | F14 |
| Level of e-commerce knowledge and familiarity in our company | F15 |
| Management attitude towards e-commerce in our company | F16 |
| The current business priorities of our company | F17 |
| Our company’s level of trust in online selling processes | F18 |
| Setup and maintenance costs of e-commerce for our company | F19 |

Figure 7.1 shows the responses of the sample in each stage of e-commerce maturity to the factors. The number ‘0’ refers to the neutral value and the responses show the distance/difference from the natural value. From the figure below, it can be seen that retailers classified in stages 1 and 2 tended to respond negatively, which means they rate these factors as barriers for them; whereas retailers in stages 3 and 4 tended to be more positive, rating more factors as incentives. In particular, the highest differences can be noted when looking at organization related factors. This means that retailers in stages 3 and 4 see their organizations as internally ready to use e-commerce whereas retailers in stages 1 and 2 do not. Companies in stages 1 and 2 also look quite different from companies in stages 3 and 4 for the consumer related factors. Interestingly, it looks as though there is more agreement between respondents in companies at all stages of maturity in term of rating the environment related factors.

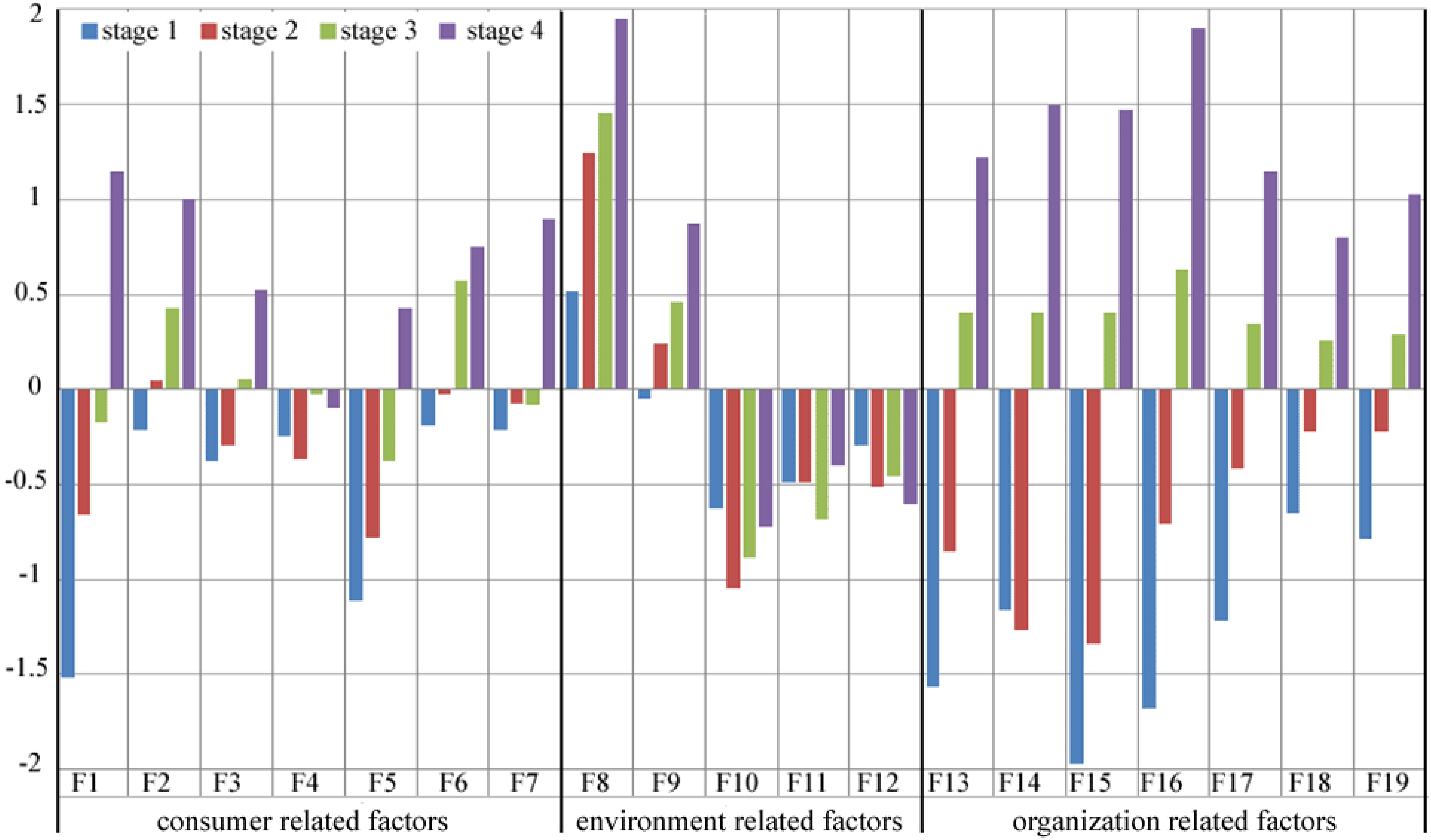

Figure 7.1: Value of the mean difference for all the 19 factors for each of the 4 groups

Figure 7.1 presents a very interesting picture. It is necessary, though, to determine the significant factors for each stage of e-commerce maturity. Thus, statistical analysis using single sample t-tests for each stage of e-commerce maturity were performed. Based on the use of the transformed 7-point scale from 1 - 7, there would be no effect for each factor if it has a mean of ‘4’ (that is, a mean of 0 in the survey data). Factors were considered significant if the t-test indicated the responses were significantly different from this neutral value, with a 2-tailed test and a minimum confidence level of 95%. Full statistical t-test results can be found in Appendix N, page 280.

Table 7.3 presents the statistical results of the t-tests for each stage of maturity for the 19 factors. All factors which have a *p* value of 0.05 or less are significant. For convenience, the significant values are in given bold. All of the 19 factors were found to be significant for at least one of the four stages of e-commerce maturity. A negative value indicates a significant barrier and a positive value (lack of a negative) indicates a significant facilitator. From the table below, it can be seen that 16 factors (1 incentive and 15 barriers) were statically significant for retailers in stage 1; 13 factors (1 incentive and 12 barriers) were statistically significant for retailers in stage 2; 6 factors (4 incentives and 2 barriers) were statistically significant for retailers in stage 3; and 16 factors (14 incentives and 2 barriers) were statically significant for retailers in stage 4.

Table 7.3: Value of the mean difference for the 19 factors for each stage of e-commerce maturity; significant factors are in bold

| | Factors | Code | Value of the mean difference for significant factors | | | | | | | |
|---|---|---|---|---|---|---|---|---|---|---|
| | | | Stage 1 no website | | Stage 2 static website | | Stage 3 interactive website | | Stage 4 E-commerce website | |
| | | | *p* | Mean Diff | *p* | Mean Diff | *p* | Mean Diff | *p* | Mean Diff |
| Consumer related factors | People's current attitudes and habits regarding shopping online | F1 | **.000** | **-1.514** | **.025** | **-.659** | .560 | -.171 | **.000** | **1.150** |
| | The current level of consumer recognition of e-commerce benefits | F2 | .058 | -.216 | .710 | .049 | **.034** | **.429** | **.000** | **1.000** |
| | The current level of consumer trust in online shopping | F3 | **.006** | **-.378** | **.038** | **-.293** | .751 | .057 | **.022** | **.525** |
| | The current level of consumer usage of credit cards for online shopping | F4 | **.027** | **-.243** | **.007** | **-.366** | .865 | -.029 | .634 | -.100 |
| | The current level of consumer demand for buying online | F5 | **.000** | **-1.108** | **.000** | **-.780** | .151 | -.371 | .130 | .425 |
| | The current level of consumer knowledge of buying online | F6 | .090 | -.189 | .830 | -.024 | **.000** | **.571** | **.001** | **.750** |
| | Level of willingness of consumers to pay delivery fees | F7 | **.030** | **-.216** | .498 | -.073 | .646 | -.086 | **.000** | **.900** |
| Environment related factors | The percentage of people using the internet | F8 | **.005** | **.514** | **.000** | **1.244** | **.000** | **1.457** | **.000** | **1.950** |
| | Level of availability of wi-fi and broadband services | F9 | .701 | -.054 | .168 | .244 | .054 | .457 | **.000** | **.875** |
| | Current level of e-commerce legislation for online shopping | F10 | **.001** | **-.622** | **.000** | **-1.049** | **.002** | **-.886** | **.004** | **-.725** |
| | Current availability of online payment options | F11 | **.004** | **-.486** | **.005** | **-.488** | **.007** | **-.686** | .118 | -.400 |
| | The current mailing and addressing system in Saudi Arabia | F12 | **.026** | **-.297** | **.016** | **-.512** | .062 | -.457 | **.007** | **-.600** |
| Organization related factors | Level of ease for our company to do e-commerce | F13 | **.000** | **-1.568** | **.000** | **-.854** | .165 | .400 | **.000** | **1.225** |
| | Nature of the products our company sells | F14 | **.000** | **-1.162** | **.000** | **-1.268** | .289 | .400 | **.000** | **1.500** |
| | Level of e-commerce knowledge and familiarity in our company | F15 | **.000** | **-1.973** | **.000** | **-1.341** | .156 | .400 | **.000** | **1.475** |
| | Management attitude towards e-commerce in our company | F16 | **.000** | **-1.676** | **.006** | **-.707** | **.026** | **.629** | **.000** | **1.900** |
| | The current business priorities of our company | F17 | **.000** | **-1.216** | **.042** | **-.415** | .183 | .343 | **.000** | **1.150** |
| | Our company's level of trust in online selling processes | F18 | **.001** | **-.649** | .221 | -.220 | .255 | .257 | **.003** | **.800** |
| | Setup and maintenance costs of e-commerce for our company | F19 | **.000** | **-.784** | .173 | -.220 | .263 | .286 | **.000** | **1.025** |

Significant codes: * significant at the 0.05 level, ** significant at the 0.001 level, *** significant at the 0.000 level

For convenience, Figure 7.2, given below, is a modified version of the previous figure, Figure 7.1, in that it presents only the significant factors as given in Table 7.3. Thus, Figure 7.2 clearly shows the significant factors for each of the stages of maturity and whether they were seen as negative (barriers) or positive (incentives).

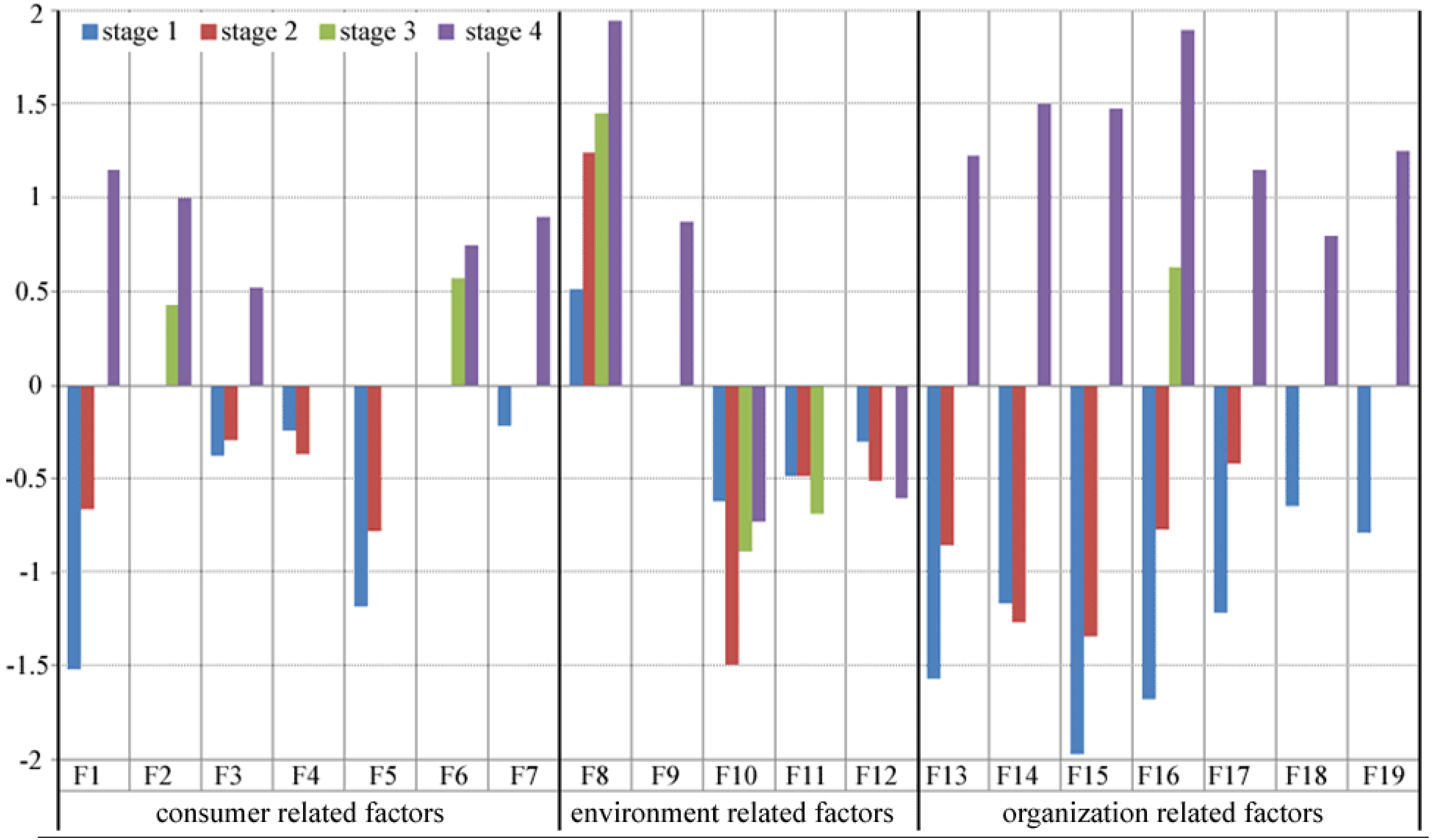

Figure 7.2: Value of the mean difference for the significant factors for each of the four stages

## 7.3. Comparing the significant incentives and barriers for each stage of e-commerce maturity

Having identified the significant factors for each stage of e-commerce maturity in the previous section, this section presents a more detailed comparison for the significant factors (incentives and barriers) for each stage of e-commerce maturity.

Table 7.4 presents all the significant factors which were considered incentives for all the stages of e-commerce maturity. Businesses in stage 1 and 2 have only one incentive, there are 4 incentives for businesses in stage 3, and for businesses in stage 4 there were 14 factors considered as incentives. It can also be seen that the incentives at the lower stages are a subset of the incentives of the higher stages.

Table 7.4: The significant factors which were considered incentives as a function of stage of e-commerce maturity

| Stage 1 | Stage 2 | Stage 3 | Stage 4 |
|---|---|---|---|
| F8: The percentage of people using the internet | F8: The percentage of people using the internet | F8: The percentage of people using the internet | F8: The percentage of people using the internet |
| | | F2: The current level of consumer recognition of e-commerce benefits | F2: The current level of consumer recognition of e-commerce benefits |
| | | F6: The current level of consumer knowledge of buying online | F6: The current level of consumer knowledge of buying online |
| | | F16: Management attitude towards e-commerce in our company | F16: Management attitude towards e-commerce in our company |
| | | | F1: People's current attitudes and habits regarding shopping online |
| | | | F3: The current level of consumer trust in online shopping |
| | | | F7: Level of willingness of consumers to pay delivery fees |
| | | | F9: Level of availability of wi-fi and broadband services |
| | | | F13: Level of ease for our company to do e-commerce |
| | | | F14: Nature of the products our company sells |
| | | | F15: Level of e-commerce knowledge and familiarity in our company |
| | | | F17: The current business priorities of our company |
| | | | F18: Our company’s level of trust in online selling processes |
| | | | F19: Setup and maintenance costs of e-commerce for our company |

Interestingly, all the four groups share incentive F8, that is, they all see the percentage of people using the Internet in Saudi Arabia as an incentive. They recognize that the population of the Internet users in Saudi Arabia is high and possibly important for businesses in e-commerce. It is the only incentive for stages 1 and 2. Participating businesses in stages 3 and 4 also share three significant incentives F2, the current level of consumer recognition of e-commerce benefits; F6, the current level of consumer knowledge of buying online; and F16, management attitude towards e-commerce in their organizations. It looks as though, beside the positive attitude of these organizations towards consumers, management attitude towards e-commerce in their organizations is a key factor. This is especially true for businesses that already use e-commerce and sell online (i.e. stage 4).

Respondents in businesses in stage 4 considered a further 10 factors as incentives compared with respondents from the lower stages.

*Three consumer related factors*

- F1: People's current attitudes and habits regarding shopping online
- F3: The current level of consumer trust in online shopping
- F7: Level of willingness of consumers to pay delivery fees

*One environment related factor*

- F9: Level of availability of wi-fi and broadband services

*Six organization related factors*

- F13: Level of ease for our company to do e-commerce
- F14: Nature of the products our company sells
- F15: Level of e-commerce knowledge and familiarity in our company
- F17: The current business priorities of our company
- F18: Our company’s level of trust in online selling processes
- F19: Setup and maintenance costs of e-commerce for our company

Now let’s consider barriers. Table 7.5 presents the factors that had a significant negative value for the mean difference from neutral and which are thus barriers. Retailers in stage 1 and 2 consider many factors as barriers, whereas retailers in stage 3 considered only one factor as a barrier and retailers in stage 4 saw two barriers.

Table 7.5: The significant factors that were considered barriers as a function of stage of e-commerce maturity

| Stage 1 | Stage 2 | Stage 3 | Stage 4 |
|---|---|---|---|
| F11: Current availability of online payment options | F11: Current availability of online payment options | F11: Current availability of online payment options | |
| F10: Current level of e-commerce legislation for online shopping | F10: Current level of e-commerce legislation for online shopping | | F10: Current level of e-commerce legislation for online shopping |
| F12: The current mailing and addressing system in Saudi Arabia | F12: The current mailing and addressing system in Saudi Arabia | | F12: The current mailing and addressing system in Saudi Arabia |
| F1: People's current attitudes and habits regarding shopping online | F1: People's current attitudes and habits regarding shopping online | | |
| F3: The current level of consumer trust in online shopping | F3: The current level of consumer trust in online shopping | | |
| F4: The current level of consumer usage of credit cards for online shopping | F4: The current level of consumer usage of credit cards for online shopping | | |
| F5: The current level of consumer demand for buying online | F5: The current level of consumer demand for buying online | | |
| F13: Level of ease for our company to do e-commerce | F13: Level of ease for our company to do e-commerce | | |
| F14: Nature of the products our company sells | F14: Nature of the products our company sells | | |
| F15: Level of e-commerce knowledge and familiarity in our company | F15: Level of e-commerce knowledge and familiarity in our company | | |
| F16: Management attitude towards e-commerce in our company | F16: Management attitude towards e-commerce in our company | | |
| F17: The current business priorities of our company | F17: The current business priorities of our company | | |
| F7: Level of willingness of consumers to pay delivery fees | | | |
| F18: Our company's level of trust in online selling processes | | | |
| F19: Setup and maintenance costs of e-commerce for our company | | | |

The only significant factor considered a barrier for retailers in stage 3 is F11, current availability of online payment options. The same factor is considered as a barrier for retailers in stages 2 and 1. However, this factor was considered as neutral for businesses in stage 4.

Interestingly, retailers in stages 1, 2 and 4 share the concern regarding F10, current level of e-commerce legislation for online shopping. They all considered this factor as a barrier. Similarly with F12, the current mailing and addressing system in Saudi Arabia; all the retailers in the three stages of e-commerce maturity 1, 2 and 4 share the same concern that this represents an inhibitor for businesses to sell online.

Retailers in stage 1 and 2 share a further 9 factors rated as barriers for them. These factors are:

*Four consumer related factors*

- F1: People's current attitudes and habits regarding shopping online
- F3: The current level of consumer trust in online shopping
- F4: The current level of consumer usage of credit cards for online shopping
- F5: The current level of consumer demand for buying online

*Five organization related factors*

- F13: Level of ease for our company to do e-commerce
- F14: Nature of the products our company sells
- F15: Level of e-commerce knowledge and familiarity in our company
- F16: Management attitude towards e-commerce in our company
- F17: The current business priorities of our company

## 7.4. Differences between stages of e-commerce maturity for the significant factors

Section 7.3 presents a list of significant incentives and barriers for each stage of maturity, but does not consider whether a significant incentive or barrier was stronger for different stages of maturity. It can be seen from section 7.3, particularly Figure 7.2, that significant factors for some stages were rated as incentives while for other stages they were rated as barriers. However, in the cases where two or more stages rate a factor in the same direction, do they differ in strength? To find out, pairwise t-tests using Bonferroni correction for multiple tests were performed.

Results showed that although respondents in stages 3 and 4 saw F2, the current level of consumer recognition of e-commerce, as an incentive, the responses were significantly lower for stage 3 respondents, with $p = 0.027$. Further, while respondents for all stages saw F8, the percentage of people using the internet, as an incentive, respondents for stage 1 rated it as less of an incentive than respondents in stage 3, with $p = 0.010$, and less than those in stage 4, with $p = 0.000$. Finally, respondents in businesses in stage 3 and 4 differed significantly for F16, management attitude towards e-commerce in their company, with respondents in businesses in stage 4 rating it even more positively than those in stage 3.

## 7.5. Do other attributes of the organizations matter?

The above analyses considered the responses to factors in terms of stage of maturity. However, it might be argued that other attributes of the organizations, e.g. organization's size or age might be important in determining responses to the factors. This would be unlikely, as the companies in the four stages came from a spread of ages and sizes, as shown in Table 6.4 page 113. However, as a check, a multiple linear regression was performed for each factor to consider responses as a function not only of stage of maturity, but also of organization size and organization age. The results confirmed that it is e-commerce maturity that is driving the results. Organization size only matters for two factors; F5, the current level of consumer demand for buying online ($p = 0.035$), and F13, level of ease for our company to do e-commerce ($p = 0.008$). However, for these factors stage of e-commerce maturity was also significant, and at a higher levels ($p =$ 0.000 and 0.000, respectively). Organization age only matters for two factors: F7, level of willingness of consumers to pay delivery fees ($p = 0.035$), and F13, level of ease for our company to do e-commerce ($p = 0.011$). Again, though, stage of e-commerce maturity was also significant for these factors, and at higher levels ($p = 0.000$ and 0.000 respectively). The multiple regressions also showed that stage of e-commerce maturity was significant for all factors except one consumer factor (F4, the current level of consumer usage of credit cards for online shopping) and for three environmental factors (F10, current level of e-commerce legislation for online shopping; F11, current availability of online payment options; and F12, the current mailing and addressing system in Saudi Arabia). Again, this is not surprising: as stated on page 121 Figure 7.2, there is more agreement between the respondents in companies at different stages in term of rating the environment related factors than for other factors. In sum, then, it is clear that stage of e-commerce maturity is the primary driving factor in the responses. Appendix P, page 291 presents the full results of the multiple linear regression.

# CHAPTER 8: DISCUSSION OF THE QUANTITATIVE FINDINGS

This chapter presents a summary and discussion of the findings. It starts with a discussion of the four stages of e-commerce maturity and the responses of respondents from each stage to the identified factors. The factors were structured around three themes: consumer, environment and organization. Each theme is discussed in terms of responses from retailers in companies at the different stages of maturity. Based on the analysis and discussion of the factors influencing retailers' decisions on whether or not to adopt e-commerce, directions and recommendations are provided to contribute to the e-commerce development in Saudi Arabia.

## 8.1. Summary of the findings and discussion

Four stages of e-commerce maturity based on the SOG-e model of Mckay, Marshall and Pranato (2000) were used in this study to classify the sample of respondents into 4 groups. It was found that there were significant differences in how retailers responded, given the stage of maturity of their company. By looking at each group separately, the study was able to determine the incentives and barriers for each stage. The use of the stage of maturity model led to findings showing ways in which views progressed from more pessimistic to more optimistic and in what ways retailers in companies at different stages were alike or not alike.

The qualitative study led to structuring data around three themes; consumer related factors, environment related factors, and organization related factors. In the next sections responses of the different groups to the factors in the different themes will be discussed.

### 8.1.1. Consumer related factors

For the convenience of the reader, Table 8.1 presents the consumer related factors that were presented in the survey and Figure 8.1 shows the consumer related factors that were found to be statistically significant (taken from Figure 7.2).

Table 8.1: consumers related factors

| | |
|---|---|
| People's current attitudes and habits regarding shopping online | F1 |
| The current level of consumer recognition of e-commerce benefits | F2 |
| The current level of consumer trust in online shopping | F3 |
| The current level of consumer usage of credit cards for online shopping | F4 |
| The current level of consumer demand for buying online | F5 |
| The current level of consumer knowledge of buying online | F6 |
| Level of willingness of consumers to pay delivery fees | F7 |

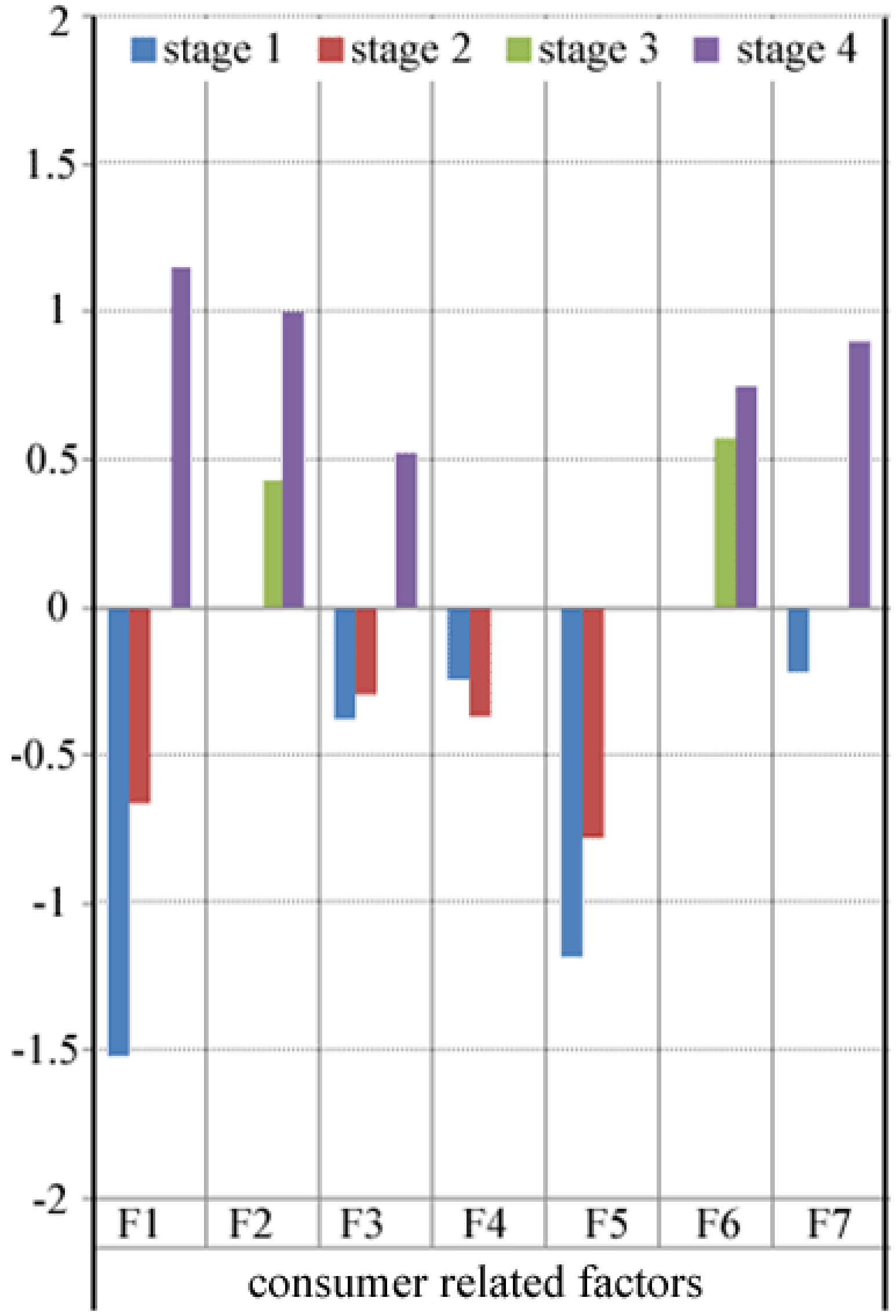

Figure 8.1: value of the mean difference for the significant consumer related factors for each of the four stages

It is quite clear that there is a stark difference between retailers in stages 1 and 2 and 3 and 4. Retailers in e-commerce maturity stages 1 and 2 had negative perceptions about consumers shopping online in Saudi Arabia; whereas retailers in stages 3 and 4 of e-commerce maturity had positive perceptions. Starting with F1 (people's current attitudes and habits regarding shopping online) retailers in stages 1 and 2 saw that the current attitude and habits of consumers shopping online as a barrier whereas retailers in stage 4 saw it as an incentive. The question which may be asked here is; on what basis do retailers in stages 1 and 2 evaluate their consumers' attitude and habits regarding online shopping? It is probable that personal feelings or experience play an important role in this evaluation. For example, Talal whose company is in stage 2 and who has no experience buying online commented in evaluating consumer's attitude toward online shopping, "*when the culture of people changed and accept to buy online we will apply e-commerce system*". He has no experience buying online and seems to think others are like him. Stage 4 retailers sell online, so they have more experience. The real question here is, given that people do buy online; why are the stage 1 and 2 retailers so negative? One possible reason is lack of knowledge and experience. This may require educational and awareness programs. Educational and awareness programs may work here for these types of retailers and enhance their knowledge regarding the online marketplace in Saudi Arabia as seen by stage 4 retailers.

In regard to F2 (the current level of consumer recognition of e-commerce benefits) only stages 3 and 4 rate it as a significant incentive, though stage 4 respondents rate it higher than those in stage 3. Respondents in the other stages responded neutrally. Again, companies in stages 3 and 4 interact with their consumers online and stage 4 retailers have experience selling online. They build their evaluation based on experience. It is as though the stage 1 and 2 retailers are not aware of the real situation, compared with retailers in stages 3 and 4. For F3 (the current level of consumer trust in online shopping) retailers in stages 1 and 2 saw a significant barrier, retailers in stage 3 were neutral, while stage 4 retailers saw it as a significant incentive. Again, we see very different responses, though all retailers are in the one country. Their perceptions are very much related to the stage of maturity of their company. Let's now consider F4 (the current level of consumer usage of credit cards for online shopping) and F5 (the current level of consumer demand for buying online) together. These two factors were found to be significant barriers only for retailers in stages 1 and 2 whereas retailers in stages 3 and 4 gave neutral responses. It is interesting here that all these retailers are in the same country but see things very differently. The question which may be asked here is, how do stage 1 and 2 retailers evaluate the online consumers' behaviour? It seems the same answer again; personal feelings and experience probably play important roles in this evaluation. For these companies, which have not interacted with online consumers, their evaluation regarding the experiences of online consumers does not fit with the real situation at least as far as retailers who do interact with online customers are concerned, but rather with their personal experience. This may suggest again that educational and awareness programs for these types of companies would help.

F6 (the current level of consumer knowledge of buying online) was found to be a significant incentive for retailers in stages 3 and 4, and neutral for stages 1 and 2. It seems that for stage 1 and 2 retailers it is more a case of them having no strong feeling about consumers' level of knowledge, rather than being hindered in adopting e-commerce because they believe consumers do not have knowledge about buying online. Finally, F7 (level of willingness of consumers to pay delivery fees) is seen as significant barrier for stage 1 but an incentive for stage 4. To know whether online consumers are willing or not to pay for delivery fees, experience (or education) is needed. Retailers in stage 4 confirm that online consumers are willing to pay for delivery fees; they have no issue with this factor at all but rather they considered such willingness a significant incentive. This suggests that the evaluation of retailers in stage 1 that consumers are not willing to pay for online orders delivery is unfounded.

It is clear based on the above discussion that although all these retailers in companies in the four stages of e-commerce maturity are in the same country they have very different opinions of consumers. Retailers in lower stages evaluate consumers based on little or no experience. Educational and awareness programs may work here for these types of retailers. Educational and awareness programs are needed to provide them with basic information about online consumers and e-commerce activities in Saudi Arabia. Let's turn now to environment related factors.

### 8.1.2. Environment related factors

For the reader's convenience, Table 8.2 presents the environment related factors and Figure 8.2 repeats the relevant section of Figure 7.2

Table 8.2: Environment related factors

| | |
|---|---|
| The percentage of people using the internet | F8 |
| Level of availability of wi-fi and broadband services | F9 |
| Current level of e-commerce legislation for online shopping | F10 |
| Current availability of online payment options | F11 |
| The current mailing and addressing system in Saudi Arabia | F12 |

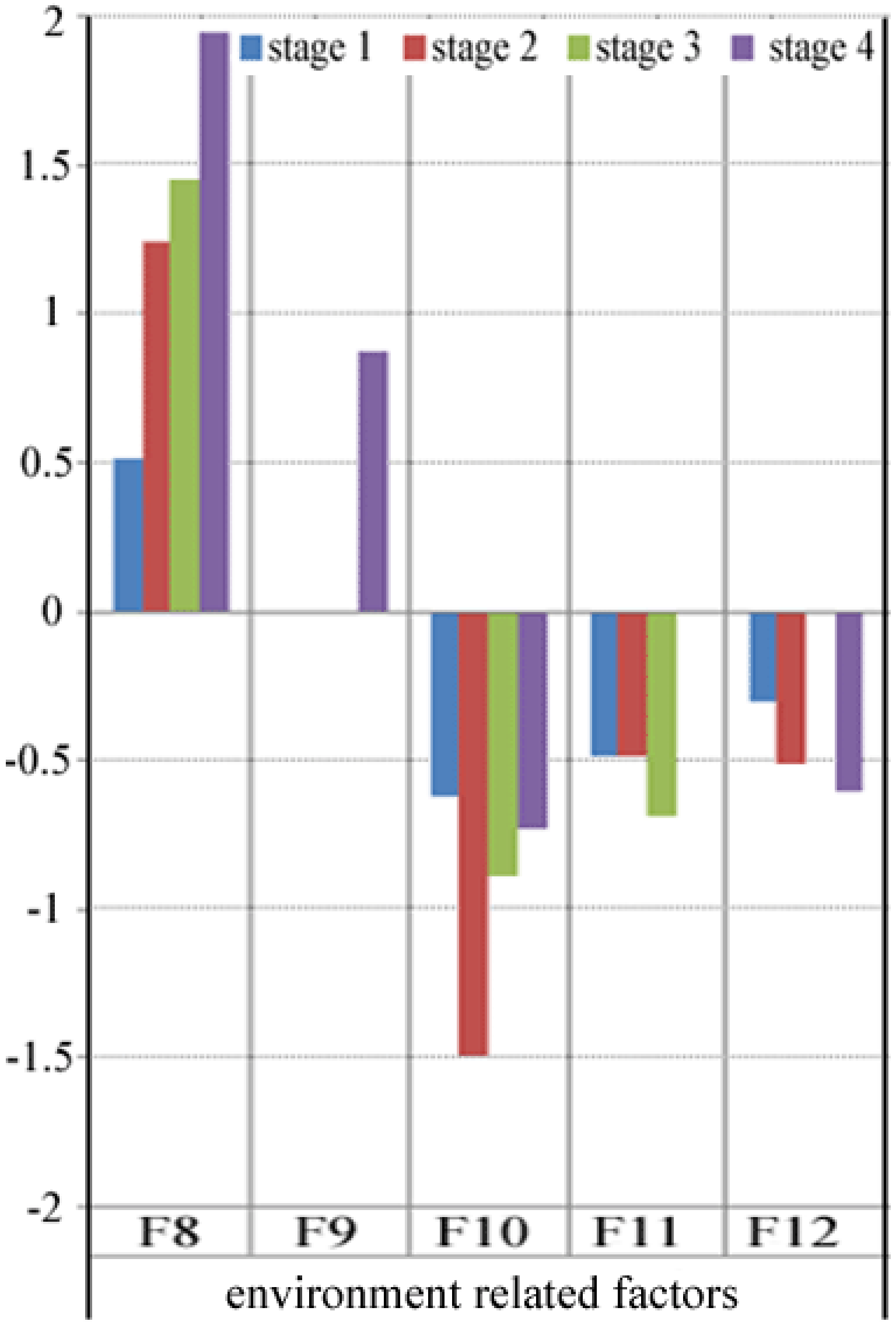

Figure 8.2: value of the mean difference for the significant environment related factors for each of the four stages

The study highlighted five environmental factors that might be significant in influencing businesses' decisions on whether or not to adopt e-commerce. Interestingly, retailers in the four stages showed much more agreement in rating the environment related factors than they did with the consumer related factors. There is no factor where one group saw it as an incentive and another saw it as a barrier (in contrast to the consumer factors). In fact, for two of the factors (F8 and F10) all four stages saw them as a significant incentive (F8) or a significant barrier (F10).

Thus, considering F8, all the stages recognized that the percentage of Internet users in Saudi Arabia is a significant incentive to online entry. Even retailers in businesses in stage 1 recognised it as incentive, though their responses were significantly lower when compared to stages 3 and 4. Similarly, considering, F10 all the groups agreed that the current level of e-commerce legislation for online shopping is Saudi Arabia is a significant barrier. The fact that even stage 4 respondents are negative shows that this is a very important problem in Saudi Arabia. It may be that retailers want better legislation so that people will be more trustful. The qualitative analysis suggested that those businesses that already sell online are looking for a means to develop trust with their consumers to increase online purchases. For example, Salem (4) suggested that "*it should be there is a certification body from the government itself to say that this company is a certified company by local government and you can buy from them. This is good to build the customers trust with the certified companies as the government trust them*". Existence of clear e-commerce legislation for online shopping helps sellers develop trust with their consumers. In addition, the existence of e-commerce legislation contributes to reducing the security concerns of non-adopters. The literature (Kraemer, Dedrick, & Melville 2006) indicated that concerns become greater in an ineffective legal and regulatory environment. In ineffective legal and regulatory environments for e-commerce privacy and security issues are rated the number one inhibitor for businesses on whether or not to adopt e-commerce in eight countries (Kraemer, Dedrick, & Melville 2006).

The current availability of online payment options, F11, was seen as a barrier for all the groups of e-commerce maturity stages except with businesses in stage 4. It was seen neutral by companies in stage 4, which have experience selling online. Several online payment options such as credit cards, SADAD, CashU, SafePay, and PayPal, are available to use in Saudi Arabia and companies which sell online in Saudi Arabia use them. The problem with businesses which considered the current available online payment systems a barrier may occur due to lack of knowledge or the perceptions about consumers that they do not accept to pay using these methods. With the consumers' reluctance to use credit cards in online purchases in Saudi Arabia and concerns of businesses regarding online payments security, other secure options may be necessary. The national payment system SADAD may be a good solution here to overcome this problem. As discussed in the qualitative analysis chapter, the national payment system SADAD may represent a solution to these issues. However, evidence from the survey results showed that a very high number of the participating businesses (about 86%) were not aware of the recent update that took place on the SADAD system, where capacity has been expanded to 20,000 billers. Further discussion will be provided later on this in the suggestions and recommendations section. It seems more practical to consider the opinion of businesses in stage 4, that the current online payment options are not an issue, but rather and incentive, because they already have experience dealing with that. On the other hand, businesses in other stages perceived that this is an issue. This may be related again to other factors such as lack of knowledge and experience.

F12 (the current mailing and addressing system in Saudi Arabia) was seen as an inhibitor for businesses in stages 1 and 2 of e-commerce maturity. Retailers in businesses in stage 4, which already use e-commerce, agree. It was discussed in the background chapter that there have been strong developments in the Saudi mailing and addressing systems since 2005. Before that there was no mailing to homes in Saudi Arabia and streets were not named. Now, each building has a physical address with number, street name, name of suburb and postcode (Saudi Post 2011). Consumers in Saudi Arabia used to receive their orders purchased online using P. O. Boxes of their own or of their relatives, friends, or work. Or orders may be picked up from the offices of shipping companies. So, the habit of delivering and receiving purchased online orders using home mailboxes has not been developed well yet. Consumers are not aware or do not trust using their homes' mailboxes. The evidence of the qualitative section of this study showed that about 36% of the businesses were not aware of the recent development of the Saudi mailing and addressing system that each building has a number and each street is named. Further efforts might be needed in order to build the awareness and trust with the Saudi mailing system. The Saudi Post has created an e-mall which is an e-commerce platform for businesses to use and sell online (E-mall 2010). One of the main purposes of e-mall is to build trust and awareness of the Saudi mailing system. The Saudi post offers either competitive or free delivery services using the Saudi mailing and addressing system. Further discussion of this will be provided later in the suggestions and recommendations section.

### 8.1.3. Organization related factors

Table 8.3 presents the organization related factors and Figure 8.3 repeats the relevant part of Figure 7.2.

Table 8.3: Organization related factors

| | |
|---|---|
| Level of ease for our company to do e-commerce | F13 |
| Nature of the products our company sells | F14 |
| Level of e-commerce knowledge and familiarity in our company | F15 |
| Management attitude towards e-commerce in our company | F16 |
| The current business priorities of our company | F17 |
| Our company's level of trust in online selling processes | F18 |
| Setup and maintenance costs of e-commerce for our company | F19 |

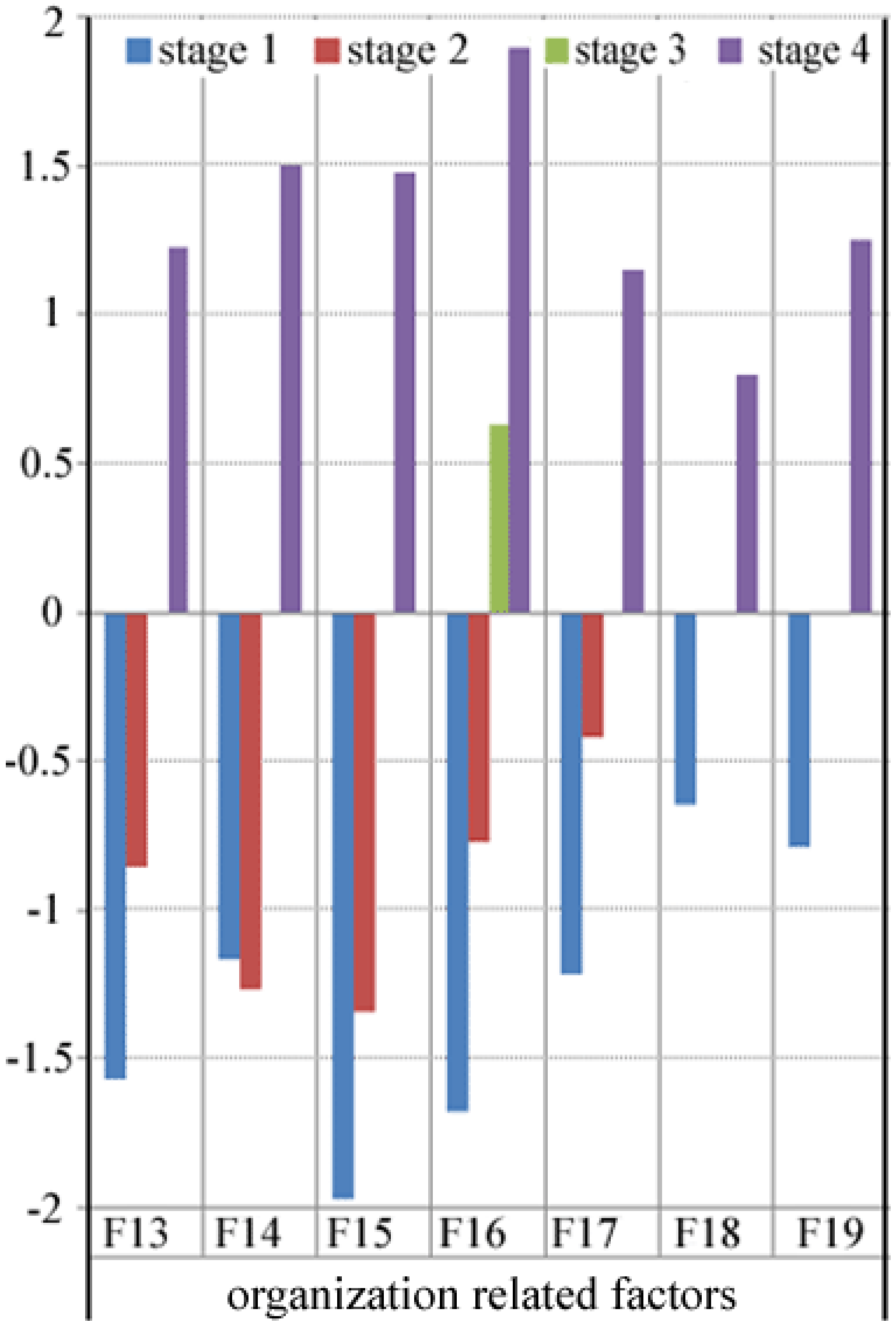

Figure8.3: value of the mean difference for the significant environment related factors for each of the four stages

Unlike the environment related factors, but like the consumer related factors, we see a striking difference between the four groups of respondents. Stage 1 respondents are negative about all the factors, while stage 4 is positive for all factors. Stage 2 respondents are negative for five factors, while stage 3 are positive for one factor though significantly less positive than stage 4 respondents.

The question which may be asked is: what makes the companies in the four stages of e-commerce maturity rate these factors so differently? This may relate to an organization's effective electronic readiness (e-readiness). An organization's e-readiness refers to the ability of the ICT infrastructure and relevant information systems to meet the business' electronic needs and abilities (EIU 2010). This includes the availability of IT staff or consultants with web-skills (Wymer & Regan 2005). For example, F13, level of ease for our company to do e-commerce, and F15, level of e-commerce knowledge and familiarity in our company, involve experience and the need for skilled staff to deal with e-commerce. For retailers in e-commerce maturity stage 4 doing e-commerce is easy, and their e-commerce knowledge and familiarity is good. Therefore they consider these factors (F13 and F15) incentives whereas for retailers in stages 1 and 2; doing e-commerce is not an easy task, and they do not have the enough knowledge to do e-commerce and familiarity.

The management attitude towards using e-commerce (F16) was considered a significant incentive for both retailers in stages 3 and 4 whereas businesses in stage 1 and stage 2 considered it a significant barrier. This means, the top management support is an important key for whether to accept or reject changes. The literature strongly supports this statement that the top management support is a key for change (Aguila-Obra & Padilla-Mele´ndez 2006; Liu 2008; Yu-hui 2008; Chong et al. 2009; Wang, Wang & Yang 2010). Rogers (2003) identified that the leadership attitude towards change is an important factor to organizational innovativeness.

While some businesses find using e-commerce costly, it is not an issue for businesses in stage 4 but instead they see the cheapness of it as incentive. Businesses in stage 4 considered that setup and maintenance costs of e-commerce, F19, is cheap and therefore they considered the cheapness as an incentive. As discussed earlier, this may refer to an organization e-readiness. Organizations with strong e-readiness find using additional technologies easier and cheaper compared to other organizations with weak e-readiness.

F18, company's level of trust in online selling processes, was only mentioned as a significant positive by businesses in stage 4 and negatively by businesses in stage 1. With regards to F17, the current business priorities of our company, similar to other factors it was rated as an incentive for businesses in stage 4 and a barrier for businesses in stages 1 and 2. This means using e-commerce has been a business priority for organizations in stage 4 whereas it has not been considered as such for organizations in stages 1 and 2.

Finally, businesses in stage 4 considered F14 a significant incentive that their products are suitable to be sold online whereas businesses in stages 1 and 2 considered their products are not suitable. This is an interesting point for discussion. The interesting thing here, as Table 6.4 in page 113 shows, is that what retailers sell is spread out throughout the 4 stages. The literature indicated that some businesses may be hindered to sell online because they think that their products are not suitable to be sold online. The outcome of this analysis shows that some businesses in Saudi Arabia sell products online that other businesses say are not suitable for selling online. This raises the question of what makes one business consider products suitable to be sold online whereas another business considers selling the same products as not suitable to be sold online. Although the businesses sell the same products, they see things differently. For example, as the qualitative analysis showed, a business considered that their beauty and perfume products are not suitable to be sold online because it was thought, people want to seeing, touching and smell the perfumes. For example, Ahmed's (2) business involves selling beauty products. He commented that it is difficult for them to sell online, consumers will not accept it because "*our products related to smell, shape which need to be physically seen*". However, there are many successful businesses around the world selling the same products online without considering this is an issue. Also seven of the stage 4 businesses, which covered in the sample of this study, reported they sell beauty and perfume products, see Table 6.4 in page 113. Again, it seems retailers need better knowledge and more experience with ways to encourage consumers to buy their product online, including answering enquiries, providing information.

## 8.2. Stage of growth and e-commerce development

From the analysis and discussion of responses from retailers in companies at different stages of e-commerce maturity, the study was able to identify that for environment factors responses differed little as a function of stage of maturity, but that for consumer and organization factors there were stark differences between the four stages. This suggests that while improvements in the environment would help all, to get the lower to higher stages it is important to carefully consider retailer perceptions of consumers and also organization factors. The question which may be asked here is: what needs to be done for businesses in the lower stages to move them to higher stages and for businesses in higher stage to continue selling online and move forward to an integrated e-business? The following discussion presents recommendations based on the study findings for businesses in the four stages. The discussion considers internal factors within an organisation, but also factors that are beyond the organization's boundary to control, which would need a third party, such as the government, to enhance and facilitate further development.

### 8.2.1. Internal changes

In this section, a discussion is presented regarding changes that need to be made internally by organizations to move from a lower stage to the next one. To help companies to make these changes, there would need to be external help, as discussed in Section 8.2.2.

For businesses in stage 1, the first step to move forward is to establish an online presence for their organizations. The real question, though, is how these companies can be encouraged to do this. Making these businesses aware of the available options to establish an online presence especially the free ones would contribute to a way of moving forward, that is, taking the first step of establishing online presence. It is most likely that external efforts are needed to show them the opportunities and to give them success stories, and to correct misconceptions in order to encourage them to move forward using the Internet for business. Further discussion of this will be provided later in the external support section.

Businesses in stage 2 are already online, but they are static and need to provide a means of interacting with consumers online. Social media such as Twitter and Facebook can be integrated to companies' websites to provide active and live interaction with customers. Saudis have been identified as having the highest percentage of their population using social networks in the Middle East and North Africa (Askool 2013). Therefore, businesses could take advantage of this by using free, wider access, and more active interaction tools. The second issue that needs to be considered for businesses in stage 2 to move forward is to use staff with IT expertise to help the organization in setting up the required IT. Again, to get companies to move, there will need to be external efforts, to be considered below.

Businesses in stage 3 are almost ready to sell online. Their websites are interactive and show product details and possibly prices, but they do not receive orders or accept payment. To move to the next stage, these companies need to use staff with technical expertise in conjunction with business oriented staff. This is because selling online is not only a technical problem; it involves products to be delivered to consumers. Again, they will need external help to achieve higher maturity, as discussed below.

Businesses in e-commerce maturity stage 4 are the highest stage of e-commerce maturity considered here; however there are further two higher stages of maturity which lead a company to move from e-commerce to e-business. As discussed in the literature review chapter, e-business is a more general term than e-commerce. E-commerce is regarded as a subset of e-business (Davis & Benamati 2003). Therefore, businesses classified in stage 4 of e-commerce maturity need to keep moving forward. Again, external help is probably needed to encourage movement. Section 8.2.2 give recommendations for external ways of helping all companies move to a higher level.

### 8.2.2. External support

The first thing to consider is who can provide support. Rogers (2003) suggested that the efforts of a change agency are a determinant in the rate of the technology adoption. A change agency can be a government or business associations that can support retailers in adopting e-commerce. Retailers in the qualitative study, especially stage 4, mentioned the importance of the role of government in developing trust with consumers. For example, Thamer (4) said "*citizens will have more trust if this subject sponsored by the government because we, in Saudi Arabia, have great confidence in anything that comes through the government*". Therefore, the relevant departments of the Saudi government, such as Ministry of Commerce, MICT, and CITC, and the relevant party from business associations such as chambers of commerce could play an important role as a change agency in diffusing online retailing.

What can change agencies in Saudi Arabia do? The study findings show that there is much agreement between all stages in rating environment factors (see Figure 8.2, page 132). All stages considered the current level of e-commerce legislation for online shopping as a barrier. Stages 1, 2 and 3 considered the current availability of online payment system as a barrier, with stage 4 being neutral. Stages 1, 2 and 4 considered the current mailing and addressing system in Saudi Arabia a barrier, with stage 3 being neutral. Given the strong agreement about barriers in the environment, it is clear that the three areas of e-commerce legislation, online payment systems, and the mailing and addressing system need improvement. In the following sections, these three areas will be discussed and examples of support will be suggested.

***Enhancing the legislation system for e-commerce*** - Clearly, the legislation system for e-commerce in Saudi Arabia needs to be enhanced to support the growth of e-commerce. Enhancing this system is in the hand of the government, perhaps with the cooperation of the private sector. The current study found that retailers in all the four stages of e-commerce maturity share the same concern that the current level of e-commerce legislation for e-commerce is a barrier to use e-commerce. This is an important issue which needs to be considered. It is confirmed by retailers who sell online that this is a real issue. There is little doubt that the clarification and enhancement of legislation and regulations, in this area, should be a priority for the government, as such matters are clearly part of its responsibilities. The enhancement and development of e-commerce legislation system would make e-commerce more attractive to businesses. For this reason, a government body should be allocated for e-commerce development in Saudi Arabia. This government body should then work with the chambers of commerce and private sector towards e-commerce solutions.

The qualitative study suggests why retailers are seeking a better legislation system for e-commerce? Salem (4) says "*It should be there is a certification body from the government itself to say that this company is a certified company by local government and you can buy from them. This is good to build the customers trust with the certified companies as the government trusts them*". Further, Thamer (4) believes that consumers trust government procedures and would feel more protected if they introduced better legislation. He says "*citizens will have more trust if this subject sponsored by the government because we, in Saudi Arabia, have great confidence in anything that comes through the government*". It is understandable why retailers are urging the government to provide good legislation system for e-commerce. With the lack of a good legislative system for e-commerce, trust is difficult to be build with consumers.

Therefore, the government must be urged, perhaps by businesses, to bring in good legislation. Studies such as the present one may be used to help urge the government to provide appropriate legislation. Moreover, once good legislation exists, it may be helpful to introduce a national trust mark. A good example of this has been applied in the United Arab Emirates (UAE). UAE has launched a national trust mark called 'Trustae' that aims to increase confidence in e-commerce (TRA 2011). Figure 8.4 gives the UAE national trust mark.

Figure 8.4: UAE national trust mark

‘TRUSTae’ emphasizes promoting a secure and safe environment for e-commerce in the UAE. In order for online retailers to earn the right of displaying ‘TRUSTae’, they have to meet specific requirements defined by the UAE Telecommunications Regulatory Authority (TRA 2011). Given the problems identified in the environment by retailers at all 4 stages, a logo like that in Figure 8.4,, used in conjunction with new legislation, could be extremely helpful.

***Introduction of more secure online payment systems* -** Retailers in stages 1, 2 and 3 share the same concern that there is a need for introducing more secure online payment systems. However, retailers in stage 4 do not see this as an issue and they manage to sell online using current online payment methods. But, what about retailers in the lowest three stages? It is likely that they lack of knowledge about payment systems because currently in Saudi Arabia there is, in fact, a good method for online payment. This payment method, called SADAD, is a national payment system that is used to pay for e-government services. It has been made available for businesses to use since 2011. However, the current study found that a very low percentage (14.4%) of 153 respondents to the survey were aware of the recent changes of allowing businesses to use SADAD. It is clear that relevant bodies, such as chambers of commerce, the SADAD management, the Ministry of Commerce, and CITC, must do more to make businesses aware that they can use the services of SADAD. With growing citizen satisfaction and use of SADAD, it has become the second largest electronic payments system in the world serving around 29 million citizens (Saudi Gazette 2013). Al-Mohamed (2011) has even stated that the SADAD system has been identified as the most preferred online payment method for the Saudi e-mall consumers. Retailers must be made aware of this. Since the Saudi government owns SADAD, the government must put more effort into marketing it and working in cooperation with the private sector to make businesses aware of using SADAD.

***More effort to build trust in the Saudi mailing and addressing system and better use of home mailboxes* -** For the last five years, there has been a notable development in the addressing and mailing system in Saudi Arabia (Natto 2012). However, the present study found that the current mailing and addressing system in Saudi Arabia is considered a barrier to e-commerce by retailers in stages 1, 2 and 4. The fact that even stage 4 retailers see it as a barrier shows that it is a real issue. To some extent retailers are not familiar with the addressing and home mailing system because it was only introduced recently (Natto 2012). To make organizations and people aware of these new services provided by the Saudi Post, the government has recently started to link some e-government services with the Saudi mailing system. It seems that this effort has not been enough to make people aware of or trust the new mailing services to homes. More effort by the government and Saudi Post to make people aware of the new system and also to ensure that products are delivered safely to buyers is vital. Providing services of insurance, for example, should also be made available. Overall, it is clear that a program to accelerate the allocation of house addresses, the adoption of mailboxes by citizens, and the provision of the necessary infrastructure, to support the secure delivery of goods to homes, is an urgent requirement for e-commerce growth.

So far, the fact that respondents in all four stages of growth see barriers in the environment have been considered. Let’s turn now to barriers related to perceptions about consumers. It is clear from the present study that stage 1 and 2 retailers have negative perceptions of consumers, and yet retailers from stages 3 and 4 have positive perceptions. Stage 1 and 2 retailers perceived the following as barriers: consumer attitudes and habits regarding shopping online, level of consumer trust in online shopping, level of consumer usage of credit cards for online shopping, and level of consumer demand for buying online. In contrast, retailers at stages 3 and 4, who have real experience either interacting with online consumers or selling products online, do not agree that these are barriers (see Figure 8.1, page 129). Taking this into consideration and the fact that stage 1 and 2 retailers have no experience in selling online, it is hard to accept their views as a true indication of consumer behavior and beliefs.. However, the fact does remain that stage 1 and 2 retailers do have negative perceptions about consumers purchasing online. The question to be addressed here is what needs to be done to correct these negative perceptions? It seems educational and awareness programs would contribute positively.

***Educational and awareness programs*** - Educational and awareness programs about consumer behavior, attitudes, beliefs, and buying habits are recommended for retailers in stages 1 and 2 of e-commerce maturity. As part of making these businesses aware of online consumer behaviors including their trust in online shopping, usage of payment methods, and demand for buying online; one recommendation is to bring them together with retailers from higher stages of e-commerce growth. Higher stage retailers have experience and positive perceptions regarding online consumers. Therefore, it would be useful in bringing them together to benefit. The question may be asked here is how to do that? It would be useful if the MCIT annual conference can be used in this regard. The Saudi Minister of Communication and Information Technology (MCIT) run an annual conference to discuss ICT issues in Saudi Arabia. An invitation through mass media to this conference would be useful. This mass media could even include information about the very positive aspects of consumer behavior that do exist. The MCIT conference could provide not only an educational program, but if it attracted retailers from all stages of growth, it would create interpersonal communication between lower stage retailers and higher stage ones. The interpersonal communication would provide two-way exchange of information and would help to dispel retailers' concerns regarding online consumers. The experience that higher stages retailers have would help lower stage retailers to become aware of the real world of the online market.

In addition to the negative perceptions that stage 1 and 2 retailers have about online consumers, they negatively rated organization factors. It is obvious in Figure 8.3 page 136 that retailers in stage 1 rate all the 7 organization factors as barriers whereas stage 4 retailers rate all 7 organization factors positively. For stage 2 retailers 5 organization factors were rated as barriers and the rest were neutral, and stage 3 only have one incentive from organization factors and the rest were neutral. Both stage 1 and 2 retailers share the same concerns regarding the level of ease for their companies to do e-commerce; nature of the products their companies sells; level of e-commerce knowledge and familiarity in their companies; management attitude towards e-commerce in their companies; and the current business priorities of their company. This clearly shows that retailers in the lower stages find e-commerce difficult to use and they clearly do not have enough knowledge to deal with e-commerce. Other retailers in companies selling the same products or that are the same age or size have achieved higher stages, so the question is raised why do the lower stage retailers have negative thoughts when others are positive? It seems that lack of knowledge breeds fear and a negative attitude. Again, it is clear that better educational and awareness programs must be implemented.

The above mentioned MCIT annual conference could be used in arranging for e-commerce solutions companies to meet with retailers in different stages. While retailers from higher stages can benefit from e-commerce solutions companies to keep updated about new technology and market opportunities, retailers from lower stages could hear of the experiences of higher stages retailers and at the same time meet with e-commerce solution companies. E-commerce solution companies could perhaps even be encouraged by the government to offer free e-commerce system trials. Trying products would lead to better understanding of how e-commerce works. The MCIT annual conference can be highly useful in the diffusion of online retailing when considering these suggestions. Something similar runs in Australia annually, called the Online Retailer Conference and e-Commerce Expo. It "brings together innovative content, inspiring speakers, strategic networking opportunities and top-tier solution providers all in the one place" (Reed Exhibitions Australia Pty Ltd 2012). It is made available for all interested people to attend and get involved and benefit from listening to experts, meeting successful retailers and e-commerce solution companies, discussing concerns, engaging in workshops, and so forth. Alongside of material promoting education in facts about e-commerce and consumer behavior, this would create an impressive communication channel for retailers in Saudi Arabia.

While educational and awareness programs are good in order to contribute to changing the perception of retailers about online consumers and improving retailers' knowledge about e-commerce; still other issues exist, particularly cost. As seen in Figure 8.3, page 136, Retailers in stage 1 are concerned about setup and maintenance costs of e-commerce (F19). Apart from education, these low stage retailers may need to be given assistance to get started on the growth to e-commerce maturity.

***Assistance in getting businesses online*** - The MCIT conference mentioned above would give stage 1 retailers information about easy and cheap solutions for businesses and about the benefits of e-commerce. These retailers could also be given access to free software, with perhaps some funding by the government for the initial setups and access to servers would help. The government could also hire e-commerce experts to meet with and help stage 1 retailers. Further, it is not even necessary for retailers to run a complete e-commerce website at the beginning of the moves into e-commerce and these lower level retailers need to be made aware of this. They may sell online through an electronic market provider such as an e-mall, a web-based business that uses the Internet to create a market that brings buyers and sellers together (Laudon & Traver, 2010). An e-mall exists in Saudi Arabia and is run by Saudi Post, which launched its e-mall project at the end of 2010 (E-mall 2010). Selling goods through an e-mall provides retailers with the opportunity to try e-commerce with little cost compared to establishing a company e-commerce website. Therefore, the efforts made by e-mall may need to be expanded and publicized more in order to get more retailers to use the facility.

In summary, Figure 8.4 below gives an illustration of recommendations. It shows businesses in the four stages and what is needed internally and externally for movement to higher stages.

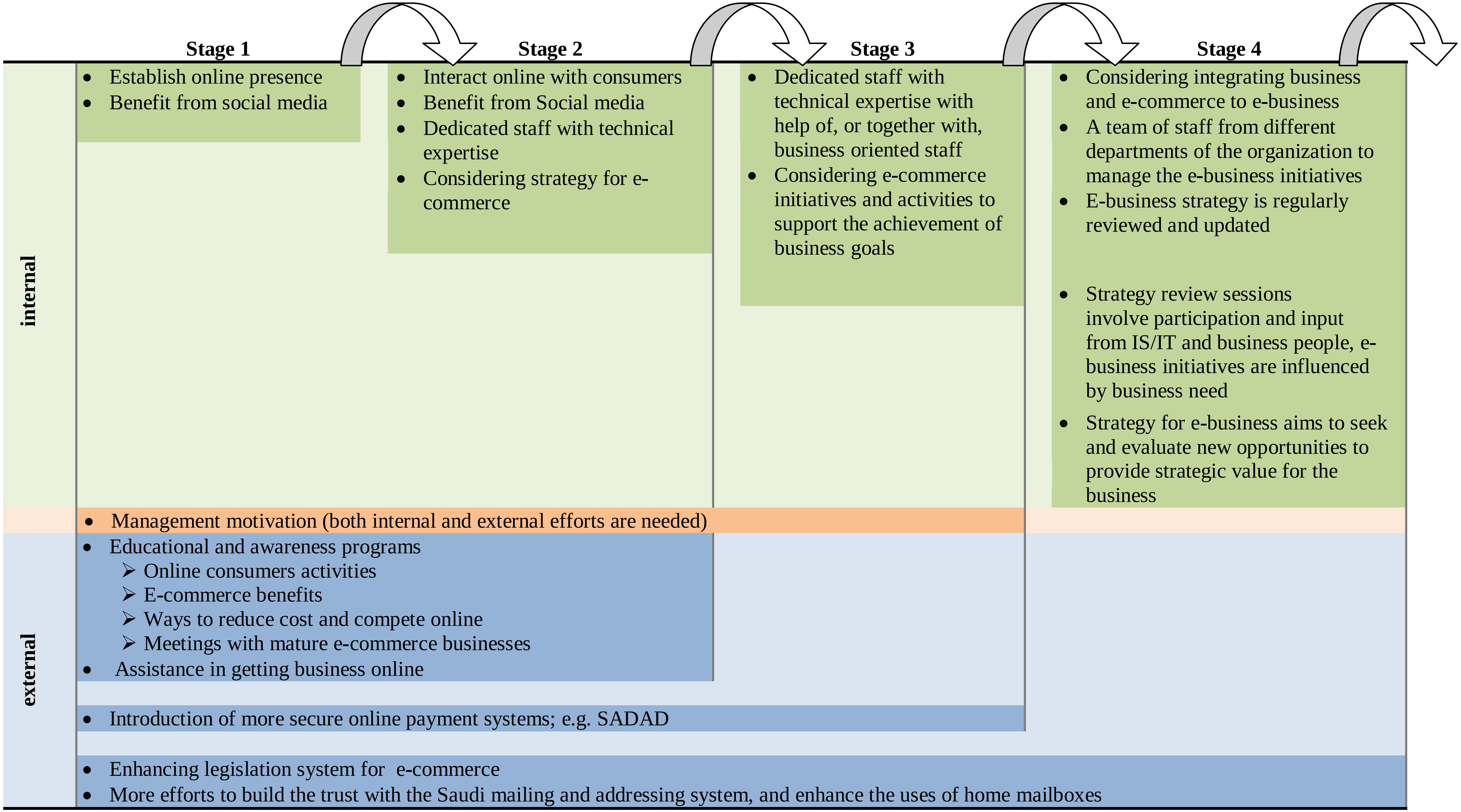

Figure 8.5: stage of growth and e-commerce development for retailers in Saudi Arabia

# CHAPTER 9: CONCLUSION

This thesis explored the factors influencing the slow progress of online retailing in Saudi Arabia. It provided rich insights into the understanding of the adoption of e-commerce based on the concerns and perceptions of retailers in different stage of e-commerce maturity. Through investigating the phenomenon of e-commerce within the context of retailers in companies at different stages of e-commerce maturity, this dissertation increases the knowledge base to find out how to diffuse online retailing in Saudi Arabia and countries with a similar background.

## 9.1. Contribution to Knowledge

The outcomes of this research have led to several achievements and novel contributions. The study investigated the factors influencing retailers' decision on whether or not to adopt e-commerce. This investigation was done in terms of four stages of e-commerce maturity. The outcomes brought out differences between the four stages in relation to consumer and organization factors while showing that there was agreement in relation to environment factors. The environment was recognized as the same by respondents in companies of all the four stages.

This investigation has helped identify the incentives and barriers for each stage of maturity. The study of retailers at different stages of e-commerce maturity is invaluable in suggesting what could be done to increase the likelihood of other company's increasing their stage of e-commerce maturity. The businesses that have reached a high stage of e-commerce maturity provide practical evidence of positive and optimistic attitudes and practices regarding the use of e-commerce, whereas the businesses that have not reached higher stages of maturity provide practical evidence of more negative and pessimistic attitudes and practices. By examining differences between retailers at different stages within the same country, much insight was gained that can be helpful for the future.

Based on the findings of the thesis, recommendations based on differences between retailers at different stages of e-commerce maturity were made, with the hope of contributing to e-commerce development in Saudi Arabia. The study, therefore, contributes to efforts leading to greater e-commerce development in Saudi Arabia. In addition, it shows the importance considering stages of e-commerce maturity in future research. When considering barriers/incentives it has been shown that there are sharp differences in opinion, depending on the stage of e-commerce maturity of retailers. While retailers at all stages were in agreement as to the Saudi Arabian environment, their responses related to consumers and organizational factors were startlingly different. This would be missed if stage of growth was not taken into account and it suggests idea to help move companies from one stage to the next.

## 9.2. Limitations and directions to future research

The current study looked at the diffusion of e-commerce in Saudi Arabia from the perspective of retailers. Since e-commerce is a business that involves other parties, e.g. regulators and consumers, this study is limited by investigating e-commerce only from the retailer side. It may be interesting to see government responses to some of the issues raised, especially those that require action by the government. Since there is agreement between businesses in all four stages of e-commerce maturity about environmental factors, e.g. legislation and online payment systems, it seems there is a need for future research to investigate these concerns with the government.

In addition, since this study showed that companies in lower stages of e-commerce maturity were pessimistic about adopting e-commerce in Saudi Arabia, a more detailed study focusing on the retailers in the higher stages could be very useful. Also, in this regard, it may be useful to consider another study covering a sample of online shoppers in Saudi Arabia, to provide retailers with detailed, objective, information about online consumers' behaviors.

# References and bibliography

# Appendices

# **Appendix A:** Ethical Approval Certificate to Conduct the Research

**Griffith UNIVERSITY**

# HUMAN RESEARCH ETHICS COMMITTEE

## ETHICAL CLEARANCE CERTIFICATE

This certificate generated on 16-02-2012.

This certificate confirms that protocol 'Diffusion of the adoption of online retailing in Saudi Arabia' (GU Protocol Number ICT/06/09/HREC) has ethical clearance from the Griffith University Human Research Ethics Committee (HREC) and has been issued with authorisation to be commenced.

The ethical clearance for this protocol runs from 28-09-2009 to 03-11-2012.

The named members of the research team for this protocol are:

Dr Steve Drew
Mr Rayed Al Ghamdi

The research team has been sent correspondence that lists the standard conditions of ethical clearance that apply to Griffith University protocols.

The HREC is established in accordance with the *National Statement on Ethical Conduct on Research Involving Humans*. The operation of this Committee is outlined in the HREC Standard Operating Procedure, which is available from www.gu.edu.au/or/ethics.

Please do not hesitate to contact me if you have any further queries about this matter.

Dr Gary Allen
Manager, Research Ethics
Office for Research
G39 room 3.55 Gold Coast Campus
Griffith University
Phone: 3735 5585
Facsimile: 07 5552 9058
Email: g.allen@griffith.edu.au

# **Appendix B:** Guidelines for the questions were asked in the interviews' sessions

## About company عن الشركة/المؤسسة بشكل عام

What are your targeted customers/ marketplace?
من هم الشريحة المستهدفة لتسويقكم؟ بمعنى: من هم الزبائن الذين تستهدفونهم؟

What channels do you use to market your products?
ماهي طرقكم الاعتيادية للإعلان والتسويق لمنتجاتكم؟

What are the favorites channels marketing products?
ماهي أفضل طريقة تسويق بالنسبة لكم؟

What do you market?
ماهي المنتجات التي تسوقونها؟

What are the most popular goods that you sell?
ماهي أشهر منتجاتكم التي تبيعونها/ أو التي عليها إقبال كبير من الزبائن؟

Are you part of commercial network?
هل لديكم تعاون (تسويقي/تجاري) مع مؤسسات تجارية أخرى؟

## Internet عن الإنترنت وخبرته

Does your organization have access to the Internet?
هل شركتكم لديها اتصال بالإنترنت؟

What kind of information does your organization's website provide?
ما المعلومات التي يوفرها موقعكم الاليكتروني؟

How do you most often communicate personally (telephone, in person, e-mail, etc)?
كيف تتواصل غالباً مع الآخرين: الهاتف ، شخصي ، بريد اليكتروني ، أخرى؟

Have you bought online?
هل سبق وأن اشتريت من الإنترنت؟

## E-commerce التصورات عن التجارة الإليكترونية

What does your social/professional group think about the risks and benefits of e-commerce?
ماذا عن اعتقاد الناس في مجتمعكم/مؤسستكم حول التجارة والتسوق من الإنترنت؟

Do you think selling online would make more profit for your organization?
هل تعتقد بأن البيع من خلال الإنترنت يحقق لكم أرباح أعلى أو أفضل؟

Do you have any friends in business that do regular online purchasing and selling of goods?
هل لديك أصدقاء يعملون في شركات تسوق وتبيع على الإنترنت؟

## E-commerce in your company عن فكرة السوق الإليكتروني بالمؤسسة

Do you find it difficult/confronting to use online system?
بشكل عام هل تتوقع أن هناك صعوبة في تطبيق نظام التسويق والبيع على الإنترنت؟ ب

What are the most difficult tasks for your organization selling products online?
ماالذي يمنعكم من تسويق منتجاتكم وبيعها على الانترنت؟

Is security of transactions an issue relevant to your decision to use e-commerce?
Is privacy of personal/financial information an issue relevant to your decision to use e-commerce?
Is trust of parties in a transaction an issue relevant to your decision to use e-commerce?
بشكل خاص: هل الأمان والخصوصية في علمليات الدفع الإليكتروني يمثل مشكلة أو عقبة نحو تطبيقكم نظام للتسويق والبيع من خلال الإنترنت؟

What makes it difficult advertising your product online?
ماالعوامل التي تُصعب استخدام موقعكم على الإنترنت لعرض كاتلوج أو نشرة بمعلومات عن منتجاتكم وأسعارها؟

If a sample of a system available to try, would you use it?
Do you want to try a sample before to apply?
هل تودون استخدام برنامج تجريبي للتسويق الإليكتروني قبل التفكير في تطبيقه في شركتكم؟

What changes needed to apply e-commerce system?
ما التغييرات المطلوبة التي تحتاجونها في منظمتكم في حال تفكيركم في تطبيق نظام تجارة إليكترونية؟

# **Appendix C:** Interview information sheet (English version)

# INTERVIEW INFORMATION SHEET

Researcher: *Rayed AlGhamdi:* ***r.alghamdi@student.griffith.edu.au***,
*School of Information and Communication Technology*
Supervisor: *Dr Steve Drew:* ***s.drew@griffith.edu.au***
*School of Information and Communication Technology*

## *Enabling e-Retail B2C eCommerce Systems in Saudi Arabia*

**Dear Sir/Madam,**
I am conducting a research project entitled "Enabling e-Retail B2C eCommerce Systems in Saudi Arabia". I appreciate your participation in this project. Please read the following information of this project.

### The Project Aims:

The project aims to explore what will facilitate the implementation of e-retail B2C e-commerce systems in Saudi Arabia. However, the online survey is for a pilot study to come up with the current situation of e-retail B2C e-commerce systems in Saudi Arabia.

### What is involved?

- Open discussion regarding e-retail B2C eCommerce Systems in your organization.
- The interview session is likely to take up to 60 minutes to complete.
- Interview session is recorded by electronic device. The recorded information will be transcribed to a document and recording will be destroyed.
- Your participation in this project is entirely voluntary.
- Report will be made available on www.shm3h.info/KSAeCommerce by Oct 2010.

### Benefits:

- Through your participation, you are contributing to identify what enable your organization implementing e-retail B2C e-commerce systems.
- In turn, this will help your organization enabling successful implementation of e-retail B2C e-commerce systems.

**For any queries regarding ethical concerns, please contact:**

***The Manager for Research Ethics,***
***Office for Research,***
***Bray Centre, Nathan Campus,***
***Griffith University***
***Ph: +61 7 3735 5585 or***
***research-ethics@griffith.edu.au***

***Computer Dep. Faculty of education,***
***King Abdulaziz University,***
***Jeddah, Saudi Arabia***
***Ph: +966 2 6916343 Ex:3284***
***Fax: +966 2 6914286***
***E-mail: computersciencedep10@gmail.com***

### Legal Privacy Statement:

The conduct of this research involves the collection, access and / or use of your identified personal information. The information collected is confidential and will not be disclosed to third parties without your consent, except to meet government, legal or other regulatory authority requirements. A de-identified copy of this data may be used for other research purposes. However, your anonymity will at all times be safeguarded. For further information consult the University's Privacy Plan at www.griffith.edu.au/ua/aa/vc/pp or telephone (+61 7 3735 5585).

# **Appendix D:** Interview information sheet (Arabic version)

# نشرة معلومات تعريفية

الباحث: رائد بن عبدالله الغامدي

r.alghamdi@student.griffith.edu.au

المشرف: د.ستيف درو

s.drew@griffith.edu.au

قسم تكنولوجيا المعلومات والاتصالات

## عنوان الدراسة:

## تفعيل تطبيق أنظمة الأسواق الإليكترونية بالمملكة العربية السعودية

تعريف السوق الإليكتروني:

يُعرف السوق الإليكتروني على أنه تسويق وبيع المنتجات التجارية من خلال الإنترنت.

### هدف الدراسة:

تهدف هذه الدراسة إلى بحث واكتشاف العوامل التي تقود إلى تسهيل تطبيق أنظمة الأسواق الإليكترونية في المملكة العربية السعودية. أحد الوسائل المستخدمة لتحقيق هذا الهدف هو إجراء مقابلات مع عينة من أصحاب/مدراء الشركات التجارية السعودية.

### ماذا يتعلق بهذه المقابلة؟

- نقاش مفتوح حول إمكانية تطبيق فكرة أنظمة الأسواق الإليكترونية بشركتكم/مؤسستكم.
- تستغرق المقابلة وقت لايزيد عن **60** دقيقة.
- يتم تسجيل المحادثات بجهاز تسجيل صوتي إليكتروني. المعلومات المسجلة سيتم تفريغها كتابيًا وسيتم التخلص من التسجيل الصوتي حفاظًا على الخصوصية.
- مشاركتك في هذه المقابلة تطوعية
- تقرير النتائج المبدئية حول نتائج هذه الدراسة سيكون مُتاح على الرابط: **www.shm3h.info/KSAeCommerce** مع نهاية شهر اكتوبر **2010** إن شاء الله.

### ماذا تعني مشاركتك؟

- مشاركتك قيِّمة في الإسهام في تحديد العوامل التي تقود إلى تفعيل تطبيق أنظمة الأسواق الإليكترونية في المملكة العربية السعودية.

- نتائج هذه الدراسة ستسهم –إن شاء الله– في تفعيل التطبيق الناجح لنظام الأسواق الإليكترونية في شركتكم/منظمتكم التجارية.

## لأية استفسارات حول الشؤون الأخلاقية لهذه الدراسة، يرجى الاتصال بـــ:

*رئيس قسم الحاسب الآلي بكلية التربية*
*جامعة الملك عبدالعزيز*
*جدة، المملكة العربية السعودية*
***Ph: +966 2 6916343 Ex:3284***
***Fax: +966 2 6914286***
***E-mail: computersciencedep10@gmail.com***

***The Manager for Research Ethics,***
***Office for Research,***
***Bray Centre, Nathan Campus,***
***Griffith University, Australia***
***Ph: +61 7 3735 5585 or***
***research-ethics@griffith.edu.au***

## بيان الخصوصية القانونية :

إجراء هذا البحث يشمل جمع والوصول و / أو استخدام المعلومات الشخصية الخاصة بك التي تم تحديدها. المعلومات التي يتم جمعها سرية ولن يتم الكشف عنها إلى أي طرف ثالث دون علمك، ما عدا لمواجهة الحكومة، والمتطلبات القانونية أو غيرها من السلطة التنظيمية. نسخة من هذه البيانات قد تستخدم لأغراض بحثية أخرى، ومع ذلك سيكون لديك ضمانات عدم الكشف عن هويتك في جميع الأوقات. لمزيد من المعلومات يُرجى الإطلاع على الخطة الجامعية للخصوصية على www.griffith.edu.au/ua/aa/vc/pp أو الاتصال على الهاتف (+61 7 3735 5585).

## Legal Privacy Statement:

The conduct of this research involves the collection, access and / or use of your identified personal information. The information collected is confidential and will not be disclosed to third parties without your consent, except to meet government, legal or other regulatory authority requirements. A de-identified copy of this data may be used for other research purposes. However, your anonymity will at all times be safeguarded. For further information consult the University's Privacy Plan at www.griffith.edu.au/ua/aa/vc/pp or telephone (+61 7 3735 5585).

# **Appendix E**: Interview consent form (English version)

## INTERVIEW CONSENT FORM

**Researcher:** ***Rayed AlGhamdi:*** **r.alghamdi@student.griffith.edu.au,**
*School of Information and Communication Technology*
**Supervisor:** ***Dr Steve Drew:*** **s.drew@griffith.edu.au**
*School of Information and Communication Technology*

# *Enabling e-Retail B2C eCommerce Systems in Saudi Arabia*

By signing below, I confirm that I have read and understood the information package and in particular that:

- I understand that my involvement in this research will involve an interview discussion of approximately 30 minutes duration.
- I have had any questions answered to my satisfaction.
- I understand that there will be no direct benefit to me from my participation in this research.
- I understand that my participation in this research is voluntary
- I understand that if I have any additional questions I can contact the research team.
- I understand that I am free to withdraw at any time, without comment or penalty.
- I agree to participate in the project.
- Report will be made available at www.shm3h.info/KSAeCommerce by Oct 2010.

Respondent, Name and Signature.........................................................................

Date....................................................................

Researcher, Name and Signature.......................................................................

Date....................................................................

# **Appendix F:** Interview consent form (Arabic version)

# نموذج موافقة على إجراء مقابلة

الباحث: رائد بن عبدالله الغامدي
r.alghamdi@student.griffith.edu.au
المشرف: د.ستيف درو
s.drew@griffith.edu.au
قسم تكنولوجيا المعلومات والاتصالات

## تفعيل تطبيق أنظمة الأسواق الإليكترونية بالمملكة العربية السعودية

بالتوقيع على هذا النموذج، أوافق بأنني اطلعت بشكل كامل على نشرة المعلومات التعريفية لهذه الدراسة، وبشكل خاص المعلومات التالية:

- مشاركتي في هذا البحث يشمل مقابلة نقاش مدتها لاتزيد عن **30** دقيقة.
- لن تكون هناك فائدة مباشرة بالنسبة لي من مشاركتي في هذا البحث.
- مشاركتي في هذه المقابلة تطوعية.
- إذا كان لدي أي أسئلة أو ملاحظات إضافية باستطاعتي الاتصال بفريق البحث.
- من حقي الانسحاب من المقابلة في أي وقت أرغب بدون تقديم أي تبرير.
- أوافق على المشاركة في هذا البحث (المقابلة).
- تقرير النتائج المبدئية حول نتائج هذه الدراسة سيكون مُتاح على الرابط:

**www.shm3h.info/KSAeCommerce** مع نهاية شهر اكتوبر **2010** إن شاء الله.

اسم وتوقيع المشارك

.......................................................................................................................................

التاريخ:   /   / 2010

اسم وتوقيع الباحث

.......................................................................................................................................

التاريخ:   /   / 2010

# **Appendix G:** Full 16 interviews transcripts

| Participant name: Ahmed | |
|---|---|
| Participant position: Marketing manger | Having website: Yes |
| Organization size: Large | Selling online: No |
| Business Category: Beauty products | E-commerce maturity stage: 2 |
| | |
| Interview form: in person interview | Language of conversation: Arabic |

*Document Created On*: 28/09/2010 6:37:58 PM
*Modified On*: 9/08/2013 2:56:00 PM
What are your targeted customers/ marketplace?
70% ladies and the rest men
This is because our products are Makeup, cosmetics and body lotions, shampoos and skin care.
What channels do you use to market your products?
Internet, SMS, and weakly & monthly magazines
What are the favorites channels marketing products?
Direct marketing in our shops
What do you market/what are your products?
Cosmetics, preparations for hair care and skin, and materials and creams for body care
What are the most popular goods that you sell?
Our products are well-known brand in worldwide; however the most popular goods are perfumes.
Does your organization have access to the Internet?
Yes
Does your company have a website?
We, as agent, have no relationship with the general website of the main company, but all the products displayed on the website of the parent company, we have shown on our website.
What kind of information does your organization's website provide?
Our website display information about our company and products but there is no sales or prices shown.
Have you bought online?
Yes, airline tickets only
What does your social/professional group think about the risks and benefits of e-commerce?
The percentage of users who browse the Internet is high, however those who sell or buy is very small. Culture of the people to accept e-shopping is not encouraging us to open the electronic market.
Do you think selling online would make more profit for your organization?
Yes of course selling online would make more profits, but we have a different marketing strategy which is having branches/shops in the main malls to be directly close to customers.
Do you find it difficult/confronting to use online system?
Recently we have bought powerful software, but unfortunately the culture of our customers is not encouraging us to go ahead with this idea. Customers do not trust to buy online and they only accept with strike conditions, because it requires credit cards payment. Recently the idea of having Internet payment cards has increased. There is International website where you can buy from and delivered to your home by International delivery companies like DHL, but internally in Saudi Arabia so far is not encouraging.
*Do you think the main problem is trust?*
Yes of course trust is the main problem. The system that protect customer right when he/she pays, he/she will receive the purchased products/services.

*Do you mean there is no clear system to protect sellers and customers rights in this field?*

Yes sure there is no clear system to protect sellers and custoemrs rights. It has to be done from early stage, customers have to be ensured there are rules and legislations protect their right. This is very important to build the trust with customers. It has to have a clear system, i.e. from government, Imposed on all involved parties. .

What makes it difficult for your company marketing and selling products online?

We have tried to market on the Internet But there is no demand, because our products related to smell, shape, and the customer wants to see physical product in front of him/her. Also the culture of people is not encouraging.

What makes it difficult advertising your product online/using online catalog?

Our products are displayed on our website but not like catalog.

What changes needed to apply e-commerce system?

We have to make contract with IT Company to build complete e-commerce system, train employees to take care of this new way of marketing and selling, offer support and after sales services. In short words, we need to have complete e-readiness.

I have a message to Saudi commerce ministry. I advice them to work harder to apply e-commerce systems. I advice the main supporter in this field is to find out a way to build the honesty/trust between customers and sellers. I suggest government department/s to take care of doing this.

I can see also one of the inhibitors is delivery fees or insurance. If the products available in the local market with the same price then the company which offer selling these products have no competitive advantage.

| | |
|---|---|
| Participant name: Ali | |
| Participant position: Owner | Having website: No |
| Organization size: Small | Selling online: No |
| Business Category: Appliances and home improvement | E-commerce maturity stage: 1 |
| | |
| Interview form: in person interview | Language of conversation: Arabic |

*Document Created On*: 28/09/2010 6:37:59 PM

*Modified On*: 9/08/2013 2:56:17 PM

What are your targeted customers/ marketplace?

All

What channels do you use to market your products?

The location of our shops

What do you market/what are your products?

Homes equipment needs: washing machines, ovens, fridges, air conditionings and TVs

What are the most popular goods that you sell?

TVs and air conditionings

Does your organization have access to the Internet?

No

Does your company have a website?

Not at all

Have you bought online?

No

What does your social/professional group think about the risks and benefits of e-commerce?

This depends on Familiarity. We are familiar with the way of normal selling where a customer has to come to our shop and see the products.

Do you think selling online would make more profit for your organization?

No, it will not be high because it cannot be compared to normal selling process. It is much better for us when a customer comes to the shop.
People here are not familiar with e-commerce and this is not something that you can apply it in a day. People here have to come to the shop to see the product and inspect it.
We prefer personal marketing it is much better for us. When you explain to a customer face to face is totally different from leaving him alone to read the feautures of a product on the Internet.
This may be benefit large and famous companies.
Do you have any friends in business that do regular online purchasing and selling of goods?
Yes I know two friends buy online but rarely.
Do you find it difficult/confronting to use online system?
We need to train our employees. Our employees are familiar with direct sales.
We do not trust online payment, the money must paid cash
The majority of similar companies, %99, do not have e-commerce websites and shops are widespread where we cannot have competitive advantage to sell online.
What makes it difficult for your company marketing and selling products online?
We will not benefit because customers are not familiar with this idea.
Is security of transactions an issue relevant to your decision to use e-commerce? Is privacy of personal/financial information an issue relevant to your decision to use e-commerce?
Yes. The majority of people in our country do not know how to buy and sell on the Internet. The ignorance of something makes it difficult.
What makes it difficult advertising your product online/using online catalog?
I say to you: specifically the nature of our business makes it difficult to go ahead with this idea. Same products are widespread in the market. Yes I agree to go ahead with this idea if there is products are not available in the local market where we can have competitive advantage. However electronic and electrical shops are almost in each street in Jeddah which makes the idea of e-commerce not useful for us.
If a sample of a system available to try, would you use it? Do you want to try a sample before to apply?
We do not think about e-commerce at all. The normal way of selling is the best for us and we do not want to sell online at all.
*In your opinion, what makes it difficult for retailers to sell inline?*
The familiarity, Sellers and customers are not familiar with online sales and purchases. This needs educational activities to remove this inhibitor at all. However in our business of selling electronic and electrical products the normal way is the best. With time may be the situation improves and people accept this idea.

| Participant name: Waleed | |
|---|---|
| Participant position: Owner | Having website: No |
| Organization size: Small | Selling online: No |
| Business Category: Audio/video | E-commerce maturity stage: 1 |
| | |
| Interview form: in person interview | Language of conversation: Arabic |

*Document Created On*: 28/09/2010 6:38:00 PM
*Modified On*: 10/08/2013 1:11:45 AM
What are your targeted customers/ marketplace?
All customers who are interested in audio recordings
What channels do you use to market your products?
SMS and brochures to advertise the latest versions
We have free service to deliver the goods. We are the exclusive company in this regard which deliver for free inside Jeddah. Even if the order is cheap we can deliver especially if the customer does not have transport. A customer can call us by phone to make an order and then we deliver the order with one of our staff if it inside Jeddah or with Alzajel shipment company if it outside Jeddah. Normally the order outside Jeddah takes 2 days to arrive.
What are the favorites channels marketing products?
The best way is taking samples of new releases and we go by, for example, some schools and give the director to display to teachers and students, and then send to us a list of orders.
What do you market/what are your products?
All Islamic audio records (cassettes and CDs), Islamic books and brochures
What are the most popular goods that you sell?
Currently Islamic brochures and Supplications cards
The sales of cassettes declined since 2003 because the cars become CDs audio systems.
*Do you have an idea to transfer from cassettes to CDs?*
No, we have customers for both. Cassettes have customers as well as CDs.
Does your organization have access to the Internet?
No
Have you bought online?
No
What does your social/professional group think about the risks and benefits of e-commerce?
In the recent years our society has received several shocks: stoke market radical decline, Sawa cards, lands and fake projects! All of these have affected the confidence and trust of people and especially on the Internet. Add to that our culture buying and selling online is very weak. Whoever sells or buys should be professional. Unaware and uneducated person find it difficult and frightening. So I can see e-commerce is limited to few International companies and those who deal with them are specific and limited customers. The Internet is Internet fertile ground for cheating and fraud. I think the acceptance to sell online will remain weak.
Do you think selling online would make more profit for your organization?
This depends on the strength of the company, and the competitive advantage offered by its products so that there are no such products on the market. Having no similar products in the local market help to achieve high profits, but if the products are available in the market, I believe that profits will be very weak.
Do you have any friends in business that do regular online purchasing and selling of goods?
No
Do you find it difficult/confronting to use online system?

You need to educate people first and spread good understanding for e-commerce. The absence of the habit to buy and sell online makes it very difficult to progress in this field. We are growing community, and the idea of e-commerce is new to us, and we need time to understand these things.
Another issue, we are a society subjected to several issues of fraud, such as stoke market radical decline, Sawa cards, lands and fake projects! We did not find a clear law to protect our rights in these issues. Now if we go ahead with selling and buying online the same problem happens in the absence of clear laws and regulations. With no clear e-commerce law and legislations applying e-commerce systems are not encouraging. In the future the environment may be encouraging, but currently the situation is not encouraging.
What makes it difficult for your company marketing and selling products online?
Financial costs involved in something that does not bring profit. As I told you earlier this type of business is not useful currently and returns very low profits.
Is security of transactions an issue relevant to your decision to use e-commerce? Is privacy of personal/financial information an issue relevant to your decision to use e-commerce?
Yes this cause a difficulty
If a sample of a system available to try, would you use it? Do you want to try a sample before to apply?
Yes we can try and this experience give us an idea how it works and how profitable will be before we involve in paying on something that is not useful.
Currently we have orders by phone. For example a customer from Algoreeat (a city 1000 km far from Jeddah) calls us to make an order. We send to him/her bank account number to deposit the money and then we send their order with Alzagel Shipment Company with 10 SR delivery fees. The order is received within 4 days. We only take the name and mobile phone number of the customer. When the order arrives in the shipment office in the customer city, the shipment company calls him/her to come to take his/her product from the office but if he/she cannot some he/she requires paying extra fees to deliver to home.
What changes needed to apply e-commerce system?
To have complete e-readiness

| Participant name: Fadi | |
|---|---|
| Participant position: Director | Having website: Yes |
| Organization size: Medium | Selling online: No |
| Business Category: Groceries | E-commerce maturity stage: 2 |
| | |
| Interview form: phone interview | Language of conversation: Arabic |

*Document Created On*: 28/09/2010 6:37:58 PM
*Modified On*: 10/08/2013 1:03:29 AM
What are your targeted customers/ marketplace?
All people
The company also has its delivery system for those who order by phones.
What channels do you use to market your products?
Newspapers, brochures, TV
What do you market/what are your products?
Pizza
Does your organization have access to the Internet?
Yes
Does your company have a website?
Yes
What kind of information does your organization’s website provide?

Information about all available types of pizza but prices and online order is not available.
All what I have to say regarding this topic, using the Internet to order our products is not our priority at the moment and we have not think about it or make study to find out what needs to be done. Thank you!

| Participant name: Hassan | |
|---|---|
| Participant position: Director | Having website: No |
| Organization size: Small | Selling online: No |
| Business Category: Kitchen utensils | E-commerce maturity stage: 1 |
| | |
| Interview form: in person interview | Language of conversation: Arabic |

*Document Created On*: 28/09/2010 6:37:57 PM
*Modified On*: 9/08/2013 2:59:38 PM
What are your targeted customers/ marketplace?
All people
What channels do you use to market your products?
The place of our shops in very important in term of advertisement which means the good location of the shop represents advertisement means for us. Also we use brochures.
What do you market/what are your products?
Household utensils, some electrical stuff, and home plastics needs.
What are the most popular goods that you sell?
Every product has its customers. Almost all our products receive same attention from customers.
Does your organization have access to the Internet?
Yes we have in big branches only.
Does your company have a website?
No
Have you bought online?
No
What does your social/professional group think about the risks and benefits of e-commerce?
This is not similar to normal way of selling where a customer comes to the shop, see a product, and make sure he/she selects the right product. Yes there are some people like the Internet.
For ourselves there is no difficulty, but we have concern regarding delivering the purchased products and its fees, and also our stuff don't know about marketing and selling on the Internet where they required training programs.
Do you think selling online would make more profit for your organization?
Using the Internet increases the chance of popularity. Today almost every home has access to the Internet, many people browse it and of course this brings customers to your shop.
Do you have any friends in business that do regular online purchasing and selling of goods?
Yes, I have friends make orders from overseas.
*Having friends buying/selling online, is not this motivating you to do so?*
Yes true, but each company has its unique system.
Do you find it difficult/confronting to use online system?
No, there is no difficulty. Nowadays people aware and as I said to you almost %90 of people use the Internet and people become well educated in this matter.

For our company we have to do training programs and sometimes to deliver the products represent an inhibitor. Also the customer maybe will not accept to pay delivering fees.
*Some companies put a price minimum limit to deliver for free, but when the order doesn't reach the price minimum limit the customer has to pay for delivering fees. Do you know this?*
Yes
What makes it difficult for your company marketing and selling products online?
For small shops/companies this is not useful. The manufacturing companies already have their products demonstrated on the Internet. For us there is no difficulty but as a manufacturing company (like Alsaef) which we import products from them they have a website. Take this example, Alsaef's products already demonstrated on the manufacturing company website, so there is no benefit for us to re invent the wheel!
Is security of transactions an issue relevant to your decision to use e-commerce? Is privacy of personal/financial information an issue relevant to your decision to use e-commerce?
Online payment maybe represents an inhibitor. However, this matter needs education for both sellers and buyers. There has to be educational programs telling what the options available are and indicating the easiness, positive, and negative aspects of each option. Educational programs are very important. When you know the positives and negatives and get the minimum limit of knowledge and awareness the matter will become easy to understand. Education plays major role in this matter.
If a sample of a system available to try, would you use it? Do you want to try a sample before to apply?
Yes sure. This is a good motivation and easy for seller and buyer.
What changes needed to apply e-commerce system?
Training programs for our stuff, maybe doing agreement with delivering/shipment companies, and the other things are easy InshAllah. Thank you

| Participant name: Moneer | |
|---|---|
| Participant position: Director | Having website: Yes |
| Organization size: Large | Selling online: Yes |
| Business Category: Telecommunications | E-commerce maturity stage: 4 |
| | |
| Interview form: in person interview | Language of conversation: Arabic |

*Document Created On*: 28/09/2010 6:37:58 PM
*Modified On*: 9/08/2013 3:00:29 PM
What are your targeted customers/ marketplace?
First we are Internet services provider and we have several departments to serve large and medium companies, and small companies and individuals. We sell to those who buy DSL cards/services. So, our targeted customer is all.
Our company has branches in Riyadh, Jeddah, and Dammam. These are our main branches in the main cities where the majority of people live in. We have also Sales Representative visiting other cities to market our products.
What channels do you use to market your products?
We use newspapers, radio, websites which have many visitors. The owner of a website who tells us they have many visitors, we advertise on their website. We also have exhibitions. We do also sponsor exhibitions, seminars and conferences.
What are the favorites channels marketing products?
It depends on the targeted customers, for individuals we prefer websites and radio. For example we select sports websites to advertise when we want to target sports lovers, and select educational websites when we want to target students and academics, we may also sponsor educational conferences. With businessmen, we use newspapers that mainly targeted for them.
What do you market/what are your products?
Our services are mainly to connect our customers with the Internet, individuals via DSL services, and companies via leaves lines. We also sell services to send sms as these used in telecommunications companies. These services usually used by marketing companies, or hospitals to send reminders for patients' appointments and so on. We also have IT projects in general networking and connections.
What are the most popular goods that you sell?
DSL services
Does your organization have access to the Internet?
Yes
Does your company have a website?
Yes
What kind of information does your organization's website provide?
E-commerce website selling DSL services
For us, we only sell services. When the customer pays directly he/she receive user name and password to use the service. We have after sale support to help our customers when they face any difficulties to set up settings and so on. So we do not have anything related to delivery, we only sell DSL services.
Do you find it difficult/confronting to use online system?
Educating people
Security issues
Connecting with strong companies' servers to ensure 24/7 access
Is security of transactions an issue relevant to your decision to use e-commerce? Is privacy of personal/financial information an issue relevant to your decision to use e-commerce?
Yes, people in general tend not to use credit cards to buy online except those who used to do. However people here afraid to buy online.
*What are the options of online payment that your company provides?*
We provide two options, credit cards and Sadad system. Sadad system is more secure system and great. With this system there is no need to entre personal payment details on our website. For those who choose to pay using this option, they receive ID number that is identified in Sadad system and they use their bank accounts to pay. It is like a bill giving to a customer when paid they can receive their products/services. Sadad is great idea and more secure than credit cards and encourage people to buy online.
*Is Sadad system encourages people to buy online?*
Yes. I personally feel more comfortable with it. I use it when I book airline tickets. When I do online purchase they provide me with a number and time to pay with Sadad system in 4 or 10 hours. I pay through my bank account and automatically they receive payment notification to complete the purchase process.
What changes needed to apply e-commerce system?
computer hardware and software
contract with a company to arrange with online payment
Employees to follow-up 24h support
Subscribe with an Internet company (ISP)
Contract with delivery companies
With delivery there is a problem if there are delivery fees, customer maybe not paying these fees except those from specific class or busy etc. This issue only with physical products but with services and software it is easy to get it only download.

| Participant name: Naif | |
|---|---|
| Participant position: Owner | Having website: Yes |
| Organization size: Medium | Selling online: No |
| Business Category: Watches | E-commerce maturity stage: 2 |
| | |
| Interview form: Phone interview | Language of conversation: Arabic |

*Document Created On*: 28/09/2010 6:37:59 PM
*Modified On*: 9/08/2013 3:05:07 PM
What are your targeted customers/ marketplace?
Individuals
What channels do you use to market your products?
Newspapers and advertisement boards on the streets
What are the favorites channels marketing products?
Newspapers
What do you market/what are your products?
Watches
What are the most popular goods that you sell?
Islamic watches that show the times of the Muslim five daily prayers in many cities around the world and show Islamic finder with Gregorian calendar
Does your organization have access to the Internet?
Yes
Does your company have a website?
Yes we have website shows information about our products in Arabic and English. Also we have e-commerce website mainly targeted for international buyers because the Purchasing power comes from overseas and that's why this website in English.
*Why not targeting Saudi marketplace?*
Our e-commerce website is not targeting consumers in Saudi Arabia and we don't have a priority to do so. This is because all our products are Plentiful in Saudi market. So, we don't have a competitive advantage in Saudi market to encourage buying online. This is why we make our main target selling online for International market. Islamic watches and clocks that we produce and sell are unique products in the international market.
Do you find it difficult/confronting to use online system?
Not really, but in Saudi market several difficulties appear: selling products online where the local market has similar plentiful and cheap products, not affording options for online payment.
Having more options for online payment other than credit cards only is an Encouraging factor to buy online. We currently use in our e-commerce system more than one option for online payment. One option is using PayPal. PayPal is a secure online payment mean for both customer and seller. It plays intermediary mean between credit card and e-commerce website. This system of online payment builds the trust also with the seller who uses it. It represents a Company of e-payment and collection that tests new companies that deal with them. For example, they have instructions not to withdraw money for the new sellers before specific period. This procedure is to ensure that the new company is serious and no complaints received from customers. If there is a complaint, they investigate and return the money to customers if the complaint is true. With this procedure customers feel happy to deal with this intermediary e-payment option which protects their rights and also build the trust with companies that deal with PayPal. So, we think if there is a similar system available in Saudi Arabia and in Arabic language will encourage to build trust which is very important factor to buy online.

| Participant name: Nasir | |
|---|---|
| Participant position: CEO | Having website: Yes |
| Organization size: Medium | Selling online: No |
| Business Category: Books | E-commerce maturity stage: 2 |
| | |
| Interview form: in person interview | Language of conversation: Arabic |

*Document Created On*: 28/09/2010 6:37:58 PM
*Modified On*: 12/08/2013 12:00:19 AM
What are your targeted customers/ marketplace?
University students
This is because most of our products targeted to them
What channels do you use to market your products?
We depend on the location of our shops. We have 5 shops in Jeddah. Sometimes we participate in internal and international Exhibitions.
What are the favorites channels marketing products?
The location of our shops is the favorite marketing channel for us. Having shops close to the common area of targeted customers (i.e. university) is the best way to market your products.
What do you market/what are your products?
Books and office and study equipments
What are the most popular goods that you sell?
Books
Does your organization have access to the Internet?
Yes we have access to the Internet in admin offices only.
What kind of information does your organization's website provide?
*Your website provides information about books and their prices. There is a possibility to do online request but there is no option for payment. When doing online request a customer has only one option to leave his contact details and a message appear he/she will be contacted soon. What happen next?*
Actually there are no sales on the Internet and honestly we do not contact customers who do such a request like this. What is on our website is display only. Sometimes some customers email us their requests. We check our e-mail every day. We distribute to students small brochures include information about books and contact details. They send to us their lists of books by Fax or e-mail and we arrange the books for them. Students refer to our website to browse information about books. We do not use the website to sell online because the online payment is not available. We only have bank account number that we give to those who would like us to deliver their orders to home. We receive their orders, *by fax or e-mail as explained before*, and we give them our bank account number to deposit the money. As soon as we receive the money we send the books with delivering companies like FedEx, DHL, or other shipment companies.
*If a customer, for example in Riyadh, sends his/her order by e-mail, do you give him/her the bank account number to deposit the money to send the books?*
Sometimes customers in Australia, Jordan and other countries contact us to ask if we have books that are not available in the market. We response as I told you before. We give them the bank account number to deposit the money and then we directly contact FedEx to send his/her order within 48 hours wherever he/she is.
Have you bought online?
No
What does your social/professional group think about the risks and benefits of e-commerce?
Honestly there is Ignorance inside our company in terms of e-commerce, and that is why our website only disply products and not receive orders. Consumers send to us their lists of books by Fax or e-mail and we arrange the books for them. Students refer to our website to browse information about books. We do not use the website to sell online because the online payment is not available. We only have bank account number that we give to those who would like us to deliver their orders to home. We receive their orders, *by fax or e-mail as explained before*, and we give them our bank account number to deposit the money. As soon as we receive the money we send the books with delivering companies like FedEx, DHL, or other shipment companies.
Also there is Ignorance in the community. Most of them don't know the meaning of e-commerce as result they will not venture. There are people when they pay online remain in doubt and not sure of receiving their purchased orders. There is fear, because ignorance breeds fear. There is a doubt to deal by money on the Internet.
Do you think selling online would make more profit for your organization?
We see the normal way of selling is the best. Selling online like Amazon.com is not useful for us in this country and all Arabic countries. This subject refers to peoples' culture of selling and buying. So the normal way of business is the best for us. Although there are more things that you can save when you transfer to sell online like tools, equipment, and rent of shops, but I don't think the Internet help us to achieve more profit. The normal way of selling remains the best for us to gain more profit.
Do you have any friends in business that do regular online purchasing and selling of goods?
No
Do you find it difficult/confronting to use online system?
No, there is no difficulty but this issue refers to the culture of people to accept and use.
What makes it difficult for your company marketing and selling products online?
Culture of people makes no benefit for us to sell online
Is security of transactions an issue relevant to your decision to use e-commerce? Is privacy of personal/financial information an issue relevant to your decision to use e-commerce?
Yes we don't trust online payment and our customers don't trust too. The ignorance of something breeds fear from it. There are thefts of Credit Cards numbers, and there are hackers penetrate your privacy. So this is a problem for the money dealing on the Internet.
If a sample of a system available to try, would you use it? Do you want to try a sample before to apply?
Yes this is good and hopeful to understand. Experiencing something is part of education.
What changes needed to apply e-commerce system?
We need computers and network, and employees' training programs in all involved e-commerce activities.
*Your company, sometimes, receive orders by phone, fax, or e-mail, money deposited into your bank account and you send with shipment companies, How do you evaluate this way of selling?*
This depends on the season. For example, we are the first company which apply this way of selling, the sales was excellent. We have started 4 years ago and the sales were high and started going down. This is because other bookshops have started to do the same thing that we do.

| Participant name: Osam | |
|---|---|
| Participant position: Regional director | Having website: Yes |
| Organization size: Large | Selling online: No |
| Business Category: Telecommunications | E-commerce maturity stage: 3 |
| | |
| Interview form: in person interview | Language of conversation: Arabic |

*Document Created On*: 28/09/2010 6:37:59 PM
*Modified On*: 9/08/2013 11:39:25 PM
What are your targeted customers/ marketplace?
We, as communications sector, aims to provide the service to anyone in need. Now phones and mobile phones has become a necessity of life. We serve 15 years old and above and we have specific products for each type of people: youth, women, VIP and so on.
What channels do you use to market your products?
Primarily we depend on the ads in the streets, roads, and television ads. As well as we advertise on newspapers and radio, and the normal way in the shops.
What do you market/what are your products?
We have more than 50 products and services; however the main services that we provide are limited in voice and data. Voice is contact service: phones lines and mobile phones SIM cards. SIM cards have different types, each type specifically made for specific class. Data is Internet services. We connect users to the Internet as we are ISP (Internet Service Provider).
Does your organization have access to the Internet?
Yes
What kind of information does your organization's website provide?
Our website was designed for two purposes: products marketing and provide the basic information that users need (e.g. calls cost …), the second purpose is to provide e-services for our customers such as paying bills, view statement and its analysis, check calls record …
We believe that most of the company's services should be available on its website to save a customer's time and effort visiting the shop. Visiting shops are required for our customers when there are physical products need to be purchased. Even physical products can be sold online but it requires us more careful thinking.
Do you find it difficult/confronting to use online system?
In general people have to become familiar with this system. Our current website provides e-services to pay bill and recharge for pre paid plans. The problem is mail address; still some people do not have mail address. Also there is another problem that people are afraid to buy online using credit cards. However, our society is classified as a youth community (most of the community are young people), this new generation will strongly accept this new technology and new type of business.
What makes it difficult for your company marketing and selling products online?
There are no inhibitors for companies to sell online but there is two points: (1) there is a need for effective and easy Internet connection, (2) well education and psychological comfort. Today almost every house has Internet connection and I believe Internet usages will increase in the future. However each new idea cause fear in the beginning. We, *Saudi community*, are still fairly new in the use of electronic commerce.
Is security of transactions an issue relevant to your decision to use e-commerce? Is privacy of personal/financial information an issue relevant to your decision to use e-commerce?
We have service to sell online but currently the response cannot be compared to the normal way. Culture of people plays a role. People are reluctant to use credit Cards. I think that this fear is exaggerated. Banks should provide easy options to have two credit cards, one with large amount of money and another one with small amount to be used in online payment. This idea makes people get used to and this will remove their fear.
We provide two options for online payment: credit cards and direct debit. However the second option is less accepted.

| Participant name: Saeed | |
|---|---|
| Participant position: Owner | Having website: No |
| Organization size: Small | Selling online: No |
| Business Category: Grocery | E-commerce maturity stage: 1 |
| | |
| Interview form: in person interview | Language of conversation: Arabic |

*Document Created On*: 28/09/2010 6:37:59 PM
*Modified On*: 9/08/2013 3:08:55 PM
What are your targeted customers/ marketplace?
All
What channels do you use to market your products?
The location of our shop
What are the favorites channels marketing products?
We are a supermarket selling food and home needs in this regard
What do you market/what are your products?
Noting specific, all our products sold every day
Does your organization have access to the Internet?
No
Does your company have a website?
No
Have you bought online?
No
What does your social/professional group think about the risks and benefits of e-commerce?
Currently it is not useful
Do you think selling online would make more profit for your organization?
No, currently it does not benefit us because people are not interested in this sort of thing.
Do you have any friends in business that do regular online purchasing and selling of goods?
No
Do you find it difficult/confronting to use online system?
Yes, you have to design a commercial website, train your employees, deliver goods. Delivering goods need special care with some of our products because they require to be stored in a specific temperature to delivered safely and healthy.
What makes it difficult for your company marketing and selling products online?
Because currently there is no rush to e-shopping in Saudi Arabia and the culture of people is not encouraging, so it is not useful for us.
Is security of transactions an issue relevant to your decision to use e-commerce? Is privacy of personal/financial information an issue relevant to your decision to use e-commerce?
Yes
What makes it difficult advertising your product online/using online catalog?
It is not useful because supermarket and mini markets are in every street.
Currently applying e-commerce system is not useful and will not benefit us. This issue requires education and training. Educated people accept this idea. Education plays key role to adopt this new technology.

| Participant name: Thamer | |
|---|---|
| Participant position: Regional Director | Having website: Yes |
| Organization size: Large | Selling online: Yes |
| Business Category: Sporting products | E-commerce maturity stage: 4 |
| | |
| Interview form: in person interview | Language of conversation: Arabic |

*Document Created On*: 28/09/2010 6:37:56 PM
*Modified On*: 10/08/2013 12:18:39 AM

What are your targeted customers/ marketplace?
We target middle class and above because our products are high quality sports equipment for homes uses and are a little expensive.
We target youth too because this group of ages concerns to have sporty bodies.
What channels do you use to market your products?
We use newspapers (every month 3 to 4 advertisements), TVs, SMS, Brochures, catalogs, banners to advertise special offers, websites and direct personal marketing. Our company has also special TV channel.
What do you market/what are your products?
Our products are sports equipment for home use. Our Mission: Sport - Fitness - Massage - Health and beauty.
What are the most popular goods that you sell?
Treadmills, exercise bikes, massage balls, and team sports games (Billiards, Tennis ...)
Does your organization have access to the Internet?
Yes
Does your company have a website?
Yes
What kind of information does your organization’s website provide?
Integrated e-commerce website
What does your social/professional group think about the risks and benefits of e-commerce?
People here used to, for long time, sell and buy directly. This means they go to shops to see and check the goods before to buy. This is the Prevailing habit of people in terms of selling and buying. This is the culture of Saudi people and Arab communities. However, from 4 years some customers have started to contact us through the website. This led us to launch an e-commerce website making it available for customers to complete their purchases online. But there is no good demand to buy online. For example, in western region of Saudi Arabia we receive 3-5 orders buying online every week and these are nothing compared to normal way of business. However we will continue providing this option (selling online) to encourage people to use it.
There is specific class in the community, according to a study we conducted 6 months ago, prefers to buy online. This class has a previous experience studying overseas in western countries and gets used to buy online. They have become comfortable with online purchases.
To use the Internet for selling and buying we must have infrastructure right. There is a movement and growth toward e-commerce but it is very slow.
*When have you launched your e-commerce website?*
Actually e-commerce website has launched since 2009 and the general website since 2007.
*You have said the online sales in your company cannot be compared to the normal way of selling, to what reasons do you refer this tendency?*
We make %10 off on online sales however the problem with people culture to accept this new way of buying. We continue sell online to encourage people to buy online.

Is security of transactions an issue relevant to your decision to use e-commerce? Is privacy of personal/financial information an issue relevant to your decision to use e-commerce?

Yes this maybe represents a difficulty confronting customers to buy online. They are afraid to put their credit cards number on the internet. I suggest for local banks to offer other easy options for online payment to encourage people to purchase online.

What changes needed to apply e-commerce system?

Before we applied e-commerce system, we have done many things: (1) build accurate networking system to Implemented the order directly without delay, (2) build internal system connecting all departments of the company: purchase, finance, warehouse, delivery and admin.

*From your view points, what prevent e-retail systems growth in KSA?*

To correctly apply e-shops, there must be pre combination of factors to be done:

Infrastructure right

Education

Regarding the education I have comment: Why we do not educate students in the classes about e-commerce. It's better to educate students how to use the Internet and benefit well rather than wasting their times on social networks and games!

*Does the delivery cause a problem for your company?*

Our company has its own delivery system by the company's cars and trucks. We deliver for free inside main cities and with delivery fees outside the main cities. We have our own Addressing system and the delivery done within 3 days. It is not an obstacle for us.

*What do you mean by Infrastructure?*

I mean the telecommunication networks inside the city and the country to be at high speed and high efficiency. As you know e-commerce systems is click and implement without delay. So the telecommunication networks should be at standard level to serve the activities of e-commerce, this is one point. (2) As much as possible no interruption. If a customer made an online purchase and he/she don't receive the order in the expected time he/she will not be satisfied and trust will be affected. The problem may be not with delivery system, it may be with the system itself not sending/making delay in the order request. (3) Having an integrated system such as Saudi Stock Market. If there is something similar for e-commerce it will be great. Having an integrated system for e-commerce is very important in terms of having people take care of the system, search for problems and solutions, and leading toward development. I believe that Saudi market is bullish and a fertile ground for investments, but –in my opinion- even we start thinking about e-commerce we need time not less than 10 years to reach the maturity in this field.

*You have said we need to have "an integrated system such as Saudi Stock Market", do you mean should be there a government department/s to be responsible about this?*

Yes this is an encouraging factor. Citizens will have more trust if this subject sponsored by the government because we, in Saudi Arabia, have great confidence in anything that comes through the government. The evidence for this Saudi Arabia was the first country that has the lowest effects of financial crises in 2009.

| Participant name: Tameem | |
|---|---|
| Participant position: Director | Having website: No |
| Organization size: Medium | Selling online: No |
| Business Category: Printing services | E-commerce maturity stage: 1 |
| | |
| Interview form: in person interview | Language of conversation: Arabic |

*Document Created On*: 28/09/2010 6:37:57 PM
*Modified On*: 10/08/2013 12:31:40 PM
What are your targeted customers/ marketplace?
Our targeted customers all people who need printings, stamps etc. Most of our customers are students, staff and families.
What channels do you use to market your products?
We advertise by Brochures, on newspapers and special newspaper for ads.
What are the favorites channels marketing products?
The favorite channel to market our products is Brochures.
What do you market/what are your products?
Our products are shields written on them letters of appreciation, stamps, business cards, wedding cards, printing services, antiques, and art paintings.
What are the most popular goods that you sell?
Popular products are shields written on them letters of appreciation, stamps.
We have also printing services but these services become less popular because almost each home has a printer and the prices of printers become more cheaper than before.
Does your organization have access to the Internet?
Yes
Does your company have a website?
No
Have you bought online?
Yes, I have bought computer hardware from America.
*How do you evaluate this experience?*
Excellent! There are many things on the Internet cheaper than local market. For example, I have bought something from the Internet costs in the local market 1500 SR, where it cost me 700-800 SR to buy from the Internet. This is great, that's right sometimes takes long time to deliver but the quality of the product is high too.
*I can understand from your talking you like online shopping and you encourage others to do so, is that right?*
I have concern with one issue. For example, before two days I had intention to buy online a computer case. After I completed the purchase processes and paid, the seller, there is also other sellers, raised the price…
*Do you mean buying from eBay?*
I'm not sure maybe eBay. Anyway, the seller raised the price after I already paid. He said this is other fees, but I said to him I do not agree because this wasn't indicated before purchasing and I select your product because it's competitive price before to add more fees! With cancelation claim there was a problem with returning back my money. Will the money be back or not! There is guarantee the money will not be lost and I had hope that the money will be back because there are a credit card record and other documentations with the seller and website that I work with but my concern is how long the process takes to return my money back.
*You mentioned that trusting to deal with the seller is very important for you…*
Yes the trust is very important. If you don't trust, you will not deal.
*For this reason, when you think to apply e-commerce system for your company, you have to seriously take care of building trust with your customers.*

Yes this is an important factor. Naturally the trust needs to be built. No one will trust you with the first impression. Selling online is difficult to show all the details of the products unless you have customers understand the details of the products because they have known this product before with normal buying and dealt with you normally. That's right I can show some details of a product online but this way is cannot be compared with the normal way of selling and buying which requires physical meeting (seller, customer, product).
What does your social/professional group think about the risks and benefits of e-commerce?
Truly, there is no trust. Even with normal dealing sometimes there is no trust.
*Do you think having experience with online selling/buying activities plays role building trust?*
Yes maybe. Clear dealing with an e-retailer from the beginning builds the trust.
*Some people say: We buy from international e-retailers from USA, UK, Australia because we trust, but dealing with an e-retailer in Saudi Arabia is impossible! Even with normal dealing there are no clear rules to protect a customer right, how about if we dealt with them online! What is your comment?*
Yes this is. Trust is required. The most important is trust, if there is no trust you cannot sell and buy.
Do you think selling online would make more profit for your organization?
I don't think so. Honestly I can notice there is no body moving toward this trend. If people know that the one who run an e-commerce website is Arabian or the company owned by Arabian, they will be unmotivated to buy.
*Do you conceder marketing and selling your products online is something like Entertainment or wasting money for building an e-commerce system?*
We may have a website and demonstrate our products, but I don't think this will increase our profits. This is because the nature of people in this country especially businessmen. We have wholesales too but I don't think businessmen, in this country, search for products on the Internet. Some businessmen came to us from far cities like Haiel and Tapok. They come to us here to see, physically, the products and make order for 300 or 400 pieces. See, their nature makes it difficult for them to buy online!
*Can I understand from your talking that the main difficulty that prevents your company not to market and sell on the Internet is because your customers don't want it?*
Yes, this is right. Let's stay on brochures and normal way. It is the same thing paid advertisements instead of paying money and efforts for e-commerce website and you don't know there is people look at it or not. The Internet is full with Ads everywhere. Personally I have not looked at Internet Ads and my friends too, only we search for the thing that we want and leave.
Do you have any friends in business that do regular online purchasing and selling of goods?
No
Do you find it difficult/confronting to use online system?
I don't know the procedures to sell online.
However, I feel it is difficult and case a headache. The normal way is better and easy. When a customer come to buy everything is finished with payment and giving him/her a product but on the Internet you have to take care of many things. Honestly sometimes I receive orders by e-mail but I don't consider them as serious orders. I give a priority to those who come to the shop to buy.
*Why you don't consider e-mails as serious orders?*
They did know our e-mail address because they have had previous deal with us, so there is a trust. However, anything comes by e-mail has a little attention and we give customers who come to our shop high priority. This trend (*not paying attention to e-mails order*) affects peoples' opinion to buy online because companies do not care.
*Some companies say that everything is available in shops, so there is no need for us to sell online. Do you think this is a right opinion? Is the need to sell online only for products that are not available in the local shops?*
It's not right. For example we have a product available in the local shops but they don't have the same quantity as we have. Others have the same products that we have but they may sell individually while we sell in wholesales. So, this is not an inhibitor for building e-commerce website. It's not logical not to build website because the product is available everywhere. That's right the product may be available everywhere but I may have a competitive advantage. For example, product X is available in the local market and you sell the same product but with something unique (*e.g. features, low-price*) makes people to buy from you.
*- Is shipping the purchased products with shipment companies represent a difficulty for your company to selling online?*
Shipment is very important stage. If a customer received his/her order in not satisfied state it will cause a problem to your business. For example, we have bought products from China, when we received them we found 25 boxes were broken out of 500 boxes. These 25 boxes cost us around 14000 SR, who will pay for this? The seller said you bought, and we shipped the products in satisfied state!
So, this is an inhibitor for us because some of our products are made from glass and need very high care in shipment.
*There are some shipments companies offer guarantees to deliver products safely, why not dealing with them?*
Yes, but who will pay for guarantees fees the seller or customer?! If you request from the customer to pay extra 10 SR for such a fees like this and if the price in total comes more than the price in the local shops, he/she may buy it from them without buy it from you online.
*Do you mean if you don't a competitive advantage, you will not benefit to sell online?*
Yes and also as I have told you before I bought a computer case online from USA while the same product is available in the local market but I purchased it online because it has a competitive advantage. The computer case that I have bought online Distinct from local products in many things: more fans with big size, laser lights etc. It maybe found the same things in the local market but will not be cheap as I bought it. At the end I get something cheap, more features and good quality.
What makes it difficult for your company marketing and selling products online?
*Has your company dealt with a shipment company?*
No, we have not dealt with well known companies like DHL, TNT, Aramix etc. However we deal with normal shipment companies (e.g., Almokarej). This company (*Almokarej*) has trucks to deliver products from city to city inside Saudi Arabia only. They deliver goods only to their offices/branches in the targeted city not to customers' homes. When the products arrived a customer will be contacted by mobile phone to come to their office/branch in his/her city to collect his/her product. There is no guarantee for the product to deliver safe.
*Is online payment represents a difficulty for your company?*
Maybe! Sometimes there is peak hours cause hang/stop for payment networks.
*Have you dealt/heard about services/websites on the Internet provides an alternative for online payment such as PayPal?*
No, I have not heard about it.
*Some companies provide several online payment services. One of these servies such is using PayPal. PayPal sometimes represents very good option to build the trust with companies' customers because PayPal has agreement with these companies not to*

*transfer the money directly to them. They keep the money for several days after payment to make sure that the customer is satisfied with his/her product and no complain received. What is your comment?*
This delays our business. The seller always thinks to have his money directly after sale to continue doing other businesses.
What makes it difficult advertising your product online/using online catalog?
We have not thought to use online catalog. This step needs a website, and the website will not received good attention from people unless you tell your customers that you have an online catalog through distributing brochures etc. It may be good idea but we have not tried to apply it.
If a sample of a system available to try, would you use it? Do you want to try a sample before to apply?
Yes, maybe!
What changes needed to apply e-commerce system?
This step reduces the cost of paying money to prepare shops. You may have a small office and market and sell your product online but you need employees to work on your business on the Internet. Instead of paying more money for the rent of the shops and their preparation, you save it to change your business online. If you think from the beginning to sell online, then you only need to pay for e-commerce website and its application and planning how to make this business attractive to customers.

| Participant name: Talal | |
|---|---|
| Participant position: Owner | Having website: Yes |
| Organization size: Small | Selling online: No |
| Business Category: Mobile phones | E-commerce maturity stage: 2 |
| | |
| Interview form: in person interview | Language of conversation: Arabic |

*Document Created On*: : 28/09/2010 6:37:59 PM
*Modified On*: 9/08/2013 10:56:36 PM
What are your targeted customers/ marketplace?
Our market is open for all
What channels do you use to market your products?
The location of our shops is very important
Personal is very important too. When when a customer asks they do not give them only answer but attract them to buy.
What do you market/what are your products?
Mobile phones ant their accessories, and SIM cards
Does your organization have access to the Internet?
Yes
Does your company have a website?
Yes
What kind of information does your organization's website provide?
General information about our company
Have you bought online?
No
What does your social/professional group think about the risks and benefits of e-commerce?
We have not reached the stage to sell and buy online yet. Most of the Internet usages are for entertainment!
Do you think selling online would make more profit for your organization?
If there is people understand the benefit of e-commerce I agree with you it help to gain more profits, but this does not work with our customers. We, currently, don't depend on new customers but loyal customers.

Do you have any friends in business that do regular online purchasing and selling of goods?
Yes I know friends buying online.
Do you find it difficult/confronting to use online system?
Nothing difficult in these days, but education is essential and plays a key role in promoting growth. It depends on your learning and understanding. There is no education, no understanding, and therefore the user feels that there is a difficulty. You must educate people first, how to buy / use / get used to purchase online. When people accept then we will do.
What makes it difficult for your company marketing and selling products online?
There is nothing specific, but the selling with customers online is not useful for us due to the habit of people. When the culture of people changed and accept to buy online we will apply e-commerce system. I repeat that the case essentially depends on education. Illiterate today is not the one who cannot read and write but the one who don't know how to deal with technologies.
Is security of transactions an issue relevant to your decision to use e-commerce? Is privacy of personal/financial information an issue relevant to your decision to use e-commerce?
The perceptions among people are that buying through the Internet cause to steal your money! ... The rumour here spreads quickly. This you can remove with education.
Is trust of parties in a transaction an issue relevant to your decision to use e-commerce?
What makes it difficult advertising your product online/using online catalog?
This became the demand now, just view is a good thing, but our customers are used to do direct view, inspect the item and checked into their own hands. So, the needs, perceptions, and acceptance of our customers are the things that lead our thinking.
What changes needed to apply e-commerce system?
Yes you need to prepare integrated system and must be sacrificed in the beginning to be fully ready for the electronic system, and also must sacrifice some of your products at the outset to gain the trust of more customers.

| Participant name: Saleh | |
|---|---|
| Participant position: Director | Having website: No |
| Organization size: Small | Selling online: No |
| Business Category: computers and accessories | E-commerce maturity stage: 1 |
| | |
| Interview form: in person interview | Language of conversation: Arabic |

*Document Created On*: 28/09/2010 6:37:58 PM
*Modified On*: 9/08/2013 6:16:25 PM
What are your targeted customers/ marketplace?
Youth
What channels do you use to market your products?
The location of our shop is the best way of marketing
What are the favorites channels marketing products?
Personal marketing
What do you market/what are your products?
Computers Hardware & Software
What are the most popular goods that you sell?
Computers Hardware like external hard disks, USB flash memory, digital cameras and mobile phones memory cards
Does your organization have access to the Internet?
Yes
Does your company have a website?

No
Have you bought online?
No
What does your social/professional group think about the risks and benefits of e-commerce?
The current Situation here is not encouraging us to go ahead with this idea.
Do you think selling online would make more profit for your organization?
I don’t think so. The normal way of our business is better, people used to see the products physically and do inspections by themselves before buying. Currently this is not benefiting us.
Do you have any friends in business that do regular online purchasing and selling of goods?
No
Do you find it difficult/confronting to use online system?
The website needs cost and even we go ahead to build an e-commerce website, it is not guaranteed to be successful. Also there are other issues like online payment, goods delivery. There might be extra fees to buy online while computer shops are widespread in Jeddah, so to go ahead with this idea is not benefiting us.
Is trust of parties in a transaction an issue relevant to your decision to use e-commerce?
Yes these issues (security and trust) cause concern to us.
What makes it difficult advertising your product online/using online catalog?
This needs to build a website while the nature of our business is personal marketing and selling. So, online catalog is not useful for us.
If a sample of a system available to try, would you use it? Do you want to try a sample before to apply?
Yes this is good. It helps us to understand how it works. Education plays very important role in this field for both sellers and buyers
What changes needed to apply e-commerce system?
Computer network, contract with delivery companies, employees training programs, arrangement with banks to get online payment systems.
What reasons do you think inhibit retailers in Saudi Arabia to apply e-commerce systems?
The current situation of customers is not encouraging. People culture to buy online discourages us to think of providing option to buy online. This subject requires education to make it easy for people understand how it works and what its benefits are. After that, when people become well educated, we can benefit from applying such idea like this. Our target, as businessmen, to gain more profit but when there is an idea is not helping us to gain this goal then there is no need to apply or think about it.

| Participant name: Mohammed | |
|---|---|
| Participant position: CEO | Having website: Yes |
| Organization size: Large | Selling online: Yes |
| Business Category: Appliances and home improvement | E-commerce maturity stage: 3 |
| | |
| Interview form: in person interview | Language of conversation: English |

*Document Created On*: 28/09/2010 6:37:59 PM
*Modified On*: 9/08/2013 2:54:33 PM

What are your targeted customers/ marketplace?
Our targeted customers are households
What channels do you use to market your products?
Advertising on newspapers and media. and also catalogs and broachers.
What are the favorites channels marketing products?
Word of mouth
What do you market/what are your products?
We sell complete home solutions and complete electronic solutions.
What are the most popular goods that you sell?
IT hardware, cameras, mobile phones, games
Does your organization have access to the Internet?
Yes
Does your company have a website?
Yes
What kind of information does your organization's website provide?
Provide information about our company, what we deal with, what type of products we sell and how to contact us as well.
Have you bought online?
Yes, many times!
What does your social/professional group think about the risks and benefits of e-commerce?
You see in this part of the world we haven't had strong e-commerce movements. I originally from the western part of the world and their day to day activities to go and check online and buy things online, it gives a comfort, convenient and gives them the ability to check prices and compared them as well. But unfortunately we have not seen strong foothold of e-commerce in this part of the world. But, InshaAllah, I'm confident because the market is so I am confident because the market is so buoyant out here. Even the economic meltdown happening in the rest of the world unfortunately, unfortunately I must say that it has not even made a dent of although here. So Inshallah I have strongly, strong advocate people will actually go into e-commerce. But we need to have the infrastructure right. In order to have the infrastructure right we must make sure that the Wi-Fi is available in better cases and more places. We must ensure that broadband is available for other people in remote places and we must also ensure that the protected plan payment??? is also taken care of, so that we do not create a system where people actually pay and do not know whether or not the goods will come or not. So it is very important to follow. So the infrastructure getting it right is very important. Another one you see how people get hooked on, people getting addicted to because of the convinence.
*I can understand from your talking that people here in KSA don't understand the benefit of e-commerce, is that right?*
No, I disagree with that, people here understand the benefit of e-commerce 100%, when I was in Riyadh I visited GYTEX (GYTEX is a fair about technologies, it is happening again in April this year you must visit GYTEX) and the time that I seen people there the indigenes original Saudi people superbly well educated and they know exactly the benefit of e-commerce and everything. But I think someone needs to kick the whole things of having the access of Wi-Fi, of having the access of broadband, having the access in every house hold. Once this done, obviously you will see the results of that very soon.

*I can understand from your talking that you have sensitively with the options available for online payment and the issue of trust how to be build …*

You are very correct, there has to be a mechanism whereby the customers feel confident and will be a %100 trusting to put their credit cards details on the Internet. In order to have that confident you must have a mechanism whereby fault doesn't happen. And also have secure networking, once we have secure network there will be no problem at all. There are secure payment systems; you just need to incorporate that into your system.

Do you find it difficult/confronting to use online system?

Absolutely not, I personally like to do move toward e-commerce system. This will actually help your customers to browse, search, check the prices for competitor, and check the quality of products in terms of features, advantages and benefits and then make the right choice. It makes comfort and less time to move around. Yes if there is specific question to particular issue they customer can contact us by phone to get help.

*I can understand from your talking that your organization is ready to apply e-commerce system, so what are you waiting for?*

We are ready to do it, but we are just waiting for few things to actually happen but it's actually in our way of thinking.

What makes it difficult for your company marketing and selling products online?

Nothing difficult but we are waiting for few things to be completed before to go ahead with this.

What changes needed to apply e-commerce system?

Not majors changes at all, only very minor changes.

As I said before we need to have products listing properly done ad also having a block to show the special offers that we have.

*From your point of view, what makes retailers in Saudi Arabia not go ahead with e-Retailing?*

Because the idea of popularity is not coming from the customers, once the businessmen or the ones who run the business in KSA get the confident that the people are more keen to go online and visit the places they will be more than happy to jump to the bandwagon and put their business in e-commerce channel.

| Participant name: Salem | |
|---|---|
| Participant position: IT Manager | Having website: Yes |
| Organization size: Medium | Selling online: Yes |
| Business Category: Grocery | E-commerce maturity stage: 4 |
| | |
| Interview form: in person interview | Language of conversation: English |

*Document Created On*: 28/09/2010 6:37:58 PM
*Modified On*: 9/08/2013 11:56:55 PM
What are your targeted customers/ marketplace?
High class customers and middle class, but before we used to serve C class but not much more.
We are Chocolate and Biscuit manufacturing company and we do have e-commerce business here
What channels do you use to market your products?
We used to market our products in the same typical way like other companies doing. We are doing door to door, flyers, newspapers, banners on the roads. Basically our session is Ramadan and we do much marketing ads during this session.
What are the favorites channels marketing products?
To be honest, our products especially locally everybody knows and it is very old company, whenever you have any good event you remember sweets and all that. Our company is not mostly depend on Ads, we do not do more ads like on TV or every where we are available, we do Ads but not much more. So, you can say Flyers and newspapers and e-mails Ads are happen. These are most commonly used channels.
*Some people says that the location of the shop itself play major role in the advertisement, what is your comment?*
Yes there are various means of doing and advertisement of your products. Of course we used to do some advertising activities inside our show rooms, so whenever people go near your show rooms that they will find what do you sell, what are the new things, attractive things, it is very known thing and we are do the same thing.
What do you market/what are your products?
Chocolate and Biscuit
What are the most popular goods that you sell?
High demand products in our company, there are 2 to 3 very high demand products like we are the sole agent in Saudi Arabia for Japanese rice cracker. Many companies selling rice cracker but you will never find the Japanese standard rice cracker than we are selling here. Second product our chocolate, our chocolate is also very high quality and selling luxury and high standards selling to customers level B. We do sell some products for C class, but basically we focus on B class.
Does your organization have access to the Internet?
Yes
Does your company have a website?
Yes
What kind of information does your organization's website provide?
We are Saudi based company but we have Complete e-commerce website open for whole world not only for Saudi Arabia.
*What encouraged your company to have e-commerce solution, is it because decision makers in your companies are familiar with these activities (i.e. used to buy , sell etc)?*
Very good question, actually we just started e-commerce website, we are having the website from long period, but we never sell on that. But from 1.5 year back we launched the e-commerce website and we do changes and now ready to sell online. We are really new in this. Our main purpose, we already know that this region doesn't give you more sales on the Internet because that habit of the people here in this region used to go outside and buy, we are not in Europe or America where people easily buy from the Internet, culturally we are different than them that's why e-commerce is not getting much more businesses. We open this channels (i.e. selling online) but we are not much successful in this field!

*You said people here culturally are not familiar to buy online, is it because they find it risky …?*

You are absolutely right; we notice that people feel difficult to put their credit card information. OK if it Amazon.com there is no problem, because Amazon for them is very trusty name but form us the will feel very difficult to buy! There is a threat feeling will they get the products or may be not getting the products!

We have to build the trust first. To build the trust in this region requires a lot of work and a lot of activities to be there. There are people from 2 to 3 years in e-commerce business like (X company) they are in this business from long time and selling very good but still very far compared to very small businesses in Europe or Aisa!

*You said there is fairness from customers to put their credit cards details on the Internet, Have you thought to provide another alternative for online payment (e.g. PayPal, Sadadسداد)?*

Sadad before was very expensive solution, it is good for large companies but it's not for middle size company like us! I agree with you that providing more online payments because not everybody has credit card, It should be there is another ways of online payments. We have another way here especially for local customers, We have call centre and the customer can call us (free calls)and make the order by phone and go to pick it up from the branch.

*Is the shipment of products cause any problem/difficulty for your company?*

We tried to do our best, we have done a small research in this area that the Fragile products sometimes losing the deals! Some customers are interested to buy from you but when they see Fragile products they will go away, it is better for them to buy from local market instead of buying online where they cannot ensured the product will shipped safely. To reduce this thing what we have done is having very strong contract with express and trusted shipment company (FedEx). We have also another shipment type, the customers who have accounts with the shipment companies they can provide us their account numbers with anyone of these companies and we will arrange with these companies to come and pick up your products and they will charge you as the same agreement you have with them.

*Is the current mailing address in Saudi Arabia case a difficulty for shipment?*

We depend on the mobile phone number; if the mobile phone number is not correct then it's difficult to deliver the product. This is not only for us but also for other companies and home delivery people they are struggling with these things. But there are certain locations (e.g. companies buildings) you can know the address clearly, you know which street and block, building, floor, flat etc. But with community houses is not very clear. So we are managing well with the mobile phone numbers actually. By calling the customer to describe to us his location we will deliver the products to him/her.

What makes it difficult for your company marketing and selling products online?

Building the awareness in the region that will take some time to adopt the culture of e-commerce. Before credit cards were not common but know become common but it took time. The awareness is coming but it takes time and we need to accelerate the development of this field. Like you are conducting this research, this is very good to highlight the difficulties to do something regard them.

*Do think Saudi government (e.g. ministry of commerce) plays major role in the development of e-commerce?*

Yes that's true. We are working now to have certificates from trusted organizations to build the trust with our customers. Like this it should be there is a certification body from the government itself to say that this company is a certified company by local government and you can buy from them. This is good to build the customers trust with the certified companies as the government trust them.

*Let's go back to our main question regarding the most difficult tasks to have e-commerce solution, what are the most difficult tasks for your company to have e-commerce website/solution?*

Building the website is very easy, You can buy cheap stuff/software from the Internet, there are e-commerce packages available cost around $100 to buy a package and publish your e-commerce website. Before it was very expensive, but know things become cheaper. You can get it for fewer amounts. The only thing that when you launce your website if you don't market it well (like go with Google ads, and all the soft communication whatever you have like highlighting your company domain name everywhere etc), you will lose your business. E-marketing is very important in this regard and requires a lot of efforts. E-marketing is like working with Google and when a user types a chocolate in Google he will find our company name there. If I didn't work hard efforts on marketing , it will never give me a $1, it will never. Because nobody knows, you have to let people know it and this doesn't require geographical location, anybody from anywhere in the world they can go and buy.

So e-marketing is very important and requires a lot of efforts to keep customers buying from you.

*What have you done with your normal customers who used to come to your company shops and buy normally? Have you advertise/show to them that you have a website and customers can go online and browse all the products and buy from their homes?*

We do certain activities before the our high profit seasons (like Ramadan) we have a campaign, in this campaign all these information written in flyers, banners ... that we have a website and you can buy online, visit it and buy from us ... and we did e-mail marketing also, we have sent more than 3000 e-mails, I'm not sure about the number but marketing people know it well. We have sent e-mails to our loyal customers. We give business cards to customers that include our website name and e-mail.

# **Appendix H:** The full coded transcripts and categories

## 1. Consumers' factors/issues/concerns

### *1.1. Culture or habit of online shopping*

Ahmed (large, 2)

Culture of the people to accept e-shopping is not encouraging us to open the electronic market.

but unfortunately the culture of our customers is not encouraging us to go ahead with this idea.

but internally in Saudi Arabia so far is not encouraging.

Nasir (medium, 2)

This subject refers to peoples' culture of selling and buying.

there is no difficulty [to sell using online system] but this issue refers to the culture of people to accept and use.

Culture of people makes no benefit for us to sell online

Osam (large, 3)

Culture of people plays a role [in discouraging online shopping].

Saeed (small, 1)

the culture of people is not encouraging, so it is not useful for us.

Saleh (small, 1)

People culture to buy online discourages us to think of providing option to buy online.

Salem (medium, 4)

we already know that this region doesn't give you more sales on the Internet because that habit of the people here in this region used to go outside and buy

Talal (small, 2)

We have not reached the stage to sell and buy online yet.

the selling with customers online is not useful for us due to the habit of people. When the culture of people changed and accept to buy online we will apply e-commerce system.

Tameem (medium, 1)

This is because the nature of people in this country especially businessmen. We have wholesales too but I don't think businessmen, in this country, search for products on the Internet.

Thamer (large, 4)

We make %10 off on online sales however the problem with people culture to accept this new way of buying.

Waleed (small, 1)

our culture buying and selling online is very weak. Whoever sells or buys should be professional.

The absence of the habit to buy and sell online makes it very difficult to progress in this field.

### *1.2. Consumers understanding of e-commerce benefits*

Mohammed (large, 1)

people here understand the benefit of e-commerce 100%

I seen people there the indigenes original Saudi people superbly well educated and they know exactly the benefit of e-commerce and everything.

Talal (small, 2)

If there is people understand the benefit of e-commerce I agree with you it help to gain more profits,

### *1.3. Consumers trust*

(ahmed (large, 2)

Customers do not trust to buy online and they only accept with strike conditions, because it requires credit cards payment.
Yes of course trust is the main problem.
This [protection system] is veryimportant to build the trust with customers.
I advice the main supporter in this field is to find out a way to build the honesty/trust between customers and sellers.
Naif (medium, 2)
With this procedure [PayPal use] customers feel happy to deal with this intermediary e-payment option which protects their rights and also build the trust with companies that deal with PayPal.
we think if there is a similar system [i.e. PayPal] available in Saudi Arabia and in Arabic language will encourage to build trust which is very important factor to buy online.
Nasir (medium, 2)
There are people when they pay online remain in doubt and not sure of receiving their purchased orders.
There is fear, because ignorance breeds fear. There is a doubt to deal by money on the Internet.
our customers don't trust too
Salem (medium, 4)
There is a threat feeling will they get the products or may be not getting the products!
We have to build the trust first. To build the trust in this region requires a lot of work and a lot of activities to be there.
We are working now to have certificates from trusted organizations to build the trust with our customers.
Tameem (medium, 1)
Naturally the trust needs to be built. No one will trust you with the first impression.
Truly, there is no trust. Even with normal dealing sometimes there is no trust.
Trust is required. The most important is trust, if there is no trust you cannot sell and buy.

### *1.4. Consumers reluctance to use credit cards*
Moneer (large, 4)
people in general tend not to use credit cards to buy online except those who used to do.
Osam (large, 3)
Also there is another problem that people are afraid to buy online using credit cards.
People are reluctant to use credit Cards.
Salem (medium, 4)
we notice that people feel difficult to put their credit card information.
Thamer (large, 4)
They are afraid to put their credit cards number on the internet.

### *1.5. Consumers preferences*
Ahmed (large, 2)
the customer wants to see and toutch physical product in front of him/her.
Ali (small, 1)
People here have to come to the shop to see the product and inspect it.
Saleh (small, 1)
people used to see the products physically and do inspections by themselves before buying.
Talal (small, 2)

this does not work with our customers. We, currently, don't depend on new customers but loyal customers.
This [online catalog] became the demand now, just view is a good thing, but our customers are used to do direct view, inspect the item and checked into their own hands. So, the needs, perceptions, and acceptance of our customers are the things that lead our thinking.
Thamer (large, 4)
People here used to, for long time, sell and buy directly. This means they go to shops to see and check the goods before to buy. This is the Prevailing habit of people in terms of selling and buying. This is the culture of Saudi people and Arab communities.

### *1.6. Consumers lived or studied in developed countries*

Thamer (large, 4)
There is specific class in the community, according to a study we conducted 6 months ago, prefers to buy online. This class has a previous experience studying overseas in western countries and gets used to buy online. They have become comfortable with online purchases.

### *1.7. Consumers level of demand of buying online*

Ahmed (large, 2)
We have tried to market on the Internet But there is no demand
Mohammed (large, 3)
we have not seen strong foothold of e-commerce in this part of the world.
Saeed (small, 1)
Because currently there is no rush to e-shopping in Saudi Arabia
Saleh (small, 1)
The current Situation here is not encouraging us to go ahead with this idea.
Salem (medium, 4)
culturally we are different than them that's why e-commerce is not getting much more businesses.
Thamer (large, 4)
there is no good demand to buy online. For example, in western region of Saudi Arabia we receive 3-5 orders buying online every week and these are nothing compared to normal way of business.

### *1.8. Consumers knowledge to buy online*

Ali (small, 1)
The majority of people in our country do not know how to buy and sell on the Internet
Nasir (medium, 2)
Also there is Ignorance in the community. Most of them don't know the meaning of e-commerce as result they will not venture.
Saleh (small, 1)
After that, when people become well educated, we can benefit from applying such idea like this.
Talal (small, 2)
The perceptions among people are that buying through the Internet cause to steal your money! ... The rumour here spreads quickly.

### *1.9. Consumers Familiarity with e-commerce*

Ali (small, 1)
People here are not familiar with e-commerce and this is not something that you can apply it in a day.
customers are not familiar with this idea

The familiarity, Sellers and customers are not familiar with online sales and purchases.
Mohammed (large, 3)
Because the idea of [e-shopping] popularity is not coming from the customers, once the businessmen or the ones who run the business in KSA get the confident that the people are more keen to go online and visit the places they will be more than happy to jump to the bandwagon and put their business in e-commerce channel.

Osam (large, 3)
In general people have to become familiar with this system [before we offer online sale system]
We, *Saudi community*, are still fairly new in the use of electronic commerce.
Talal (small, 2)
When people accept then we will do.
Waleed (small, 1)
We are growing community, and the idea of e-commerce is new to us, and we need time to understand these things.

### *1.10. Consumers distrust Arabian online businesses*
Tameem (medium, 1)
If people know that the one who run an e-commerce website is Arabian or the company owned by Arabian, they will be unmotivated to buy.

## 2. Environment factors/issues/concerns
### *2.1. Wi-fi & broadband services availability*
Mohammed (large, 3)
But we need to have the infrastructure right. In order to have the infrastructure right we must make sure that the Wi-Fi is available in better cases and more places. We must ensure that broadband is available for other people in remote places
But I think someone needs to kick the whole things of having the access of Wi-Fi, of having the access of broadband, having the access in every house hold. Once this done, obviously you will see the results of that very soon.

### *2.2. Protection system*
Ahmed (large, 2)
The system that protect customer right when he/she pays, he/she will receive the purchased products/services.
Yes sure there is no clear system to protect sellers and customers rights.
It has to be done from early stage, customers have to be ensured there are rules and legislations protect their right.
Mohammed (large, 3)
we must also ensure that the protected plan payment??? is also taken care of, so that we do not create a system where people actually pay and do not know whether or not the goods will come or not
Naif (medium, 2)
With this procedure [PayPal use] customers feel happy to deal with this intermediary e-payment option which protects their rights
Waleed (small, 1)
We did not find a clear law to protect our rights in these issues [SAWA and stockmarket]. Now if we go ahead with selling and buying online the same problem happens in the absence of clear laws and regulations. With no clear e-commerce law and legislations applying e-commerce systems are not encouraging.

### *2.3. Required action by government*

Ahmed (large, 2)

It has to have a clear [protection] system, i.e. from government, Imposed on all involved parties. .

I have a message to Saudi commerce ministry. I advice them to work harder to apply e-commerce systems. I advice the main supporter in this field is to find out a way to build the honesty/trust between customers and sellers. I suggest government department/s to take care of doing this.

Salem (medium, 4)

Like this it should be there is a certification body from the government itself to say that this company is a certified company by local government and you can buy from them. This is good to build the customers trust with the certified companies as the government trust them.

Thamer (large, 4)

Having an integrated system such as Saudi Stock Market. If there is something similar for e-commerce it will be great. Having an integrated system for e-commerce is very important in terms of having people take care of the system, search for problems and solutions, and leading toward development.

Citizens will have more trust if this subject sponsored by the government because we, in Saudi Arabia, have great confidence in anything that comes through the government. The evidence for this Saudi Arabia was the first country that has the lowest effects of financial crises in 2009.

### *2.4. online payment system*

Mohammed (large, 3)

there has to be a mechanism whereby the customers feel confident and will be a %100 trusting to put their credit cards details on the Internet.

There are secure payment systems; you just need to incorporate that into your system.

Moneer (large, 4)

We provide two options [for online payment], credit cards and Sadad system.

Naif (medium, 2)

not affording options for online payment [cause a difficulty using online system, he mentioned PayPal]

Having more options for online payment other than credit cards only is an Encouraging factor to buy online.

Osam (large, 3)

Banks should provide easy options to have two credit cards, one with large amount of money and another one with small amount to be used in online payment.

This idea [debit cards] makes people get used to and this will remove their fear.

We provide two options for online payment: credit cards and direct debit. However the second option is less accepted.

Saleh (small, 1)

Also there are other issues like online payment [difficulty]

Salem (medium, 4)

I agree with you that providing more online payments because not everybody has credit card, It should be there is another ways of online payments.

Themer (large, 4)

I suggest for local banks to offer other easy options for online payment to encourage people to purchase online.

### *2.5. SADAD_national online payment system*

Moneer (large, 4)

Sadad system is more secure system and great.
Sadad is great idea and more secure than credit cards and encourage people to buy online.

Salem (medium, 4)

Sadad before was very expensive solution, it is good for large companies but it's not for middle size company like us!

### *2.6. PayPal*

Naif (medium, 2)

We currently use in our e-commerce system more than one option for online payment. One option is using PayPal. PayPal is a secure online payment mean for both customer and seller. It plays intermediary mean between credit card and e-commerce website. This system of online payment builds the trust also with the seller who uses it. It represents a Company of e-payment and collection that tests new companies that deal with them. For example, they have instructions not to withdraw money for the new sellers before specific period. This procedure is to ensure that the new company is serious and no complaints received from customers. If there is a complaint, they investigate and return the money to customers if the complaint is true.

### *2.7. Delivery issues*

Ahmed (large, 2)

I can see also one of the inhibitors is delivery fees or insurance.

Hassan (small, 1)

we have concern regarding delivering the purchased products and its fees
sometimes to deliver the products represent an inhibitor. Also the customer maybe will not accept to pay delivering fees.

Moneer (large, 4)

With delivery there is a problem if there is deleviry fees, customer maybe not paying these fees except those from specific class or busy etc.

Saeed (small, 1)

Delivering goods need special care with some of our products because they require to be stored in a specific temperature to delivered safely and healthy.

Saleh (small, 1)

There might be extra fees to buy online

Tameem (medium, 1)

Shipment is very important stage. If a customer received his/her order in not satisfied state it will cause a problem to your business.
If you request from the customer to pay extra 10 SR for such a fees like this and if the price in total comes more than the price in the local shops, he/she may buy it from them without buy it from you online.

### *2.8. Issues relate to Saudi mailing and addressing system*

Osam (large, 3)

The problem is mail address; still some people do not have mail address.

Salem (medium, 4)

We depend on the mobile phone number; if the mobile phone number is not correct then it's difficult to deliver the product. This is not only for us but also for other companies and home delivery people they are struggling with these things. But there are certain locations (e.g. companies buildings) you can know the address clearly, you know which street and block, building, floor, flat etc. But with community houses is not very clear. So we are managing well with the mobile phone numbers actually. By calling the customer to describe to us his location we will deliver the products to him/her.

### *2.9. Need for researches*

Salem (medium, 4)

The awareness is coming but it takes time and we need to accelerate the development of this field. Like you are conducting this research, this is very good to highlight the difficulties to do something regard them.

### *2.10. Need for Educational programs*

Ali (small, 1)

This needs educational activities to remove this inhibitor at all.

Hassan (small, 1)

this matter needs education for both sellers and buyers. There has to be educational programs telling what the options available are and indicating the easiness, positive, and negative aspects of each option. Educational programs are very important. When you know the positives and negatives and get the minimum limit of knowledge and awareness the matter will become easy to understand. Education plays major role in this matter.

Saeed (small, 1)

This issue requires education and training. Educated people accept this idea. Education plays key role to adopt this new technology.

Saleh (small, 1)

This subject requires education to make it easy for people understand how it works and what its benefits are.

Salem (medium, 4)

Building the awareness in the region that will take some time to adopt the culture of e-commerce.

Talal (small, 2)

Nothing difficult in these days, but education is essential and plays a key role in promoting growth. It depends on your learning and understanding. There is no education, no understanding, and therefore the user feels that there is a difficulty. You must educate people first, how to buy / use / get used to purchase online.

I repeat that the case essentially depends on education. Illiterate today is not the one who cannot read and write but the one who don't know how to deal with technologies.

Thamer (large, 4)

Regarding the education I have comment: Why we do not educate students in the classes about e-commerce. It's better to educate students how to use the Internet and benefit well rather than wasting their times on social networks and games!

Waleed (small, 1)

You need to educate people first and spread good understanding for e-commerce.

### *2.11. Free trails for e-commerce software*

Hassan (small, 1)

Yes sure. This is a good motivation and easy for seller and buyer.

Nasir (medium, 2)

Yes this is good and hopeful to understand. Experiencing something is part of education.

Saleh (small, 1)

Yes this is good. It helps us to understand how it works. Education plays very important role in this field for both sellers and buyers

Waleed (small, 1)

Yes we can try and this experience give us an idea how it works and how profitable will be before we involve in paying on something that is not useful.

## 3. Market factors/issues/concerns

### *3.1. Products availability in local market*

Ali (small, 1)

Same products [electrical and electronics] are widespread in the market. Yes I agree to go ahead with this idea if there is products are not available in the local market where we can have competitive advantage.

Naïf (medium, 2)

selling products online where the local market has similar plentiful and cheap products [is difficult to gain consumers online]

Saeed (small, 1)

It is not useful because supermarket and mini markets are in every street.

Saleh (small, 1)

computer shops are widespread in Jeddah, so to go ahead with this idea is not benefiting us.

Tameem (medium, 1)

It's not right. For example we have a product available in the local shops but they don't have the same quantity as we have. Others have the same products that we have but they may sell individually while we sell in wholesales. So, this is not an inhibitor for building e-commerce website. It's not logical not to build website because the product is available everywhere. That's right the product may be available everywhere but I may have a competitive advantage. For example, product X is available in the local market and you sell the same product but with something unique (*e.g. features, low-price*) makes people to buy from you.

Waleed (small, 1)

Having no similar products in the local market help to achieve high profits, but if the products are available in the market, I believe that profits will be very weak.

### *3.2. level of offering competative advantage*

Naif (medium, 2)

This is because all our products are Plentiful in Saudi market. So, we don't have a competitive advantage in Saudi market to encourage buying online. This is why we make our main target selling online for International market. Islamic watches and clocks that we produce and sell are unique products in the international market.

### *3.3. E-marketing required hard efforts*

Salem (medium, 4)

The only thing that when you launce your website if you don't market it well (like go with Google ads, and all the soft communication whatever you have like highlighting your company domain name everywhere  etc), you will lose your business. E-marketing is very important in this regard and requires a lot of efforts.  E-marketing is like working with Google and when a user types a chocolate in Google he will find our company name there. If I didn't work hard efforts on marketing , it will never give me a $1, it will never. Because nobody knows, you have to let people know it and this doesn't require geographical location, anybody from anywhere in the world they can go and buy.

So e-marketing is very important and requires a lot of efforts to keep customers buying from you.

## 4. Organization factors/issues/concerns

### *4.1. E-commerce difficulty*

Saeed (small, 1)

Yes [it is difficult], you have to design a commercial website, train your employees, deliver goods.

Saleh (small, 1)

The website needs cost and even we go ahead to build an e-commerce website, it is not guaranteed to be successful. Also there are other issues like online payment [difficulty], goods delivery [difficulty].

Tameem (medium, 1)

Selling online is difficult to show all the details of the products unless you have customers understand the details of the products because they have known this product before with normal buying and dealt with you normally.

I feel it is difficult and case a headache. The normal way is better and easy.

### *4.2. experience buying online*

Ahmed (large, 2)

Yes, airline tickets only

Ali (small, 1)

No

Hassan (small, 1)

No

Mohammed (large, 3)

Yes, many times!

I originally from the western part of the world and their day to day activities to go and check online and buy things online, it gives a comfort, convenient and gives them the ability to check prices and compared them as well.

Moneer (large, 4)

Yes

Nasir (medium, 2)

No

Saeed (small, 1)

No

Saleh (small, 1)

No

Talal (small, 2)

No

Osam (large, 3)

Yes

Tameem (medium, 1)

Yes, I have bought computer hardware from America.

Waleed (small, 1)

No

### *4.3. Issues related to nature of the business*

Ali (small, 1)

The majority of similar companies, %99, do not have e-commerce websites and shops are widespread where we cannot have competitive advantage to sell online.

the nature of our business makes it difficult to go ahead with this idea.

Same products [electrical and electronics] are widespread in the market.

Saleh (small, 1)

the nature of our business is personal marketing and selling. So, online catalog is not useful for us.

### *4.4. Issues related to offering low prices*

Ahmed (large, 1)

If the products available in the local market with the same price then the company which offer selling these products have no competitive advantage.

### *4.5. Issues related to products types*

Ahmed (large, 2)

there is no demand, because our products related to smell, shape, and the customer wants to see and touch physical product

Ali (small, 1)

However electronic and electrical shops are almost on every corner in Jeddah which makes the idea of e-commerce not useful for us.

Salem (medium, 4)

Some customers are interested to buy from you but when they see Fragile products they will go away, it is better for them to buy from local market instead of buying online where they cannot ensured the product will shipped safely.

Tameem (medium, 1)

this [delivery in satisfying condition] is an inhibitor for us because some of our products are made from glass and need very high care in shipment.

### *4.6. lack of e-commerce knowledge in our company*

Hassan, (small, 1)

our stuff don't know about marketing and selling on the Internet where they required training programs.

Nasir (medium, 2)

Honestly there is Ignorance inside our company in terms of e-commerce, and that is why our website only disply products and not receive orders.

Tameem (medium, 1)

I don't know the procedures to sell online.

### *4.7. Management Attitude toward e-commerce*

Mohammed (large, 3)

I personally like to do move toward e-commerce system. This will actually help your customers to browse, search, check the prices for competitor, and check the quality of products in terms of features, advantages and benefits and then make the right choice. It makes comfort and less time to move around.

Osam (lager, 3)

We believe that most of the company's services should be available on its website to save a customer's time and effort visiting the shop. Visiting shops are required for our customers when there are physical products need to be purchased. Even physical products can be sold online but it requires us more careful thinking.

Thamer (large, 3)

However we will continue providing this option (selling online) to encourage people to use it.

### *4.8. Our business familiarity or preference*

Ali (small, 1)

This depends on Familiarity. We are familiar with the way of normal selling where a customer has to come to our shop and see the products.

It is much better for us when a customer comes to the shop.

We prefer personal marketing it is much better for us. When you explain to a customer face to face is totally different from leaving him alone to read the features of a product on the Internet.

Our employees are familiar with direct sales.

We do not think about e-commerce at all. The normal way of selling is the best for us and we do not want to sell online at all.
However in our business of selling electronic and electrical products the normal way is the best. With time may be the situation improves and people accept this idea.
Hassam (small, 1)
This is not similar to normal way of selling where a customer comes to the shop, see a product, and make sure he/she selects the right product.
Nasir (medium, 2)
We see the normal way of selling is the best.
So the normal way of business is the best for us.
The normal way of selling remains the best for us to gain more profit.
Salem (small, 1)
The normal way of our business is better
Talal (small, 2)
this does not work with our customers. We, currently, don’t depend on new customers but loyal customers.
Tameem (medium , 1)
this way is cannot be compared with the normal way of selling and buying which requires physical meeting (seller, customer, product).
When a customer come to buy everything is finished with payment and giving him/her a product but on the Internet you have to take care of many things.

### *4.9. Our business priority*

Fadi (medium, 2)
using the Internet to order our products is not our priority at the moment
Naif (medium, 2)
Our e-commerce website is not targeting consumers in Saudi Arabia and we don’t have a priority to do so.
Tameem (medium, 1)
sometimes I receive orders by e-mail but I don’t consider them as serious orders. I give a priority to those who come to the shop to buy.

### *4.10. Our marketing Strategy*

Ahmed (large, 2)
we have a different marketing strategy which is having branches/shops in the main malls to be directly close to customers.

### *4.11. Own delivery system*

Fadi (medium, 2)
The company also has its delivery system
Thamer (large, 4)
Our company has its own delivery system by the company’s cars and trucks. We deliver for free inside main cities and with delivery fees outside the main cities. We have our own Addressing system and the delivery done within 3 days. It is not an obstacle for us.

### *4.12. Name: Personal feeling or experience*

Mohmmed (large, 3)
I personally like to do move toward e-commerce system.
Moneer (large, 3)
I personally feel more comfortable with it. I use it when I book airline tickets.
Nasir (medium, 2)
There are thefts of Credit Cards numbers, and there are hackers penetrate your privacy. So this is a problem for the money dealing on the Internet.

Tameem (medium, 1)
I have concern with one issue. For example, before two days I had intention to buy online a computer case. After I completed the purchase processes and paid, the seller, there is also other sellers, raised the price…
The Internet is full with Ads everywhere. Personally I have not looked at Internet Ads and my friends too, only we search for the thing that we want and leave.
For example, we have bought products from China, when we received them we found 25 boxes were broken out of 500 boxes. These 25 boxes cost us around 14000 SR, who will pay for this? The seller said you bought, and we shipped the products in satisfied state!
So, this [delivery in satisfing condition] is an inhibitor for us because some of our products are made from glass and need very high care in shipment.
Waleed (small, 1)
The Internet is Internet fertile ground for cheating and fraud. I think the acceptance to sell online will remain weak.
Another issue, we are a society subjected to several issues of fraud, such as stoke market radical decline, Sawa cards, lands and fake projects!

### *4.13. Phone or e-mail order*

Fadi (medium, 2)
The company also has its delivery system for those who order by phones.
Nasir (medium, 2)
Sometimes some customers email us their requests. We check our e-mail every day. They send to us their lists of books by Fax or e-mail and we arrange the books for them. Students refer to our website to browse information about books.
We receive their orders, *by fax or e-mail as explained before*, and we give them our bank account number to deposit the money. As soon as we receive the money we send the books with delivering companies like FedEx, DHL, or other shipment companies.
Salem (medium, 4)
We have another way here especially for local customers, We have call centre and the customer can call us (free calls)and make the order by phone and go to pick it up from the branch.
Waleed (small, 1)
We have free service to deliver the goods. We are the exclusive company in this regard which deliver for free inside Jeddah. Even if the order is cheap we can deliver especially if the customer does not have transport. A customer can call us by phone to make an order and then we deliver the order with one of our staff if it inside Jeddah or with Alzajel shipment company if it outside Jeddah. Normally the order outside Jeddah takes 2 days to arrive.
Currently we have orders by phone. For example a customer from Algoreeat (a city 1000 km far from Jeddah) calls us to make an order. We send to him/her bank account number to deposit the money and then we send their order with Alzagel Shipment Company with 10 SR delivery fees. The order is received within 4 days. We only take the name and mobile phone number of the customer. When the order arrives in the shipment office in the customer city, the shipment company calls him/her to come to take his/her product from the office but if he/she cannot some he/she requires paying extra fees to deliver to home.

### *4.14. Planning to use e-commerce*

Mohammed (large, 3)
We are ready to do it, but we are just waiting for few things to actually happen but it’s actually in our way of thinking.

### *4.15. Research is required*

Fadi (medium, 2)
we have not think about it or make study to find out what needs to be done

### *4.16. Name: security concerns*

Saleh (small, 1)
Yes these issues [security and trust] cause concern to us.
Waleed (small, 1)
Yes this [security of online transactions] cause a difficulty

### *4.17. Sellers distrust online payment*

Ali (small, 1)
We do not trust online payment, the money must paid cash
Hassan (small, 1)
Online payment maybe represents an inhibitor.
Nasir (medium, 2)
Yes we don’t trust online payment

### *4.18. Setup cost concern*

Saleh (small, 1)
The website needs cost and even we go ahead to build an e-commerce website, it is not guaranteed to be successful.
Salem (medium, 4)
Building the website is very easy, You can buy cheap stuff/software from the Internet, there are e-commerce packages available cost around $100 to buy a package and publish your e-commerce website. Before it was very expensive, but know things become cheaper.
Tameem (medium, 4)
It is the same thing paid advertisements instead of paying money and efforts for e-commerce website and you don’t know there is people look at it or not.
Waleed (small, 1)
Financial costs involved in something that does not bring profit.

## 5. Perceptions

### *5.1. Using Internet increase seller popularity*

Hassan (small, 1)
Using the Internet increases the chance of popularity.

### *5.2. Perception about Internet users in KSA*

Ahmed (large, 2)
The percentage of users who browse the Internet is high
Hassan (small, 1)
Today almost every home has access to the Internet, many people browse it and of course this brings customers to your shop.
Nowadays people aware and as I said to you almost %90 of people use the Internet and people become well educated in this matter.
Osam (large, 3)
Today almost every house has Internet connection and I believe Internet usages will increase in the future.

### *5.3. Perception about e-commerce users*

Ahmed (large, 2)

those who sell or buy is very small.
Mohammed (large, 3)
You see in this part of the world we haven't had strong e-commerce movements.
Moneer (large, 4)
people here afraid to buy online.
Talal (small, 2)
Most of the Internet usages are for entertainment!
Thamer (large, 4)
from 4 years some customers have started to contact us through the website. This led us to launch an e-commerce website making it available for customers to complete their purchases online.

### *5.4. Perception about e-commerce profits*

Ahmed (large, 2)
Yes of course selling online would make more profits
Ali (small, 1)
it will not be high because it cannot be compared to normal selling process.
We will not benefit because customers are not familiar with this idea.
Yes I agree to go ahead with this idea if there is products are not available in the local market where we can have competitive advantage.
Hassan (small, 1)
For small shops/companies this is not useful.
Nasir (medium, 2)
Selling online like Amazon.com is not useful for us in this country and all Arabic countries.
Although there are more things that you can save when you transfer to sell online like tools, equipment, and rent of shops, but I don't think the Internet help us to achieve more profit.
Saeed (small, 1)
currently it does not benefit us because people are not interested in this sort of thing.
Saleh (small, 1)
I don't think so [selling online would make more profits].
Currently this is not benefiting us.
Our target, as businessmen, to gain more profit but when there is an idea is not helping us to gain this goal then there is no need to apply or think about it.
Talal (small, 2)
selling with customers online is not useful for us due to the habit of people.
Tameem (medium, 1)
I don't think [using e-commerce will increase our profits]so.
We may have a website and demonstrate our products, but I don't think this will increase our profits.
Waleed (small, 1)
this type of business is not useful currently and returns very low profits.

### *5.5. Perception about e-commerce*

Mohammed (large, 3)
This will actually help your customers to browse, search, check the prices for competitor, and check the quality of products in terms of features, advantages and benefits and then make the right choice. It makes comfort and less time to move around.
Saeed (small, 1)
Currently it is not useful
Tameem (medium, 1)

This step reduces the cost of paying money to prepare shops. You may have a small office and market and sell your product online but you need employees to work on your business on the Internet. Instead of paying more money for the rent of the shops and their preparation, you save it to change your business online. If you think from the beginning to sell online, then you only need to pay for e-commerce website and its application and planning how to make this business attractive to customers.

Waleed (small, 1)

I can see e-commerce is limited to few International companies and those who deal with them are specific and limited customers.

### *5.6. future of E-commerce in KSA*

Mohammed (large, 3)

I am confident because the market is so buoyant out here.

So Inshallah I have strongly, strong advocate people will actually go into e-commerce.

Osam (large, 3)

However, our society is classified as a youth community (most of the community are young people), this new generation will strongly accept this new technology and new type of business.

Tahmer (large, 4)

There is a movement and growth toward e-commerce but it is very slow.

I believe that Saudi market is bullish and a fertile ground for investments, but –in my opinion- even we start thinking about e-commerce we need time not less than 10 years to reach the maturity in this field.

## 6. General info

### *6.1. Required changes in our organization to use e-commerce*

Ahmed (large, 2)

We have to make contract with IT Company to build complete e-commerce system, train employees to take care of this new way of marketing and selling, offer support and after sales services. In short words, we need to have complete e-readiness.

Hassan (small, 1)

Training programs for our stuff, maybe doing agreement with delivering/shipment companies

Mohammed (large, 3)

And also have secure networking, once we have secure network there will be no problem at all.

Not majors changes at all, only very minor changes.

we need to have products listing properly done ad also having a block to show the special offers that we have.

Moneer (large, 4)

computer hardware and software

contract with a company to arrange with online payment

Employees to follow-up 24h support

Subscribe with an Internet company (ISP)

Contract with delivery companies

Nasir (medium2)

We need computers and network, and employees' training programs in all involved e-commerce activities.

Saleh (small, 1)

Computer network, contract with delivery companies, employees training programs, arrangement with banks to get online payment systems.

Talal (small, 1)

prepare integrated system and must be sacrificed in the beginning to be fully ready for the electronic system, and also must sacrifice some of your products at the outset to gain the trust of more customers.

Thamer (large, 4)

Before we applied e-commerce system, we have done many things: (1) build accurate networking system to Implemented the order directly without delay, (2) build internal system connecting all departments of the company: purchase, finance, warehouse, delivery and admin.

Waleed (small, 1)

To have complete e-readiness

### *6.2. Name: Purpose of the website*

Ahmed (large, 2)

Our website display information about our company and products but there is no sales or prices shown.

Fadi (medium, 2)

Information about all available types of pizza but prices and online order is not available.

Mohammed (large, 3)

Provide information about our company, what we deal with, what type of products we sell and how to contact us as well.

Moneer (large, 4)

E-commerce website selling DSL services

Naif (medium, 2)

Yes we have website shows information about our products in Arabic and English.
we have e-commerce website mainly targeted for international buyers because the Purchasing power comes from overseas and that's why this website in English.

Nasir (medium, 2)

Actually there are no sales on the Internet and honestly we do not contact customers who do such a request like this. What is on our website is display only.

Osam (large, 3)

Our website was designed for two purposes: producats marketing and provide the basic information that users need (e.g. calls cost …), the second purpose is to provide e-services for our customers such as paying bills, view statement and its analysis, check calls record

Salem (medium, 4)

We are Saudi based company but we have Complete e-commerce website open for whole world not only for Saudi Arabia.

Talal (small, 2)

General information about our company

Thamer (large, 4)

Integrated e-commerce website

### *6.3. Name: Marketing channels*

Ahmed (large, 2)

Internet, SMS, and weakly & monthly magazines

Ali (small, 1)

The [physical] location of our shops

Fadi (medium, 2)

Newspapers, brochures, TV

Hassan (small, 1)

The place of our shops in very important in term of advertisement which means the good location of the shop represents advertisement means for us. Also we use brochures.

Mohammed (large, 3)

Advertising on newspapers and media and also catalogues and broachers

Moneer (large, 4)

We use newspapers, radio, websites which have many visitors. The owner of a website who tells us they have many visitors, we advertise on their website. We also have exhibitions. We do also sponsor exhibitions, seminars and conferences.

Naif (medium, 2)

Newspapers and advertisement boards on the streets

Nasir (medium, 2)

We depend on the location of our shops. We have 5 shops in Jeddah. Sometimes we participate in internal and international Exhibitions.

Osam (large, 3)

Primarily we depend on the ads in the streets, roads, and television ads. As well as we advertise on newspapers and radio, and the normal way in the shops.

Saeed (small, 1)

The location of our shop

Saleh (small, 1)

The location of our shop is the best way of marketing

Salem (medium, 4)

We used to market our products in the same typical way like other companies doing. We are doing door to door, flyers, newspapers, banners on the roads. Basically our session is Ramadan and we do much marketing ads during this session.

Talal (medium, 2)

The location of our shops is very important

Personal is very important too. When when a customer asks they do not give them only answer but attract them to buy.

Tameem (medium, 1)

We advertise by Brochures, on newspapers and special newspaper for ads.

Thamer (large, 4)

We use newspapers (every month 3 to 4 advertisements), TVs, SMS, Brochures, catalogues, banners to advertise special offers, websites and direct personal marketing. Our company has also special TV channel.

Waleed (small, 1)

SMS and brochures to advertise the latest versions

# **Appendix I:** Approval for further extension to the research ethics application

GRIFFITH UNIVERSITY HUMAN RESEARCH ETHICS COMMITTEE

16-Aug-2013

Dear Mr Al Ghamdi

I write further to your application for a variation to your approved protocol "Enabling e-Retail B2C eCommerce Systems in Saudi Arabia" (GU Ref No: ICT/06/09/HREC). This request has been considered by the Office for Research.

The OR resolved to approve the requested variation:

Request to vary the protocol as follows:
1) Re-open the protocol for further data collection from a sample of retailers in Saudi Arabia (100-200). Identity of participants may occur in answering some questions, however, in the research report all the identities will be kept anonymous. The survey will be used in three forms: paper, online and phone interview. Copy of survey materials provided.
2) Extend the clearance date. An extension to the protocol has been granted until 31 December 2013, as the researchers did not clarify an extension date in the variation request. If the researchers require further time for data collection, please advise.

This decision is subject to ratification at the next meeting of the HREC. However, you are authorised to immediately commence the revised project on this basis. I will only contact you again about this matter if the HREC raises any additional questions or comments about this variation.

Regards

Dr Kristie Westerlaken
Policy Officer
Office for Research
Bray Centre, Nathan Campus
Griffith University
ph: +61 (0)7 373 58043
fax: +61 (07) 373 57994
email: k.westerlaken@griffith.edu.au
web:

Cc:

Researchers are reminded that the Griffith University Code for the Responsible Conduct of Research provides guidance to researchers in areas such as conflict of interest, authorship, storage of data, & the training of research students.
You can find further information, resources and a link to the University's Code by visiting http://policies.griffith.edu.au/pdf/Code%20for%20the%20Responsible%20Conduct%20of%20Research.pdf

# **Appendix J:** The letter which was sent for checking the clarity of the statements made by the researcher based on qualitative analysis

Dear Dr…
Thank you for accepting the invitation to participate evaluating my research survey clarity. Please refer to the attached document and provide your comments in the answer area based on the following:

- Do the statements I write/provide well describe the phrases provided by the participants?
- Are my statements formulated in neutral ways? These statements will be included in the survey questions and I want to make sure that the statements do not have either positive or negative indications.
- If you do not agree with any statement, please suggest better wording to describe participants' phrases.

Thank you for your assistance

1. **People's current attitudes and habits regarding shopping in person**

    - "*People culture to buy online discourages us to think of providing option to buy online*"
    - "*Our culture buying and selling online is very weak. Whoever sells or buys should be professional*"

| Your comment: |
|---|
| |

2. **Current level of consumers understanding of e-commerce benefits**

    - "*If there is people understand the benefit of e-commerce I agree with you it help to gain more profits*"
    - "*people here understand the benefit of e-commerce 100%*"

| Your comment: |
|---|
| |

3. **Current level of consumer trust in online shopping**
    - "*Customers do not trust to buy online and they only accept with strike conditions, because it requires credit cards payment*"
    - "*Naturally the trust needs to be built. No one will trust you with the first impression*"

| Your comment: |
|---|
| |

4. **Current level of consumer credit card usage**
   - "*we notice that people feel difficult to put their credit card information*"
   - "*Also there is another problem that people are afraid to buy online using credit cards*"

| Your comment: |
|---|
| |

5. **Current level of consumers' demand of buying online**
   - "*there is no good demand to buy online. For example, in western region of Saudi Arabia we receive 3-5 orders buying online every week and these are nothing compared to normal way of business*"
   - "*we have not seen strong foothold of e-commerce in this part of the world*"

| Your comment: |
|---|
| |

6. **The number of people using the internet**
   - "*Today almost every house has Internet connection and I believe Internet usages will increase in the future*"
   - "*The percentage of users who browse the Internet is high*"

| Your comment: |
|---|
| |

7. **Current level of consumer knowledge of buying online**
   - "*This subject requires education to make it easy for people understand how it works and what its benefits are*"
   - "*The majority of people in our country do not know how to buy and sell on the Internet*"

| Your comment: |
|---|
| |

8. **Current level of consumers familiarity with e-commerce**
   - "*We [Saudi community] are still fairly new in the use of electronic commerce*"
   - "*People here are not familiar with e-commerce and this is not something that you can apply it in a day*"

| Your comment: |
|---|
| |

**9. Level of willingness of consumers to pay delivery fees**

- "*With delivery there is a problem if there is delivery fees, customer maybe not paying these fees except those from specific class or busy*"
- "*we have concern regarding delivering the purchased products and its fees*"

| Your comment: |
|---|
| |

**10. Level of availability of Wi-fi and broadband services**

- "*But I think someone needs to kick the whole things of having the access of Wi-Fi, of having the access of broadband, having the access in every house hold. Once this done, obviously you will see the results of that very soon*"
- "*But we need to have the infrastructure right. In order to have the infrastructure right we must make sure that the Wi-Fi is available in better cases and more places. We must ensure that broadband is available for other people in remote places*"

| Your comment: |
|---|
| |

**11. Current level of safeguards to protect consumers when online shopping**

- "*It has to be done from early stage, customers have to be ensured there are rules and legislations protect their right*"
- "*With no clear e-commerce law and legislations applying e-commerce systems are not encouraging*"

| Your comment: |
|---|
| |

**12. Current availability of online payment options**

- "*Banks should provide easy options to have two credit cards, one with large amount of money and another one with small amount to be used in online payment*"
- "*There are secure payment systems; you just need to incorporate that into your system*"

| Your comment: |
|---|
| |

**13. The available of SADAD for online payment for internet shopping**

- "*SADAD before was very expensive solution, it is good for large companies but it's not for middle size company like us*"
- "*SADAD system is more secure system and great*"

| Your comment: |
|---|
| |

**14. The current mailing and addressing system in Saudi Arabia**

- "*We depend on the mobile phone number; if the mobile phone number is not correct then it's difficult to deliver the product. This is not only for us but also for other companies and home delivery people they are struggling with these things. But there are certain locations (e.g. companies buildings) you can know the address clearly, you know which street and block, building, floor, flat etc. But with community houses is not very clear.*"
- "*The problem is mail address; still some people do not have mail address*"

Your comment:

**15. Level of easiness of doing e-commerce**

- "*I feel it is difficult and case a headache. The normal way is better and easy*"
- "*Selling online is difficult to show all the details of the products unless you have customers understand the details of the products because they have known this product before with normal buying and dealt with you normally*"

Your comment:

**16. Nature of our products we sell**

- "*Same products [electrical and electronics] are widespread in the market*"
- "*Some customers are interested to buy from you but when they see Fragile products they will go away, it is better for them to buy from local market instead of buying online where they cannot ensured the product will shipped safely*"

Your comment:

**17. Level of e-commerce knowledge and familiarity in our company**

- "*Our employees are familiar with direct sales*"
- "*our staff don't know about marketing and selling on the Internet where they required training programs*"

Your comment:

**18. Management attitude towards e-commerce**

- "*we will continue providing this option (selling online) to encourage people to use it*"

- "*I personally like to do move toward e-commerce system. This will actually help your customers to browse, search, check the prices for competitor, and check the quality of products in terms of features, advantages and benefits and then make the right choice. It makes comfort and less time to move around*"

| Your comment: |
|---|
| |

**19. Our current business priority**

- "*using the Internet to order our products is not our priority at the moment*"
- "*sometimes I receive orders by e-mail but I don't consider them as serious orders. I give a priority to those who come to the shop to buy*"

| Your comment: |
|---|
| |

**20. Our company's level of trust in online selling processes**

- "*We do not trust online payment, the money must paid cash*"
- "*this [security of online transactions] cause a difficulty*"

| Your comment: |
|---|
| |

**21. Setup and maintenance cost**

- "*Financial costs involved in something that does not bring profit*"
- "*Building the website is very easy, you can buy cheap stuff/software from the Internet, there are e-commerce packages available cost around $100 to buy a package and publish your e-commerce website. Before it was very expensive, but know things become cheaper*"

| Your comment: |
|---|
| |

# **Appendix K:** Survey Design (English Version)

*School of Information and Communication Technology*
Researcher: Rayed AlGhamdi
***Research Title: Diffusion of the Adoption of Online Retailing Systems in Saudi Arabia***

***Research purpose:***
The purpose of this survey is to collect information from retailers in KSA that will assist the researcher to provide empirical evidence for this academic research in relation to the diffusion of the adoption of online retailing systems in Saudi Arabia.

***The expected benefit of the research:***
This research will assist the researcher completing his Griffith University PhD requirements. Data collected from this survey and the outcomes of its analysis will support further academic research in the field of e-commerce in general and in the same field in Saudi Arabia specifically. The outcomes of this study are expected to contribute to the development of e-commerce in Saudi Arabia.

***Voluntary participation:***
Your participation is voluntary. You do not need to answer every question unless you wish to do so; however the answers you provide in this survey will help the researcher to complete the analysis of data. I highly appreciate your time answering all the questions.

***The ethical conduct of this research:***
This research is being conducted in accordance with the National Statement on Ethical Conduct in Human Research and all the ethical standards that are applied by Griffith University in social research (Ethics Protocol Number ICT/06/09/HREC).

***Feedback to you:***
Participants will be offered a copy of the result of the research if they completed and returned the questionnaire.

***Legal Privacy Statement:***
The conduct of this research involves the collection, access and / or use of your identified personal information. The information collected is confidential and will not be disclosed to third parties without your consent, except to meet government, legal or other regulatory authority requirements. A de-identified copy of this data may be used for other research purposes. However, your anonymity will at all times be safeguarded. For further information consult the University's Privacy Plan at www.griffith.edu.au/ua/aa/vc/pp or telephone (+61 7 3735 5585).

Any questions or comments regarding this survey should be addressed to:

**Rayed AlGhamdi**
*School of ICT*
*Griffith University*
*Mobile: +61 411787195*
*r.alghamdi@griffith.edu.au*

**Prof. Kamal Mansour Jambi**
*Faculty of Computing and IT*
*King Abdulaziz University,*
*Ph: +966 2 6400000 Ex: 52926*
*Fax: +966 2 6405845*
*kjambi@kau.edu.sa*

**The Manager for Research Ethics**
*Office for Research*
*Bray Centre, Nathan Campus,*
*Griffith University*
*Ph: +61 7 3735 5585 or*
*research-ethics@griffith.edu.au*

1. **What is your position in this organization?**
   - ○ Managing Director/General Manager/CEO
   - ○ Deputy Managing Director/Deputy General manager
   - ○ Department Manager
   - ○ Other....................................................... (please specify your position)

2. **How old is your organization?**
   - ○ Less than 10 years
   - ○ 10 – 20 years
   - ○ 21 – 30 years
   - ○ More than 30 years

3. **Which one of the following best describes your organization's primary business?**
   - ○ Apparel, accessories, and footwear
   - ○ Books and school needs
   - ○ Building materials
   - ○ Cars, auto parts, and accessories
   - ○ Computers and accessories
   - ○ Electronics
   - ○ Furniture
   - ○ Groceries
   - ○ Jewelry
   - ○ Home appliances
   - ○ Medicine and medical equipment
   - ○ Optical products
   - ○ Perfumes and beauty products
   - ○ Printing equipment and/or services
   - ○ Sporting goods
   - ○ Telecommunications services
   - ○ Toys and video games
   - ○ Travel and tourism
   - ○ Others.................................................................. (please specify)

4. **How many employees work in your organization?**
   - ○ 1-5
   - ○ 6-50
   - ○ 51-200
   - ○ More than 200

5. **How much is the annual revenue for your organization?**
   - ○ Up to SAR 20 million
   - ○ SAR 20-50 million
   - ○ More than SAR 50 million

6. **Which one of the following best describes the impact of e-commerce on your organization?**
   - ○ E-commerce has no impact on our business
   - ○ E-commerce has little impact on our business
   - ○ E-commerce has considerable impact on our business
   - ○ E-commerce is critical to our business

**7. If you sell online, when did you start doing this?**

- O 1 year or less
- O 2 years
- O 3 years
- O 4 years
- O Others.................................................................. (please specify)

**8. If you sell online, what percentage of total sales comes from your website?**

- O Less than 10%
- O 10 to 29%
- O 30-50%
- O More than 50%
- O Others.................................................................. (please specify)

9. **When does your organization expect to have the following e-commerce capabilities?** For each of the items please indicate by putting a tick (✓)in the appropriate column.

| E-commerce Capabilities | Have it now | Within 3 years | Probably never | No definite plan |
|---|---|---|---|---|
| Publish information about our organization on our own website | | | | |
| Use of the internet for marketing and advertisements | | | | |
| Publish product descriptions including prices on our own website | | | | |
| Communicate with consumers through our own website | | | | |
| Communicate with consumers using social networks (e.g. twitter and facebook). | | | | |
| Get consumer feedback or ratings of products on our website | | | | |
| Allow consumers to place orders on our website using a shopping cart | | | | |
| Accept payment from consumers using online payment systems | | | | |
| Have dedicated IT staff and business-oriented staff to manage e-commerce activities | | | | |
| Place orders with our suppliers over the Internet | | | | |

**10. For each of the factors below, please rate the impact that each has on the decision to use e-commerce in your organization using this 7 point -3 to +3 scale:**

| -3 | -2 | -1 | 0 | +1 | +2 | +3 |
|---|---|---|---|---|---|---|
| Major barrier | | | No influence | | | Strong incentive |

| *Factor* | Level of Impact *Barrier* -------------------------------------- *Incentive* | | | | | | |
|---|---|---|---|---|---|---|---|
| | -3 | -2 | -1 | 0 | +1 | +2 | +3 |
| People's current attitudes and habits regarding shopping online | ○ | ○ | ○ | ○ | ○ | ○ | ○ |
| The current level of consumer recognition of e-commerce benefits | ○ | ○ | ○ | ○ | ○ | ○ | ○ |
| The current level of consumer trust in online shopping | ○ | ○ | ○ | ○ | ○ | ○ | ○ |
| The current level of consumer usage of credit cards for online shopping | ○ | ○ | ○ | ○ | ○ | ○ | ○ |
| The current level of consumer demand for buying online | ○ | ○ | ○ | ○ | ○ | ○ | ○ |
| The current level of consumer knowledge of buying online | ○ | ○ | ○ | ○ | ○ | ○ | ○ |
| Level of willingness of consumers to pay delivery fees | ○ | ○ | ○ | ○ | ○ | ○ | ○ |
| The percentage of people using the internet | ○ | ○ | ○ | ○ | ○ | ○ | ○ |
| Level of availability of wi-fi and broadband services | ○ | ○ | ○ | ○ | ○ | ○ | ○ |
| Current level of e-commerce legislation for online shopping | ○ | ○ | ○ | ○ | ○ | ○ | ○ |
| Current availability of online payment options | ○ | ○ | ○ | ○ | ○ | ○ | ○ |
| The current mailing and addressing system in Saudi Arabia | ○ | ○ | ○ | ○ | ○ | ○ | ○ |
| Level of ease for our company to do e-commerce | ○ | ○ | ○ | ○ | ○ | ○ | ○ |
| Nature of the products our company sells | ○ | ○ | ○ | ○ | ○ | ○ | ○ |
| Level of e-commerce knowledge and familiarity in our company | ○ | ○ | ○ | ○ | ○ | ○ | ○ |
| Management attitude towards e-commerce in our company | ○ | ○ | ○ | ○ | ○ | ○ | ○ |
| The current business priorities of our company | ○ | ○ | ○ | ○ | ○ | ○ | ○ |
| Our company's level of trust in online selling processes | ○ | ○ | ○ | ○ | ○ | ○ | ○ |
| Setup and maintenance costs of e-commerce for our company | ○ | ○ | ○ | ○ | ○ | ○ | ○ |

11. **Please write below if you have more to say about factors that inhibit or motivate your organization to adopt and use e-commerce.**
    ....................................................................................................................
    ....................................................................................................................
    ....................................................................................................................
    ....................................................................................................................

12. **Are you aware of the recent developments regarding the mailing and addressing system carried out by the Saudi Post?**
    - ○ Yes
    - ○ No

13. **Are you aware of the update that took place on the SADAD system, where capacity has been expanded to 20,000 billers?**
    - ○ Yes
    - ○ No

14. **Have you read the latest report released by the Saudi Communication and Information Technology Commission (CITC) regarding Internet usage and e-commerce activities?**
    - ○ Yes
    - ○ No

15. **If you want to suggest or say anything regarding e-commerce in Saudi Arabia, please write below**
    ....................................................................................................................
    ....................................................................................................................
    ....................................................................................................................
    ....................................................................................................................
    ....................................................................................................................

**Thank you for your time and participation in completing this questionnaire.**

Please tick any box that applies to you

- ❒ I am interested in the e-commerce development in Saudi Arabia and would be willing to be contacted in the future for further participation
- ❒ I would like to receive a summary of the survey results.

If you ticked any of the above, please provide your preferred contact details

Contact name: ......................................................................................
Address: ......................................................................................
......................................................................................
Contact No./E-mail: ......................................................................................

# **Appendix L:** Survey Design (Arabic Version)

الباحث: رائد بن عبدالله الغامدي

**عنوان البحث: تبني نظم البيع عن طريق الانترنت لتجار التجزئة في المملكة العربية السعودية**

**هدف الاستبيان:**

يهدف هذا الاستبيان لجمع معلومات تتعلق بموضوع الدراسة من الشركات والمؤسسات التجارية (قطاع تجارة التجزئة) العاملة في السعودية؛ لمعرفة مدى انتشار البيع على الإنترنت والتحديات التي تواجهه.

**الفائدة المتوقعة لهذا البحث:**

جميع المعلومات التي تجمع في هذا الاستبيان غرضها الاستخدام البحثي، وهذا الاستبيان هو جزء من دراسة دكتوراه يقوم بها الباحث لدعم تطبيق أنظمة التسوق الإليكتروني في المملكة العربية السعودية.

**المشاركة تطوعية:**

المشاركة في تعبئة هذا الاستبيان تطوعية، كل إجابة تقدمها في هذا الاستبيان لها قيمة عالية جدًا، فهي تساعد الباحث على التحليل الدقيق والوصول لنتائج موثقة.

**المعايير الأخلاقية لهذا البحث:**

إن هذا البحث وجميع مايتعلق به من جمع معلومات خاضع للشروط والمعايير الأخلاقية لإجراء البحوث الاجتماعية المطبقة في جامعة Griffith (Ethics Protocol Number ICT/06/09/HREC) ، وهو ما يضمن أن جميع المعلومات التي تُجمع من قبل المشاركين لن تستخدم إلا للأغراض البحثية.

**الحصول على نتائج هذا الاستبيان:**

المشارك الذي يكمل جميع إجابات أسئلة هذه الاستبانة يحق له الحصول على ملخص لنتيجة البحث ، يُرسل على بريده الالكتروني إذا رغب ذلك.

**بيان الخصوصية:**

إن جميع المعلومات المتعلقة بتحديد هوية المشارك في هذا الاستبيان سرية ولن يتم الكشف عنها إلى أطراف ثالثة دون موافقتك، باستثناء تلبية متطلبات الحكومة أو غيرها من متطلبات الهيئات التنظيمية التي تثبت بأن هذا البحث يسير وفق الشؤون الأخلاقية. البيانات التي تجمع من هذا الاستبيان قد تستخدم لأغراض بحثية أخرى، ومع ذلك وفي جميع الأحوال فإن بيانات تحديد هوية المشارك ستبقى سرية. وللمزيد من المعلومات حول خطة الجامعة بشأن الخصوصية يُرجى الاطلاع على www.griffith.edu.au/ua/aa/vc/pp أو الاتصال على الهاتف (+61 7 3735 5585).

للحصول على المزيد من الأسئلة أو الاستفسارات المتعلقة بهذا الاستبيان يرجى الاتصال بـ

الباحث : رائد بن عبدالله الغامدي
تقنية المعلومات والاتصالات
جامعة Griffith – أستراليا
هاتف: 558441379 966+
r.alghamdi@griffith.edu.au

أ.د. كمال منصور جمبي
كلية الحاسبات وتقنية المعلومات
جامعة الملك عبدالعزيز
هاتف: +966 2 6400000 Ex: 52926
kjambi@kau.edu.sa

The Manager for Research Ethics,
Bray Centre, Office for Research,
Griffith University Nathan Campus,
Ph: +61 7 3735 5585 or
research-ethics@griffith.edu.au

**16. ما هو مركزك الوظيفي في هذه الشركة/المؤسسة؟**
- O المدير العام / الرئيس التنفيذي
- O نائب المدير العام/ نائب الرئيس التنفيذي
- O مدير إدارة أو قسم
- O أخرى.................................................................. (من فضلك حدد)

**17. كم هو عمر شركتكم/مؤسستكم؟**
- O أقل من 10 سنوات
- O 10 – 20 سنة
- O 21 – 30 سنة
- O أكثر من 30 سنة

**18. أي الخيارات التالية يصف النشاط الرئيسي لشركتكم/مؤسستكم؟**
- O ملابس وإكسسوارات، وأحذية
- O كتب واحتياجات مدراس
- O مواد البناء
- O سيارات، قطع غيار السيارات، وإكسسوارات
- O أجهزة كمبيوتر وملحقاتها
- O إلكترونيات
- O أثاث منزلي
- O مواد غذائية وتموينات منزلية
- O مجوهرات
- O أجهزة وأدوات المنزلية
- O الأدوية والمعدات الطبية
- O المنتجات البصرية
- O عطور ومستحضرات تجميل
- O معدات وخدمات الطباعة
- O أدوات ومنتجات رياضية
- O خدمات الاتصالات السلكية واللاسلكية
- O ألعاب الأطفال
- O سفر وسياحة
- O أخرى .............................................................. (من فضلك حدد)

**19. كم عدد الموظفين العاملين بشركتكم/مؤسستكم؟**
- O 1- 5
- O 6 - 50
- O 51 - 200
- O أكثر من 200

**20. كم هو العائد السنوي لشركتكم/مؤسستكم؟**
- O إلى 20 مليون ريال سعودي
- O 20 – 50 مليون ريال سعودي
- O أكثر من 50 مليون ريال سعودي

**21. أي مما يلي يصف درجة تأثير التجارة الإلكترونية على شركتكم/مؤسستكم؟**
- O التجارة الإلكترونية ليس لها أي تأثير على أعمالنا
- O التجارة الإلكترونية لها تأثير ضئيل على أعمالنا
- O التجارة الإلكترونية لها تأثير كبير على عملنا
- O التجارة الإلكترونية المحدد الرئيس لعملنا

**22. إذا كانت شركتكم/مؤسستكم تبيع عن طريق الإنترنت؛ منذ متى بدأتم البيع بهذه الطريقة؟**
- O سنة أو اقل
- O سنتان
- O 3 سنوات
- O 4 سنوات فأكثر

○ أخرى ........................................................ (من فضلك حدد)

**23. إذا كانت شركتكم/مؤسستكم تبيع عن طريق الإنترنت؛ كم هي نسبة المبيعات عن طريق استخدام هذه الطريقة من إجمالي البيع الكلي؟**

○ أقل من 10 %
○ 10 – 29 %
○ 30 – 50 %
○ أكثر من 50 %
○ أخرى ........................................................ (من فضلك حدد)

**24. متى تتوقع أن يكون لشركتكم/مؤسستكم قدرات التجارة الإلكترونية التالية؟**
لكل من البنود يرجى الإشارة عن طريق وضع علامة (✓) علامة في خانة العمود المناسب.

| قدرات التجارة الإلكترونية | لدينا الآن | خلال 3 سنوات | لن نستخدم | لاتوجد خطة محددة |
| --- | --- | --- | --- | --- |
| نشر معلومات حول شركتنا/مؤسستنا على موقعنا على الانترنت | | | | |
| استخدام شبكة الإنترنت لأغراض التسويق والإعلانات | | | | |
| نشر وصف لمنتجاتنا وأسعارها على موقعنا على الإنترنت | | | | |
| التواصل مع عملاءنا عن طريق موقعنا على الانترنت | | | | |
| التواصل مع عملاءنا باستخدام الشبكات الاجتماعية (مثل تويتر و فيسبوك و يوتيوب). | | | | |
| موقعنا على الإنترنت يتيح للعملاء إضافة تعليقاتهم وتقييم المنتجات | | | | |
| بإمكان عملاءنا الشراء عن طريق موقعنا على الإنترنت باستخدام عربة التسوق shopping cart | | | | |
| بإمكان عملاءنا إجراء عمليات الدفع باستخدام أحد خيارات الدفع الالكتروني (مثل:البطاقات الائتمانية – نظام سداد– PayPal) | | | | |
| مجموعة موظفين ذوي تخصصات تقنية معلومات IT وتخصصات تجارية لإدارة أنشطة التجارة الإلكترونية | | | | |
| إمكانية الشراء عن طريق الإنترنت من الموردين | | | | |

**25. لكل من العوامل أدناه، يرجى تحديد معدل تأثيرها على قرار استخدام التجارة الإلكترونية في شركتكم/مؤسستكم**

| -3 | -2 | -1 | 0 | +1 | +2 | +3 |
|---|---|---|---|---|---|---|
| عائق قوي | | | ليس له تأثير | | | حافز قوي |

| العوامل المؤثرة | درجة التأثير<br>عائق------------------------------ حافز<br>-3 | -2 | -1 | 0 | +1 | +2 | +3 |
|---|---|---|---|---|---|---|---|
| عادات وتوجهات الناس الحالية نحو التسوق عن طريق الإنترنت | ○ | ○ | ○ | ○ | ○ | ○ | ○ |
| المستوى الحالي لمعرفة المستهلكين بفوائد التجارة الإلكترونية | ○ | ○ | ○ | ○ | ○ | ○ | ○ |
| المستوى الحالي لثقة المستهلكين في التسوق عن طريق الانترنت | ○ | ○ | ○ | ○ | ○ | ○ | ○ |
| المستوى الحالي لاستخدام المستهلكين بطاقات الائتمان للتسوق عن طريق الإنترنت | ○ | ○ | ○ | ○ | ○ | ○ | ○ |
| المستوى الحالي لإقبال المستهلكين على الشراء عن طريق الانترنت | ○ | ○ | ○ | ○ | ○ | ○ | ○ |
| المستوى الحالي لمعرفة وخبرة المستهلكين بالشراء عن طريق الانترنت | ○ | ○ | ○ | ○ | ○ | ○ | ○ |
| مستوى استعداد المستهلكين لدفع رسوم توصيل الطلبات | ○ | ○ | ○ | ○ | ○ | ○ | ○ |
| نسبة عدد السكان الذين يستخدمون الإنترنت | ○ | ○ | ○ | ○ | ○ | ○ | ○ |
| المستوى الحالي لتوفر خدمات النطاق العريض والإنترنت السريع | ○ | ○ | ○ | ○ | ○ | ○ | ○ |
| المستوى الحالي لتشريعات التجارة الإلكترونية للتسوق عن طريق الإنترنت | ○ | ○ | ○ | ○ | ○ | ○ | ○ |
| الخيارات الحالية المتوفرة للدفع عن طريق الإنترنت | ○ | ○ | ○ | ○ | ○ | ○ | ○ |
| النظام البريدي الحالي ونظام العنونة في السعودية | ○ | ○ | ○ | ○ | ○ | ○ | ○ |
| مستوى سهولة التجارة الإلكترونية بالنسبة لشركتنا/مؤسستنا | ○ | ○ | ○ | ○ | ○ | ○ | ○ |
| طبيعة المنتجات التي تبيعها شركتنا/مؤسستنا | ○ | ○ | ○ | ○ | ○ | ○ | ○ |
| مستوى المعرفة والخبرة لشركتنا بالتجارة الإلكترونية | ○ | ○ | ○ | ○ | ○ | ○ | ○ |
| توجهات الإدارة في شركتنا/مؤسستنا تجاه التجارة الإلكترونية | ○ | ○ | ○ | ○ | ○ | ○ | ○ |
| أولويات العمل الحالية لشركتنا/مؤسستنا | ○ | ○ | ○ | ○ | ○ | ○ | ○ |
| مستوى ثقتنا في عمليات البيع عن طريق الانترنت | ○ | ○ | ○ | ○ | ○ | ○ | ○ |
| تكاليف إنشاء وصيانة نظام التجارة الإلكترونية بالنسبة لشركتنا/مؤسستنا | ○ | ○ | ○ | ○ | ○ | ○ | ○ |

**26. يُرجى الكتابة أدناه إذا كان لديك المزيد لتقوله عن عوائق أو محفزات استخدام التجارة الإلكترونية في شركتكم/مؤسستكم.**

..........................................................................................................................................

..........................................................................................................................................

..........................................................................................................................................

..........................................................................................................................................

..........................................................................................................................................

**27. هل لديك علم بالتطورات الحديثة لنظام العنونة البريدية التي نفذها البريد السعودي حيث أصبح لكل مسكن عنوان؟**

- ○ نعم
- ○ لا

**28. هل لديك علم بالتحديثات التي جرت على نظام سداد للمدفوعات الالكترونية ؛ حيث تم توسيع القدرة الاستيعابية للمفوترين إلى 20 ألف مفوتر؟**

- ○ نعم
- ○ لا

29. **هل اطلعت على أحدث تقرير صادر عن هيئة الاتصالات وتقنية المعلومات بشأن استخدامات الإنترنت وأنشطة التجارة الإلكترونية بالسعودية؟**

- ○ نعم
- ○ لا

**30. إذا كان لديك ما ترغب إضافته أو قوله بخصوص التسوق اليكتروني في السعودية، من فضلك اكتبه هنا**

..........................................................................................................................................

..........................................................................................................................................

..........................................................................................................................................

..........................................................................................................................................

**شكرا لك على المشاركة في الإجابة على أسئلة هذه الاستبانة**

---

هذا الجزء اختياري ، من فضلك ضع علامة (✓) أمام ما ترغبه

- ❐ لدي اهتمام بموضوع تطوير التجارة الإلكترونية في السعودية، ويسعدني المشاركة في أبحاث مستقبلية ذات العلاقة
- ❐ أرغب في الحصول على ملخص لنتائج هذا البحث

إذا قمت باختيار أيًا من البنود أعلاه ، من فضلك اكتب تفاصيل التواصل التالية:

الاسم: ..........................................................................................................................

العنوان: ..........................................................................................................................

..........................................................................................................................................

رقم الاتصال / البريد الإلكتروني: ..........................................................................................

# **Appendix M:** Full statistical descriptive analysis using SPSS v.20

**Statistics**

| | N | |
|---|---|---|
| | Valid | Missing |
| What is your position in this organization? | 153 | 0 |
| [Other]  What is your position in this organization? | 153 | 0 |
| How old is your organization? | 152 | 1 |
| Which one of the following best describes your organization’s primary business? | 153 | 0 |
| [Other]  Which one of the following best describes your organization’s primary business? | 153 | 0 |
| How many employees work in your organization? | 153 | 0 |
| How much is the annual revenue for your organization? | 143 | 10 |
| Which one of the following best describes the impact of e-commerce on your organization? | 153 | 0 |
| When did your organization start selling online? | 45 | 108 |
| What percentage of total sales comes from your website? | 45 | 108 |
| [Other]  What percentage of total sales comes from your website? | 153 | 0 |
| [Publish information about our organization on our own website] | 153 | 0 |
| [Use of the internet for marketing and advertisements] | 153 | 0 |
| [Publish product descriptions including prices on our own website] | 153 | 0 |
| [Communicate with consumers through our own website] | 153 | 0 |
| [Communicate with consumers using social networks (e.g. twitter and facebook).] | 153 | 0 |
| [Get consumer feedback or ratings of products on our website ] | 153 | 0 |
| [Allow consumers to place orders on our website using a shopping cart ] | 153 | 0 |
| [Accept payment from consumers using online payment systems] | 153 | 0 |
| [Have dedicated IT staff  and business-oriented staff to manage e-commerce activities] | 153 | 0 |
| [Place orders with our suppliers over the Internet ] | 153 | 0 |
| E-commerce Maturity Level | 153 | 0 |
| [People's current attitudes and habits regarding shopping online] | 153 | 0 |
| [The current level of consumer recognition of e-commerce benefits] | 153 | 0 |
| [The current level of consumer trust in online shopping] | 153 | 0 |
| [The current level of consumer usage of credit cards for online shopping] | 153 | 0 |
| [The current level of consumer demand for buying online] | 153 | 0 |
| [The current level of consumer knowledge of buying online] | 153 | 0 |
| [Level of willingness of consumers to pay delivery fees] | 153 | 0 |
| [The percentage of people using the internet] | 153 | 0 |
| [Level of availability of wi-fi and broadband services] | 153 | 0 |
| [Current level of e-commerce legislation for online shopping] | 153 | 0 |
| [Current availability of online payment options] | 153 | 0 |
| [The current mailing and addressing system in Saudi Arabia] | 153 | 0 |
| [Level of ease for our company to do e-commerce] | 153 | 0 |
| [Nature of the products our company sells] | 153 | 0 |
| [Level of e-commerce knowledge and familiarity in our company] | 153 | 0 |
| [Management attitude towards e-commerce in our company] | 153 | 0 |
| [The current business priorities of our company] | 153 | 0 |
| [Our company’s level of trust in online selling processes] | 153 | 0 |
| [Setup and maintenance costs of e-commerce for our company] | 153 | 0 |
| Please write below if you have more to say about factors that inhibit or motivate your organization to adopt and use e-commerce. | 153 | 0 |
| Are you aware of the recent developments regarding the mailing and addressing system carried out by the Saudi Post? | 153 | 0 |
| Are you aware of the update that took place on the SADAD system, where capacity has been expanded to 20,000 billers? | 153 | 0 |
| Have you read the latest report released by the Saudi Communication and Information Technology Commission (CITC) regarding Internet usage and e-commerce activities? | 153 | 0 |
| If you want to suggest or say anything regarding e-commerce in Saudi Arabia, please write below | 153 | 0 |

## Frequency Table

**What is your position in this organization?**

| | | Frequency | Percent | Valid Percent | Cumulative Percent |
|---|---|---|---|---|---|
| Valid | Managing Director/General Manager/CEO/Owner | 63 | 41.2 | 41.2 | 41.2 |
| | Deputy Managing Director/Deputy General manager | 38 | 24.8 | 24.8 | 66.0 |
| | Department Manager | 43 | 28.1 | 28.1 | 94.1 |
| | Other | 9 | 5.9 | 5.9 | 100.0 |
| | Total | 153 | 100.0 | 100.0 | |

**[Other] What is your position in this organization?**

| | | Frequency | Percent | Valid Percent | Cumulative Percent |
|---|---|---|---|---|---|
| Valid | | 148 | 96.7 | 96.7 | 96.7 |
| | ???? | 1 | .7 | .7 | 97.4 |
| | ????? ??? | 1 | .7 | .7 | 98.0 |
| | ??????? ??????????? | 1 | .7 | .7 | 98.7 |
| | Only Recommendations | 1 | .7 | .7 | 99.3 |
| | Test | 1 | .7 | .7 | 100.0 |
| | Total | 153 | 100.0 | 100.0 | |

**How old is your organization?**

| | | Frequency | Percent | Valid Percent | Cumulative Percent |
|---|---|---|---|---|---|
| Valid | Less than 10 years | 31 | 20.3 | 20.4 | 20.4 |
| | 10-20 years | 61 | 39.9 | 40.1 | 60.5 |
| | 21-30 years | 38 | 24.8 | 25.0 | 85.5 |
| | more than 30 years | 22 | 14.4 | 14.5 | 100.0 |
| | Total | 152 | 99.3 | 100.0 | |
| Missing | System | 1 | .7 | | |
| Total | | 153 | 100.0 | | |

**Which one of the following best describes your organization’s primary business?**

| | | Frequency | Percent | Valid Percent | Cumulative Percent |
|---|---|---|---|---|---|
| Valid | Apparel, accessories, and footwear | 15 | 9.8 | 9.8 | 9.8 |
| | Books and school needs | 9 | 5.9 | 5.9 | 15.7 |
| | Building materials | 5 | 3.3 | 3.3 | 19.0 |
| | Cars, auto parts, and accessories | 7 | 4.6 | 4.6 | 23.5 |
| | Computers and accessories | 9 | 5.9 | 5.9 | 29.4 |
| | Electronics | 12 | 7.8 | 7.8 | 37.3 |
| | Furniture | 10 | 6.5 | 6.5 | 43.8 |
| | Groceries | 14 | 9.2 | 9.2 | 52.9 |
| | Jewellry & watches | 7 | 4.6 | 4.6 | 57.5 |
| | Home appliances | 7 | 4.6 | 4.6 | 62.1 |
| | Medicine and medical equipment | 7 | 4.6 | 4.6 | 66.7 |
| | Perfumes and beauty products | 14 | 9.2 | 9.2 | 75.8 |
| | Printing equipment and/or services | 3 | 2.0 | 2.0 | 77.8 |
| | Sporting goods | 5 | 3.3 | 3.3 | 81.0 |
| | Telecommunications services | 6 | 3.9 | 3.9 | 85.0 |
| | Toys and video games | 8 | 5.2 | 5.2 | 90.2 |
| | Travel and tourism | 5 | 3.3 | 3.3 | 93.5 |
| | othes | 10 | 6.5 | 6.5 | 100.0 |
| | Total | 153 | 100.0 | 100.0 | |

**[Other] Which one of the following best describes your organization’s primary business?**

| | | Frequency | Percent | Valid Percent | Cumulative Percent |
|---|---|---|---|---|---|
| Valid | | 144 | 94.1 | 94.1 | 94.1 |
| | ???? ??????? | 1 | .7 | .7 | 94.8 |
| | ????? ?????? ?????? ????? ????? ???? ?????? | 1 | .7 | .7 | 95.4 |
| | ????? ??????? ????????? ? ???? ??????? ?????? ???? | 1 | .7 | .7 | 96.1 |
| | ????? ???????? | 1 | .7 | .7 | 96.7 |
| | ????? ???????????? | 1 | .7 | .7 | 97.4 |
| | ?????? ????? | 1 | .7 | .7 | 98.0 |
| | ??????? ????? ????? ?????????? ??? ?????? ???? ??? | 1 | .7 | .7 | 98.7 |
| | Central Air conditioning for buildings | 1 | .7 | .7 | 99.3 |
| | Decorat ??????? | 1 | .7 | .7 | 100.0 |
| | Total | 153 | 100.0 | 100.0 | |

**How many employees work in your organization?**

| | | Frequency | Percent | Valid Percent | Cumulative Percent |
|---|---|---|---|---|---|
| Valid | 1-5 | 21 | 13.7 | 13.7 | 13.7 |
| | 6-50 | 42 | 27.5 | 27.5 | 41.2 |
| | 51-200 | 53 | 34.6 | 34.6 | 75.8 |
| | more than 200 | 37 | 24.2 | 24.2 | 100.0 |
| | Total | 153 | 100.0 | 100.0 | |

**How much is the annual revenue for your organization?**

| | | Frequency | Percent | Valid Percent | Cumulative Percent |
|---|---|---|---|---|---|
| Valid | Up to SAR 20 million | 62 | 40.5 | 43.4 | 43.4 |
| | SAR 20-50 million | 44 | 28.8 | 30.8 | 74.1 |
| | More than SAR 50 million | 37 | 24.2 | 25.9 | 100.0 |
| | Total | 143 | 93.5 | 100.0 | |
| Missing | System | 10 | 6.5 | | |
| Total | | 153 | 100.0 | | |

**Which one of the following best describes the impact of e-commerce on your organization?**

| | | Frequency | Percent | Valid Percent | Cumulative Percent |
|---|---|---|---|---|---|
| Valid | E-commerce has no impact on our business | 65 | 42.5 | 42.5 | 42.5 |
| | E-commerce has little impact on our business | 37 | 24.2 | 24.2 | 66.7 |
| | E-commerce has considerable impact on our business | 39 | 25.5 | 25.5 | 92.2 |
| | E-commerce is critical to our business | 12 | 7.8 | 7.8 | 100.0 |
| | Total | 153 | 100.0 | 100.0 | |

**When did your organization start selling online?**

| | | Frequency | Percent | Valid Percent | Cumulative Percent |
|---|---|---|---|---|---|
| Valid | 1 year or less | 10 | 6.5 | 22.2 | 22.2 |
| | 2 years | 5 | 3.3 | 11.1 | 33.3 |
| | 3 years | 9 | 5.9 | 20.0 | 53.3 |
| | 4 years or more | 21 | 13.7 | 46.7 | 100.0 |
| | Total | 45 | 29.4 | 100.0 | |
| Missing | System | 108 | 70.6 | | |
| Total | | 153 | 100.0 | | |

**What percentage of total sales comes from your website?**

| | | Frequency | Percent | Valid Percent | Cumulative Percent |
|---|---|---|---|---|---|
| Valid | Less than 10% | 12 | 7.8 | 26.7 | 26.7 |
| | 10 to 29% | 10 | 6.5 | 22.2 | 48.9 |
| | 30-50% | 10 | 6.5 | 22.2 | 71.1 |
| | More than 50% | 10 | 6.5 | 22.2 | 93.3 |
| | others | 3 | 2.0 | 6.7 | 100.0 |
| | Total | 45 | 29.4 | 100.0 | |
| Missing | System | 108 | 70.6 | | |
| Total | | 153 | 100.0 | | |

**[Other] What percentage of total sales comes from your website?**

| | | Frequency | Percent | Valid Percent | Cumulative Percent |
|---|---|---|---|---|---|
| Valid | | 150 | 98.0 | 98.0 | 98.0 |
| | 100 | 1 | .7 | .7 | 98.7 |
| | 100% | 1 | .7 | .7 | 99.3 |
| | 100% ?????? ?? ???? ???????? | 1 | .7 | .7 | 100.0 |
| | Total | 153 | 100.0 | 100.0 | |

**[Publish information about our organization on our own website]**

| | | Frequency | Percent | Valid Percent | Cumulative Percent |
|---|---|---|---|---|---|
| Valid | have it now | 113 | 73.9 | 73.9 | 73.9 |
| | within 3 years | 8 | 5.2 | 5.2 | 79.1 |
| | probably never | 8 | 5.2 | 5.2 | 84.3 |
| | no definite plan | 24 | 15.7 | 15.7 | 100.0 |
| | Total | 153 | 100.0 | 100.0 | |

**[Use of the internet for marketing and advertisements]**

| | | Frequency | Percent | Valid Percent | Cumulative Percent |
|---|---|---|---|---|---|
| Valid | have it now | 88 | 57.5 | 57.5 | 57.5 |
| | within 3 years | 7 | 4.6 | 4.6 | 62.1 |
| | probably never | 13 | 8.5 | 8.5 | 70.6 |
| | no definite plan | 45 | 29.4 | 29.4 | 100.0 |
| | Total | 153 | 100.0 | 100.0 | |

**[Publish product descriptions including prices on our own website]**

| | | Frequency | Percent | Valid Percent | Cumulative Percent |
|---|---|---|---|---|---|
| Valid | have it now | 81 | 52.9 | 52.9 | 52.9 |
| | within 3 years | 7 | 4.6 | 4.6 | 57.5 |
| | probably never | 20 | 13.1 | 13.1 | 70.6 |
| | no definite plan | 45 | 29.4 | 29.4 | 100.0 |
| | Total | 153 | 100.0 | 100.0 | |

**[Communicate with consumers through our own website]**

| | | Frequency | Percent | Valid Percent | Cumulative Percent |
|---|---|---|---|---|---|
| Valid | have it now | 80 | 52.3 | 52.3 | 52.3 |
| | within 3 years | 5 | 3.3 | 3.3 | 55.6 |
| | probably never | 19 | 12.4 | 12.4 | 68.0 |
| | no definite plan | 49 | 32.0 | 32.0 | 100.0 |
| | Total | 153 | 100.0 | 100.0 | |

**[Communicate with consumers using social networks (e.g. twitter and facebook).]**

| | | Frequency | Percent | Valid Percent | Cumulative Percent |
|---|---|---|---|---|---|
| Valid | have it now | 73 | 47.7 | 47.7 | 47.7 |
| | within 3 years | 12 | 7.8 | 7.8 | 55.6 |
| | probably never | 14 | 9.2 | 9.2 | 64.7 |
| | no definite plan | 54 | 35.3 | 35.3 | 100.0 |
| | Total | 153 | 100.0 | 100.0 | |

**[Get consumer feedback or ratings of products on our website ]**

| | | Frequency | Percent | Valid Percent | Cumulative Percent |
|---|---|---|---|---|---|
| Valid | have it now | 63 | 41.2 | 41.2 | 41.2 |
| | within 3 years | 12 | 7.8 | 7.8 | 49.0 |
| | probably never | 27 | 17.6 | 17.6 | 66.7 |
| | no definite plan | 51 | 33.3 | 33.3 | 100.0 |
| | Total | 153 | 100.0 | 100.0 | |

**[Allow consumers to place orders on our website using a shopping cart ]**

| | | Frequency | Percent | Valid Percent | Cumulative Percent |
|---|---|---|---|---|---|
| Valid | have it now | 38 | 24.8 | 24.8 | 24.8 |
| | within 3 years | 17 | 11.1 | 11.1 | 35.9 |
| | probably never | 60 | 39.2 | 39.2 | 75.2 |
| | no definite plan | 38 | 24.8 | 24.8 | 100.0 |
| | Total | 153 | 100.0 | 100.0 | |

**[Accept payment from consumers using online payment systems]**

| | | Frequency | Percent | Valid Percent | Cumulative Percent |
|---|---|---|---|---|---|
| Valid | have it now | 40 | 26.1 | 26.1 | 26.1 |
| | within 3 years | 15 | 9.8 | 9.8 | 35.9 |
| | probably never | 61 | 39.9 | 39.9 | 75.8 |
| | no definite plan | 37 | 24.2 | 24.2 | 100.0 |
| | Total | 153 | 100.0 | 100.0 | |

**[Have dedicated IT staff and business-oriented staff to manage e-commerce activities]**

| | | Frequency | Percent | Valid Percent | Cumulative Percent |
|---|---|---|---|---|---|
| Valid | have it now | 46 | 30.1 | 30.1 | 30.1 |
| | within 3 years | 16 | 10.5 | 10.5 | 40.5 |
| | probably never | 53 | 34.6 | 34.6 | 75.2 |
| | no definite plan | 38 | 24.8 | 24.8 | 100.0 |
| | Total | 153 | 100.0 | 100.0 | |

**[Place orders with our suppliers over the Internet ]**

| | | Frequency | Percent | Valid Percent | Cumulative Percent |
|---|---|---|---|---|---|
| Valid | have it now | 44 | 28.8 | 28.8 | 28.8 |
| | within 3 years | 10 | 6.5 | 6.5 | 35.3 |
| | probably never | 43 | 28.1 | 28.1 | 63.4 |
| | no definite plan | 56 | 36.6 | 36.6 | 100.0 |
| | Total | 153 | 100.0 | 100.0 | |

**E-commerce Maturity Level**

| | | Frequency | Percent | Valid Percent | Cumulative Percent |
|---|---|---|---|---|---|
| Valid | No website | 37 | 24.2 | 24.2 | 24.2 |
| | Static website | 41 | 26.8 | 26.8 | 51.0 |
| | Interactive website | 35 | 22.9 | 22.9 | 73.9 |
| | E-commerce website | 40 | 26.1 | 26.1 | 100.0 |
| | Total | 153 | 100.0 | 100.0 | |

**[People's current attitudes and habits regarding shopping online]**

| | | Frequency | Percent | Valid Percent | Cumulative Percent |
|---|---|---|---|---|---|
| Valid | -3 major barrier | 23 | 15.0 | 15.0 | 15.0 |
| | -2 barrier | 27 | 17.6 | 17.6 | 32.7 |
| | -1 minor barrier | 24 | 15.7 | 15.7 | 48.4 |
| | 0 no influence | 15 | 9.8 | 9.8 | 58.2 |
| | +1 weak incentive | 31 | 20.3 | 20.3 | 78.4 |
| | +2 incentive | 26 | 17.0 | 17.0 | 95.4 |
| | +3 strong incentive | 7 | 4.6 | 4.6 | 100.0 |
| | Total | 153 | 100.0 | 100.0 | |

**[The current level of consumer recognition of e-commerce benefits]**

| | | Frequency | Percent | Valid Percent | Cumulative Percent |
|---|---|---|---|---|---|
| Valid | -3 major barrier | 3 | 2.0 | 2.0 | 2.0 |
| | -2 barrier | 6 | 3.9 | 3.9 | 5.9 |
| | -1 minor barrier | 12 | 7.8 | 7.8 | 13.7 |
| | 0 no influence | 74 | 48.4 | 48.4 | 62.1 |
| | +1 weak incentive | 37 | 24.2 | 24.2 | 86.3 |
| | +2 incentive | 18 | 11.8 | 11.8 | 98.0 |
| | +3 strong incentive | 3 | 2.0 | 2.0 | 100.0 |
| | Total | 153 | 100.0 | 100.0 | |

**[The current level of consumer trust in online shopping]**

| | | Frequency | Percent | Valid Percent | Cumulative Percent |
|---|---|---|---|---|---|
| Valid | -3 major barrier | 6 | 3.9 | 3.9 | 3.9 |
| | -2 barrier | 9 | 5.9 | 5.9 | 9.8 |
| | -1 minor barrier | 18 | 11.8 | 11.8 | 21.6 |
| | 0 no influence | 81 | 52.9 | 52.9 | 74.5 |
| | +1 weak incentive | 29 | 19.0 | 19.0 | 93.5 |
| | +2 incentive | 8 | 5.2 | 5.2 | 98.7 |
| | +3 strong incentive | 2 | 1.3 | 1.3 | 100.0 |
| | Total | 153 | 100.0 | 100.0 | |

**[The current level of consumer usage of credit cards for online shopping]**

| | | Frequency | Percent | Valid Percent | Cumulative Percent |
|---|---|---|---|---|---|
| Valid | -3 major barrier | 7 | 4.6 | 4.6 | 4.6 |
| | -2 barrier | 6 | 3.9 | 3.9 | 8.5 |
| | -1 minor barrier | 24 | 15.7 | 15.7 | 24.2 |
| | 0 no influence | 93 | 60.8 | 60.8 | 85.0 |
| | +1 weak incentive | 19 | 12.4 | 12.4 | 97.4 |
| | +2 incentive | 3 | 2.0 | 2.0 | 99.3 |
| | +3 strong incentive | 1 | .7 | .7 | 100.0 |
| | Total | 153 | 100.0 | 100.0 | |

**[The current level of consumer demand for buying online]**

| | | Frequency | Percent | Valid Percent | Cumulative Percent |
|---|---|---|---|---|---|
| Valid | -3 major barrier | 17 | 11.1 | 11.1 | 11.1 |
| | -2 barrier | 28 | 18.3 | 18.3 | 29.4 |
| | -1 minor barrier | 26 | 17.0 | 17.0 | 46.4 |
| | 0 no influence | 36 | 23.5 | 23.5 | 69.9 |
| | +1 weak incentive | 32 | 20.9 | 20.9 | 90.8 |
| | +2 incentive | 10 | 6.5 | 6.5 | 97.4 |
| | +3 strong incentive | 4 | 2.6 | 2.6 | 100.0 |
| | Total | 153 | 100.0 | 100.0 | |

**[The current level of consumer knowledge of buying online]**

| | | Frequency | Percent | Valid Percent | Cumulative Percent |
|---|---|---|---|---|---|
| Valid | -3 major barrier | 5 | 3.3 | 3.3 | 3.3 |
| | -2 barrier | 2 | 1.3 | 1.3 | 4.6 |
| | -1 minor barrier | 9 | 5.9 | 5.9 | 10.5 |
| | 0 no influence | 84 | 54.9 | 54.9 | 65.4 |
| | +1 weak incentive | 36 | 23.5 | 23.5 | 88.9 |
| | +2 incentive | 17 | 11.1 | 11.1 | 100.0 |
| | Total | 153 | 100.0 | 100.0 | |

**[Level of willingness of consumers to pay delivery fees]**

| | | Frequency | Percent | Valid Percent | Cumulative Percent |
|---|---|---|---|---|---|
| Valid | -3 major barrier | 3 | 2.0 | 2.0 | 2.0 |
| | -2 barrier | 7 | 4.6 | 4.6 | 6.5 |
| | -1 minor barrier | 9 | 5.9 | 5.9 | 12.4 |
| | 0 no influence | 100 | 65.4 | 65.4 | 77.8 |
| | +1 weak incentive | 16 | 10.5 | 10.5 | 88.2 |
| | +2 incentive | 16 | 10.5 | 10.5 | 98.7 |
| | +3 strong incentive | 2 | 1.3 | 1.3 | 100.0 |
| | Total | 153 | 100.0 | 100.0 | |

**[The percentage of people using the internet]**

| | | Frequency | Percent | Valid Percent | Cumulative Percent |
|---|---|---|---|---|---|
| Valid | -3 major barrier | 3 | 2.0 | 2.0 | 2.0 |
| | -2 barrier | 2 | 1.3 | 1.3 | 3.3 |
| | -1 minor barrier | 6 | 3.9 | 3.9 | 7.2 |
| | 0 no influence | 26 | 17.0 | 17.0 | 24.2 |
| | +1 weak incentive | 47 | 30.7 | 30.7 | 54.9 |
| | +2 incentive | 36 | 23.5 | 23.5 | 78.4 |
| | +3 strong incentive | 33 | 21.6 | 21.6 | 100.0 |
| | Total | 153 | 100.0 | 100.0 | |

**[Level of availability of wi-fi and broadband services]**

| | | Frequency | Percent | Valid Percent | Cumulative Percent |
|---|---|---|---|---|---|
| Valid | -3 major barrier | 4 | 2.6 | 2.6 | 2.6 |
| | -2 barrier | 5 | 3.3 | 3.3 | 5.9 |
| | -1 minor barrier | 19 | 12.4 | 12.4 | 18.3 |
| | 0 no influence | 53 | 34.6 | 34.6 | 52.9 |
| | +1 weak incentive | 51 | 33.3 | 33.3 | 86.3 |
| | +2 incentive | 14 | 9.2 | 9.2 | 95.4 |
| | +3 strong incentive | 7 | 4.6 | 4.6 | 100.0 |
| | Total | 153 | 100.0 | 100.0 | |

**[Current level of e-commerce legislation for online shopping]**

| | | Frequency | Percent | Valid Percent | Cumulative Percent |
|---|---|---|---|---|---|
| Valid | -3 major barrier | 33 | 21.6 | 21.6 | 21.6 |
| | -2 barrier | 12 | 7.8 | 7.8 | 29.4 |
| | -1 minor barrier | 19 | 12.4 | 12.4 | 41.8 |
| | 0 no influence | 76 | 49.7 | 49.7 | 91.5 |
| | +1 weak incentive | 10 | 6.5 | 6.5 | 98.0 |
| | +2 incentive | 3 | 2.0 | 2.0 | 100.0 |
| | Total | 153 | 100.0 | 100.0 | |

**[Current availability of online payment options]**

| | | Frequency | Percent | Valid Percent | Cumulative Percent |
|---|---|---|---|---|---|
| Valid | -3 major barrier | 17 | 11.1 | 11.1 | 11.1 |
| | -2 barrier | 14 | 9.2 | 9.2 | 20.3 |
| | -1 minor barrier | 27 | 17.6 | 17.6 | 37.9 |
| | 0 no influence | 75 | 49.0 | 49.0 | 86.9 |
| | +1 weak incentive | 14 | 9.2 | 9.2 | 96.1 |
| | +2 incentive | 4 | 2.6 | 2.6 | 98.7 |
| | +3 strong incentive | 2 | 1.3 | 1.3 | 100.0 |
| | Total | 153 | 100.0 | 100.0 | |

**[The current mailing and addressing system in Saudi Arabia]**

| | | Frequency | Percent | Valid Percent | Cumulative Percent |
|---|---|---|---|---|---|
| Valid | -3 major barrier | 21 | 13.7 | 13.7 | 13.7 |
| | -2 barrier | 8 | 5.2 | 5.2 | 19.0 |
| | -1 minor barrier | 13 | 8.5 | 8.5 | 27.5 |
| | 0 no influence | 95 | 62.1 | 62.1 | 89.5 |
| | +1 weak incentive | 12 | 7.8 | 7.8 | 97.4 |
| | +2 incentive | 4 | 2.6 | 2.6 | 100.0 |
| | Total | 153 | 100.0 | 100.0 | |

**[Level of ease for our company to do e-commerce]**

| | | Frequency | Percent | Valid Percent | Cumulative Percent |
|---|---|---|---|---|---|
| Valid | -3 major barrier | 19 | 12.4 | 12.4 | 12.4 |
| | -2 barrier | 19 | 12.4 | 12.4 | 24.8 |
| | -1 minor barrier | 31 | 20.3 | 20.3 | 45.1 |
| | 0 no influence | 26 | 17.0 | 17.0 | 62.1 |
| | +1 weak incentive | 26 | 17.0 | 17.0 | 79.1 |
| | +2 incentive | 26 | 17.0 | 17.0 | 96.1 |
| | +3 strong incentive | 6 | 3.9 | 3.9 | 100.0 |
| | Total | 153 | 100.0 | 100.0 | |

**[Nature of the products our company sells]**

| | | Frequency | Percent | Valid Percent | Cumulative Percent |
|---|---|---|---|---|---|
| Valid | -3 major barrier | 33 | 21.6 | 21.6 | 21.6 |
| | -2 barrier | 12 | 7.8 | 7.8 | 29.4 |
| | -1 minor barrier | 17 | 11.1 | 11.1 | 40.5 |
| | 0 no influence | 31 | 20.3 | 20.3 | 60.8 |
| | +1 weak incentive | 20 | 13.1 | 13.1 | 73.9 |
| | +2 incentive | 21 | 13.7 | 13.7 | 87.6 |
| | +3 strong incentive | 19 | 12.4 | 12.4 | 100.0 |
| | Total | 153 | 100.0 | 100.0 | |

**[Level of e-commerce knowledge and familiarity in our company]**

| | | Frequency | Percent | Valid Percent | Cumulative Percent |
|---|---|---|---|---|---|
| Valid | -3 major barrier | 35 | 22.9 | 22.9 | 22.9 |
| | -2 barrier | 15 | 9.8 | 9.8 | 32.7 |
| | -1 minor barrier | 27 | 17.6 | 17.6 | 50.3 |
| | 0 no influence | 18 | 11.8 | 11.8 | 62.1 |
| | +1 weak incentive | 20 | 13.1 | 13.1 | 75.2 |
| | +2 incentive | 27 | 17.6 | 17.6 | 92.8 |
| | +3 strong incentive | 11 | 7.2 | 7.2 | 100.0 |
| | Total | 153 | 100.0 | 100.0 | |

**[Management attitude towards e-commerce in our company]**

| | | Frequency | Percent | Valid Percent | Cumulative Percent |
|---|---|---|---|---|---|
| Valid | -3 major barrier | 19 | 12.4 | 12.4 | 12.4 |
| | -2 barrier | 19 | 12.4 | 12.4 | 24.8 |
| | -1 minor barrier | 28 | 18.3 | 18.3 | 43.1 |
| | 0 no influence | 24 | 15.7 | 15.7 | 58.8 |
| | +1 weak incentive | 20 | 13.1 | 13.1 | 71.9 |
| | +2 incentive | 19 | 12.4 | 12.4 | 84.3 |
| | +3 strong incentive | 24 | 15.7 | 15.7 | 100.0 |
| | Total | 153 | 100.0 | 100.0 | |

**[The current business priorities of our company]**

| | | Frequency | Percent | Valid Percent | Cumulative Percent |
|---|---|---|---|---|---|
| Valid | -3 major barrier | 14 | 9.2 | 9.2 | 9.2 |
| | -2 barrier | 18 | 11.8 | 11.8 | 20.9 |
| | -1 minor barrier | 12 | 7.8 | 7.8 | 28.8 |
| | 0 no influence | 65 | 42.5 | 42.5 | 71.2 |
| | +1 weak incentive | 13 | 8.5 | 8.5 | 79.7 |
| | +2 incentive | 20 | 13.1 | 13.1 | 92.8 |
| | +3 strong incentive | 11 | 7.2 | 7.2 | 100.0 |
| | Total | 153 | 100.0 | 100.0 | |

**[Our company’s level of trust in online selling processes]**

| | | Frequency | Percent | Valid Percent | Cumulative Percent |
|---|---|---|---|---|---|
| Valid | -3 major barrier | 12 | 7.8 | 7.8 | 7.8 |
| | -2 barrier | 6 | 3.9 | 3.9 | 11.8 |
| | -1 minor barrier | 5 | 3.3 | 3.3 | 15.0 |
| | 0 no influence | 105 | 68.6 | 68.6 | 83.7 |
| | +1 weak incentive | 4 | 2.6 | 2.6 | 86.3 |
| | +2 incentive | 6 | 3.9 | 3.9 | 90.2 |
| | +3 strong incentive | 15 | 9.8 | 9.8 | 100.0 |
| | Total | 153 | 100.0 | 100.0 | |

**[Setup and maintenance costs of e-commerce for our company]**

| | | Frequency | Percent | Valid Percent | Cumulative Percent |
|---|---|---|---|---|---|
| Valid | -3 major barrier | 13 | 8.5 | 8.5 | 8.5 |
| | -2 barrier | 7 | 4.6 | 4.6 | 13.1 |
| | -1 minor barrier | 5 | 3.3 | 3.3 | 16.3 |
| | 0 no influence | 99 | 64.7 | 64.7 | 81.0 |
| | +1 weak incentive | 4 | 2.6 | 2.6 | 83.7 |
| | +2 incentive | 8 | 5.2 | 5.2 | 88.9 |
| | +3 strong incentive | 17 | 11.1 | 11.1 | 100.0 |
| | Total | 153 | 100.0 | 100.0 | |

**Are you aware of the recent developments regarding the mailing and addressing system carried out by the Saudi Post?**

| | | Frequency | Percent | Valid Percent | Cumulative Percent |
|---|---|---|---|---|---|
| Valid | no | 54 | 35.3 | 35.3 | 35.3 |
| | yes | 99 | 64.7 | 64.7 | 100.0 |
| | Total | 153 | 100.0 | 100.0 | |

**Are you aware of the update that took place on the SADAD system, where capacity has been expanded to 20,000 billers?**

| | | Frequency | Percent | Valid Percent | Cumulative Percent |
|---|---|---|---|---|---|
| Valid | no | 131 | 85.6 | 85.6 | 85.6 |
| | yes | 22 | 14.4 | 14.4 | 100.0 |
| | Total | 153 | 100.0 | 100.0 | |

**Have you read the latest report released by the Saudi Communication and Information Technology Commission (CITC) regarding Internet usage and e-commerce activities?**

| | | Frequency | Percent | Valid Percent | Cumulative Percent |
|---|---|---|---|---|---|
| Valid | no | 97 | 63.4 | 63.4 | 63.4 |
| | yes | 56 | 36.6 | 36.6 | 100.0 |
| | Total | 153 | 100.0 | 100.0 | |

# **Appendix N:** Full statistical results of the t-test

**t-test results for participating businesses classified in stage 1 (*n=37*)**

| | N | Mean | Std. Deviation | Std. Error Mean |
|---|---|---|---|---|
| F1 | 37 | 2.49 | 1.239 | .204 |
| F2 | 37 | 3.78 | .672 | .111 |
| F3 | 37 | 3.62 | .794 | .131 |
| F4 | 37 | 3.76 | .641 | .105 |
| F5 | 37 | 2.89 | 1.220 | .201 |
| F6 | 37 | 3.81 | .660 | .108 |
| F7 | 37 | 3.78 | .584 | .096 |
| F8 | 37 | 4.51 | 1.044 | .172 |
| F9 | 37 | 3.95 | .848 | .139 |
| F10 | 37 | 3.38 | 1.037 | .170 |
| F11 | 37 | 3.51 | .961 | .158 |
| F12 | 37 | 3.70 | .777 | .128 |
| F13 | 37 | 2.43 | 1.168 | .192 |
| F14 | 37 | 2.84 | 1.236 | .203 |
| F15 | 37 | 2.03 | 1.093 | .180 |
| F16 | 37 | 2.32 | 1.056 | .174 |
| F17 | 37 | 2.78 | 1.182 | .194 |
| F18 | 37 | 3.35 | 1.111 | .183 |
| F19 | 37 | 3.22 | 1.228 | .202 |

| | Test Value = 4 | | | | | |
|---|---|---|---|---|---|---|
| | t | df | Sig. (2-tailed) | Mean Difference | 95% Confidence Interval of the Difference | |
| | | | | | Lower | Upper |
| F1 | -7.432 | 36 | .000 | -1.514 | -1.93 | -1.10 |
| F2 | -1.956 | 36 | .058 | -.216 | -.44 | .01 |
| F3 | -2.898 | 36 | .006 | -.378 | -.64 | -.11 |
| F4 | -2.307 | 36 | .027 | -.243 | -.46 | -.03 |
| F5 | -5.526 | 36 | .000 | -1.108 | -1.51 | -.70 |
| F6 | -1.744 | 36 | .090 | -.189 | -.41 | .03 |
| F7 | -2.253 | 36 | .030 | -.216 | -.41 | -.02 |
| F8 | 2.992 | 36 | .005 | .514 | .17 | .86 |
| F9 | -.388 | 36 | .701 | -.054 | -.34 | .23 |
| F10 | -3.647 | 36 | .001 | -.622 | -.97 | -.28 |
| F11 | -3.079 | 36 | .004 | -.486 | -.81 | -.17 |
| F12 | -2.328 | 36 | .026 | -.297 | -.56 | -.04 |
| F13 | -8.166 | 36 | .000 | -1.568 | -1.96 | -1.18 |
| F14 | -5.718 | 36 | .000 | -1.162 | -1.57 | -.75 |
| F15 | -10.984 | 36 | .000 | -1.973 | -2.34 | -1.61 |
| F16 | -9.657 | 36 | .000 | -1.676 | -2.03 | -1.32 |
| F17 | -6.260 | 36 | .000 | -1.216 | -1.61 | -.82 |
| F18 | -3.551 | 36 | .001 | -.649 | -1.02 | -.28 |
| F19 | -3.883 | 36 | .000 | -.784 | -1.19 | -.37 |

**t-test results for participating businesses classified in stage 2 (*n=41*)**

| | N | Mean | Std. Deviation | Std. Error Mean |
|---|---|---|---|---|
| F1 | 41 | 3.34 | 1.811 | .283 |
| F2 | 41 | 4.05 | .835 | .130 |
| F3 | 41 | 3.71 | .873 | .136 |
| F4 | 41 | 3.63 | .829 | .130 |
| F5 | 41 | 3.22 | 1.314 | .205 |
| F6 | 41 | 3.98 | .724 | .113 |
| F7 | 41 | 3.93 | .685 | .107 |
| F8 | 41 | 5.24 | .943 | .147 |
| F9 | 41 | 4.24 | 1.113 | .174 |
| F10 | 41 | 2.95 | 1.303 | .203 |
| F11 | 41 | 3.51 | 1.052 | .164 |
| F12 | 41 | 3.49 | 1.306 | .204 |
| F13 | 41 | 3.15 | 1.195 | .187 |
| F14 | 41 | 2.73 | 1.703 | .266 |
| F15 | 41 | 2.66 | 1.591 | .248 |
| F16 | 41 | 3.29 | 1.553 | .243 |
| F17 | 41 | 3.59 | 1.264 | .197 |
| F18 | 41 | 3.78 | 1.129 | .176 |
| F19 | 41 | 3.78 | 1.013 | .158 |

| | Test Value = 4 | | | | | |
|---|---|---|---|---|---|---|
| | t | df | Sig. (2-tailed) | Mean Difference | 95% Confidence Interval of the Difference | |
| | | | | | Lower | Upper |
| F1 | -2.328 | 40 | .025 | -.659 | -1.23 | -.09 |
| F2 | .374 | 40 | .710 | .049 | -.21 | .31 |
| F3 | -2.147 | 40 | .038 | -.293 | -.57 | -.02 |
| F4 | -2.825 | 40 | .007 | -.366 | -.63 | -.10 |
| F5 | -3.804 | 40 | .000 | -.780 | -1.20 | -.37 |
| F6 | -.216 | 40 | .830 | -.024 | -.25 | .20 |
| F7 | -.684 | 40 | .498 | -.073 | -.29 | .14 |
| F8 | 8.447 | 40 | .000 | 1.244 | .95 | 1.54 |
| F9 | 1.403 | 40 | .168 | .244 | -.11 | .60 |
| F10 | -5.154 | 40 | .000 | -1.049 | -1.46 | -.64 |
| F11 | -2.970 | 40 | .005 | -.488 | -.82 | -.16 |
| F12 | -2.511 | 40 | .016 | -.512 | -.92 | -.10 |
| F13 | -4.574 | 40 | .000 | -.854 | -1.23 | -.48 |
| F14 | -4.768 | 40 | .000 | -1.268 | -1.81 | -.73 |
| F15 | -5.400 | 40 | .000 | -1.341 | -1.84 | -.84 |
| F16 | -2.916 | 40 | .006 | -.707 | -1.20 | -.22 |
| F17 | -2.100 | 40 | .042 | -.415 | -.81 | -.02 |
| F18 | -1.244 | 40 | .221 | -.220 | -.58 | .14 |
| F19 | -1.388 | 40 | .173 | -.220 | -.54 | .10 |

**t-test results for participating businesses classified in stage 3 (*n=35*)**

| | N | Mean | Std. Deviation | Std. Error Mean |
|---|---|---|---|---|
| F1 | 35 | 3.83 | 1.723 | .291 |
| F2 | 35 | 4.43 | 1.145 | .194 |
| F3 | 35 | 4.06 | 1.056 | .178 |
| F4 | 35 | 3.97 | .985 | .166 |
| F5 | 35 | 3.63 | 1.497 | .253 |
| F6 | 35 | 4.57 | .850 | .144 |
| F7 | 35 | 3.91 | 1.095 | .185 |
| F8 | 35 | 5.46 | 1.442 | .244 |
| F9 | 35 | 4.46 | 1.358 | .230 |
| F10 | 35 | 3.11 | 1.568 | .265 |
| F11 | 35 | 3.31 | 1.409 | .238 |
| F12 | 35 | 3.54 | 1.400 | .237 |
| F13 | 35 | 4.40 | 1.666 | .282 |
| F14 | 35 | 4.40 | 2.199 | .372 |
| F15 | 35 | 4.40 | 1.631 | .276 |
| F16 | 35 | 4.63 | 1.592 | .269 |
| F17 | 35 | 4.34 | 1.494 | .253 |
| F18 | 35 | 4.26 | 1.314 | .222 |
| F19 | 35 | 4.29 | 1.487 | .251 |

| | Test Value = 4 | | | | | |
|---|---|---|---|---|---|---|
| | t | df | Sig. (2-tailed) | Mean Difference | 95% Confidence Interval of the Difference | |
| | | | | | Lower | Upper |
| F1 | -.589 | 34 | .560 | -.171 | -.76 | .42 |
| F2 | 2.214 | 34 | .034 | .429 | .04 | .82 |
| F3 | .320 | 34 | .751 | .057 | -.31 | .42 |
| F4 | -.172 | 34 | .865 | -.029 | -.37 | .31 |
| F5 | -1.468 | 34 | .151 | -.371 | -.89 | .14 |
| F6 | 3.977 | 34 | .000 | .571 | .28 | .86 |
| F7 | -.463 | 34 | .646 | -.086 | -.46 | .29 |
| F8 | 5.979 | 34 | .000 | 1.457 | .96 | 1.95 |
| F9 | 1.992 | 34 | .054 | .457 | -.01 | .92 |
| F10 | -3.343 | 34 | .002 | -.886 | -1.42 | -.35 |
| F11 | -2.878 | 34 | .007 | -.686 | -1.17 | -.20 |
| F12 | -1.931 | 34 | .062 | -.457 | -.94 | .02 |
| F13 | 1.420 | 34 | .165 | .400 | -.17 | .97 |
| F14 | 1.076 | 34 | .289 | .400 | -.36 | 1.16 |
| F15 | 1.451 | 34 | .156 | .400 | -.16 | .96 |
| F16 | 2.336 | 34 | .026 | .629 | .08 | 1.18 |
| F17 | 1.358 | 34 | .183 | .343 | -.17 | .86 |
| F18 | 1.158 | 34 | .255 | .257 | -.19 | .71 |
| F19 | 1.137 | 34 | .263 | .286 | -.22 | .80 |

**t-t results for participating businesses classified in stage 4 (*n=40*)**

| | N | Mean | Std. Deviation | Std. Error Mean |
|---|---|---|---|---|
| F1 | 40 | 5.15 | 1.545 | .244 |
| F2 | 40 | 5.00 | 1.261 | .199 |
| F3 | 40 | 4.53 | 1.396 | .221 |
| F4 | 40 | 3.90 | 1.317 | .208 |
| F5 | 40 | 4.43 | 1.738 | .275 |
| F6 | 40 | 4.75 | 1.335 | .211 |
| F7 | 40 | 4.90 | 1.194 | .189 |
| F8 | 40 | 5.95 | 1.467 | .232 |
| F9 | 40 | 4.88 | 1.305 | .206 |
| F10 | 40 | 3.28 | 1.502 | .237 |
| F11 | 40 | 3.60 | 1.582 | .250 |
| F12 | 40 | 3.40 | 1.336 | .211 |
| F13 | 40 | 5.23 | 1.387 | .219 |
| F14 | 40 | 5.50 | 1.519 | .240 |
| F15 | 40 | 5.48 | 1.467 | .232 |
| F16 | 40 | 5.90 | 1.446 | .229 |
| F17 | 40 | 5.15 | 1.545 | .244 |
| F18 | 40 | 4.80 | 1.604 | .254 |
| F19 | 40 | 5.03 | 1.593 | .252 |

| | Test Value = 4 | | | | | |
|---|---|---|---|---|---|---|
| | t | df | Sig. (2-tailed) | Mean Difference | 95% Confidence Interval of the Difference | |
| | | | | | Lower | Upper |
| F1 | 4.707 | 39 | .000 | 1.150 | .66 | 1.64 |
| F2 | 5.016 | 39 | .000 | 1.000 | .60 | 1.40 |
| F3 | 2.379 | 39 | .022 | .525 | .08 | .97 |
| F4 | -.480 | 39 | .634 | -.100 | -.52 | .32 |
| F5 | 1.547 | 39 | .130 | .425 | -.13 | .98 |
| F6 | 3.553 | 39 | .001 | .750 | .32 | 1.18 |
| F7 | 4.767 | 39 | .000 | .900 | .52 | 1.28 |
| F8 | 8.408 | 39 | .000 | 1.950 | 1.48 | 2.42 |
| F9 | 4.242 | 39 | .000 | .875 | .46 | 1.29 |
| F10 | -3.053 | 39 | .004 | -.725 | -1.21 | -.24 |
| F11 | -1.599 | 39 | .118 | -.400 | -.91 | .11 |
| F12 | -2.841 | 39 | .007 | -.600 | -1.03 | -.17 |
| F13 | 5.588 | 39 | .000 | 1.225 | .78 | 1.67 |
| F14 | 6.245 | 39 | .000 | 1.500 | 1.01 | 1.99 |
| F15 | 6.357 | 39 | .000 | 1.475 | 1.01 | 1.94 |
| F16 | 8.308 | 39 | .000 | 1.900 | 1.44 | 2.36 |
| F17 | 4.707 | 39 | .000 | 1.150 | .66 | 1.64 |
| F18 | 3.153 | 39 | .003 | .800 | .29 | 1.31 |
| F19 | 4.069 | 39 | .000 | 1.025 | .52 | 1.53 |

# **Appendix O:** Full statistical results of the pairwise test

```
> pairwise.t.test (f1Data$Response, f1Data$grp, p.adjust="bonf")

        Pairwise comparisons using t tests with pooled SD

data:  f1Data$Response and f1Data$grp

  a       b       c
b 0.14633 -       -
c 0.00225 0.83095 -
d 7.9e-12 5.1e-07 0.00078

P value adjustment method: bonferroni
> pairwise.t.test (f2Data$Response, f2Data$grp, p.adjust="bonf")

        Pairwise comparisons using t tests with pooled SD

data:  f2Data$Response and f2Data$grp

  a       b       c
b 1.000   -       -
c 0.032   0.338   -
d 1.8e-07 9.7e-06 0.027

P value adjustment method: bonferroni
> pairwise.t.test (f3Data$Response, f3Data$grp, p.adjust="bonf")

        Pairwise comparisons using t tests with pooled SD

data:  f3Data$Response and f3Data$grp

  a       b       c
b 1.00000 -       -
c 0.43940 0.83832 -
d 0.00012 0.00034 0.08060

P value adjustment method: bonferroni
> pairwise.t.test (f4Data$Response, f4Data$grp, p.adjust="bonf")

        Pairwise comparisons using t tests with pooled SD

data:  f4Data$Response and f4Data$grp

  a    b    c
b 1.00 -    -
c 1.00 0.50 -
d 1.00 0.33 1.00

P value adjustment method: bonferroni
```

```
> pairwise.t.test (f5Data$Response, f5Data$grp, p.adjust="bonf")

        Pairwise comparisons using t tests with pooled SD

data:  f5Data$Response and f5Data$grp

  a       b       c
b 1.00000 -       -
c 0.18292 1.00000 -
d 1.3e-05 0.00053 0.05621

P value adjustment method: bonferroni
> pairwise.t.test (f6Data$Response, f6Data$grp, p.adjust="bonf")

        Pairwise comparisons using t tests with pooled SD

data:  f6Data$Response and f6Data$grp

  a       b       c
b 1.0000  -       -
c 0.0022  0.0235  -
d 4.7e-06 9.9e-05 1.0000

P value adjustment method: bonferroni
> pairwise.t.test (f7Data$Response, f7Data$grp, p.adjust="bonf")

        Pairwise comparisons using t tests with pooled SD

data:  f7Data$Response and f7Data$grp

  a       b       c
b 1       -       -
c 1       1       -
d 4.0e-08 6.8e-07 1.4e-06

P value adjustment method: bonferroni
> pairwise.t.test (f8Data$Response, f8Data$grp, p.adjust="bonf")

        Pairwise comparisons using t tests with pooled SD

data:  f8Data$Response and f8Data$grp

  a       b     c
b 0.083   -     -
c 0.010   1.000 -
d 8.1e-06 0.057 0.544

P value adjustment method: bonferroni
```

```
> pairwise.t.test (f9Data$Response, f9Data$grp, p.adjust="bonf")

        Pairwise comparisons using t tests with pooled SD

data:  f9Data$Response and f9Data$grp

  a      b      c
b 1.0000 -      -
c 0.3708 1.0000 -
d 0.0023 0.0683 0.5930

P value adjustment method: bonferroni
> pairwise.t.test (f10Data$Response, f10Data$grp, p.adjust="bonf")

        Pairwise comparisons using t tests with pooled SD

data:  f10Data$Response and f10Data$grp

  a b c
b 1 - -
c 1 1 -
d 1 1 1

P value adjustment method: bonferroni
> pairwise.t.test (f11Data$Response, f11Data$grp, p.adjust="bonf")

        Pairwise comparisons using t tests with pooled SD

data:  f11Data$Response and f11Data$grp

  a       b       c
b 1       -       -
c 1       1       -
d 4.3e-07 9.2e-08 1.1e-08

P value adjustment method: bonferroni
> pairwise.t.test (f12Data$Response, f12Data$grp, p.adjust="bonf")

        Pairwise comparisons using t tests with pooled SD

data:  f12Data$Response and f12Data$grp

  a b c
b 1 - -
c 1 1 -
d 1 1 1

P value adjustment method: bonferroni
```

```
> pairwise.t.test (f13Data$Response, f13Data$grp, p.adjust="bonf")

        Pairwise comparisons using t tests with pooled SD

data:  f13Data$Response and f13Data$grp

  a       b       c
b 0.10746 -       -
c 1.5e-08 0.00034 -
d < 2e-16 2.6e-11 0.01299

P value adjustment method: bonferroni
> pairwise.t.test (f14Data$Response, f14Data$grp, p.adjust="bonf")

        Pairwise comparisons using t tests with pooled SD

data:  f14Data$Response and f14Data$grp

  a       b       c
b 1.00000 -       -
c 0.00057 0.00013 -
d 3.9e-11 2.6e-12 0.00803

P value adjustment method: bonferroni
> pairwise.t.test (f15Data$Response, f15Data$grp, p.adjust="bonf")

        Pairwise comparisons using t tests with pooled SD

data:  f15Data$Response and f15Data$grp

  a       b       c
b 0.3055  -       -
c 2.7e-10 2.0e-06 -
d < 2e-16 1.2e-15 0.0027

P value adjustment method: bonferroni
> pairwise.t.test (f16Data$Response, f16Data$grp, p.adjust="bonf")

        Pairwise comparisons using t tests with pooled SD

data:  f16Data$Response and f16Data$grp

  a       b       c
b 0.01412 -       -
c 3.1e-10 0.00027 -
d < 2e-16 2.6e-14 0.00026

P value adjustment method: bonferroni
```

```
> pairwise.t.test (f17Data$Response, f17Data$grp, p.adjust="bonf")

        Pairwise comparisons using t tests with pooled SD

data:  f17Data$Response and f17Data$grp

  a       b       c
b 0.086   -       -
c 1.3e-05 0.059   -
d 1.3e-12 2.9e-07 0.024

P value adjustment method: bonferroni
> pairwise.t.test (f18Data$Response, f18Data$grp, p.adjust="bonf")

        Pairwise comparisons using t tests with pooled SD

data:  f18Data$Response and f18Data$grp

  a       b       c
b 0.83071 -       -
c 0.01766 0.63102 -
d 3e-06   0.00095 0.22391

P value adjustment method: bonferroni
> pairwise.t.test (f19Data$Response, f19Data$grp, p.adjust="bonf")

        Pairwise comparisons using t tests with pooled SD

data:  f19Data$Response and f19Data$grp

  a       b       c
b 0.4232  -       -
c 0.0041  0.4821  -
d 1.7e-08 4.8e-05 0.0476

P value adjustment method: bonferroni
```

# **Appendix P:** Full statistical results of multiple linear regression

**Coefficients[a]**

| Model | Unstandardized Coefficients | | Standardized Coefficients | t | p |
|---|---|---|---|---|---|
| | B | Std. Error | Beta | | |
| (Constant) | 2.316 | .479 | | 4.833 | .000 |
| How old is your organization? | -.208 | .162 | -.108 | -1.285 | .201 |
| How many employees work in your organization? | -.097 | .158 | -.052 | -.617 | .538 |
| E-commerce Maturity Level | .855 | .116 | .517 | 7.386 | .000 |

a. Dependent Variable: [F1: People's current attitudes and habits regarding shopping online]

**Coefficients[a]**

| Model | Unstandardized Coefficients | | Standardized Coefficients | t | p |
|---|---|---|---|---|---|
| | B | Std. Error | Beta | | |
| (Constant) | 3.400 | .305 | | 11.139 | .000 |
| How old is your organization? | -.030 | .103 | -.027 | -.294 | .769 |
| How many employees work in your organization? | -.011 | .101 | -.010 | -.107 | .915 |
| E-commerce Maturity Level | .406 | .074 | .414 | 5.514 | .000 |

a. Dependent Variable: [F2: The current level of consumer recognition of e-commerce benefits]

**Coefficients[a]**

| Model | Unstandardized Coefficients | | Standardized Coefficients | t | p |
|---|---|---|---|---|---|
| | B | Std. Error | Beta | | |
| (Constant) | 3.127 | .321 | | 9.741 | .000 |
| How old is your organization? | -.062 | .109 | -.053 | -.566 | .572 |
| How many employees work in your organization? | .085 | .106 | .076 | .800 | .425 |
| E-commerce Maturity Level | .307 | .078 | .310 | 3.965 | .000 |

a. Dependent Variable: [F3: The current level of consumer trust in online shopping]

**Coefficients[a]**

| Model | Unstandardized Coefficients | | Standardized Coefficients | t | p |
|---|---|---|---|---|---|
| | B | Std. Error | Beta | | |
| (Constant) | 3.245 | .295 | | 10.998 | .000 |
| How old is your organization? | .038 | .100 | .037 | .381 | .704 |
| How many employees work in your organization? | .114 | .097 | .116 | 1.176 | .242 |
| E-commerce Maturity Level | .067 | .071 | .077 | .941 | .348 |

a. Dependent Variable: [F4: The current level of consumer usage of credit cards for online shopping]

**Coefficients[a]**

| Model | Unstandardized Coefficients | | Standardized Coefficients | t | p |
|---|---|---|---|---|---|
| | B | Std. Error | Beta | | |
| (Constant) | 3.053 | .433 | | 7.054 | .000 |
| How old is your organization? | -.002 | .147 | -.001 | -.015 | .988 |
| How many employees work in your organization? | -.303 | .143 | -.193 | -2.125 | .035 |
| E-commerce Maturity Level | .523 | .105 | .377 | 5.004 | .000 |

a. Dependent Variable: [F5: The current level of consumer demand for buying online]

**Coefficients[a]**

| Model | Unstandardized Coefficients | | Standardized Coefficients | t | p |
|---|---|---|---|---|---|
| | B | Std. Error | Beta | | |
| (Constant) | 3.076 | .281 | | 10.945 | .000 |
| How old is your organization? | -.025 | .095 | -.024 | -.265 | .791 |
| How many employees work in your organization? | .156 | .093 | .153 | 1.682 | .095 |
| E-commerce Maturity Level | .336 | .068 | .373 | 4.946 | .000 |

a. Dependent Variable: [F6: The current level of consumer knowledge of buying online]

**Coefficients[a]**

| Model | Unstandardized Coefficients | | Standardized Coefficients | t | p |
|---|---|---|---|---|---|
| | B | Std. Error | Beta | | |
| (Constant) | 3.994 | .278 | | 14.344 | .000 |
| How old is your organization? | -.201 | .094 | -.189 | -2.132 | .035 |
| How many employees work in your organization? | -.097 | .092 | -.094 | -1.051 | .295 |
| E-commerce Maturity Level | .349 | .067 | .383 | 5.196 | .000 |

a. Dependent Variable: [F7: Level of willingness of consumers to pay delivery fees]

**Coefficients[a]**

| Model | Unstandardized Coefficients | | Standardized Coefficients | t | p |
|---|---|---|---|---|---|
| | B | Std. Error | Beta | | |
| (Constant) | 4.420 | .376 | | 11.763 | .000 |
| How old is your organization? | -.110 | .127 | -.079 | -.863 | .390 |
| How many employees work in your organization? | .002 | .124 | .001 | .013 | .990 |
| E-commerce Maturity Level | .451 | .091 | .379 | 4.973 | .000 |

a. Dependent Variable: [F8: The percentage of people using the internet]

**Coefficients[a]**

| Model | Unstandardized Coefficients | | Standardized Coefficients | t | p |
|---|---|---|---|---|---|
| | B | Std. Error | Beta | | |
| (Constant) | 4.376 | .341 | | 12.845 | .000 |
| How old is your organization? | -.353 | .115 | -.281 | -1.693 | .091 |
| How many employees work in your organization? | .028 | .112 | .023 | .245 | .806 |
| E-commerce Maturity Level | .303 | .082 | .280 | 3.676 | .000 |

a. Dependent Variable: [F9: Level of availability of wi-fi and broadband services]

**Coefficients[a]**

| Model | Unstandardized Coefficients | | Standardized Coefficients | t | p |
|---|---|---|---|---|---|
| | B | Std. Error | Beta | | |
| (Constant) | 3.554 | .411 | | 8.648 | .000 |
| How old is your organization? | -.098 | .139 | -.069 | -.703 | .483 |
| How many employees work in your organization? | -.063 | .136 | -.046 | -.464 | .644 |
| E-commerce Maturity Level | .014 | .099 | .012 | .141 | .888 |

a. Dependent Variable: [F10: Current level of e-commerce legislation for online shopping]

**Coefficients[a]**

| Model | Unstandardized Coefficients | | Standardized Coefficients | t | p |
|---|---|---|---|---|---|
| | B | Std. Error | Beta | | |
| (Constant) | 3.214 | .381 | | 8.431 | .000 |
| How old is your organization? | -.153 | .129 | -.117 | -1.188 | .237 |
| How many employees work in your organization? | .234 | .126 | .184 | 1.864 | .064 |
| E-commerce Maturity Level | .005 | .092 | .005 | .056 | .955 |

a. Dependent Variable: [F11: Current availability of online payment options]

**Coefficients[a]**

| Model | Unstandardized Coefficients | | Standardized Coefficients | t | p |
|---|---|---|---|---|---|
| | B | Std. Error | Beta | | |
| (Constant) | 4.069 | .369 | | 11.026 | .000 |
| How old is your organization? | -.143 | .125 | -.113 | -1.145 | .254 |
| How many employees work in your organization? | .009 | .122 | .008 | .076 | .939 |
| E-commerce Maturity Level | -.096 | .089 | -.088 | -1.079 | .282 |

a. Dependent Variable: [F12: The current mailing and addressing system in Saudi Arabia]

**Coefficients[a]**

| Model | Unstandardized Coefficients | | Standardized Coefficients | t | p |
|---|---|---|---|---|---|
| | B | Std. Error | Beta | | |
| (Constant) | 1.293 | .401 | | 3.223 | .002 |
| How old is your organization? | -.348 | .136 | -.193 | -2.563 | .011 |
| How many employees work in your organization? | .356 | .132 | .203 | 2.690 | .008 |
| E-commerce Maturity Level | .943 | .097 | .610 | 9.731 | .000 |

a. Dependent Variable: [F13: Level of ease for our company to do e-commerce]

**Coefficients[a]**

| Model | Unstandardized Coefficients | | Standardized Coefficients | t | p |
|---|---|---|---|---|---|
| | B | Std. Error | Beta | | |
| (Constant) | 2.144 | .516 | | 4.159 | .000 |
| How old is your organization? | -.370 | .175 | -.174 | -1.080 | .284 |
| How many employees work in your organization? | .041 | .170 | .020 | .239 | .812 |
| E-commerce Maturity Level | .989 | .125 | .541 | 7.939 | .000 |

a. Dependent Variable: [F14: Nature of the products our company sells]

**Coefficients[a]**

| Model | Unstandardized Coefficients | | Standardized Coefficients | t | p |
|---|---|---|---|---|---|
| | B | Std. Error | Beta | | |
| (Constant) | 1.439 | .437 | | 3.296 | .001 |
| How old is your organization? | -.181 | .148 | -.087 | -1.223 | .223 |
| How many employees work in your organization? | -.178 | .144 | -.088 | -1.233 | .220 |
| E-commerce Maturity Level | 1.240 | .106 | .692 | 11.749 | .000 |

a. Dependent Variable: [F15: Level of e-commerce knowledge and familiarity in our company]

**Coefficients[a]**

| Model | Unstandardized Coefficients | | Standardized Coefficients | t | p |
|---|---|---|---|---|---|
| | B | Std. Error | Beta | | |
| (Constant) | 1.640 | .427 | | 3.837 | .000 |
| How old is your organization? | -.144 | .145 | -.070 | -.993 | .322 |
| How many employees work in your organization? | -.123 | .141 | -.062 | -.876 | .383 |
| E-commerce Maturity Level | 1.227 | .103 | .699 | 11.888 | .000 |

a. Dependent Variable: [F16: Management attitude towards e-commerce in our company]

**Coefficients[a]**

| Model | Unstandardized Coefficients | | Standardized Coefficients | t | p |
|---|---|---|---|---|---|
| | B | Std. Error | Beta | | |
| (Constant) | 2.562 | .412 | | 6.224 | .000 |
| How old is your organization? | -.093 | .139 | -.055 | -.668 | .505 |
| How many employees work in your organization? | -.144 | .136 | -.088 | -1.062 | .290 |
| E-commerce Maturity Level | .807 | .099 | .555 | 8.117 | .000 |

a. Dependent Variable: [F17: The current business priorities of our company]

**Coefficients[a]**

| Model | Unstandardized Coefficients | | Standardized Coefficients | t | p |
|---|---|---|---|---|---|
| | B | Std. Error | Beta | | |
| (Constant) | 3.201 | .393 | | 8.154 | .000 |
| How old is your organization? | -.151 | .133 | -.103 | -1.135 | .258 |
| How many employees work in your organization? | -.010 | .130 | -.007 | -.077 | .939 |
| E-commerce Maturity Level | .492 | .095 | .392 | 5.190 | .000 |

a. Dependent Variable: [F18: Our company's level of trust in online selling processes]

**Coefficients[a]**

| Model | Unstandardized Coefficients | | Standardized Coefficients | t | p |
|---|---|---|---|---|---|
| | B | Std. Error | Beta | | |
| (Constant) | 2.561 | .405 | | 6.317 | .000 |
| How old is your organization? | -.113 | .137 | -.073 | -.825 | .411 |
| How many employees work in your organization? | .111 | .134 | .074 | .832 | .407 |
| E-commerce Maturity Level | .596 | .098 | .447 | 6.081 | .000 |

a. Dependent Variable: [F19: Setup and maintenance costs of e-commerce for our company]